# Evolving Chaos: Identifying New Attractors of the Generalised Lorenz Family

Indranil Pan[1] and Saptarshi Das[2,*]

1. *Department of Earth Science and Engineering, Imperial College London, Exhibition Road, SW7 2AZ, United Kingdom.*
2. *Department of Mathematics, College of Engineering, Mathematics and Physical Sciences, University of Exeter, Penryn Campus, Cornwall TR10 9FE, United Kingdom*

**Email:**

i.pan11@imperial.ac.uk (I. Pan)

s.das3@exeter.ac.uk, saptarshi@pe.jusl.ac.in (S. Das*)

**Corresponding author's phone number:** +44(0)7448572598

**Abstract:**

In a recent paper, we presented an intelligent evolutionary search technique through genetic programming (GP) for finding new analytical expressions of nonlinear dynamical systems, similar to the classical Lorenz attractor's which also exhibit chaotic behaviour in the phase space. In this paper, we extend our previous finding to explore yet another gallery of new chaotic attractors which are derived from the original Lorenz system of equations. Compared to the previous exploration with sinusoidal type transcendental nonlinearity, here we focus on only cross-product and higher-power type nonlinearities in the three state equations. We here report over 150 different structures of chaotic attractors along with their one set of parameter values, phase space dynamics and the Largest Lyapunov Exponents (LLE). The expressions of these new Lorenz-like nonlinear dynamical systems have been automatically evolved through multi-gene genetic programming (MGGP). In the past two decades, there have been many claims of designing new chaotic attractors as an incremental extension of the Lorenz family. We provide here a large family of chaotic systems whose structure closely resemble the original Lorenz system but with drastically different phase space dynamics. This advances the state of the art knowledge of discovering new chaotic systems which can find application in many real-world problems. This work may also find its archival value in future in the domain of new chaotic system discovery.

## 1. Introduction

Investigation of new chaotic attractors showing rich phase space dynamics has been widely researched in many studies particularly in the field of cryptography and secure communication [1] and explaining naturally occurring complex systems in biology, economics, chemistry and physics [2]. There have been several claims of inventing new chaotic attractors in recent years as a derivative of the celebrated Lorenz system [3], [4] to the Lorenz family of systems [5], [3] e.g. Rossler, Rucklidge, Chen, Genesio-Tesi, Shimizu-Morioka [6], Chen [7], Lu [8], Liu [9], Qi [10], Sprott [11]–[14] etc. Given a highly complex time series, there have been very few successful results to find out the structure of the underlying chaotic system, especially when the order and exact functional complexity of the dynamical system is not precisely known. In most cases, this inverse problem is infeasible but given a pool of chaotic attractor structures, it is rather easy to simulate the state variables and compare it with the observation. Using the Takens' embedding theorem [15], [16] and recorded time series of just one state variable of finite length, it is possible to reconstruct the original phase space dynamics of the underlying attractor without precisely knowing its mathematical structure. Under such a scenario and with a proper choice of embedding delay, the LLE of the reconstructed attractors closely approach that of the original one.

Our previous study in [17] report more than 100 chaotic attractors having at least a single sinusoidal nonlinearity in one of the state equations. It was also argued in [17] that the rich phase space dynamics may be an effect of increased number of equilibrium points due to the transcendental terms e.g. sinusoidal function of the state variables. Here, we explore a different family of attractors having rather a much simpler cross-product and higher power type nonlinearity involving the three state variables. We here show that even with simple algebraic expressions without the previously explored transcendental terms e.g. sinh, cosh, sin, cos, exp, as shown in [17] and Sprott [11]–[14], quite complex phase space dynamics can be generated using an evolutionary search with genetic programming. During automatic evolution of chaotic system expressions, the state time series based time delay embedding method has been employed to calculate the Lyapunov exponent [16] for the initial screening. However, numerical calculation of the LLE based on finite length of only one state time series could lead to false discovery of many nonlinear dynamical systems as chaotic, if a positive LLE criteria were imposed on the values calculated on these set of simulations. Therefore, after the initial time series based evolution of chaotic systems and LLE computation with time delay embedding, the true LLEs have been recalculated using a symbolic differentiation scheme for each of the newly evolved expression of the state equation through genetic programming, while calculating the Jacobian matrix of these new nonlinear dynamical system. We believe that these new chaotic attractors are going to serve as a useful archival reference for future designers of chaotic cryptography and many natural scientists as these extends the generalised Lorenz family with cross-product nonlinearity to a much wider library of Lorenz like attractors.

## 2. Genetic programming to evolve new chaotic attractors
### 2.1. Search method and objective function

Single and multigene GP have been used to search for the chaotic system expressions as reported in our previous exploration [17]. The single gene GP helps evolving a single state equation while keeping the other two as that of the Lorenz attractor, whereas the multigene GP simultaneously evolves two or all the three state equations to find a completely new chaotic system compared to the structures of the classical Lorenz system [3]. The single or multi-gene GP evolves with the objective of maximising the LLE, thus starting with a moderately good attractor (original Lorenz system) as an initial guess to find the good attractors while discarding expressions for the bad ones, with a low value of LLE.

The GP algorithm usually fits a symbolic regression problem while evolving explicit mathematical expression [18]. However compared to the original implementation in [18] the GP algorithm has been modified as an unsupervised learning or maximisation problem instead of a supervised input-output regression type data fitting problem. Within the GP algorithm, the temporal evolution of each dynamical system was calculated on a simulation time length of 50 seconds using the ode3 Bogacki-Shampine solver, with a fixed step size of 0.01 sec. The population size for the GP algorithm is taken as 25, with 50 generation and plain lexicographic tournament selection with size 3. The termination criterion was the maximum number of iterations. The maximum depth of the GP tree was taken as 5 and 2 particularly for mutation. The ratio of mathematical operators to numeric values in the GP evolved expression were kept as 0.5 and all four basic arithmetic operations were used in the evolutionary search i.e. $\{+,-,\times,\div\}$.

### 2.2. Complexity handling in the GP algorithm

One important point for our implementation of the GP algorithm is the mechanism to control the complexity and complete the simulations in a realistic time frame. We did a code profiling and observed that the maximum time required by the algorithm was in simulating the temporal evolution of each GP generated expression and calculating the Lyapunov exponent based on time delay embedding. On further inspection we found that some of the expressions that the GP algorithm evaluated in the later iterations had been evaluated before in some previous generation. Therefore, to get rid of these redundant calculations, we made an archive mechanism which stores each unique expression and its corresponding Lyapunov exponent. The logic of evaluating each GP candidate is then modified, such that the expression is first matched with the archive and if it exists in the archive, the LLE value is returned reading off from the archive table. In case it is a new expression that does not exist in the archive, the dynamical evolutions of the states are carried out along with computation of the Lyapunov exponents and these are added to the archive. This modification alone made the code approximately 27 times faster and helped in completing the runs in a realistic time frame.

Also, there are a few other modifications that we did to the expression evaluation to speed up the computation. In some of the cases, the GP algorithm was trying to evaluate only one gene (i.e. only the *x* or *y* expression instead of both). This meant that the dynamical system would have only one or two state equations instead of three. We kept a check for such cases and assigned a high constant penalty value to the fitness function (without evaluating it) in such cases. Also in a few cases we found that the GP algorithm was trying to evaluate expression which did not have any state variables (on the right-hand side of the equations) and had only constants. This meant that either of $\{\dot{x},\dot{y},\dot{z}\}$ were constants in the dynamical system. Almost all of these expressions did not give rise to chaotic attractors in our initial runs, so we also implemented a check to assign a high constant penalty value to the fitness function for such unwanted cases.

Some of the expressions did not have the above-mentioned problems, but they were unstable systems, i.e. when the time evolutions of the states were simulated, some of the states diverged to infinity. Since we used Simulink to integrate the state variables, there are two cases which can happen. In the first case, within the prescribed simulation time, the solution becomes infinity and simulation stops. In such cases, the length of the state vector (i.e. values of x, y, z at each time instant) is less than what would be obtained in a normal simulation, as all the time steps could not be simulated. In the second case, the whole simulation is completed, and the values of the state variables become very large (but do not go to infinity). This indicates that if we would have simulated it for some more time, these state variables would probably had diverged to infinity. For both of these cases, we did not invoke the expensive Lyapunov exponent calculation. We assigned a fixed high penalty value to the fitness function. We used the same penalty assignment technique when the last 20 time-steps of the state evolution did not show any appreciable change (the mean of the difference was less than 0.01). This implied that the dynamical system had converged to a fixed point and therefore would not exhibit chaos. We did not subsequently evaluate the Lyapunov exponent for these cases. The objective of assigning high penalty values for such cases is that this mechanism helps in steering the GP search away from such areas of the objective function space where it is less likely to find chaotic attractors.

### 2.3. Lorenz family of attractors with product nonlinearity revisited

As opposed to relatively complex state equations involving transcendental functions e.g. sine, cosine and hyperbolic tangent terms as in [13], [14], [17], we here show that even simple algebraic operations like multiplication, division, and higher powers (cross product type nonlinearity as a whole) can generate sufficiently complex and innovative phase space patterns in third order nonlinear dynamical systems as compared the classical Lorenz system and its family of attractors [19]. The state equations of the classical Lorenz system (1) have been modified in four different categories as shown in (2) by modifying two/three state equations together as an extension of the Lorenz system structure.

$$\dot{x} = 10(y-x), \dot{y} = 28x - xz - y, \dot{z} = xy - (8/3)z \qquad (1)$$

For convenience of nomenclature the four categories are named as Lorenz-XY, Lorenz-YZ, Lorenz-XZ and Lorenz-XYZ family (2), indicating which state equation has been searched using the multigene GP algorithm.

$$\begin{aligned}
\text{Lorenz-XY family:} \quad & \dot{x} = g(x,y,z), \dot{y} = f(x,y,z), \dot{z} = xy - (8/3)z \\
\text{Lorenz-YZ family:} \quad & \dot{x} = 10(y-x), \dot{y} = f(x,y,z), \dot{z} = h(x,y,z) \\
\text{Lorenz-XZ family:} \quad & \dot{x} = g(x,y,z), \dot{y} = 28x - xz - y, \dot{z} = h(x,y,z) \\
\text{Lorenz-XYZ family:} \quad & \dot{x} = g(x,y,z), \dot{y} = f(x,y,z), \dot{z} = h(x,y,z)
\end{aligned} \qquad (2)$$

In (2), the first three cases retains one equation same as that of the Lorenz attractor while modifying the others with the GP algorithm. The fourth case in (2), modifies all the three state equations thus yielding any possible complex structure with a choice of search functions like multiplication, division etc., acting as the cause of nonlinearity in the system. The search starts with the classical Lorenz system as the initial guess and evolves through GP with an objective of maximising the LLE. We also report the intermediate results during such evolutionary search, since many similar expressions may yield completely different phase space characteristics. Therefore, it is also important to look at the intermediate attractors and not just at the final converged result given by the GP algorithm.

### 2.4. Symbolic Jacobian calculation of the new expressions evolved by GP algorithm

Previous exploration reveals that simple structures of chaotic systems with only product nonlinearity (without an explicit transcendental term), as a natural successor of Lorenz family of attractors [6], [20]–[22], is capable of generating sufficiently complex chaotic motions in the phase space. The GP searches for new chaotic system state expressions while maximising the LLE obtained from time delay embedding of a single time series [23] as a candidate solution. But this LLE is not absolutely reliable to finally report if the embedding delay is wrongly chosen.

Therefore in the post processing stage, after the GP based initial search of probable candidate solutions, we adopted a symbolic Jacobian computation to obtain the exact algebraic structure based LLEs [24], for all the attractors. We read each expression and reconstructed the dynamical system using the Symbolic Math Toolbox (SMT) in Matlab. We used variable precision arithmetic up to five digits for the conversion of floating point constants to symbolic constants. The Jacobian of each set of dynamical system was calculated symbolically using the functions in SMT of Matlab. For both the LLE computation and phase space trajectory calculation, we used a fixed step size of 0.05 sec and time span of $10^4$ sec with the Dormand-Prince algorithm of numerical integration in Matlab programming platform (as outlined in [24] and an initial condition of $\{x_0 = 2, y_0 = 3, z_0 = 1\}$.

### *2.5. Post-processing of evolved expressions*

Reporting hundreds of attractors manually is a hugely time-consuming work as well. Therefore, we automated the post-processing part too. For all the expressions with positive LLE, we again reconstructed the dynamical system using the SMT of Matlab. We used the symbolic *simplify()* function to make the expressions as simple as possible. Then we replaced all the constant co-efficients in the expressions using variables like $\{a, b, c, \cdots\}$ etc. and added another string which shows the values of these variables (e.g. $a = 2, b = 3$ etc.). Finally we used the *latex()* function of SMT to report the expressions in a format which can be read by a TeX system and therefore facilitate converting them in the form of equations. For the final plots of the phase portraits as presented in this paper and the supplementary material, we simulated each attractor for 200 seconds with an ode3 Bogacki-Shampine solver and a fixed step size of 0.01 secs.

### 3. Results

The GP algorithm to search for the four cases in (2) were run 100 times, while the final converged expressions as well as the intermediate expressions have been reported here that produces a positive LLE. The simulations were run on a 64-bit Windows desktop with 16 GB memory and I7, 3.4 GHz processor and the run time for each of the four cases being approximately 2 days. The expression for the state equations and the symbolically computed LLE values with a sampling time of 0.05 and time length of $10^4$ sec have been reported in the subsequent sections in

Table 1-Table 4 respectively. Each of the four tables report the new expressions evolved by the GP algorithm along with the coefficient values and LLE. Most of the phase portraits are variants of the two-wing butterfly structure of the original Lorenz family of systems with some difference and has been reported in the supplementary material. Few interesting phase space patterns worth noticing have been reported here in Figure 1-Figure 11. These interesting phase space behaviours have been given a few names like uneven wings, island, sea shell, land snail, wink, binocular, tangled string etc. due to their resemblance with the real-world objects. Also, few entries of the following tables have the same attractors with different constants but completely different phase space behaviour as well as significantly different LLE which are not differentiated.

### 3.1. Extended Lorenz-XY family with cross-product nonlinearity

Table 1: Expressions of the generalized Lorenz-XY family of attractors

| Name | $\dot{x}$ expression | $\dot{y}$ expression | Coefficient values | LLE |
|---|---|---|---|---|
| Lorenz-XY1 | $ay-bx$ | $cx-y-xz$ | $a=10, b=10, c=28.058$ | 0.9037 |
| Lorenz-XY2 | $ay-bx$ | $cx-xz+d$ | $a=10, b=10, c=28, d=8.531$ | 0.9661 |
| Lorenz-XY3 | $ay-bx$ | $-x(z-c)$ | $a=10, b=10, c=27$ | 0.9771 |
| Lorenz-XY4 | $ay-bx$ | $cx-xz-d$ | $a=10, b=10, c=28, d=0.62818$ | 0.9989 |
| Lorenz-XY5 | $ay-bx$ | $-cx(dz-e)$ | $a=10, b=10, c=1.7764, d=0.56295, e=19.619$ | 1.1293 |
| Lorenz-XY6 | $ay-bx$ | $cx-y-xz+d$ | $a=10, b=10, c=28, d=1.3961$ | 0.8936 |
| Lorenz-XY7 | $ay-bx$ | $-x(z^2+z-c)$ | $a=10, b=10, c=28$ | 1.1035 |
| Lorenz-XY8 | $ay-bx$ | $-xz^2+cx-y$ | $a=10, b=10, c=28$ | 1.2097 |
| Lorenz-XY9 | $ay-bx$ | $x(z-c)(z-d)$ | $a=10, b=10, c=28, d=1$ | 0.9800 |
| Lorenz-XY10 | $ay-bx$ | $-x(cz-d)$ | $a=10, b=10, c=11, d=29$ | 1.0197 |
| Lorenz-XY11 | $ay-bx+c$ | $dx-y-xz$ | $a=10, b=10, c=8.1133, d=28$ | 0.8716 |
| Lorenz-XY12 | $ay-bx-c$ | $dx-xz-e$ | $a=6.2914, b=7.2914, c=3.0003, d=28.0, e=12.792$ | 0.8051 |
| Lorenz-XY13 | $ay-bx+c$ | $-x(z-d)$ | $a=10, b=10, c=34.008, d=28$ | 0.9103 |
| Lorenz-XY14 | $ay-bx$ | $-xz^2+cx+d$ | $a=10, b=10, c=28, d=1.7497$ | 1.0508 |
| Lorenz-XY15 | $ay-bx$ | $cx-xz^2-(x+d)(x+z)$ | $a=10, b=10, c=28, d=2.6022$ | 0.9762 |
| Lorenz-XY16 | $ay-bx$ | $-x(z^2-c)$ | $a=10, b=10, c=26$ | 1.1272 |
| Lorenz-XY17 | $ay-bx$ | $-cx(dz^2-e)$ | $a=10, b=10, c=4.4409, d=0.22518, e=5.5028$ | 1.0848 |
| Lorenz-XY18 | $ay - bx$ | $-xz^2 - cz + dx$ | $a = 10, b = 10, c = 3.5625, d = 28$ | 1.2823 |
| Lorenz-XY19 | $ay - bx$ | $-xz^2 + z + cx + d$ | $a = 10, b = 10, c = 28, d = 9.3871$ | 1.2869 |
| Lorenz-XY20 | $ay - bx$ | $-xz^2 + cx - y + d$ | $a = 10, b = 10, c = 28, d = 5.9679$ | 1.2308 |
| Lorenz-XY21 | $x^2 - zx - y - z$ | $-x(z - a)$ | $a = 27$ | 0.0031 |
| Lorenz-XY22 | $ay - bx$ | $-xz^2 + cx - d$ | $a = 10, b = 10, c = 28, d = 5.938$ | 1.1280 |
| Lorenz-XY23 | $ay - bx + c$ | $dx - xz - e$ | $a = 10, b = 10, c = 0.19889, d = 29, e = 7.9648$ | 1.0024 |
| Lorenz-XY24 | $ay - bx$ | $cx + y - xz + d$ | $a = 10, b = 10, c = 26, d = 2.7398$ | 1.0320 |
| Lorenz-XY25 | $ay - bx$ | $cx + y - dxz$ | $a = 10, b = 10, c = 27, d = 2$ | 1.0509 |
| Lorenz-XY26 | $ay - bx$ | $-x(cz - d)(z + e)$ | $a = 10, b = 10, c = 2, d = 7, e = 4$ | 1.1450 |
| Lorenz-XY27 | $-ax - z$ | $bx - xz - c$ | $a = 9.8541, b = 18.545, c = 499.52$ | 0.0017 |
| Lorenz-XY28 | $ay - bx - c$ | $dx - y - xz$ | $a = 10, b = 10, c = 5.7251, d = 28$ | 0.8865 |
| Lorenz-XY29 | $ay - bx$ | $cx + dy - xz - e$ | $a = 10, b = 20, c = 18, d = 10, e = 1.5421$ | 1.6578 |
| Lorenz-XY30 | $ay - bx$ | $cx + dy - xz$ | $a = 10, b = 20, c = 18, d = 9$ | 1.2971 |
| Lorenz-XY31 | $ay - bx$ | $cx + dy - xz + e$ | $a = 10, b = 20, c = 17, d = 10, e = 7.6208$ | 1.3788 |
| Lorenz-XY32 | $ay - bx$ | $-xz^2 + cx + dy - e$ | $a = 10, b = 30, c = 18, d = 10, e = 8.1933$ | 0.7552 |
| Lorenz-XY33 | $ay - bx$ | $x(z^2 - cz + d)$ | $a = 10, b = 10, c = 28, d = 28$ | 0.9780 |
| Lorenz-XY34 | $ay - bx + cz$ | $dx - y - xz$ | $a = 9, b = 10, c = 0.027707, d = 28$ | 0.8254 |
| Lorenz-XY35 | $ay - bx$ | $cx + y - xz - d$ | $a = 10, b = 10, c = 27, d = 28$ | 0.6964 |
| Lorenz-XY36 | $ay - bx - c$ | $dx - xz + e$ | $a = 10, b = 10, c = 25.812, d = 28, e = 2.5322$ | 0.9646 |
| Lorenz-XY37 | $ay - bx - c$ | $-x(z - d)$ | $a = 10, b = 10, c = 25.812, d = 28$ | 0.9519 |
| Lorenz-XY38 | $ay - bx - c$ | $-dx(ez - f)$ | $a = 10, b = 10, c = 25.812,$ $d = 3.5527, e = 0.28147, f = 7.8956$ | 0.9532 |

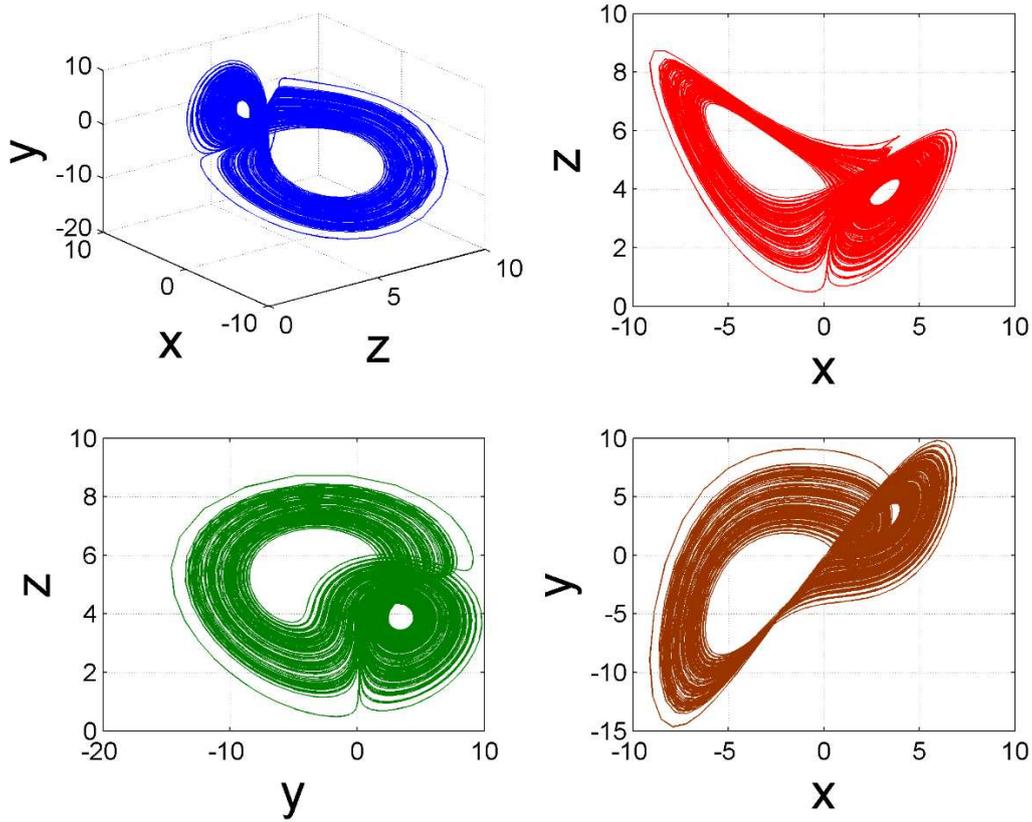

*Figure 1: Phase space dynamics of Lorenz-XY15 (uneven wings)*

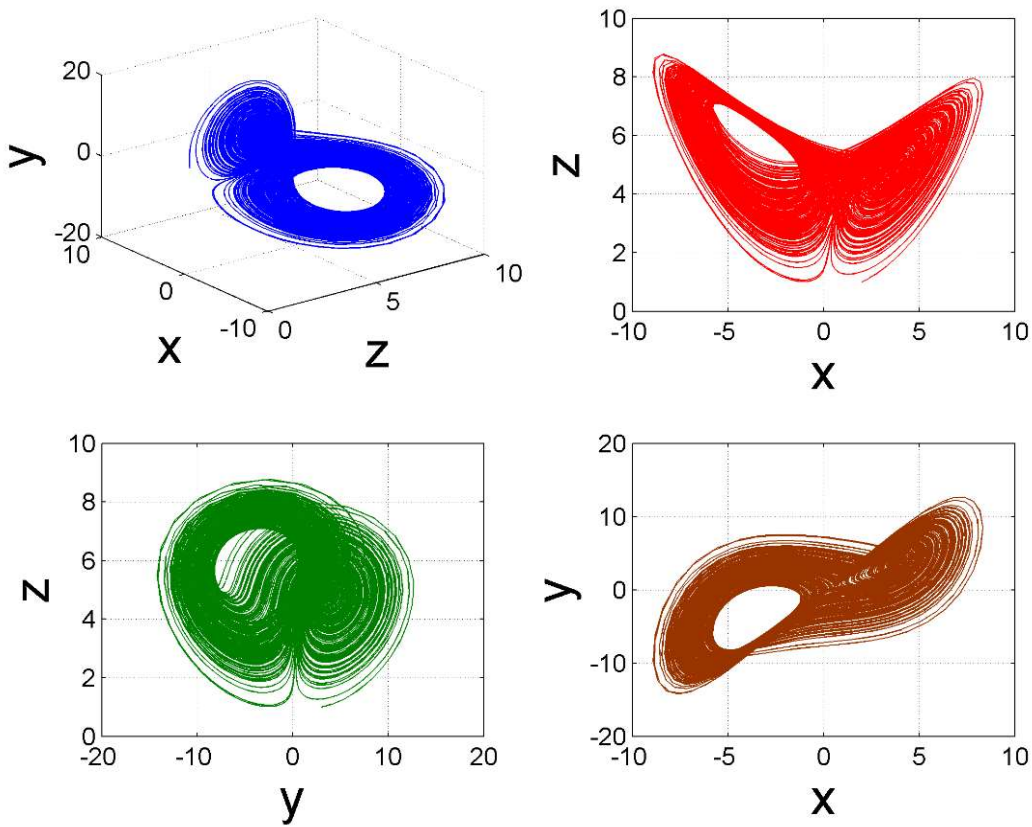

*Figure 2: Phase space dynamics of Lorenz-XY18 (pirate's eye patch)*

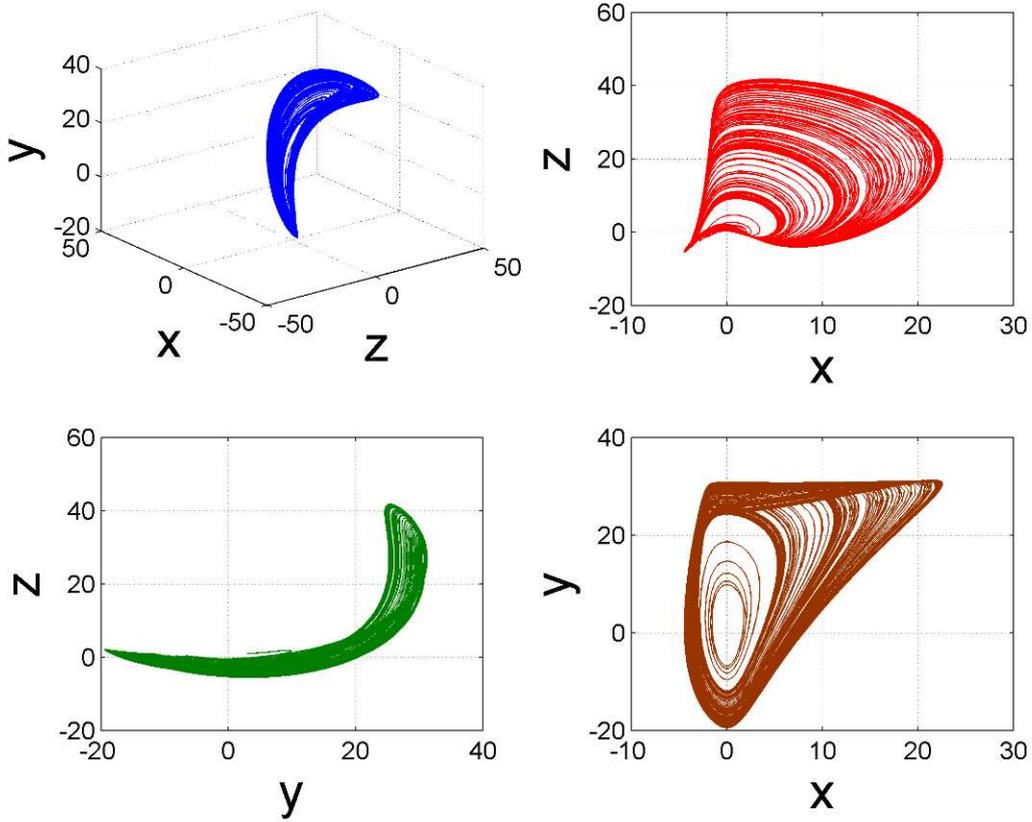

*Figure 3: Phase space dynamics of Lorenz-XY21 (wafer)*

### 3.2. Extended Lorenz-YZ family with cross-product nonlinearity

Table 2: Expressions of the generalized Lorenz-YZ family of attractors

| Name | $\dot{y}$ expression | $\dot{z}$ expression | Coefficient values | LLE |
|---|---|---|---|---|
| Lorenz-YZ1 | $ax - y - xz$ | $xy - bz$ | $a = 28, b = 2.6667$ | 0.9019 |
| Lorenz-YZ2 | $ax - y - xz$ | $xy - bz + c$ | $a = 28, b = 2.6667, c = 4.0252$ | 0.8637 |
| Lorenz-YZ3 | $-x(z - a)$ | $xy - bz$ | $a = 27, b = 2.6667$ | 0.9774 |
| Lorenz-YZ4 | $-ax(z - b)$ | $x^2 - cz$ | $a = 2, b = 14, c = 2.6667$ | 1.2641 |
| Lorenz-YZ5 | $ax - xz - b$ | $x^2 - cz$ | $a = 28, b = 2.6667, c = 2.6667$ | 1.2185 |
| Lorenz-YZ6 | $ax - xz - b$ | $xy - cz$ | $a = 28, b = 8.9901, c = 2.6667$ | 0.9639 |
| Lorenz-YZ7 | $-ax(bz - c)$ | $xy - dz$ | $a = 3.5527, b = 0.28147, c = 7.6082, d = 2.6667$ | 0.9787 |
| Lorenz-YZ8 | $ax - y - xz$ | $xy - b$ | $a = 28, b = 74.746$ | 1.3999 |
| Lorenz-YZ9 | $-xz$ | $xy - a$ | $a = 74.746$ | 1.6303 |
| Lorenz-YZ10 | $-xz$ | $y^2 - a$ | $a = 74.746$ | 1.3638 |
| Lorenz-YZ11 | $x - y - z - xz$ | $xy - a$ | $a = 74.746$ | 1.3995 |
| Lorenz-YZ12 | $-z(x+a)$ | $xy - b$ | $a = 1, b = 74.746$ | 1.5515 |
| Lorenz-YZ13 | $-x(z-a)$ | $by^2 - c$ | $a = 1, b = 27.03, c = 74.746$ | 1.3767 |
| Lorenz-YZ14 | $y - x - xz$ | $xy - a$ | $a = 74.746$ | 1.8763 |
| Lorenz-YZ15 | $ax - y - xz$ | $bx^2 - cz$ | $a = 28, b = 2, c = 2.6667$ | 0.4151 |
| Lorenz-YZ16 | $-x(z-a)$ | $-bz - x^2(x-c)$ | $a = 28, b = 2.6667, c = 28$ | 1.2885 |
| Lorenz-YZ17 | $-ax(bz-c)$ | $x^2 - d$ | $a = 7.1054, b = 0.14074, c = 3.9453, d = 2.6667$ | 1.2698 |
| Lorenz-YZ18 | $-ax(bz-c)$ | $x^2 - dz$ | $a = 1.3333, b = 2, c = 21, d = 2.6667$ | 1.2646 |
| Lorenz-YZ19 | $-x(z-a)$ | $bx^2 - cz$ | $a = 28, b = 28, c = 2.6667$ | 1.2637 |
| Lorenz-YZ20 | $-ax(bz-c)$ | $dx^2 - ez$ | $a = 0.00032, b = 3125, c = 87246, d = 28, e = 2.6667$ | 1.2693 |
| Lorenz-YZ21 | $-xz^2 + ax - y$ | $xy - bz$ | $a = 28, b = 2.6667$ | 1.2106 |
| Lorenz-YZ22 | $ax - y - xz$ | $x^2 - bz$ | $a = 27.909, b = 2.6667$ | 0.0007 |
| Lorenz-YZ23 | $ax - xz - b$ | $xy - c$ | $a = 27, b = 14.771, c = 74.667$ | 1.3358 |
| Lorenz-YZ24 | $-xz - a$ | $xy - b$ | $a = 14.771, b = 74.667$ | 1.3372 |
| Lorenz-YZ25 | $-x(z+a)$ | $xy - b$ | $a = 1, b = 74.667$ | 1.6417 |
| Lorenz-YZ26 | $az - bx - xz + c$ | $xy - d$ | $a = 2.6667, b = 3.6667, c = 2.6667, e = 74.667$ | 0.0100 |
| Lorenz-YZ27 | $-x(z-a)$ | $xy - b$ | $a = 27, b = 74.667$ | 1.6349 |
| Lorenz-YZ28 | $-xz^2 + ax - y$ | $xy - z - y$ | $a = 28$ | 0.8249 |
| Lorenz-YZ29 | $ax + y - z - xz$ | $xy - bz$ | $a = 28, b = 2.6667$ | 0.5543 |
| Lorenz-YZ30 | $ax - y - xz$ | $bxy - cz$ | $a = 28, b = 28, c = 2.6667$ | 0.9035 |
| Lorenz-YZ31 | $ax - y - xz$ | $xy(z+b) - cz$ | $a = 28, b = 1, c = 2.6667$ | 1.5476 |
| Lorenz-YZ32 | $ax - y - xz$ | $by(cxz + d) - ez$ | $a = 28, b = 0.33333, c = 3, d = 8, e = 2.6667$ | 1.2596 |
| Lorenz-YZ33 | $-xz^2 + ax - y$ | $xy - b$ | $a = 28, b = 7.1111$ | 1.5321 |
| Lorenz-YZ34 | $-xz^2 + ax - b$ | $xy - c$ | $a = 28, b = 0.92202, c = 7.1111$ | 1.7466 |
| Lorenz-YZ35 | $z - y - xz$ | $xy - a$ | $a = 74.667$ | 1.3432 |
| Lorenz-YZ36 | $-y-xz-a$ | $xy-b$ | $a=9.6885, b=74.667$ | 1.2938 |
| Lorenz-YZ37 | $a-xz-y$ | $xy-b$ | $a=2.6667, b=74.667$ | 1.4171 |
| Lorenz-YZ38 | $ax-xz-b$ | $xy-cz$ | $a=28, b=1.3014, c=2.6667$ | 0.9989 |
| Lorenz-YZ39 | $ax-y-bxz$ | $x^2-cz-d$ | $a=28, b=2.6667, c=2.6667, d=2.6667$ | 1.0485 |
| Lorenz-YZ40 | $-ax(bz-c)$ | $xy-dz$ | $a=0.33333, b=3, c=76, d=2.6667$ | 0.9390 |
| Lorenz-YZ41 | $-x(z-a)$ | $x(y-b)-cz$ | $a=28, b=2.3958, c=2.6667$ | 0.9594 |
| Lorenz-YZ42 | $-ax(z-b)$ | $cxy-dz$ | $a=2, b=14, c=28, d=2.6667$ | 1.0001 |
| Lorenz-YZ43 | $-x(z^2+z-a)$ | $bx^2-cz$ | $a=28, b=28, c=2.6667$ | 2.4157 |
| Lorenz-YZ44 | $ax-xz+b$ | $cx^2-dz$ | $a=28, b=1.1561, c=28, d=2.6667$ | 1.1838 |
| Lorenz-YZ45 | $-x(z^2+z-a)$ | $bxy-cz$ | $a=28, b=28, c=2.6667$ | 1.1041 |
| Lorenz-YZ46 | $-x(az-b)$ | $cxy-dz$ | $a=29, b=28, c=28, d=2.6667$ | 0.9991 |
| Lorenz-YZ47 | $-xz^2+ax-y$ | $x-z+xy$ | $a=28$ | 0.9478 |
| Lorenz-YZ48 | $ax-y-xz$ | $bx+xy-c$ | $a=28, b=2.6667, c=74.667$ | 1.5134 |
| Lorenz-YZ49 | $-ax-y-xz$ | $xy-b$ | $a=5.7244, b=74.667$ | 1.3908 |
| Lorenz-YZ50 | $ax-y-xz-b$ | $xy-c$ | $a=28, b=0.9829, c=74.667$ | 1.2503 |
| Lorenz-YZ51 | $y-xz-a$ | $xy-b$ | $a=2.6667, b=74.667$ | 1.8804 |
| Lorenz-YZ52 | $y-xz-a$ | $xy-by-c$ | $a=2.6667, b=2.6667, c=62.169$ | 1.4222 |
| Lorenz-YZ53 | $ax-y-xz$ | $x(x+y-b)-cz$ | $a=28, b=7.1699, c=2.6667$ | 0.8510 |
| Lorenz-YZ54 | $ax-y-xz$ | $x(bx-y)-cz$ | $a=28, b=29, c=2.6667$ | 1.0104 |
| Lorenz-YZ55 | $ax-y-xz$ | $y(bx-y)-cz$ | $a=28, b=28, c=2.6667$ | 0.9189 |
| Lorenz-YZ56 | $ax-y-xz$ | $bxy-c$ | $a=28, b=2.6667, c=74.667$ | 1.3748 |
| Lorenz-YZ57 | $ax-y-xz$ | $x(bx-c)-dz$ | $a=28, b=27, c=7.6956, d=2.6667$ | 0.8891 |
| Lorenz-YZ58 | $x-y-xz$ | $xy-a$ | $a=74.667$ | 1.3832 |
| Lorenz-YZ59 | $-x(x+z-a)$ | $bx^2-cz$ | $a=28, b=28, c=2.6667$ | 1.2990 |
| Lorenz-YZ60 | $ax - y - xz$ | $xy - z - b$ | $a = 28, b = 32.11$ | 1.0011 |
| Lorenz-YZ61 | $ax - y - bxz - c$ | $xy - dz - e$ | $a = 28, b = 4.2232, c = 3.4967, d = 2.6667, e = 18.54$ | 1.3681 |
| Lorenz-YZ62 | $x - y - axz$ | $xy - bz - c$ | $a = 4.2232, b = 2.6667, c = 18.54$ | 0.9577 |
| Lorenz-YZ63 | $ax - bxz - c$ | $xy - dz - e$ | $a = 27, b = 4.2232, c = 3.4967, d = 2.6667, e = 18.54$ | 1.3622 |
| Lorenz-YZ64 | $-y - axz - b$ | $xy - cz - d$ | $a = 4.2232, b = 0.83007, c = 2.6667, d = 18.54$ | 0.9290 |
| Lorenz-YZ65 | $a - bxz$ | $xy - cz - d$ | $a = 2.6667, b = 4.2232, c = 2.6667, d = 18.54$ | 1.0190 |
| Lorenz-YZ66 | $ax + by - z - xz$ | $xy - cz$ | $a = 27, b = 2, c = 2.6667$ | 0.6472 |
| Lorenz-YZ67 | $x^2 - zxa + by$ | $xy - cz$ | $a = 2, b = 7.0951, c = 2.6667$ | 0.0004 |
| Lorenz-YZ68 | $ax + by - xz$ | $xy - cz$ | $a = 28, b = 2.0743, c = 2.6667$ | 1.0599 |
| Lorenz-YZ69 | $ax - y - xz$ | $xz(y - b) - cz$ | $a = 28, b = 1, c = 2.6667$ | 1.2912 |
| Lorenz-YZ70 | $-xz^2 - z + ax$ | $xy - bz$ | $a = 28, b = 2$ | 1.3159 |
| Lorenz-YZ71 | $-ax(bz^2 - c)$ | $xy - dz$ | $a = 1.3323, b = 2.2243, c = 20.266, d = 2$ | 1.1954 |
| Lorenz-YZ72 | $-ax(bz^2 - c)$ | $xy - z - x$ | $a = 1.3323, b = 2.2243, c = 20.266$ | 0.8326 |
| Lorenz-YZ73 | $ax - y - xz$ | $x(bx + y) - cz$ | $a = 28, b = 2, c = 2.6667$ | 0.8956 |
| Lorenz-YZ74 | $ax + bz - xz$ | $xy - cz$ | $a = 28, b = 0.57681, c = 2.6667$ | 0.6138 |
| Lorenz-YZ75 | $-ax(z - b)$ | $x(x + y) - cz$ | $a = 2, b = 14, c = 2.6667$ | 1.2037 |
| Lorenz-YZ76 | $-x(z - a)$ | $x(x + b) - cz$ | $a = 28, b = 4.1059, c = 2.6667$ | 1.0411 |

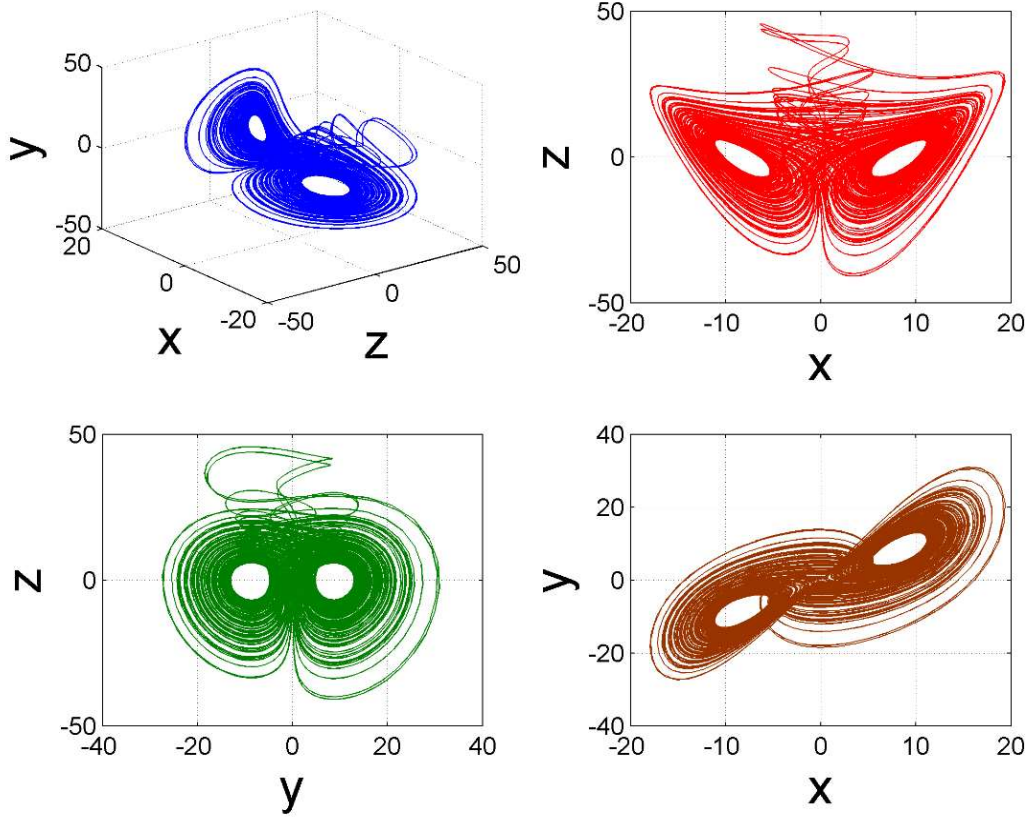

*Figure 4: Phase space dynamics of Lorenz-YZ10 (Mask of Zoro)*

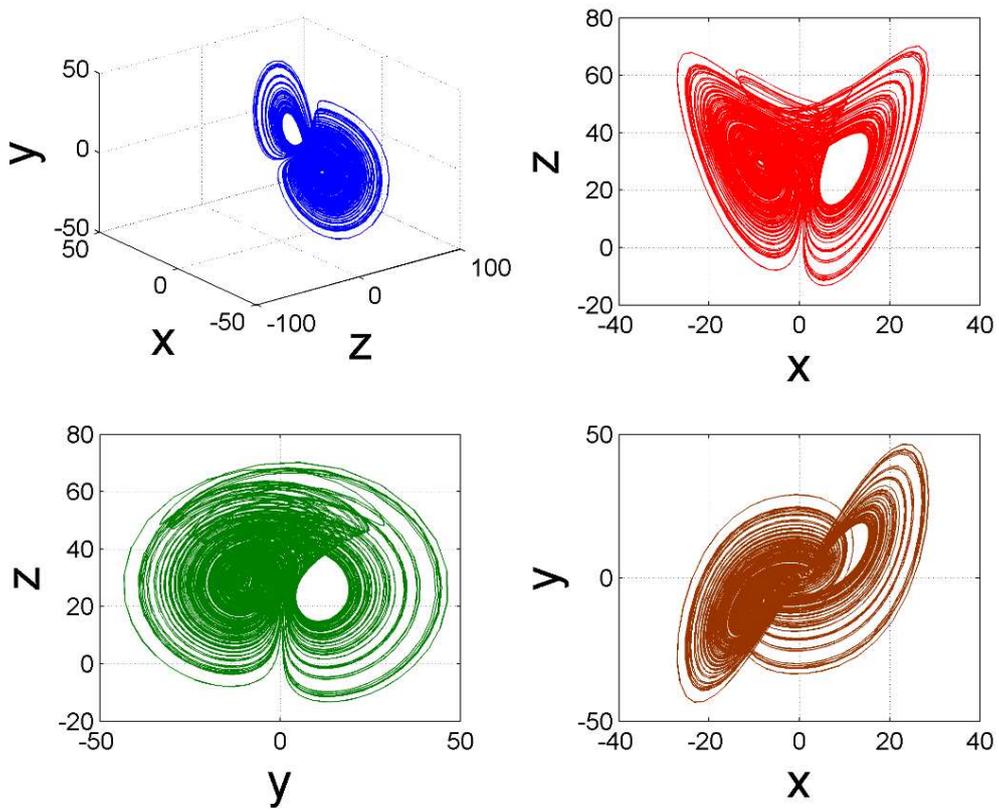

*Figure 5: Phase space dynamics of Lorenz-YZ23 (island)*

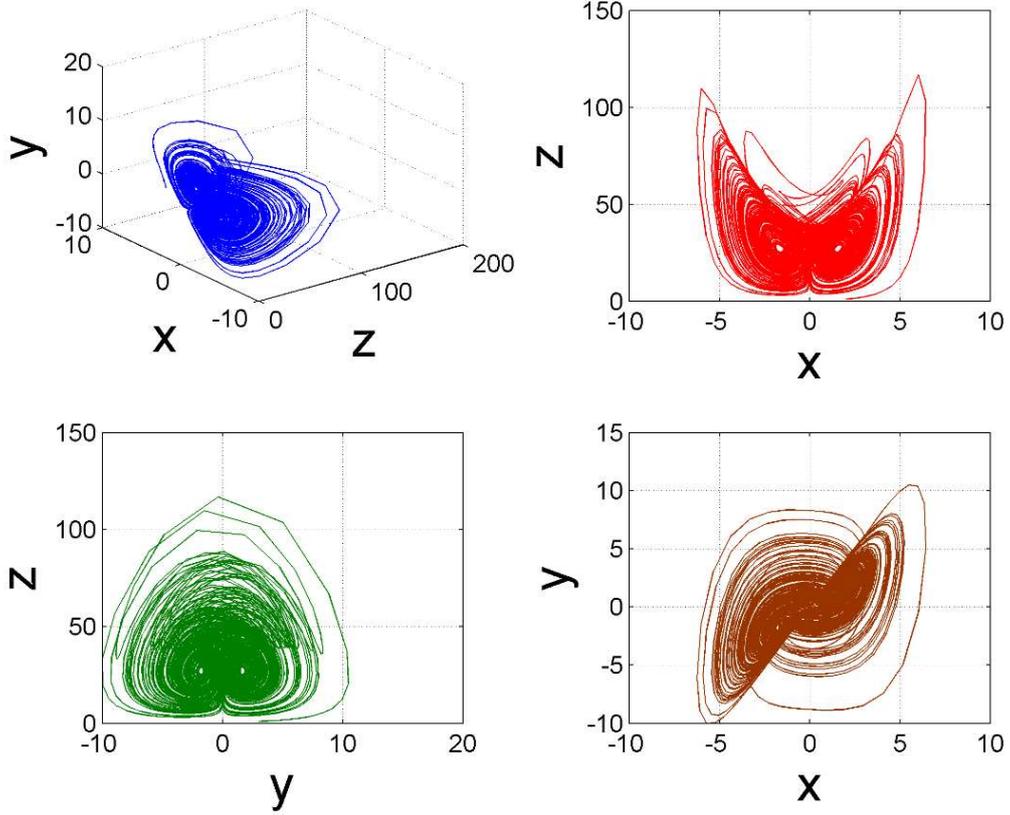

*Figure 6: Phase space dynamics of Lorenz-YZ32 (angry birds)*

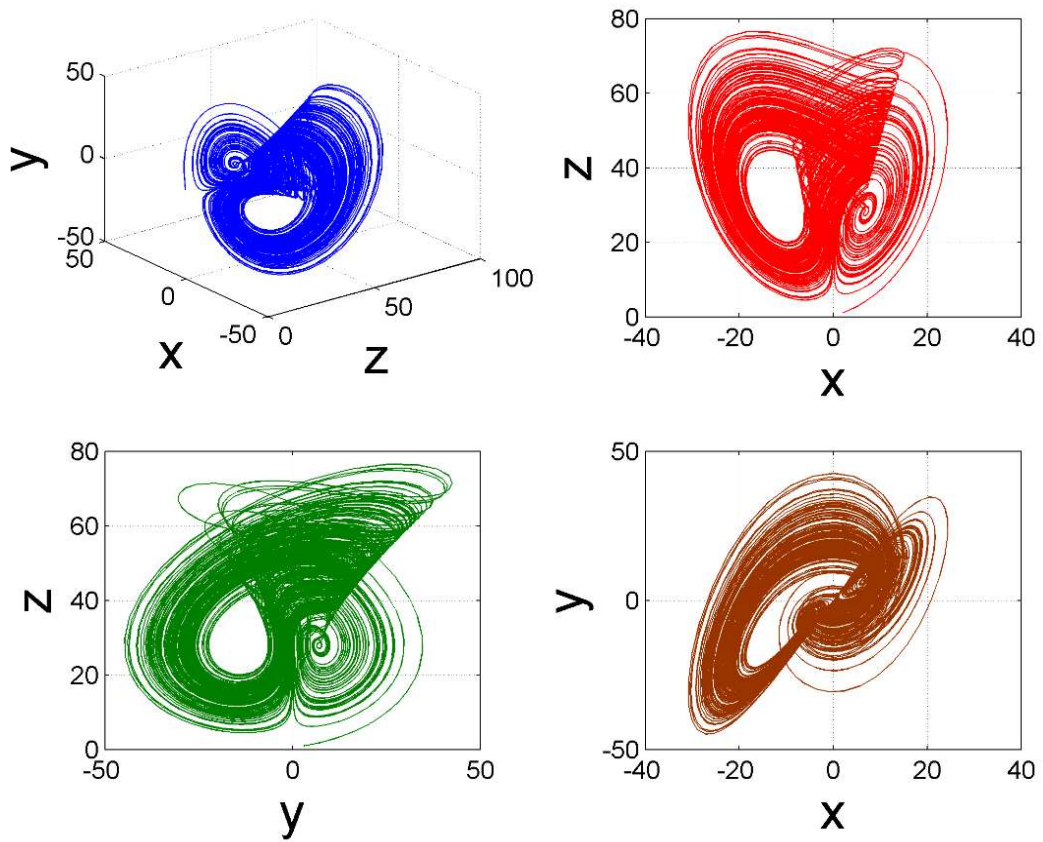

*Figure 7: Phase space dynamics of Lorenz-YZ76 (sea shell)*

### 3.3. Extended Lorenz-XZ family with cross-product nonlinearity

Table 3: Expressions of the generalized Lorenz-XZ family of attractors

| Name | $\dot{x}$ expression | $\dot{z}$ expression | Coefficient values | LLE |
|---|---|---|---|---|
| Lorenz-XZ1 | $ay - bx + c$ | $xy - dz$ | $a = 10, b = 10, c = 0.63329, d = 2.6667$ | 0.9031 |
| Lorenz-XZ2 | $ay - bx - c$ | $xy - dz$ | $a = 10, b = 9, c = 2.6667, d = 2.6667$ | 0.9547 |
| Lorenz-XZ3 | $ay - bx$ | $xy - cz$ | $a = 10, b = 10, c = 2.6667$ | 0.9019 |
| Lorenz-XZ4 | $ay - bx$ | $cx^2 - dz$ | $a = 10, b = 10, c = 2, d = 2.6667$ | 0.4151 |
| Lorenz-XZ5 | $ay - bx$ | $x(x+y) - cz$ | $a = 10, b = 10, c = 2.6667$ | 0.9000 |
| Lorenz-XZ6 | $ay - bx$ | $cxy - dz$ | $a = 10, b = 10, c = 10, d = 2.6667$ | 0.9027 |
| Lorenz-XZ7 | $ay - bx$ | $(x+y)^2 - cz$ | $a = 10, b = 10, c = 2.6667$ | 0.8827 |
| Lorenz-XZ8 | $ay - bx$ | $cx(x+y) - dz$ | $a = 10, b = 10, c = 2, d = 2.6667$ | 0.8935 |
| Lorenz-XZ9 | $ay - bx$ | $xy - cz - x$ | $a = 10, b = 10, c = 1.6667$ | 0.7530 |
| Lorenz-XZ10 | $ay - bx$ | $xy - cz - d$ | $a = 10, b = 10, c = 2.6667, d = 5.7345$ | 0.9502 |
| Lorenz-XZ11 | $-az - by(x-y)$ | $xy - cz$ | $a = 1.2154, b = 0.45577, c = 2.6667$ | 0.4518 |
| Lorenz-XZ12 | $ay - bx$ | $xy - y - cz - dx$ | $a = 10, b = 10, c = 2.6667, d = 1.7544$ | 0.7349 |
| Lorenz-XZ13 | $ay - bx$ | $xy - cz - y + x(x-d)$ | $a = 10, b = 10, c = 2.6667, d = 2.3465$ | 0.9197 |
| Lorenz-XZ14 | $ay - bx$ | $xy - cz - dx - e$ | $a = 10, b = 10, c = 2.6667, d = 1.7544, e = 2.2835$ | 0.8479 |
| Lorenz-XZ15 | $ay - bx$ | $xy - cz + d$ | $a = 10, b = 10, c = 2.6667, d = 1.1615$ | 0.8932 |
| Lorenz-XZ16 | $ay - bx$ | $x(x+c) - dz$ | $a = 10, b = 10, c = 3.8174, d = 2.6667$ | 0.8312 |
| Lorenz-XZ17 | $ay - bx$ | $cx(dx+e) - fz$ | $a = 10, b = 10, c = 0.66667, d = 15, e = 4, f = 2.6667$ | 0.8999 |
| Lorenz-XZ18 | $ay - bx - c$ | $x^2 - dz$ | $a = 10, b = 10, c = 2.6667, d = 2.6667$ | 0.9819 |

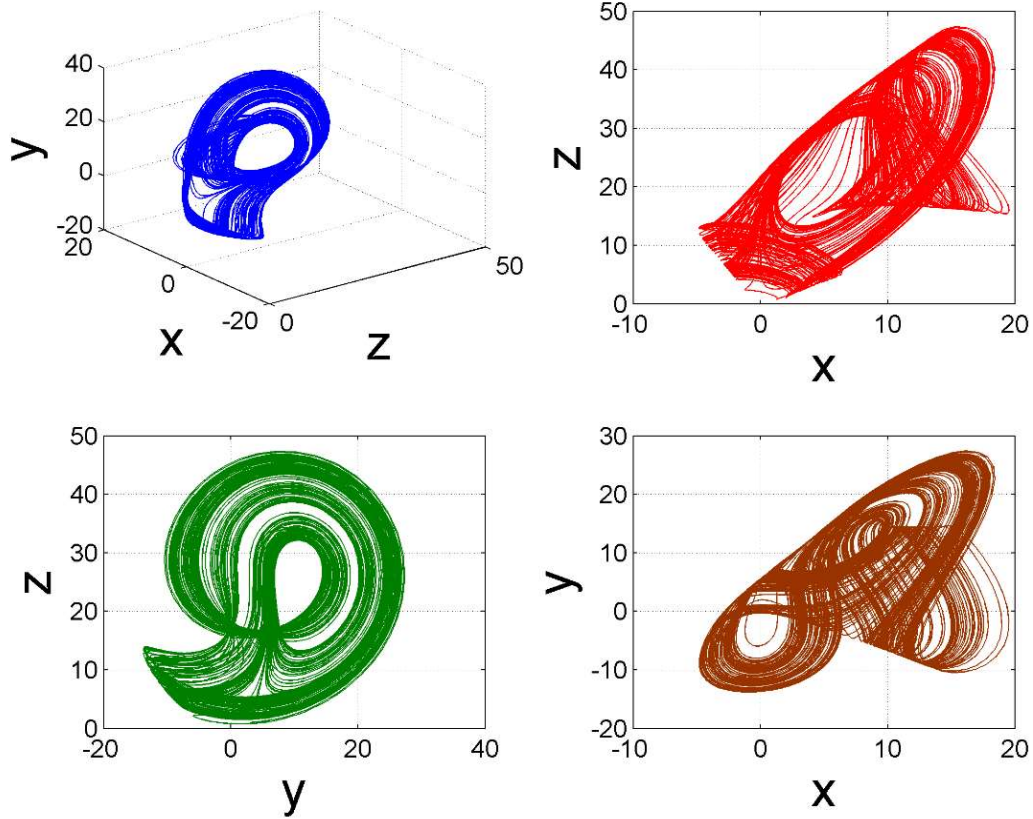

*Figure 8: Phase space dynamics of Lorenz-XZ11 (land snail)*

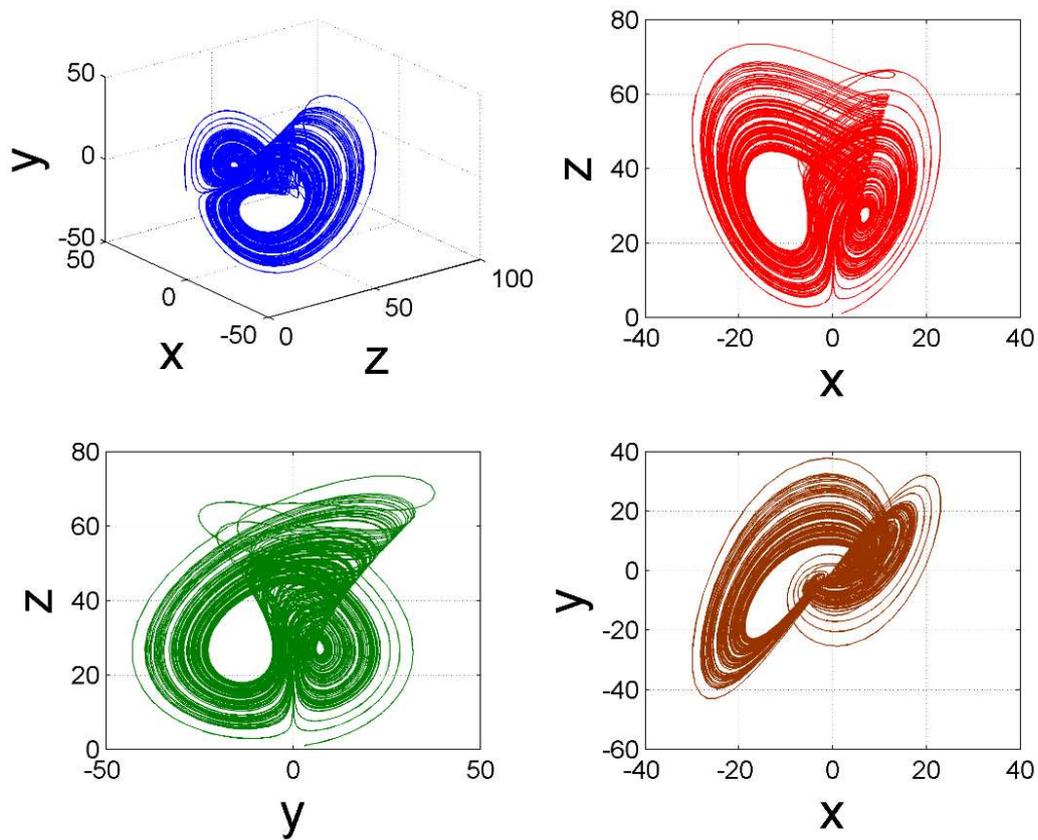

*Figure 9: Phase space dynamics of Lorenz-ZX16 (wink)*

## 3.4. Extended Lorenz-XYZ family with cross-product nonlinearity

Table 4: Expressions of the generalized Lorenz-XYZ family of attractors

| Name | $\dot{x}$ expression | $\dot{y}$ expression | $\dot{z}$ expression | Coefficient values | LLE |
|---|---|---|---|---|---|
| Lorenz-XYZ1 | $ay-bx$ | $-xz^2+cx-y$ | $xy-dz$ | $a=10, b=10, c=28, d=2.6667$ | 1.210579 |
| Lorenz-XYZ2 | $ay-bx$ | $-cx(z-d)$ | $xy-ez$ | $a=10, b=10, c=2, d=14, e=2.6667$ | 0.999431 |
| Lorenz-XYZ3 | $ay-bx$ | $-cx(dz-e)$ | $xy-fz$ | $a=10, b=10, c=0.33333, d=3, e=76, f=2.6667$ | 0.938996 |
| Lorenz-XYZ4 | $-az(x-y)$ | $x$ | $x^2-bx+z-c$ | $a=7.1111, b=1.6667, c=28$ | 0.890715 |
| Lorenz-XYZ5 | $-az(x-y)$ | $-bx$ | $x^2-cx+z-d$ | $a=7.1111, b=1.1549, c=1.6667, d=28$ | 0.013961 |
| Lorenz-XYZ6 | $ay-bx$ | $cx-y-xz$ | $xy-d$ | $a=10, b=10, c=28, d=74.667$ | 1.386173 |
| Lorenz-XYZ7 | $ay-bx$ | $-xz$ | $xy-c$ | $a=10, b=10, c=74.667$ | 1.648004 |
| Lorenz-XYZ8 | $ay-bx$ | $cx-xz+d$ | $xy-ez$ | $a=10, b=10, c=28, d=4.9295, e=2.6667$ | 0.991365 |
| Lorenz-XYZ9 | $ay-bx$ | $cx+y-xz+d$ | $x^2-ez$ | $a=10, b=10, c=28, d=2.651, e=2.6667$ | 1.600634 |
| Lorenz-XYZ10 | $ay-bx$ | $cx-y-dxz$ | $xy-ez$ | $a=10, b=10, c=28, d=28, e=2.6667$ | 0.902656 |
| Lorenz-XYZ11 | $ay-bx$ | $x-y-xz$ | $xy-c$ | $a=10, b=10, c=74.667$ | 1.383194 |
| Lorenz-XYZ12 | $ay-bx$ | $-cx(dz-e)$ | $xy-fz$ | $a=10, b=10, c=1.7764, d=0.56295, e=11.883, f=2.6667$ | 0.834024 |
| Lorenz-XYZ13 | $ay-bx$ | $cx-y-xz$ | $(d-y)x^2-ez$ | $a=10, b=10, c=28, d=28, e=2.6667$ | 0.950368 |
| Lorenz-XYZ14 | $ay-bx$ | $cx-xz-d$ | $xy-ez$ | $a=10, b=10, c=27, d=4.6119, e=2.6667$ | 0.967698 |
| Lorenz-XYZ15 | $ay-bx$ | $cx-y-xz$ | $dz(exy-f)$ | $a=10, b=10, c=28, d=2.9606, e=1.5348, f=0.90072$ | 1.573857 |
| Lorenz-XYZ16 | $ay-bx$ | $-x(z-c)$ | $dz(exy-f)$ | $a=10, b=10, c=27, d=2.9606, e=1.5348, f=0.90072$ | 1.797136 |
| Lorenz-XYZ17 | $ay-bx$ | $cx-y-xz$ | $dxy-ez$ | $a=10, b=10, c=28, d=12.117, e=2.6667$ | 0.902901 |
| Lorenz-XYZ18 | $ay-bx$ | $-x(z-c)$ | $x(x+d)-ez$ | $a=10, b=10, c=27, d=1.2624, e=2.6667$ | 1.118337 |
| Lorenz-XYZ19 | $ay-bx$ | $cx-y-xz$ | $dx^2-ez$ | $a=10, b=10, c=28, d=10, e=2.6667$ | 0.350399 |
| Lorenz-XYZ20 | $ay-bx$ | $cx-y-xz$ | $x(x+y)-dz$ | $a=10, b=10, c=28, d=2.6667$ | 0.899956 |
| Lorenz-XYZ21 | $ay-bx$ | $cx-y-xz$ | $x(x+dy)-ez$ | $a=10, b=10, c=28, d=2, e=2.6667$ | 0.921967 |
| Lorenz-XYZ22 | $ay-bx$ | $-x(z-c)$ | $dz(exy-f)$ | $a=10, b=10, c=27, d=0.33333, e=3, f=8$ | 1.799491 |
| Lorenz-XYZ23 | $ay-bx$ | $-x(z-c)$ | $dxy-ez$ | $a=10, b=10, c=27, d=5.6746, e=2.6667$ | 0.978276 |
| Lorenz-XYZ24 | $ay-bx$ | $-x(z-c)$ | $dxy-ez+f$ | $a=10, b=10, c=27, d=5.6746,$ $e=2.6667, f=11.946$ | 0.870589 |
| Lorenz-XYZ25 | $ay-bx$ | $-x(z-c)$ | $x^2-dz$ | $a=10, b=10, c=28, d=2.6667$ | 1.263459 |
| Lorenz-XYZ26 | $ay-bx$ | $cx-z(dx-z)$ | $xy-ez$ | $a=10, b=10, c=28, d=28, e=2.6667$ | 0.944587 |
| Lorenz-XYZ27 | $ay-bx$ | $-x(z-c)$ | $dx^2-ez$ | $a=10, b=10, c=28, d=28, e=2.6667$ | 1.263673 |
| Lorenz-XYZ28 | $ay-bx$ | $cx-xz-d$ | $x^2-ez$ | $a=10, b=10, c=28, d=1.3852, e=2.6667$ | 1.26158 |
| Lorenz-XYZ29 | $ay-bx+c$ | $dx-xz-e$ | $x^2-fz$ | $a=10, b=10, c=19.366, d=28,$ $e=1.3852, f=2.6667$ | 1.254098 |
| Lorenz-XYZ30 | $ay-bx+c$ | $-dx(ez-f)$ | $x^2-gz$ | $a=10, b=10, c=19.366, d=4.4409,$ $e=0.22518, f=5.5252, g=2.6667$ | 1.129121 |
| Lorenz-XYZ31 | $ay-bx+c$ | $dx-xz-e$ | $xy-fz$ | $a=10, b=10, c=19.366, d=28,$ $e=1.3852, f=2.6667$ | 0.978701 |
| Lorenz-XYZ32 | $ay-bx$ | $cx-xz-d$ | $xy-e$ | $a=10, b=10, c=28, d=2.6667, e=74.667$ | 1.584448 |
| Lorenz-XYZ33 | $ay-bx$ | $-x(z-c)$ | $xy-d$ | $a=10, b=10, c=27, d=74.667$ | 1.634896 |
| Lorenz-XYZ34 | $ay-bx$ | $cx+y-xz+d$ | $xy-e$ | $a=10, b=10, c=28, d=2.5796, e=74.667$ | 1.878482 |
| Lorenz-XYZ35 | $ay-bx$ | $cx+y-xz+d$ | $xy-ez$ | $a=10, b=10, c=28, d=2.5796, e=2.6667$ | 1.069293 |
| Lorenz-XYZ36 | $ay-bx$ | $cx-xz+d$ | $xy-e$ | $a=10, b=10, c=28.995, d=2.5796, e=74.667$ | 1.591306 |
| Lorenz-XYZ37 | $ay-bx$ | $cx+dy-xz$ | $xy-ez$ | $a=10, b=10, c=28, d=0.013854, e=2.6667$ | 1.0003 |

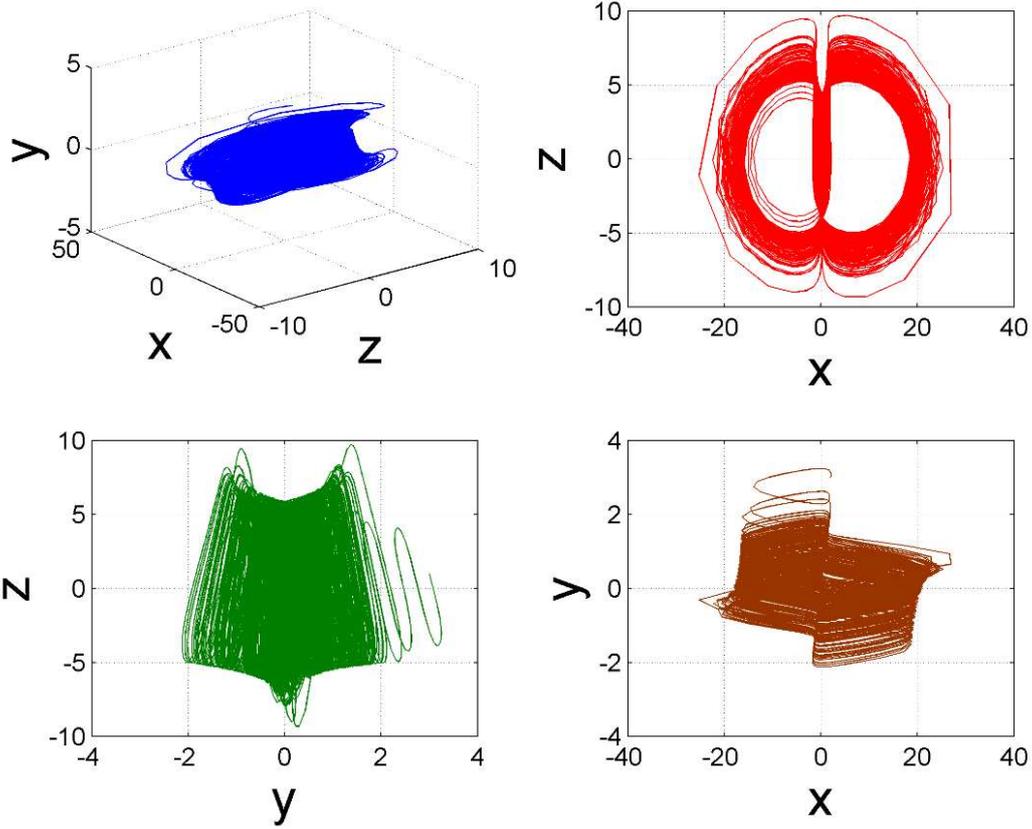

*Figure 10: Phase space dynamics of Lorenz-XYZ4 (binocular)*

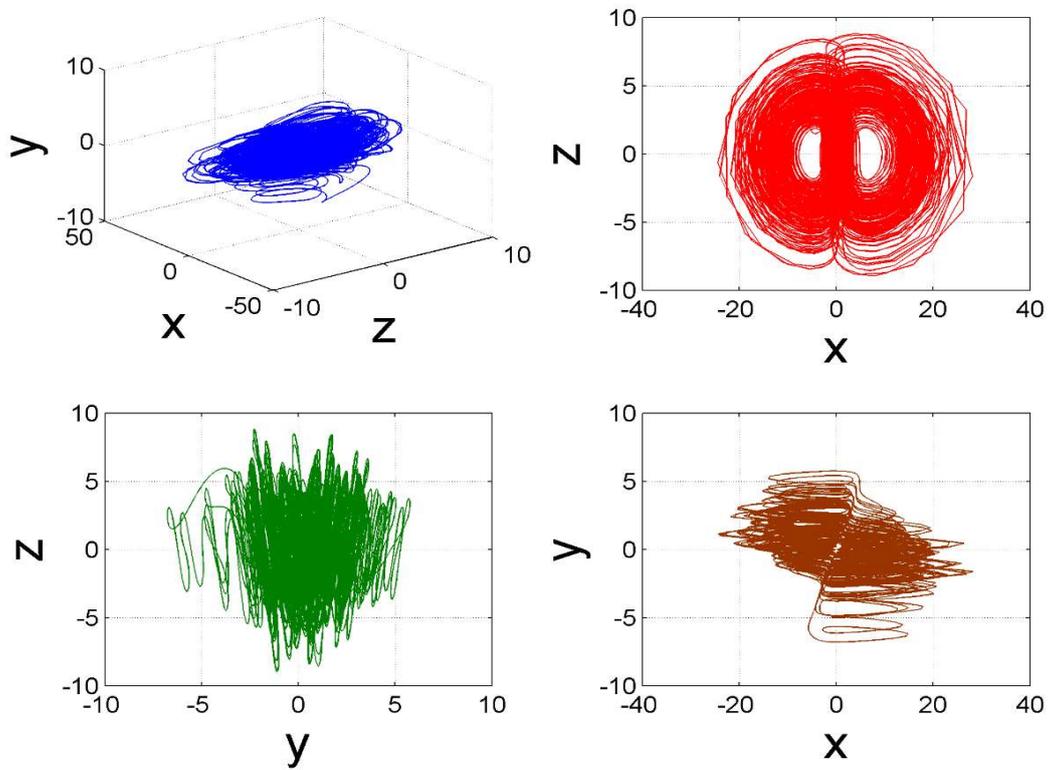

*Figure 11: Phase space dynamics of Lorenz-XYZ5 (tangled string)*

## 4. Discussions

Here we have focussed only on simple operators (like addition, subtraction, multiplication, division and higher powers of state variables) for evolving new chaotic systems with GP, as opposed to other complex nonlinear functions e.g. sinusoids introduced in [17]. The objective here is to find out similar attractors within the Lorenz family, with similar expressions but which exhibit drastically different looking phase portraits. Though we were successful in finding quite rich and novel phase portrait characteristics from these relatively simple algebraic expressions, a vast majority of the chaotic attractors (as shown in the supplementary materials) have very similar phase space dynamics to the original Lorenz system. Also, due to the use of such simple operators, it might happen that some of these chaotic attractors actually belong to other generalised family of attractors which have been studied before in the vast amount of available literature, claiming invention of yet another chaotic attractor, as an extension of the Lorenz family. Also, in the past it has been seen that many claims of such findings of algebraically unique chaotic attractors are indeed a result of transformation of the state variables which make them identical e.g. please see the discussions, controversies and replies on the claim of Chen [25][26][27], Lu [28] and generalised Lorenz system [29], [30] as special classes of Lorenz system. This needs to be analysed on a case by case basis for the attractors reported in this paper and the present paper shows that a GP based search can automatically evolve such hundreds of simple algebraic expressions for new unexplored family of chaotic systems. However, we do not claim that the new structures reported here cannot be reduced to some special cases of already existing vast zoo or gallery of 3D chaotic flows. It is also worth noting that our list does not claim to encompass all possible attractors but shows only few hundreds of candidate solutions, as there can be many more possible solutions depending on increasing complexity of the algebraic state equations. Other types of nonlinearities (like hyperbolic tangent, exponential, logarithmic etc.) can also be followed up as a sequel of the presently reported results.

Recently Gao *et al.* [31] claimed to improve over the GP based chaos evolution results in [17] for designing chaotic system corresponding to the global optima. However, it is worth noting that even local minima found by the evolutionary search with different expressions and dynamical behaviours can indeed be useful to discover new gallery of chaotic systems. With such an aim, even sub-optimal intermediate solutions with slightly positive Lyapunov exponents may show rich phase space dynamics as the complexity does not necessarily always correlate with high LLE. This is the reason why we here report even the intermediate search results with LLE > 0, similar to the earlier explorations in [17], during the evolutionary search process and not only at the final converged results of GP algorithm, which has been misinterpreted in Gao *et al.* [31]. Also, in this paper, we have listed the newly found chaotic system as the GP search progressed and not using the calculated LLE. Only the interesting phase space characteristics are shown here and the rest of them resembling the Lorenz system have been reported in the supplementary material. A ranking and categorization of the newly found attractors are not attempted here similar to the studies on different symmetry, equilibria and multi-stability [32]–[34]. This is left as a scope of future work as the complexity of the chaotic attractors does not always correlate with the LLE and as such there are many atypical behaviours reported here with a low LLE value which needs further investigation.

Sprott first introduced a similar concept of maximizing the chaoticity or complexity (strangeness or fractal structure) using either the Lyapunov exponent or the Kaplan-Yorke dimension ($D_{KY} = 1 - L_{max}/L_{min}$, where $L_{max}/L_{min}$ is the ratio of maximum and minimum Lyapunov exponents) [35] which we also adopt in this study using only the LLE criteria in the GP based combined structure and parameter optimization. It is also argued in [35] that the Lyapunov exponents can be made arbitrarily high by suitable choice of the constants within the same structure of the nonlinear dynamical system. However, our genetic programming based search attempts to find new algebraic structures and does not search for the constants of a fixed algebraic structure by maximizing the LLE using an optimizer unlike [35]. This is the fundamental difference between searching for new structures as well as constants (using GP) compared to searching for constants in a fixed structure (e.g. using genetic algorithm or other evolutionary or swarm algorithms) with an objective of maximizing a cost function. Since the GP does crossover and mutation of the subtrees of symbolic expressions rather than the variables, it can produce novel algebraic structures which fixed structure parameter optimizers cannot achieve.

In the GP algorithm, the ratio of mathematical operators to the numeric values are kept as 0.5 as also reported in [17]. This enables the GP algorithm to explore new mathematical expressions, rather than searching different constants within the same structure which could have been easily done with simpler versions like genetic algorithms and other evolutionary optimizers for fixed structure parameter optimization problems. Therefore, starting with the Lorenz system, the GP algorithm give equal emphasis on searching new constants within the same structure as well as evolving drastically new expressions which might lead to retaining the similar terms like part of the original Lorenz system structure e.g. constants like 8/3 = 2.6667 in the third state equation, 28 in the second state equation or 10 in the first state equation etc. Since the GP algorithm evolves tree like structures, an already found expression yielding good chaotic behaviour may be copied in the next generation as a part of the tree while other parts are evolved independently through crossover and mutation. In fact, there may be many other possible evolved structures starting from different chaotic systems with different constants which is beyond the scope of the present paper.

During the GP based search method as well as during the LLE computation and obtaining the phase portraits, the initial conditions were kept fixed as mentioned before. There is a possibility that this initial condition of the state variables are outside the basin of attraction as discussed in [35], for the candidate solutions in a particular run of GP. These solutions automatically get rejected as the LLE does not reach a positive value. A more exhaustive albeit computationally heavy way can be to search with various initial conditions within the GP algorithm to obtain yet more new attractors.

Moreover, the estimation of LLE can be sensitive to the choice of step-size, initial condition and the total time span. For a longer time-span of $10^4$ sec, a sampling time of 0.05 sec yields reasonable computational time. Therefore, according to the previous literature especially in the case of Lorenz system, our LLE calculation seems to be correct up to two decimal places whereas an accurate LLE up to five decimal places would typically need $10^{10}$ simulation steps with the same sampling time as reported in [35] which is left as a scope of future work due to massive computational burden for the present large family of newly found chaotic attractors. In future, a further study is needed to explore the sensitivity of the LLE calculation for these large family of chaotic systems by varying initial conditions, step size and total time for simulation.

## 5. Conclusions

This paper reports further extension of Lorenz family of chaotic attractors using simple cross product type nonlinearity. The structure of the chaotic systems, one set of parameter values exhibiting chaotic behaviour and the rich phase space dynamics have also been reported for more than 150 new nonlinear dynamical systems. Future work may include using these new chaotic system structures in explaining naturally occurring complex dynamics [36], [37], like the use of Lorenz system in prediction of atmospheric dynamics as well as in the applied field like secure communication and cryptography [1].

**Appendix**

Phase portraits of all the 4 classes of extended Lorenz family of systems, reported in the Tables are shown in the supplementary material for brevity.

# Supplementary Material

## 1. Phase portraits of generalized Lorenz-XY family of attractors

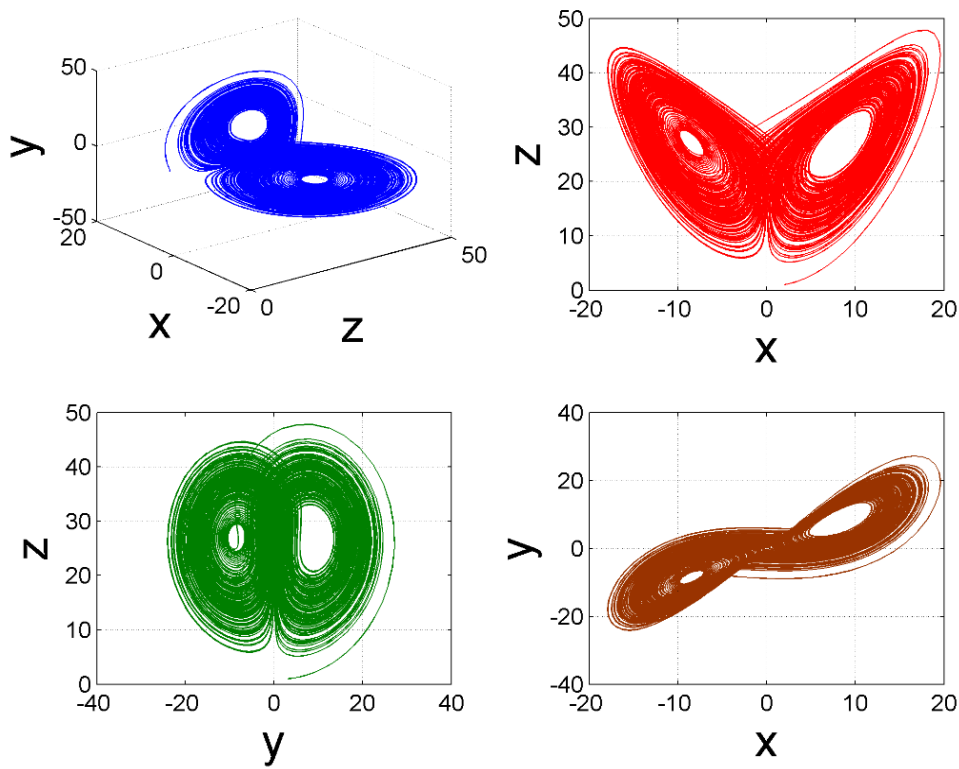

*Figure 1: Phase space dynamics of Lorenz-XY1*

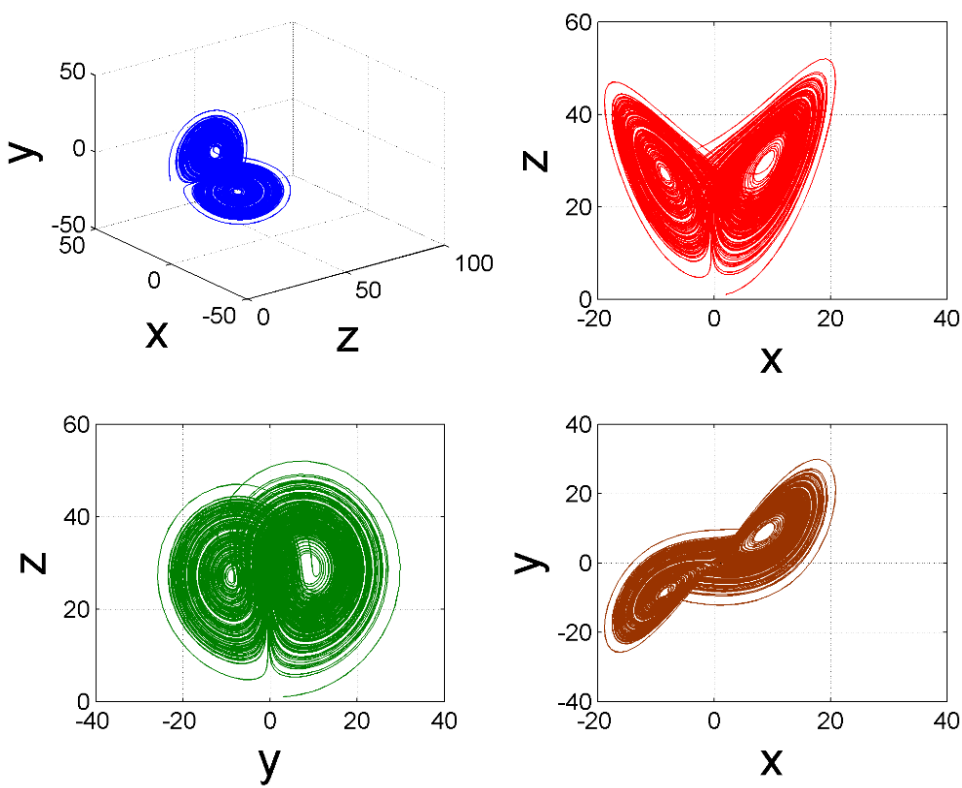

*Figure 2: Phase space dynamics of Lorenz-XY2*

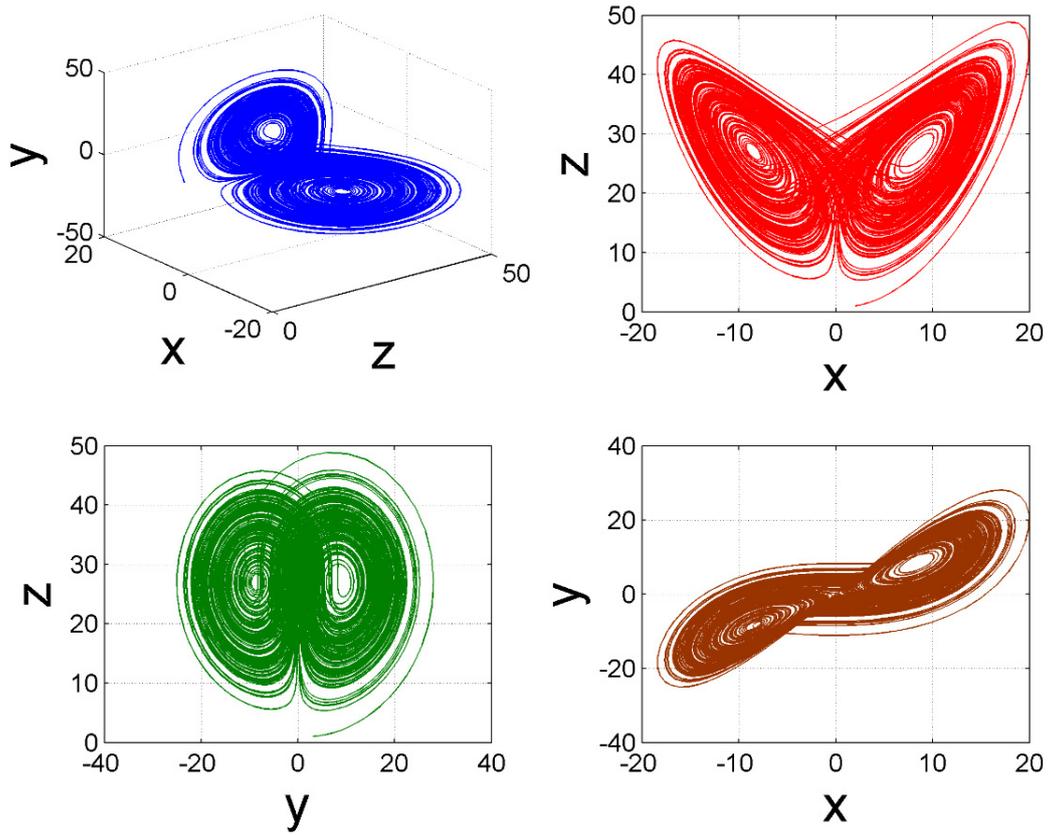

*Figure 3: Phase space dynamics of Lorenz-XY3*

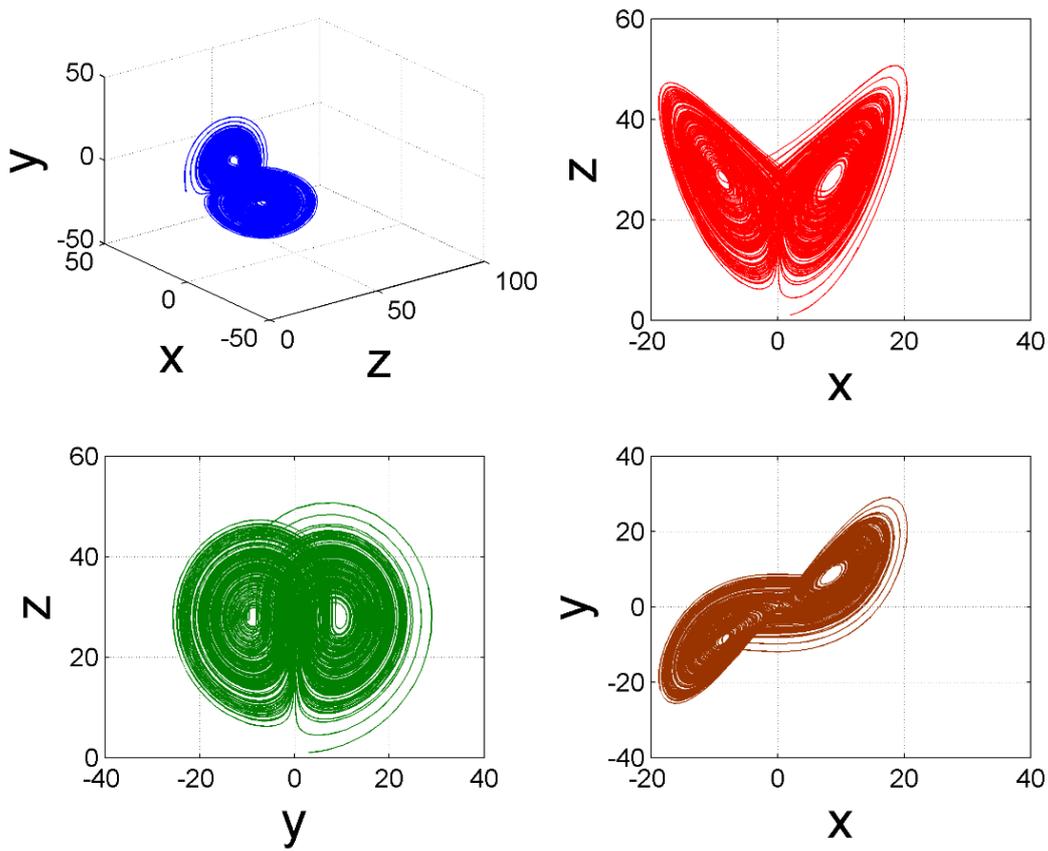

*Figure 4: Phase space dynamics of Lorenz-XY4*

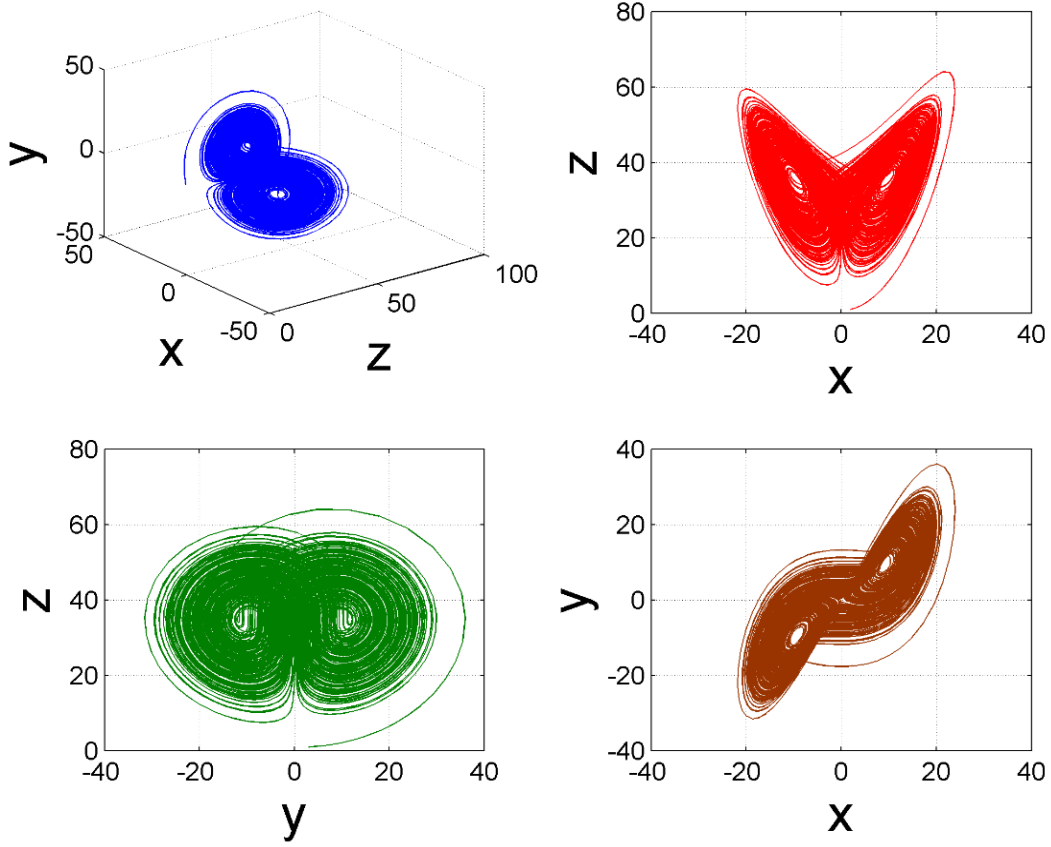

*Figure 5: Phase space dynamics of Lorenz-XY5*

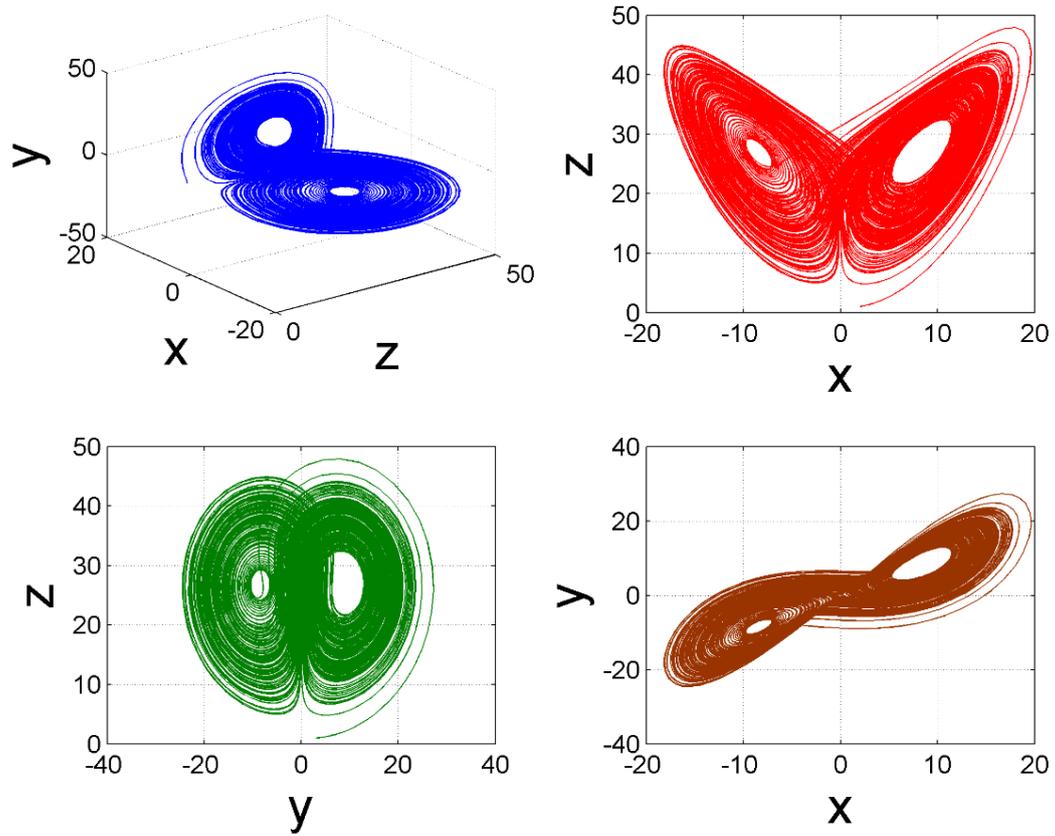

*Figure 6: Phase space dynamics of Lorenz-XY6*

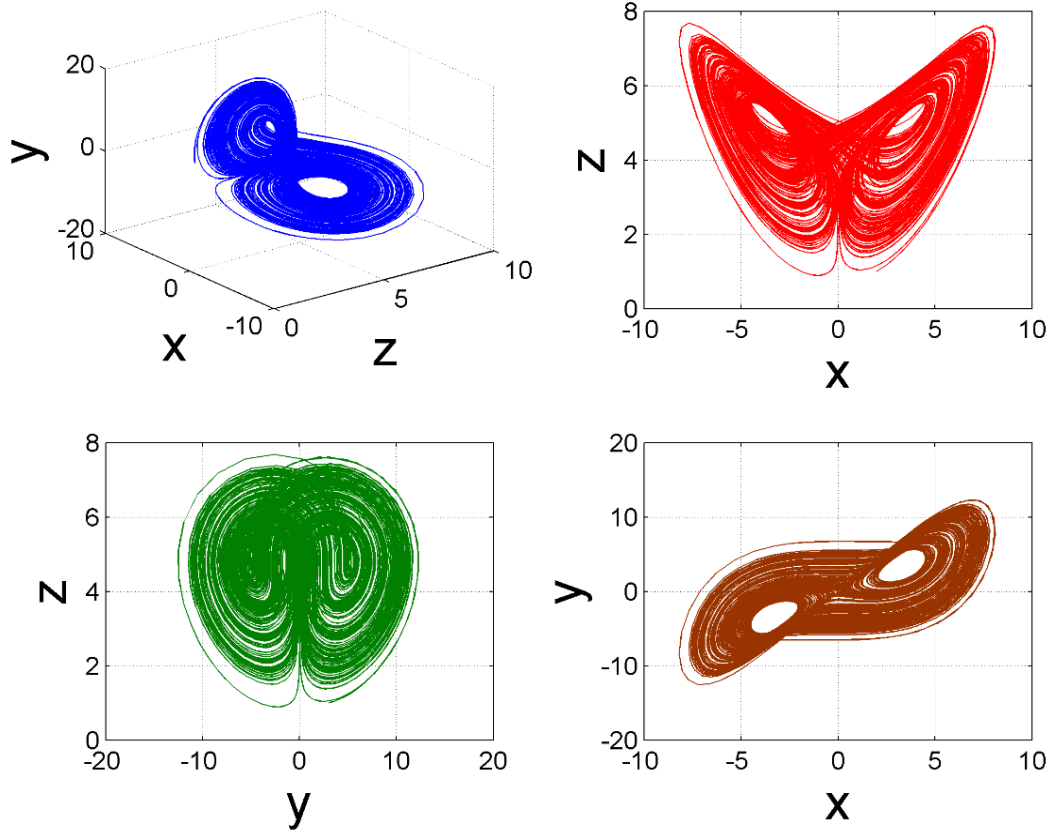

*Figure 7: Phase space dynamics of Lorenz-XY7*

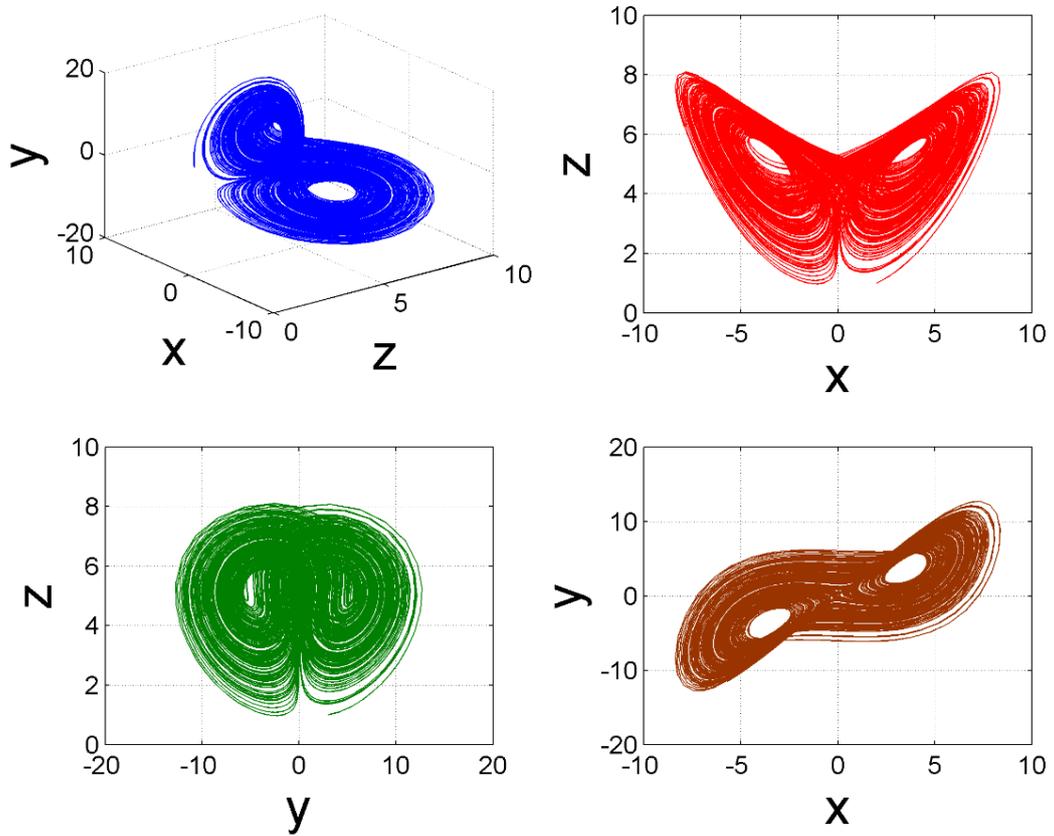

*Figure 8: Phase space dynamics of Lorenz-XY8*

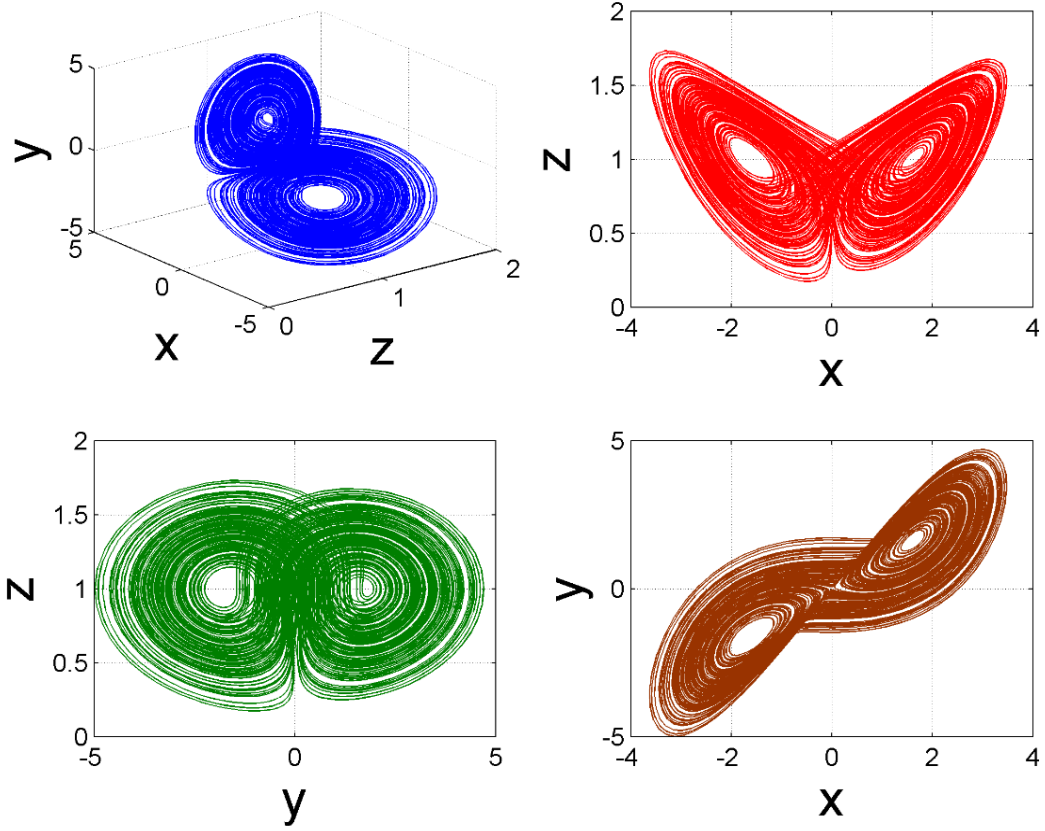

*Figure 9: Phase space dynamics of Lorenz-XY9*

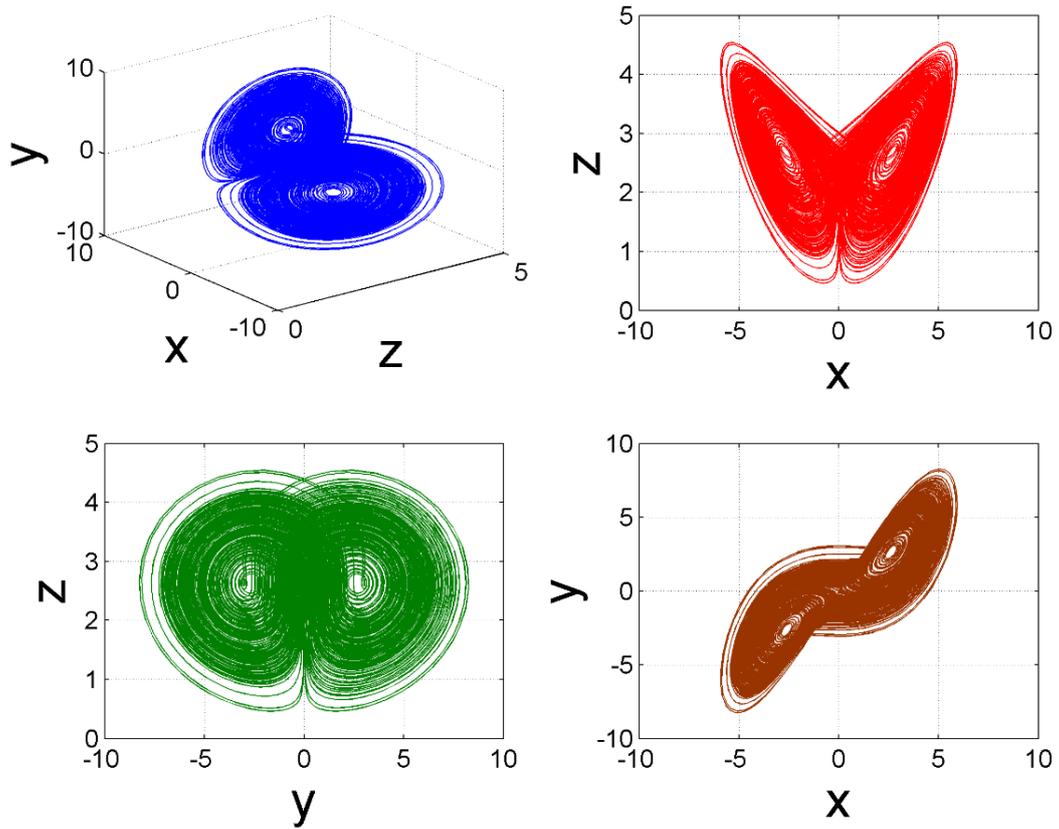

*Figure 10: Phase space dynamics of Lorenz-XY10*

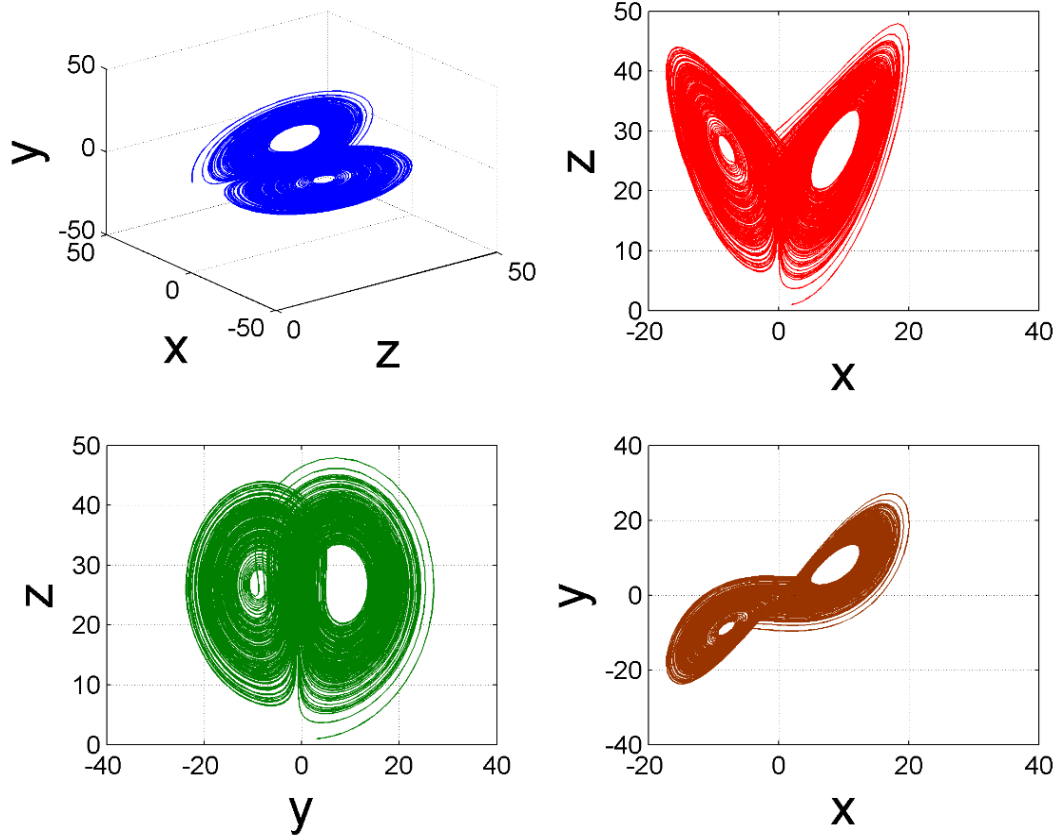

*Figure 11: Phase space dynamics of Lorenz-XY11*

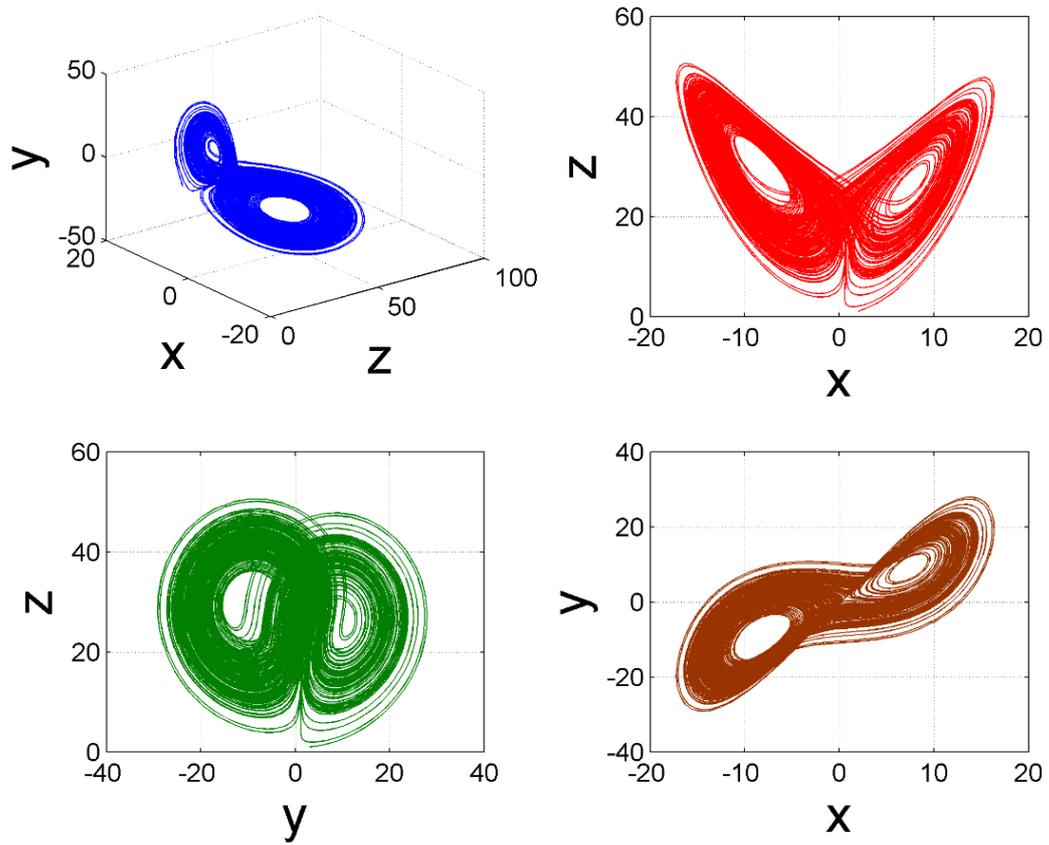

*Figure 12: Phase space dynamics of Lorenz-XY12*

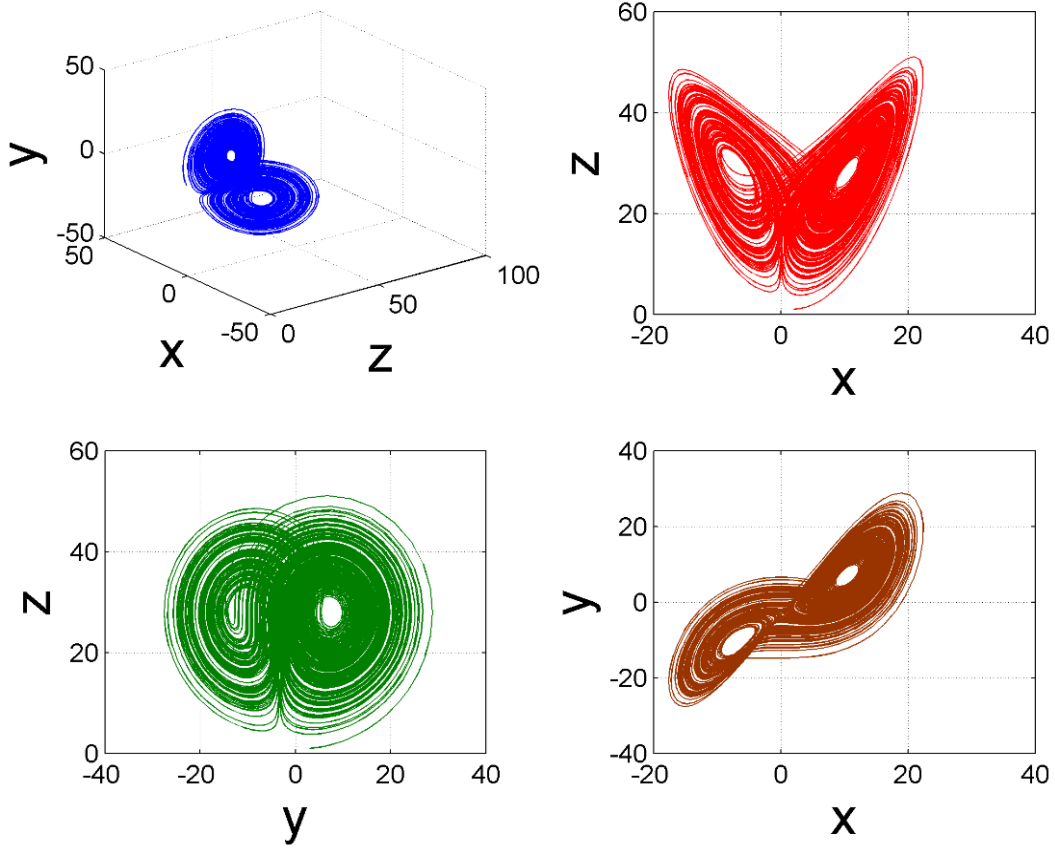

*Figure 13: Phase space dynamics of Lorenz-XY13*

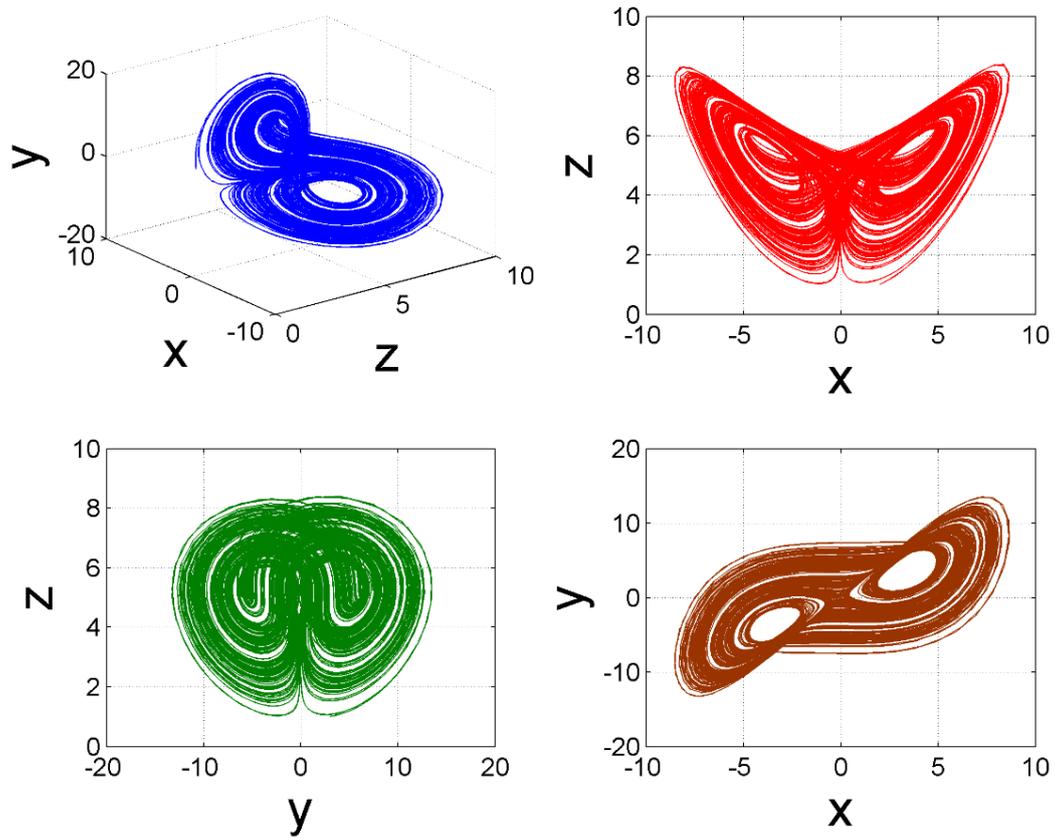

*Figure 14: Phase space dynamics of Lorenz-XY14*

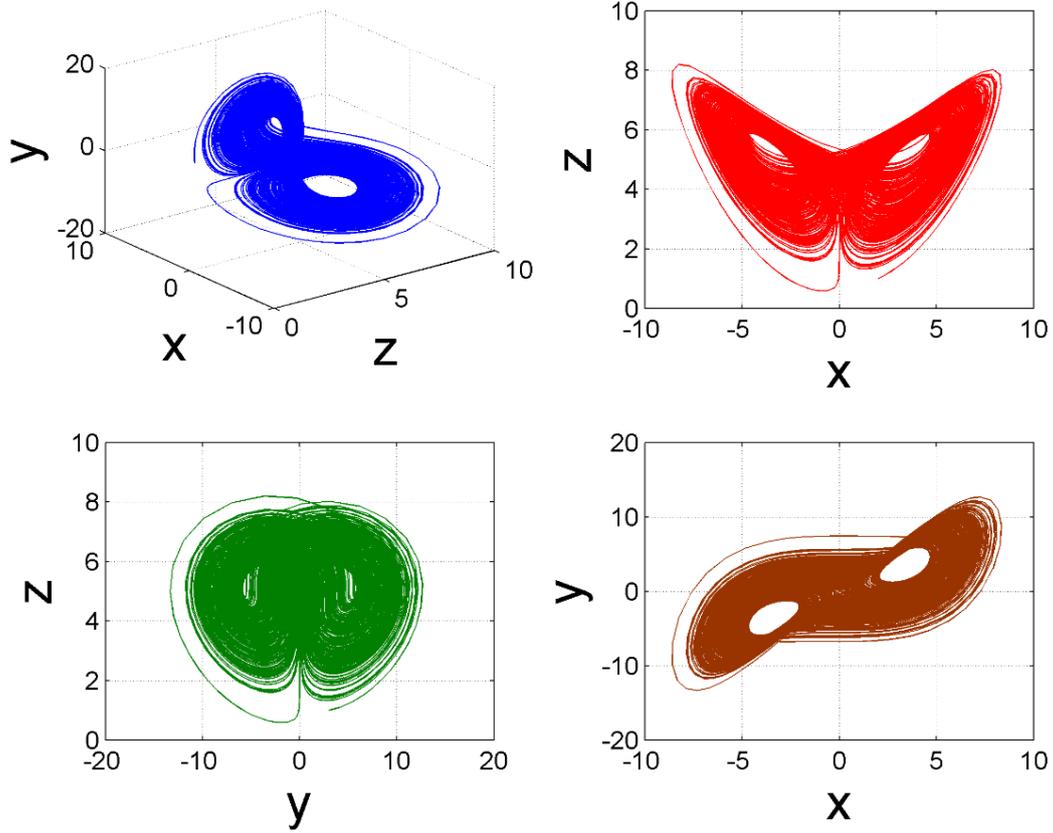

*Figure 15: Phase space dynamics of Lorenz-XY16*

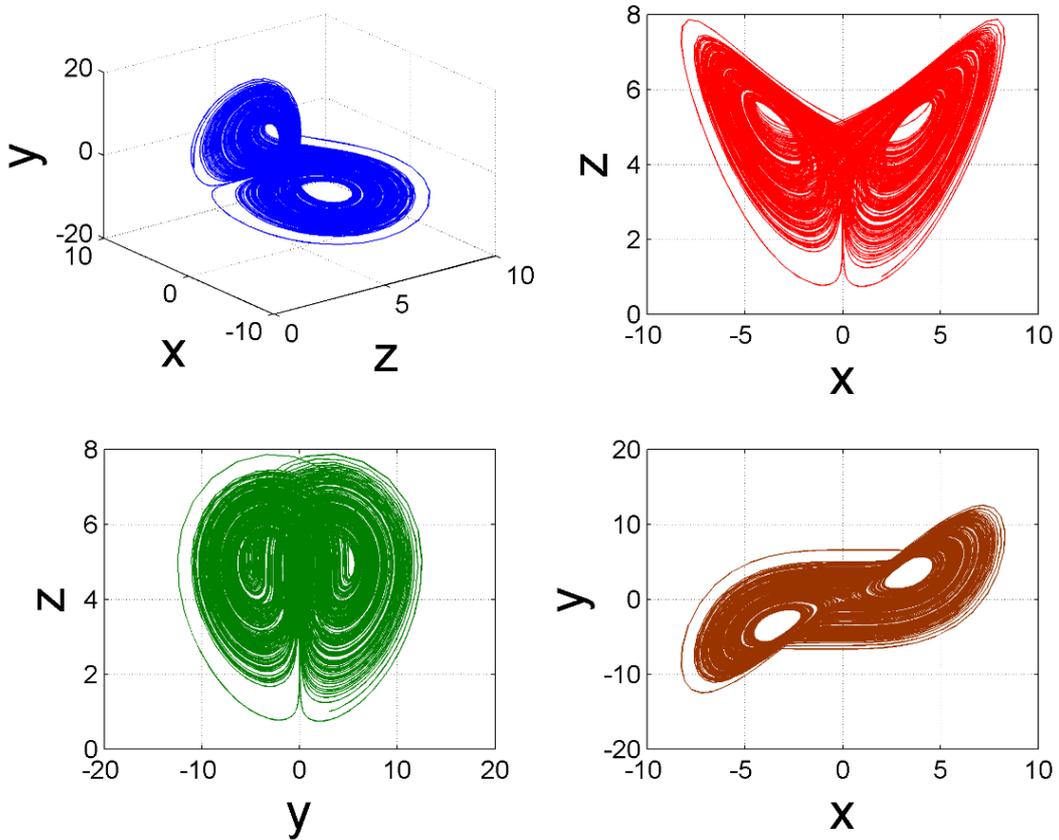

*Figure 16: Phase space dynamics of Lorenz-XY17*

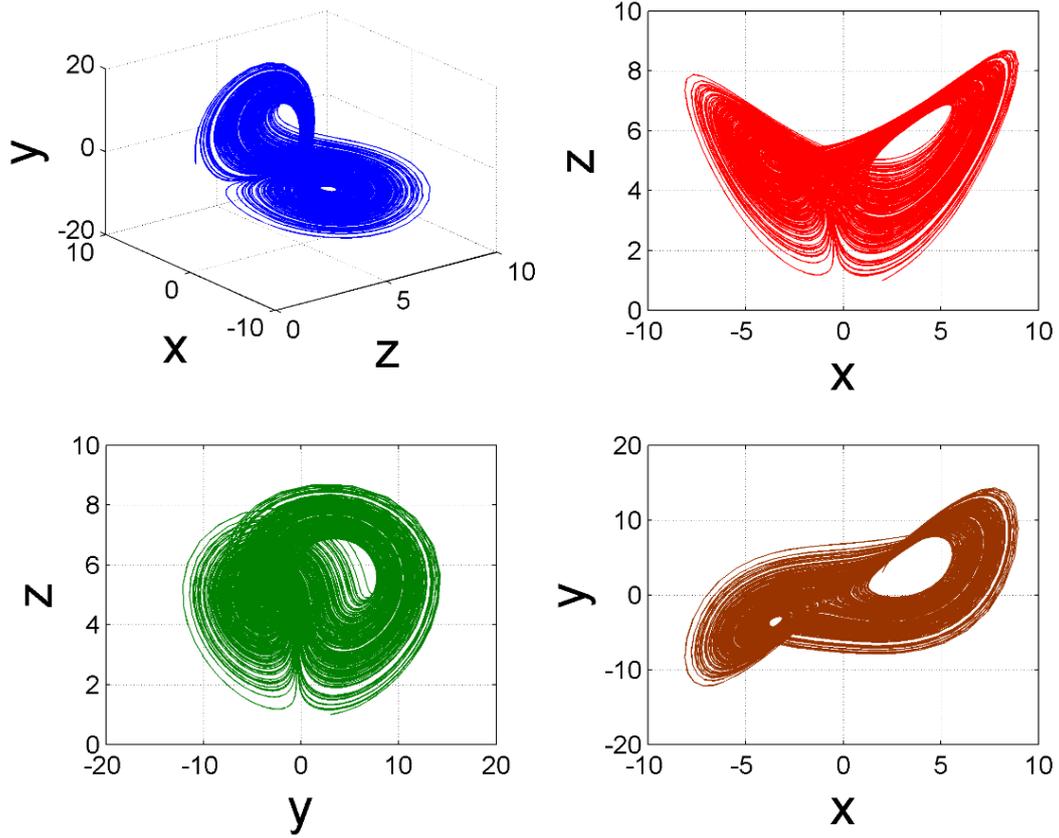

*Figure 17: Phase space dynamics of Lorenz-XY19*

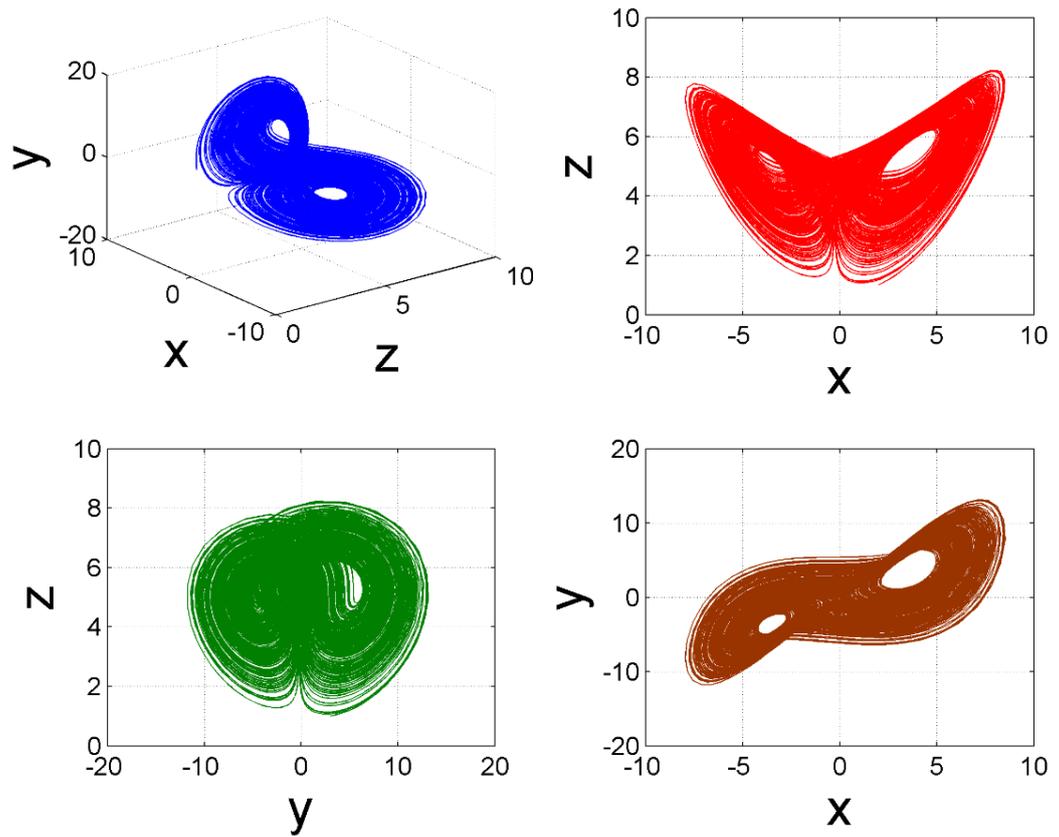

*Figure 18: Phase space dynamics of Lorenz-XY20*

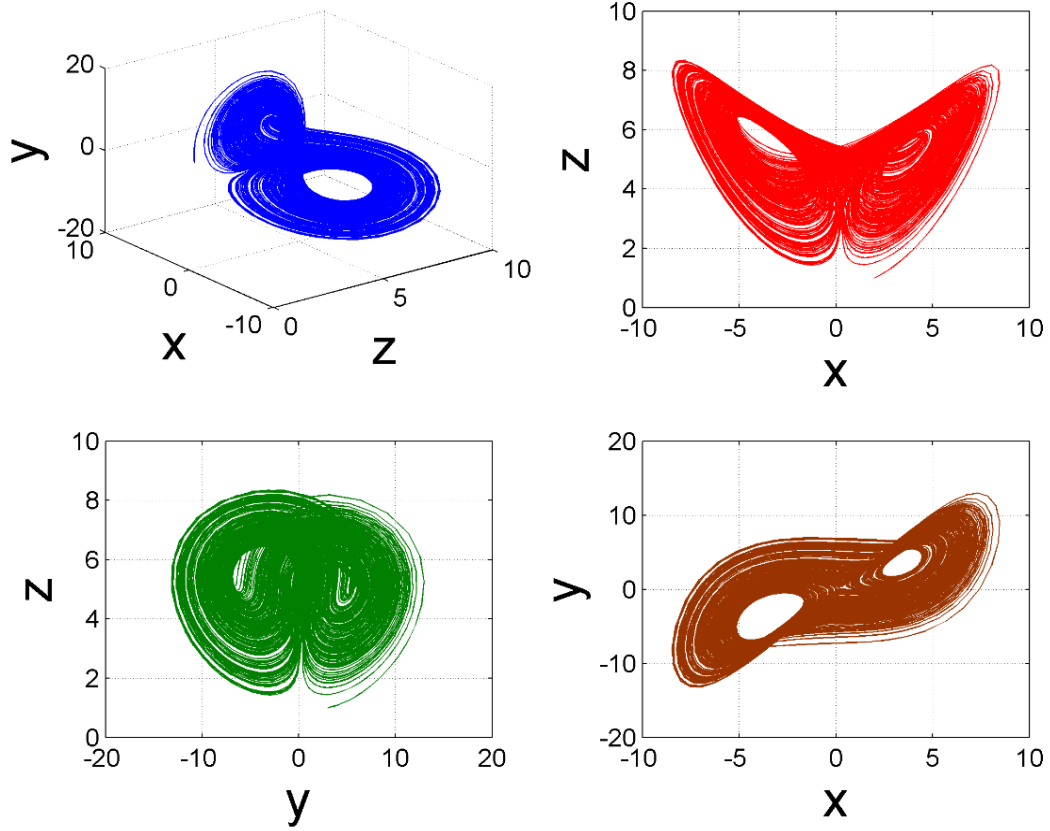

*Figure 19: Phase space dynamics of Lorenz-XY22*

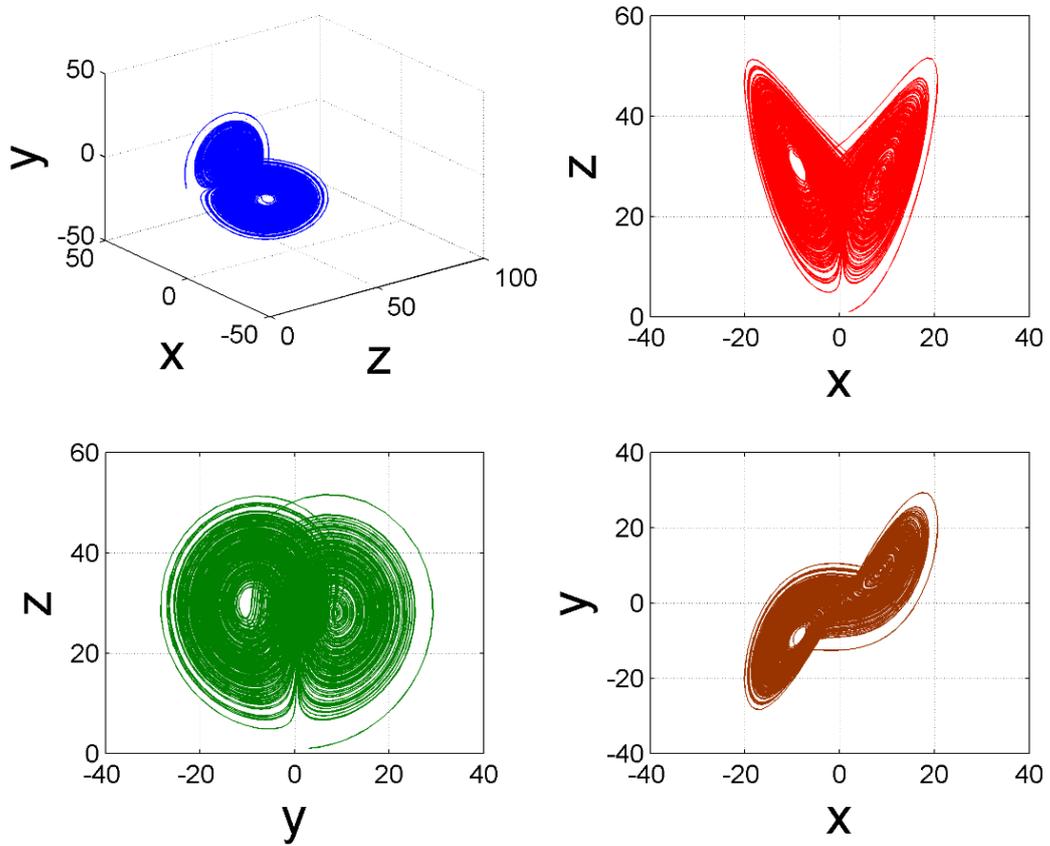

*Figure 20: Phase space dynamics of Lorenz-XY23*

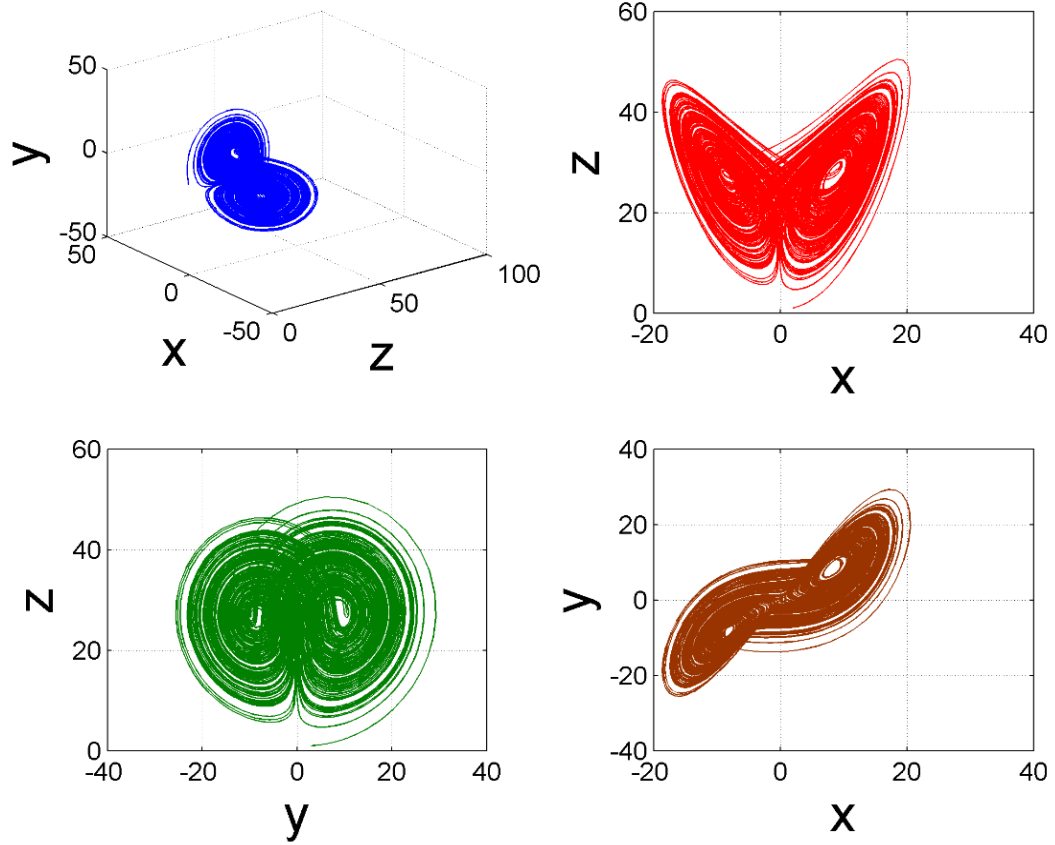

*Figure 21: Phase space dynamics of Lorenz-XY24*

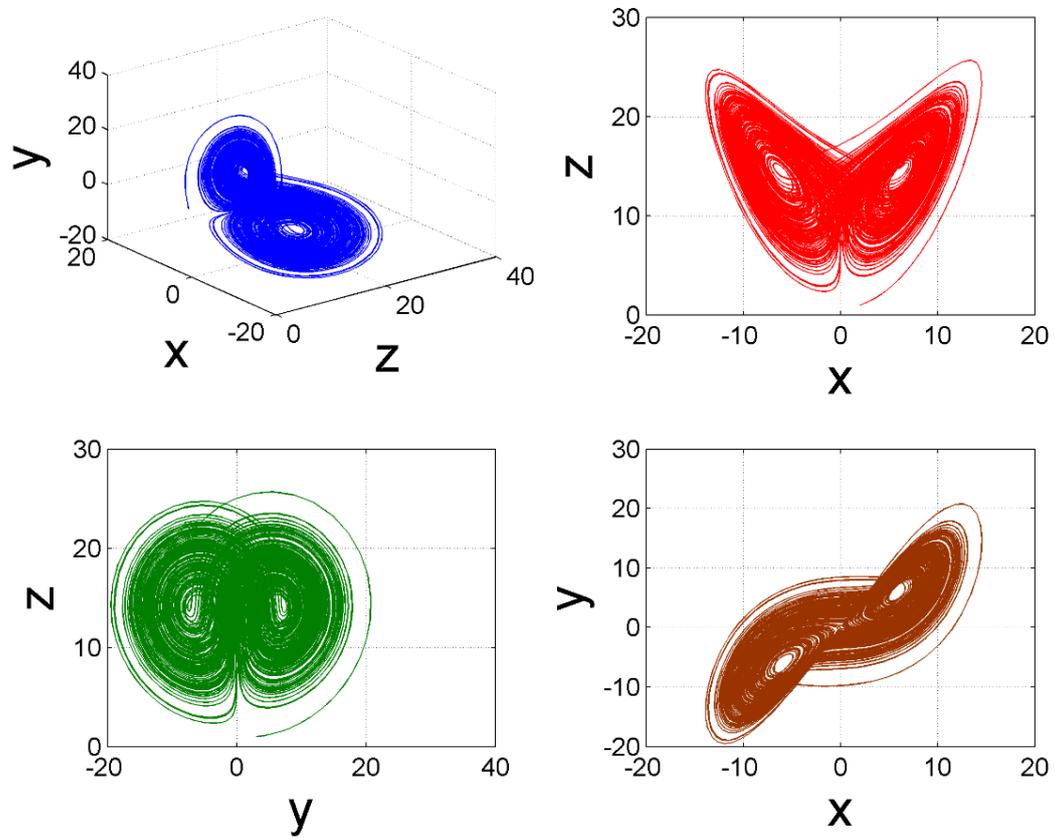

*Figure 22: Phase space dynamics of Lorenz-XY25*

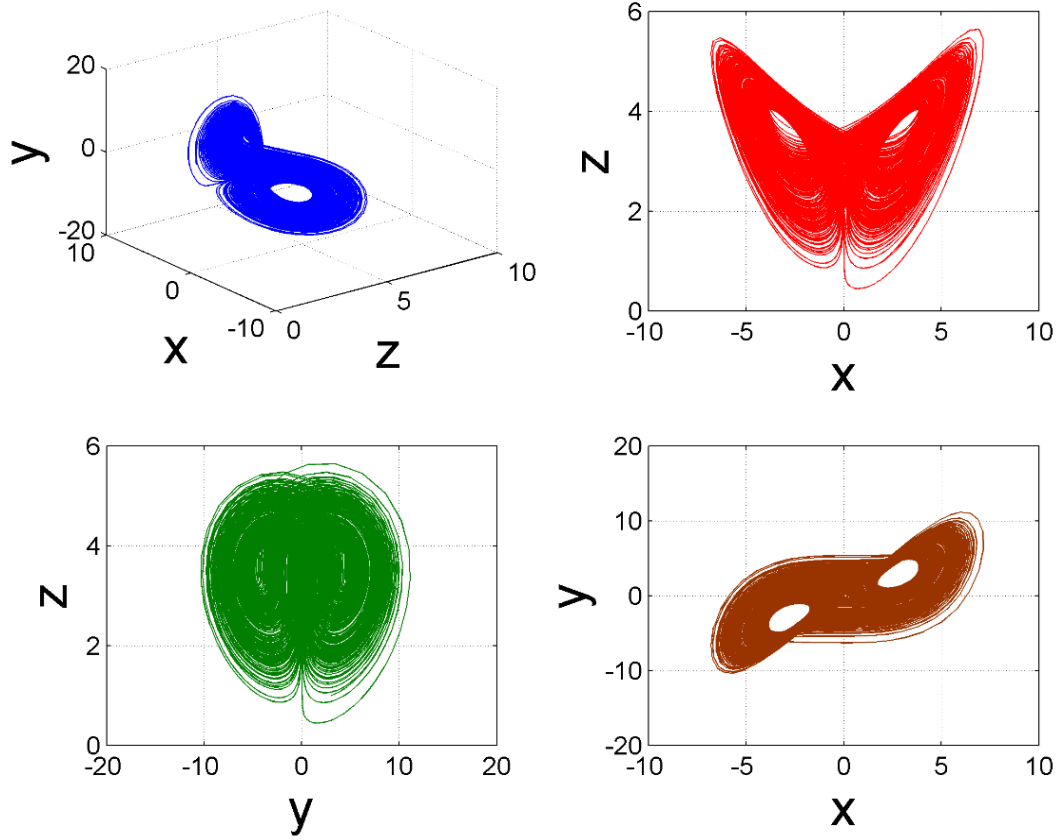

*Figure 23: Phase space dynamics of Lorenz-XY26*

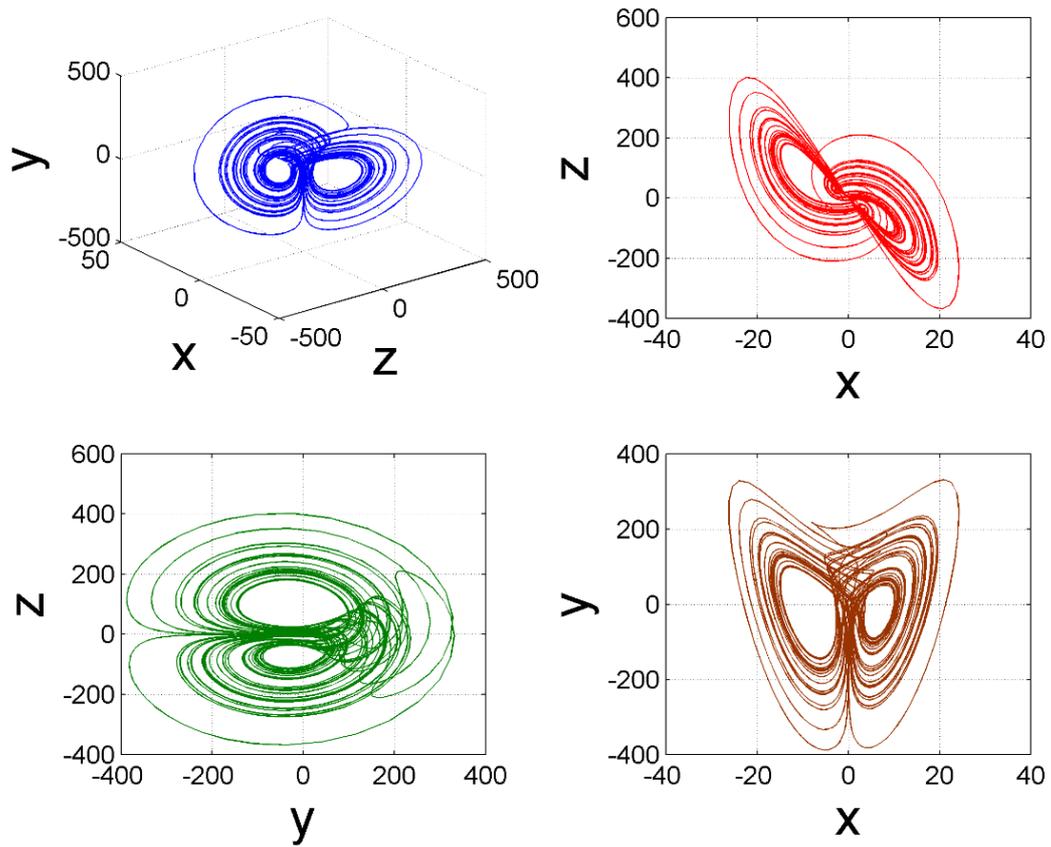

*Figure 24: Phase space dynamics of Lorenz-XY27*

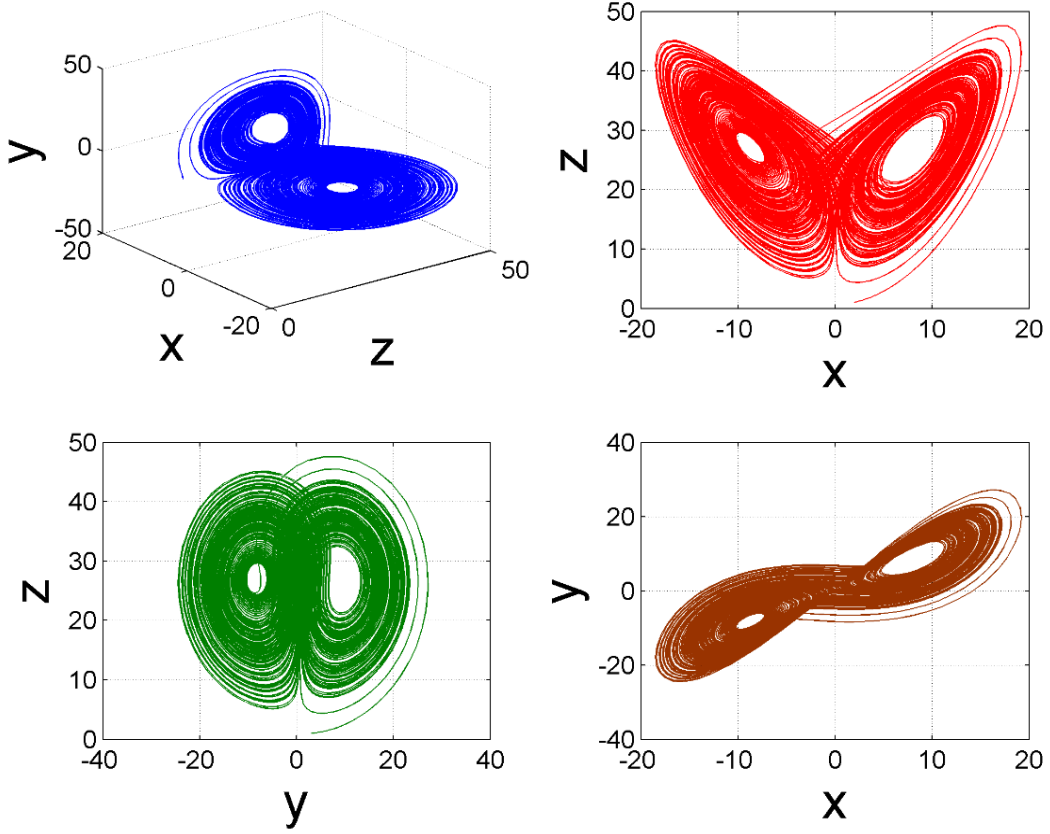

*Figure 25: Phase space dynamics of Lorenz-XY28*

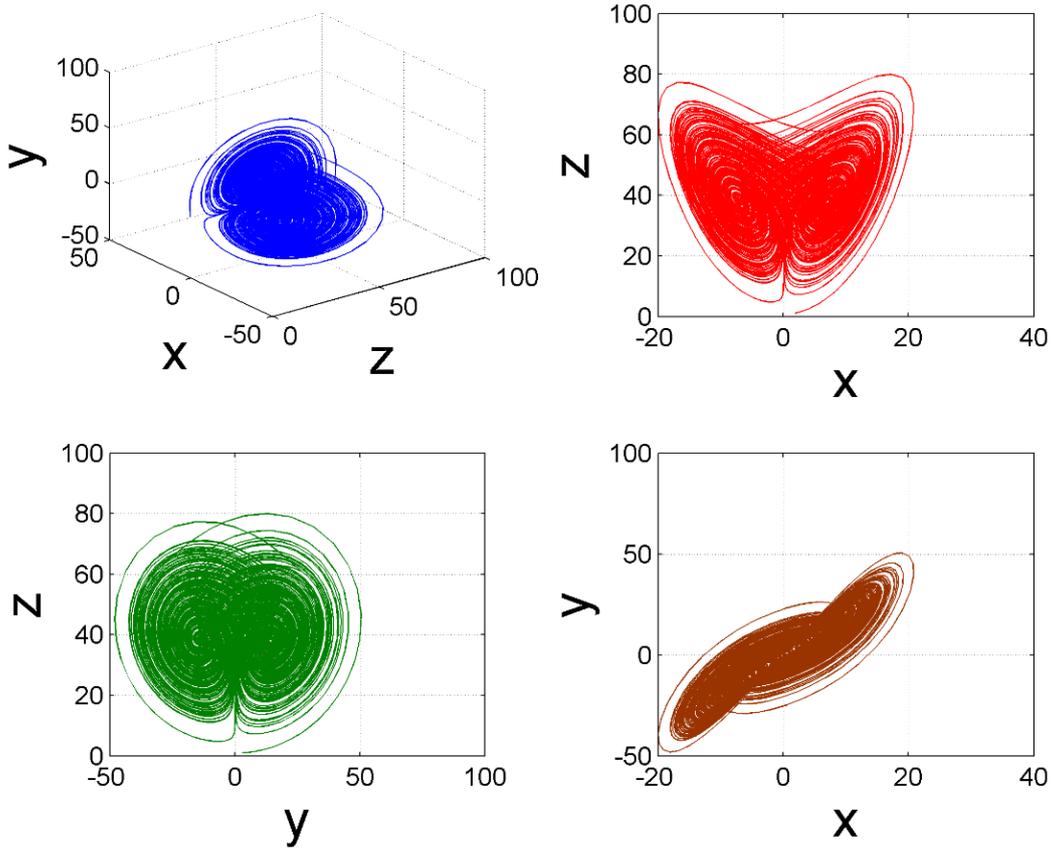

*Figure 26: Phase space dynamics of Lorenz-XY29*

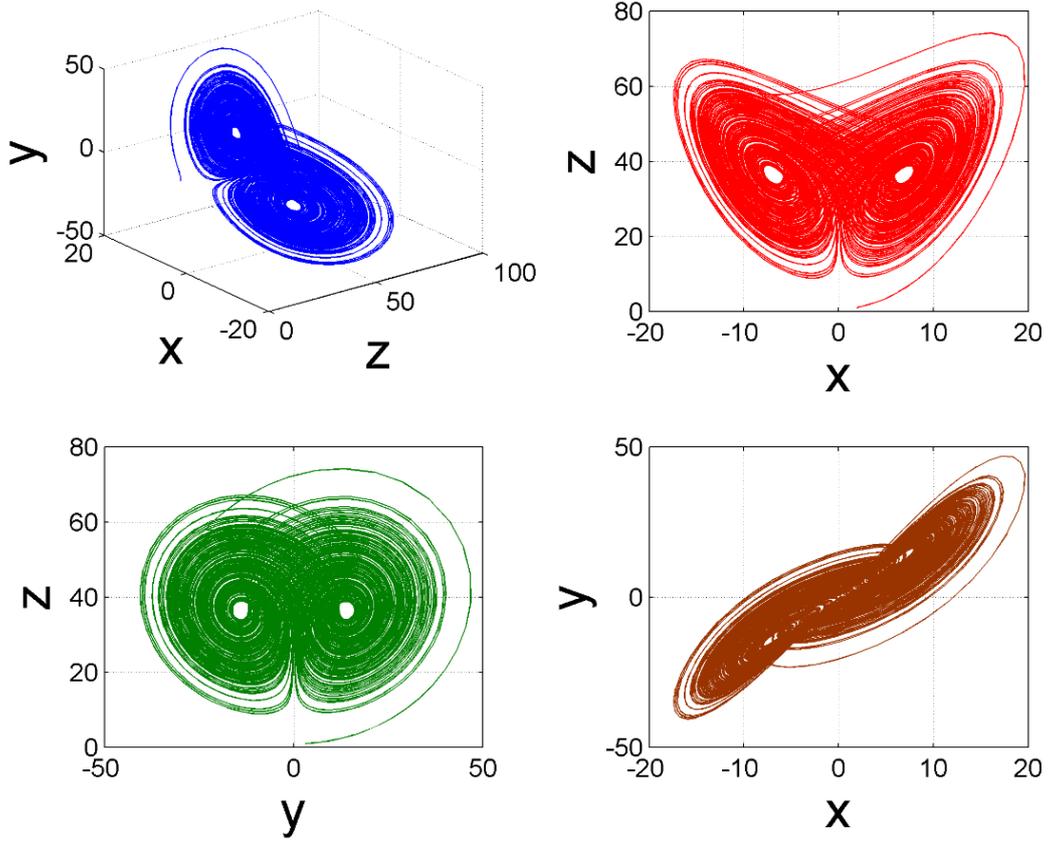

*Figure 27: Phase space dynamics of Lorenz-XY30*

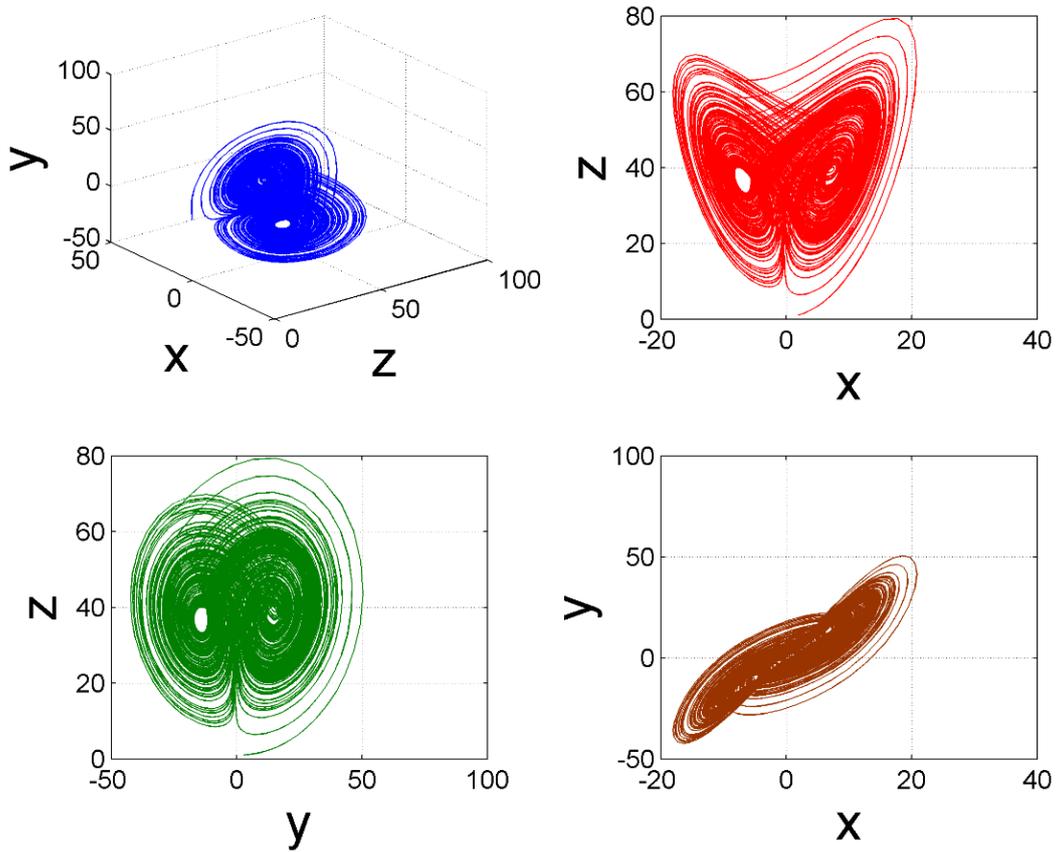

*Figure 28: Phase space dynamics of Lorenz-XY31*

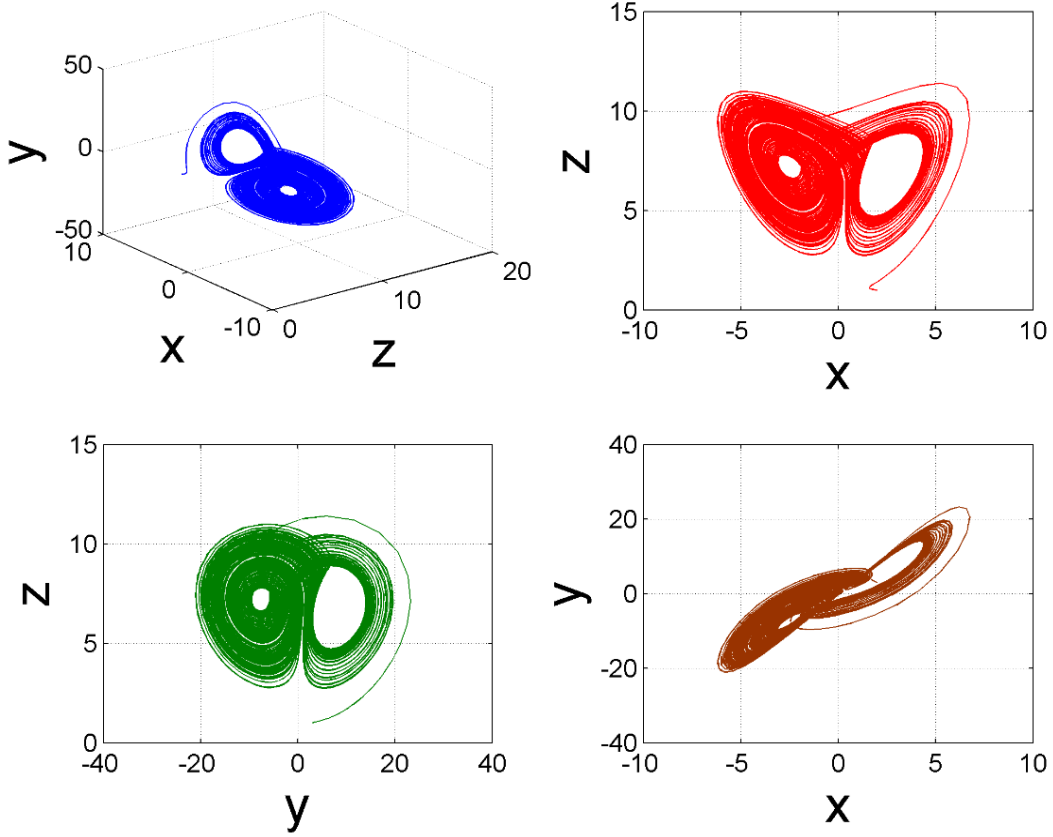

*Figure 29: Phase space dynamics of Lorenz-XY32*

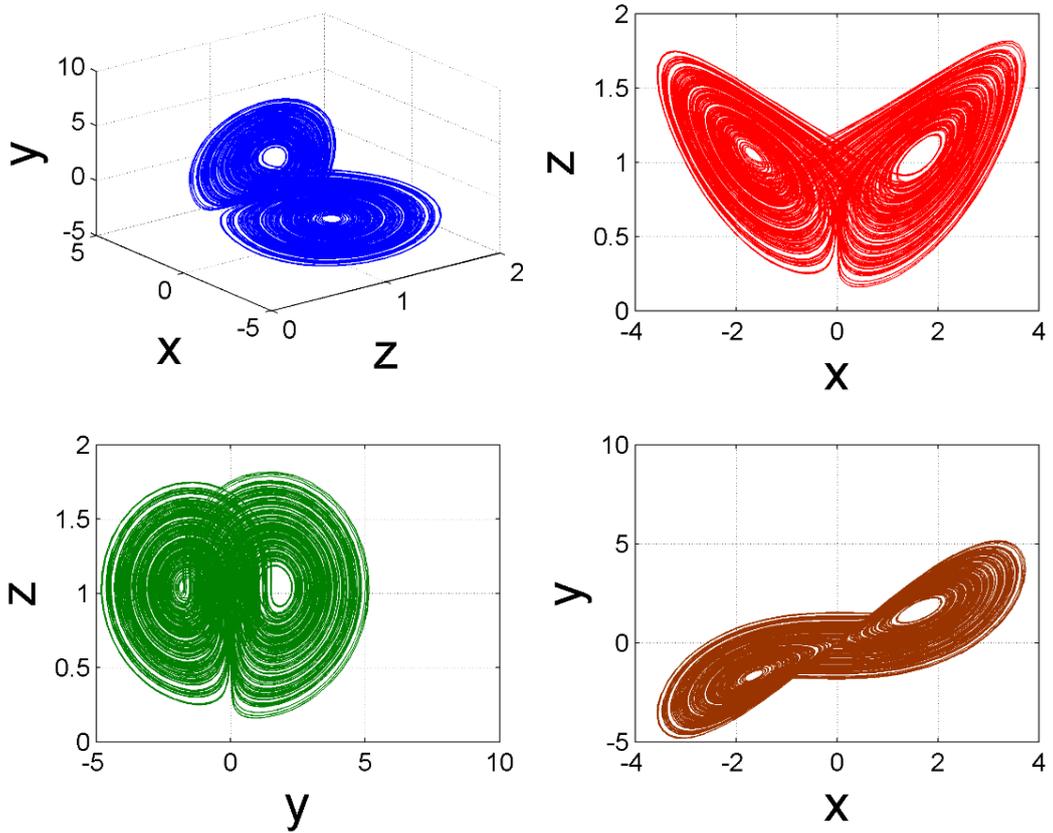

*Figure 30: Phase space dynamics of Lorenz-XY33*

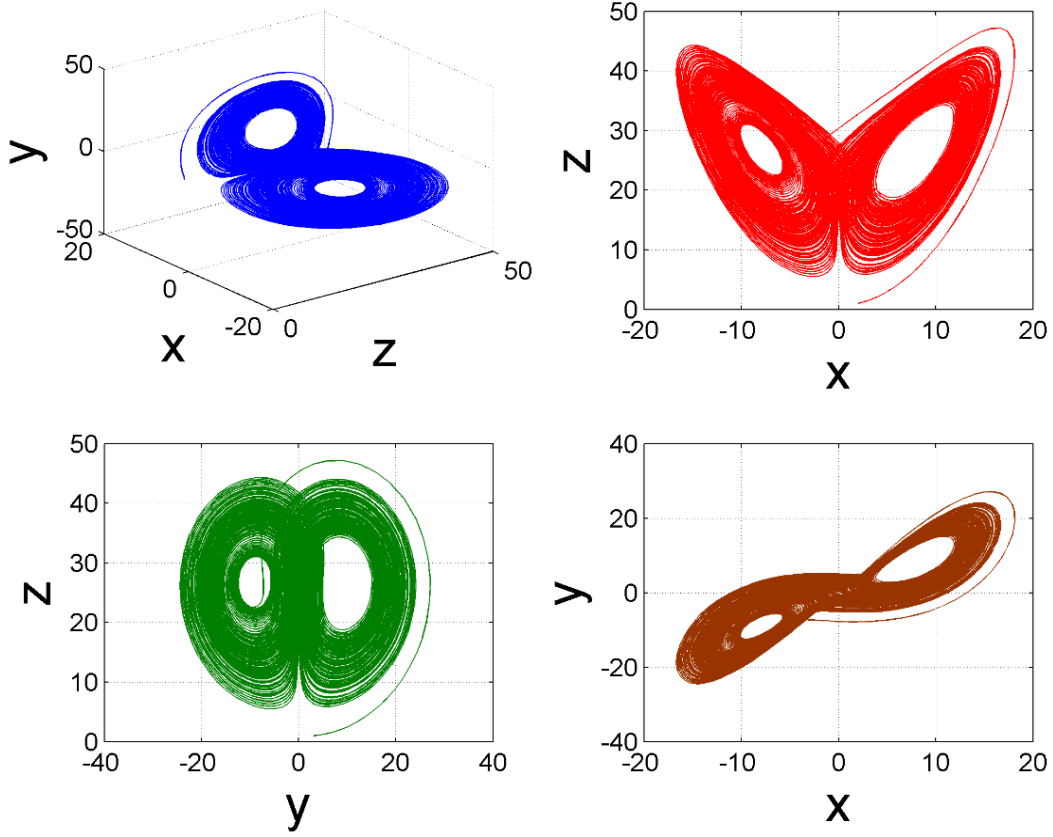

*Figure 31: Phase space dynamics of Lorenz-XY34*

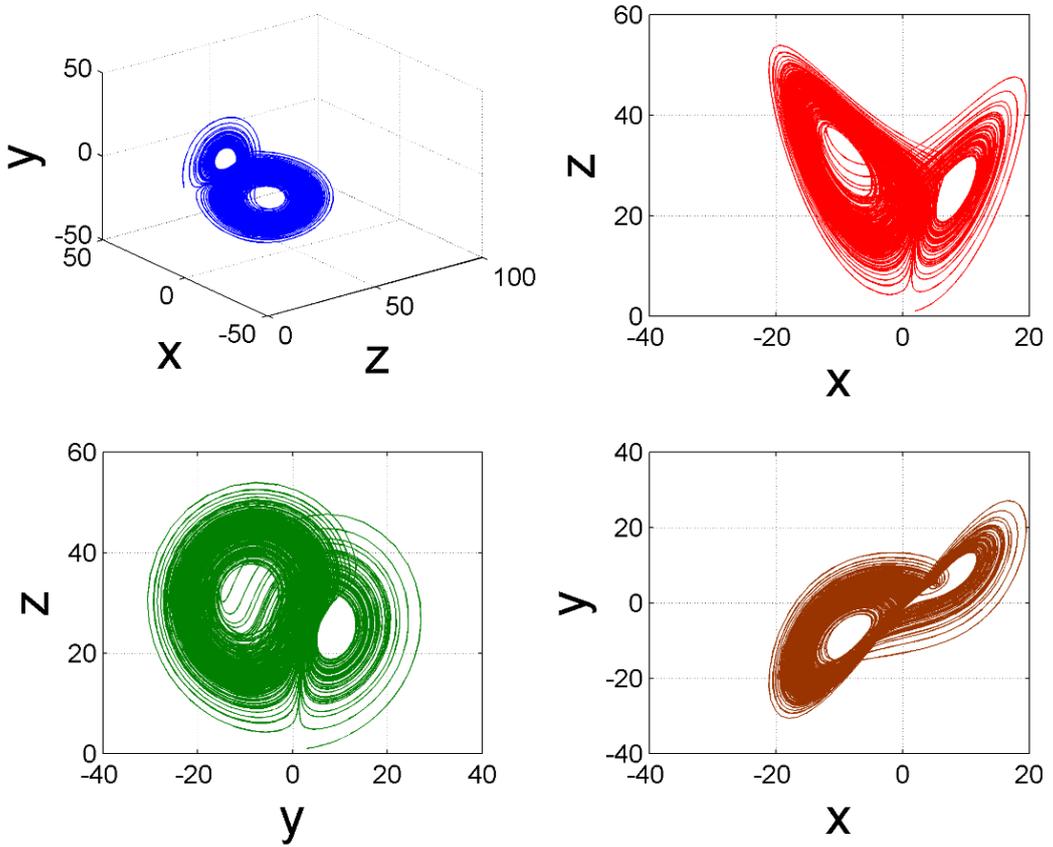

*Figure 32: Phase space dynamics of Lorenz-XY35*

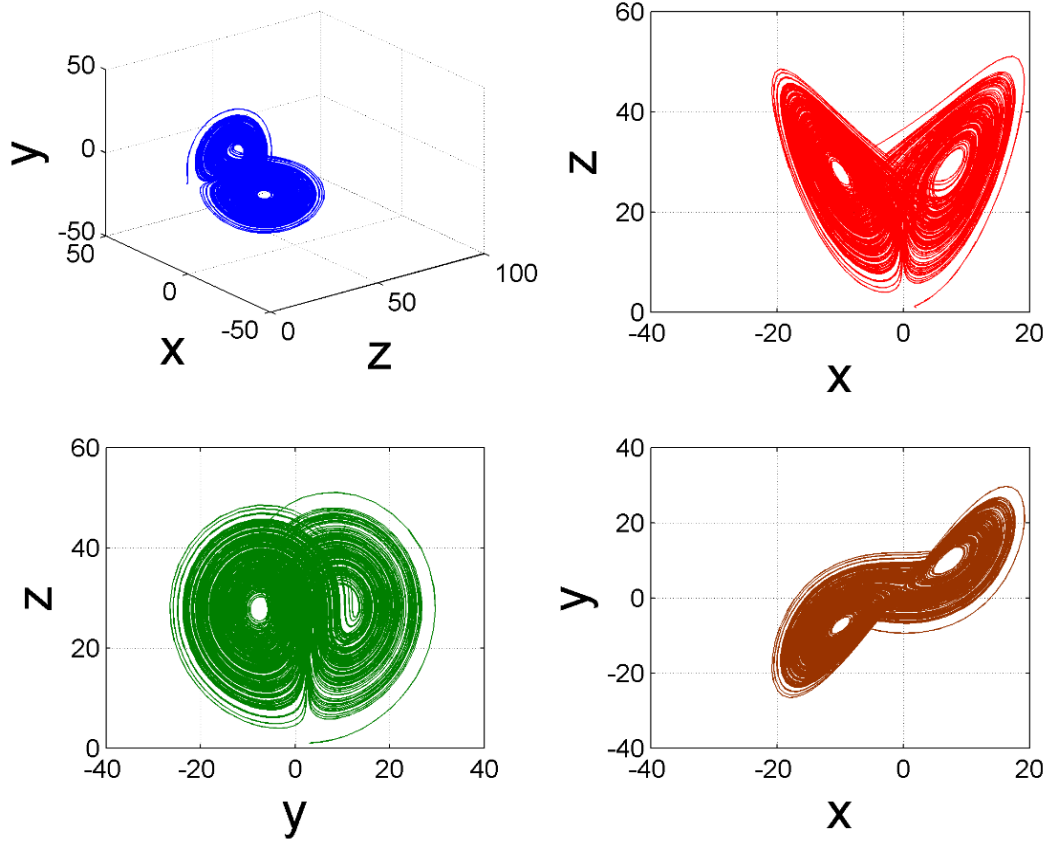

*Figure 33: Phase space dynamics of Lorenz-XY36*

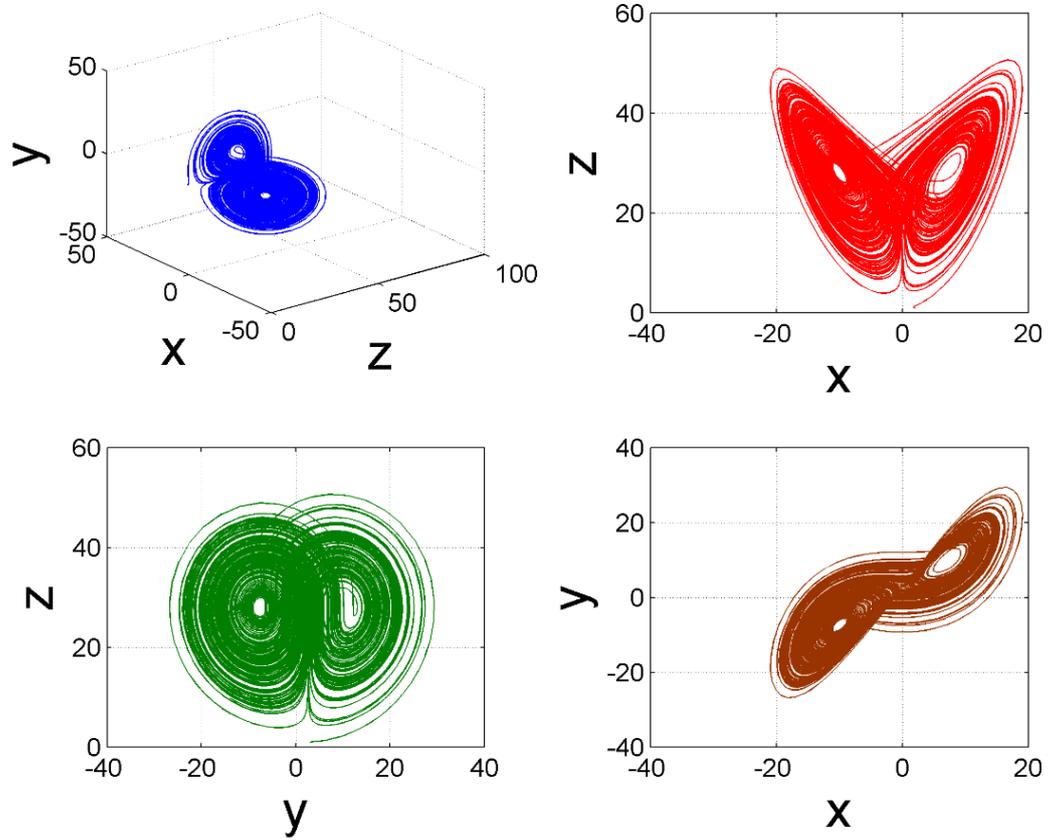

*Figure 34: Phase space dynamics of Lorenz-XY37*

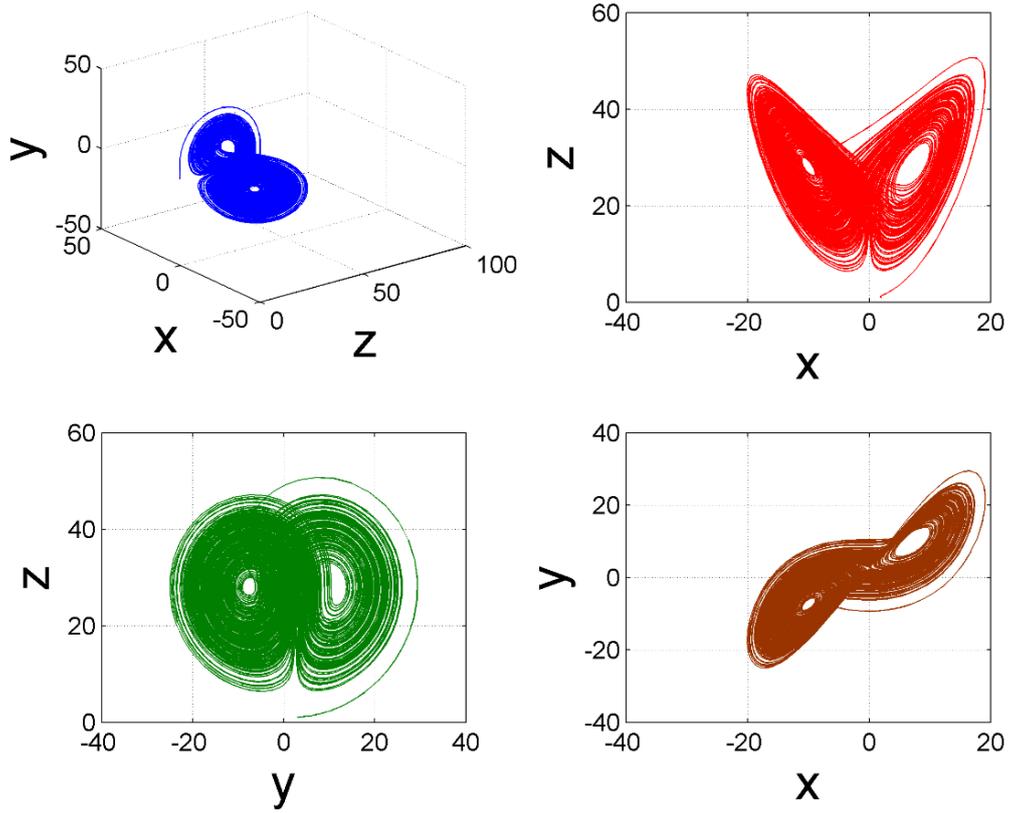

*Figure 35: Phase space dynamics of Lorenz-XY38*

## 2. Phase portraits of generalized Lorenz-YZ family of attractors

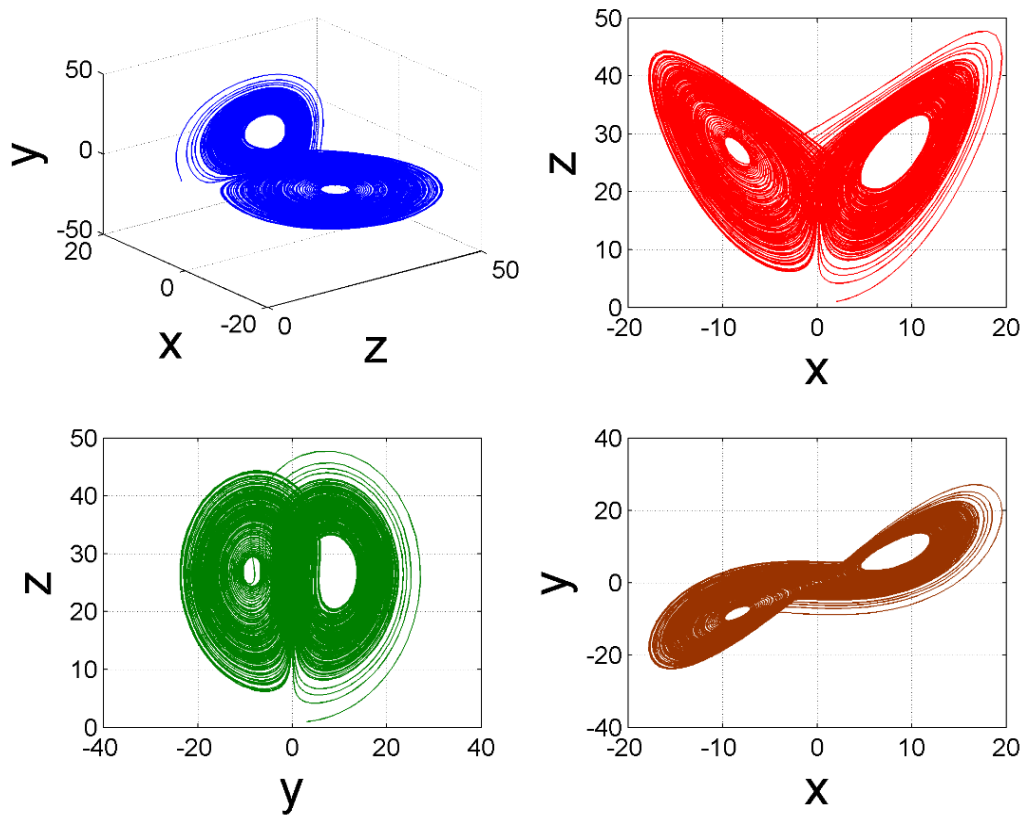

*Figure 36: Phase space dynamics of Lorenz-YZ1*

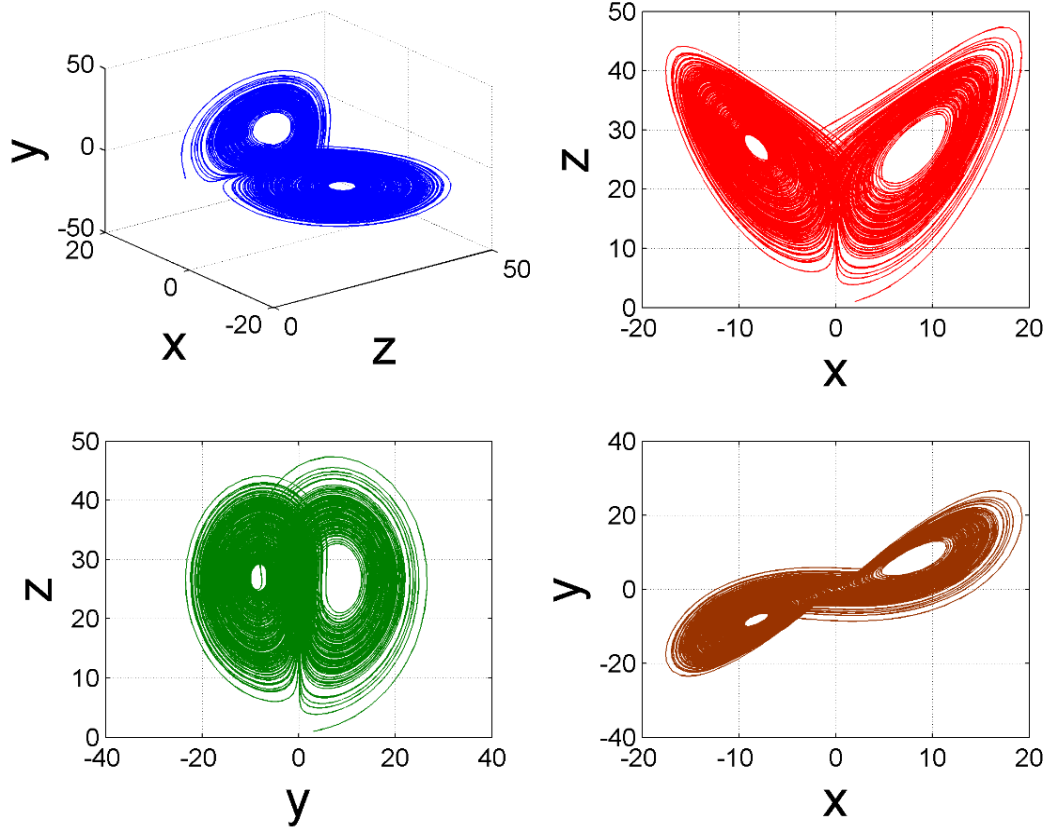

*Figure 37: Phase space dynamics of Lorenz-YZ2*

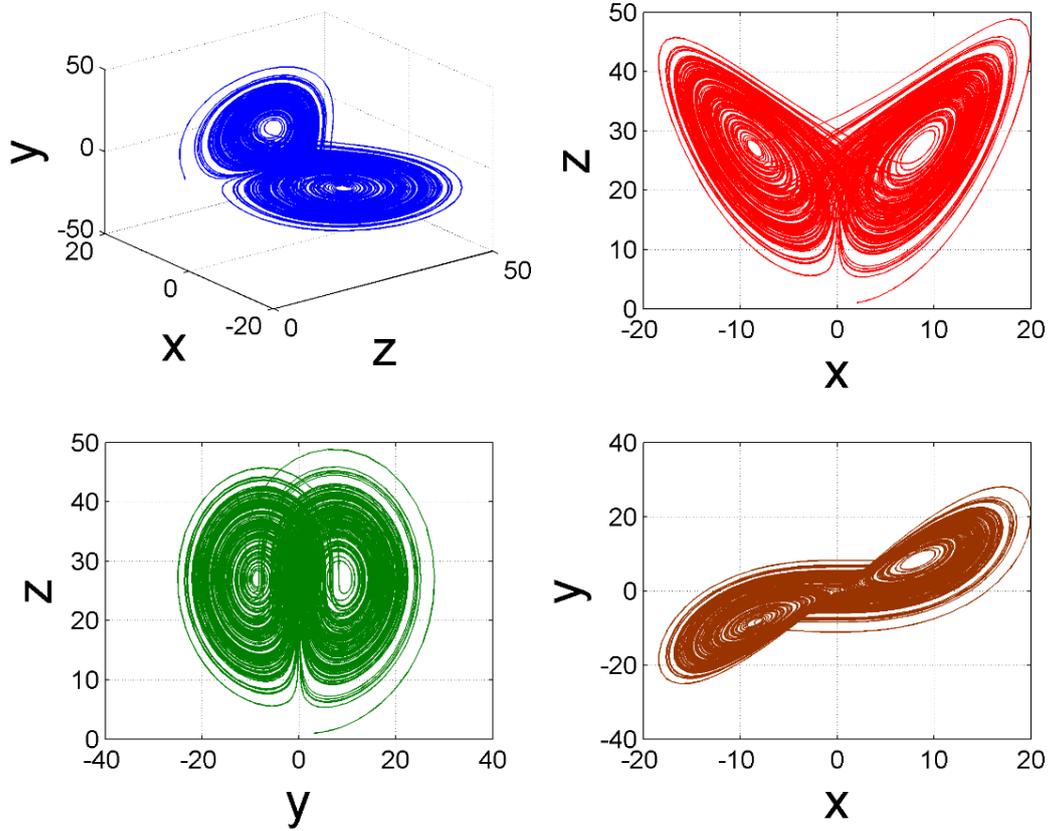

*Figure 38: Phase space dynamics of Lorenz-YZ3*

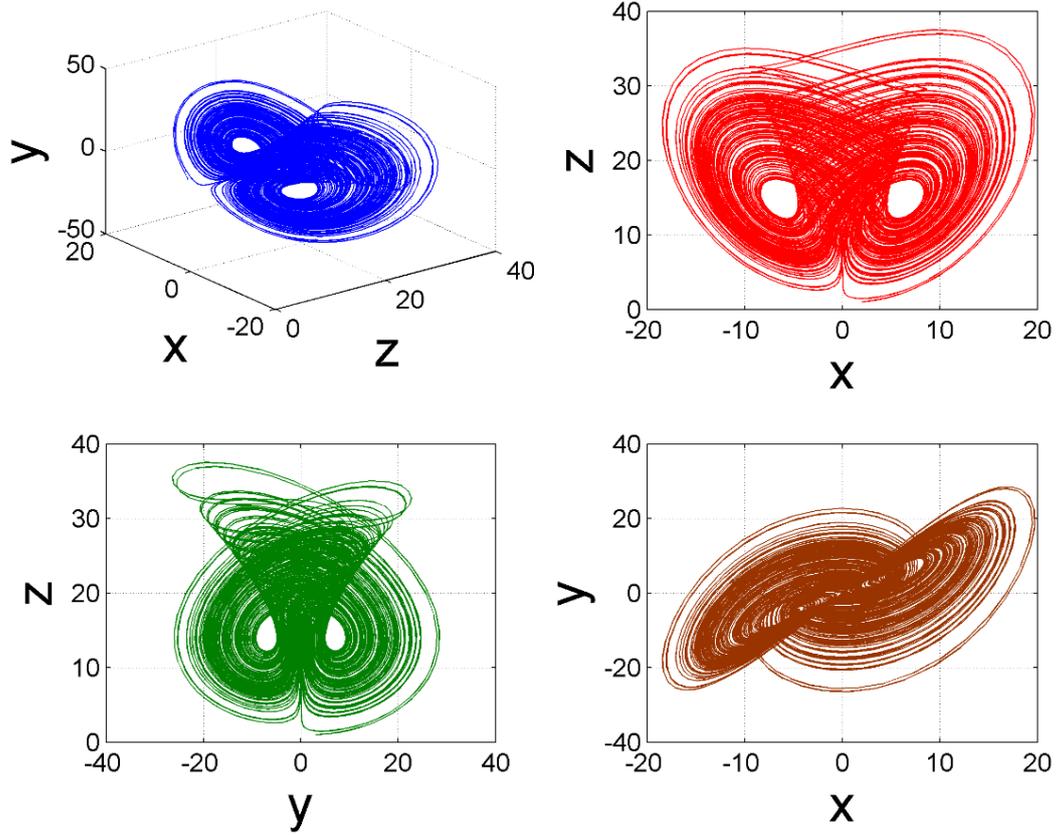

*Figure 39: Phase space dynamics of Lorenz-YZ4*

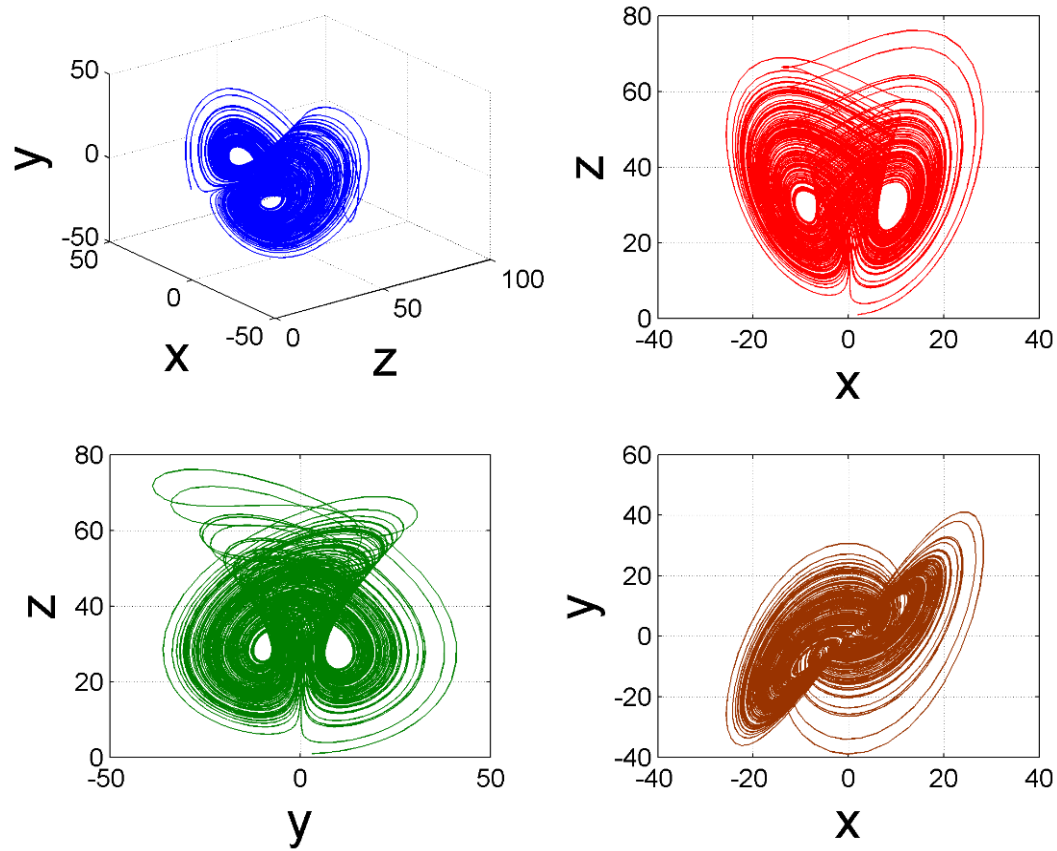

*Figure 40: Phase space dynamics of Lorenz-YZ5*

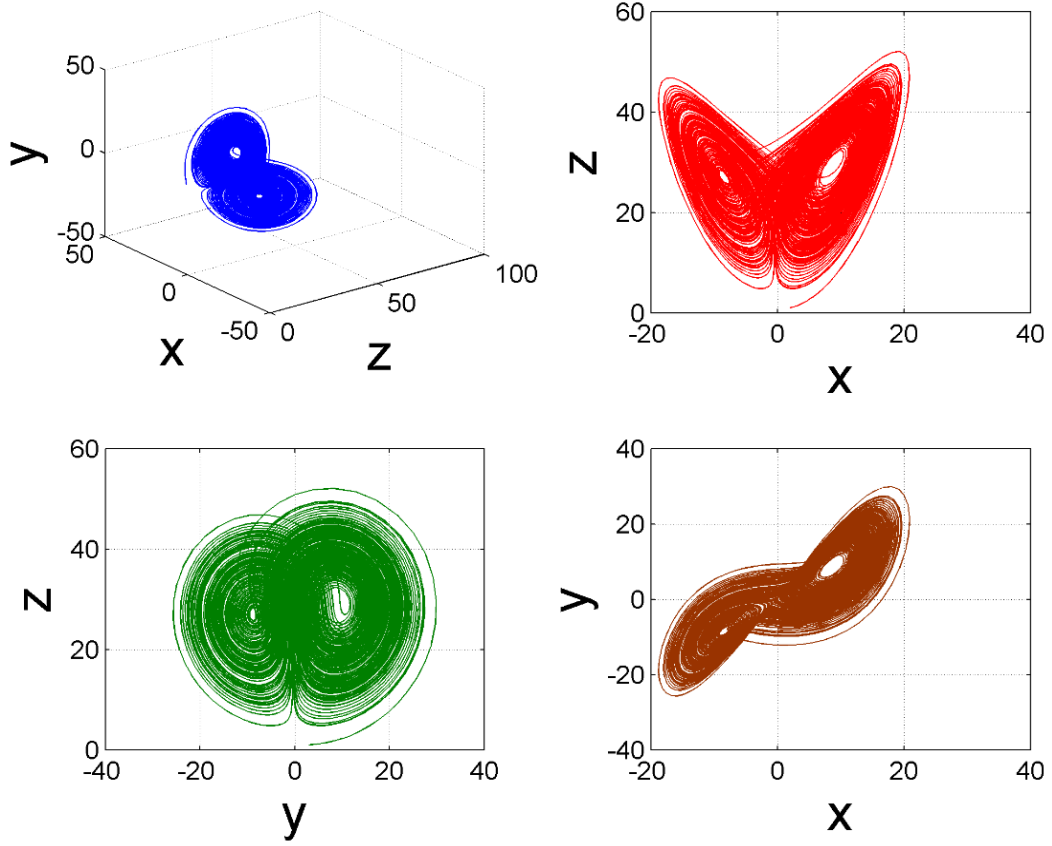

*Figure 41: Phase space dynamics of Lorenz-YZ6*

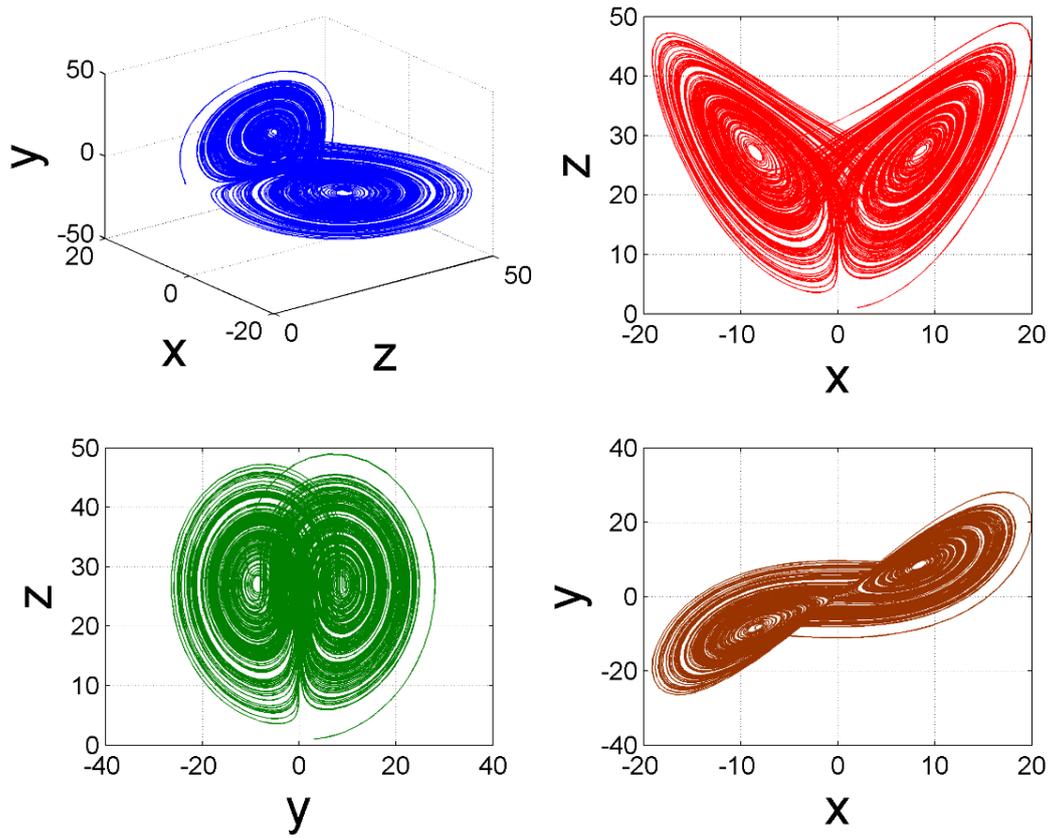

*Figure 42: Phase space dynamics of Lorenz-YZ7*

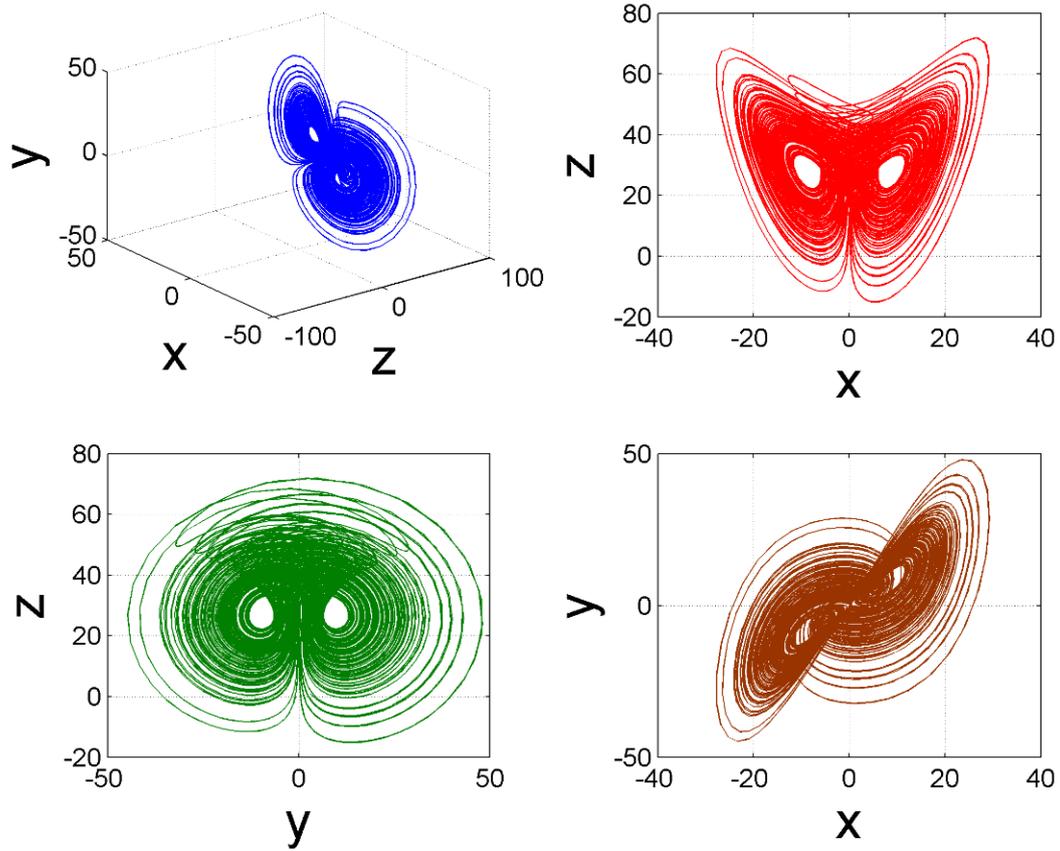

*Figure 43: Phase space dynamics of Lorenz-YZ8*

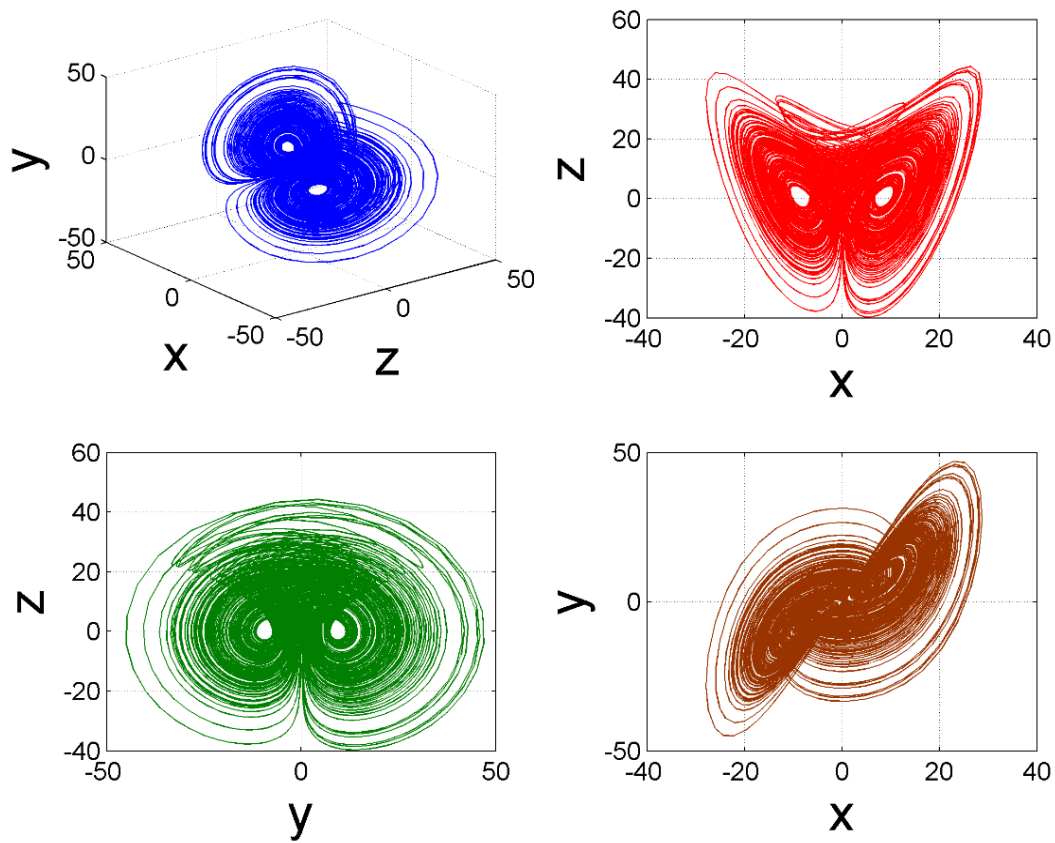

*Figure 44: Phase space dynamics of Lorenz-YZ9*

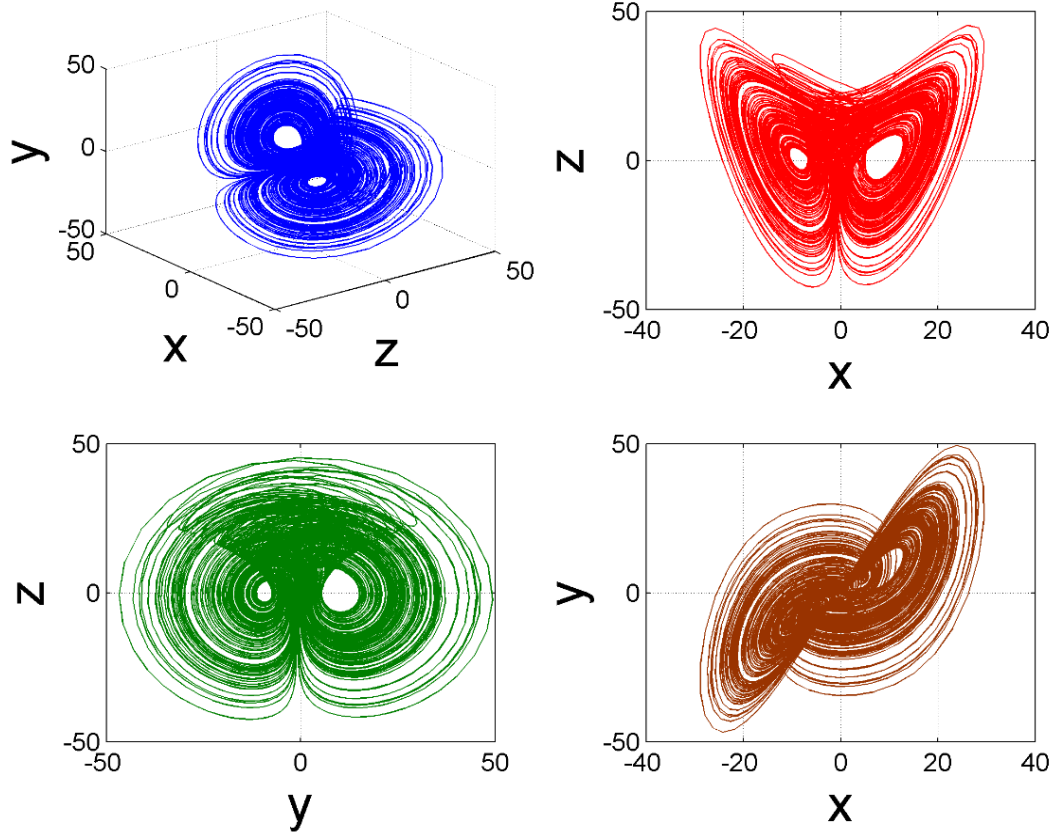

*Figure 45: Phase space dynamics of Lorenz-YZ11*

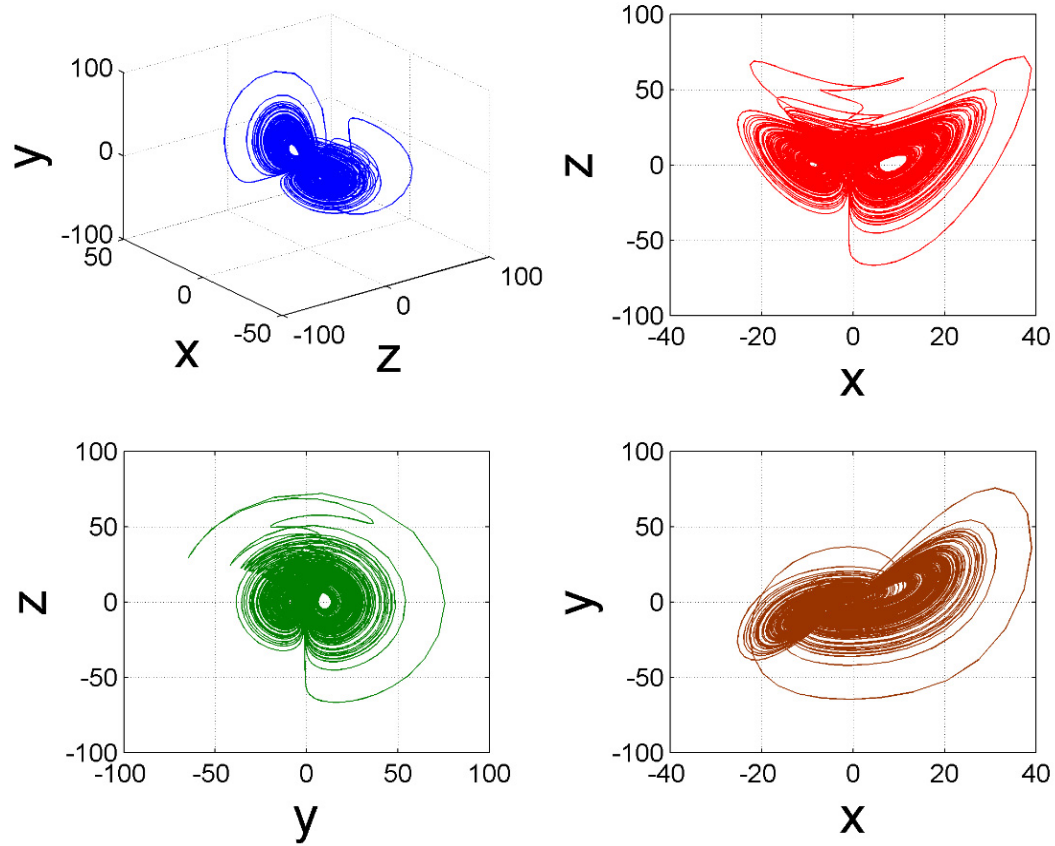

*Figure 46: Phase space dynamics of Lorenz-YZ12*

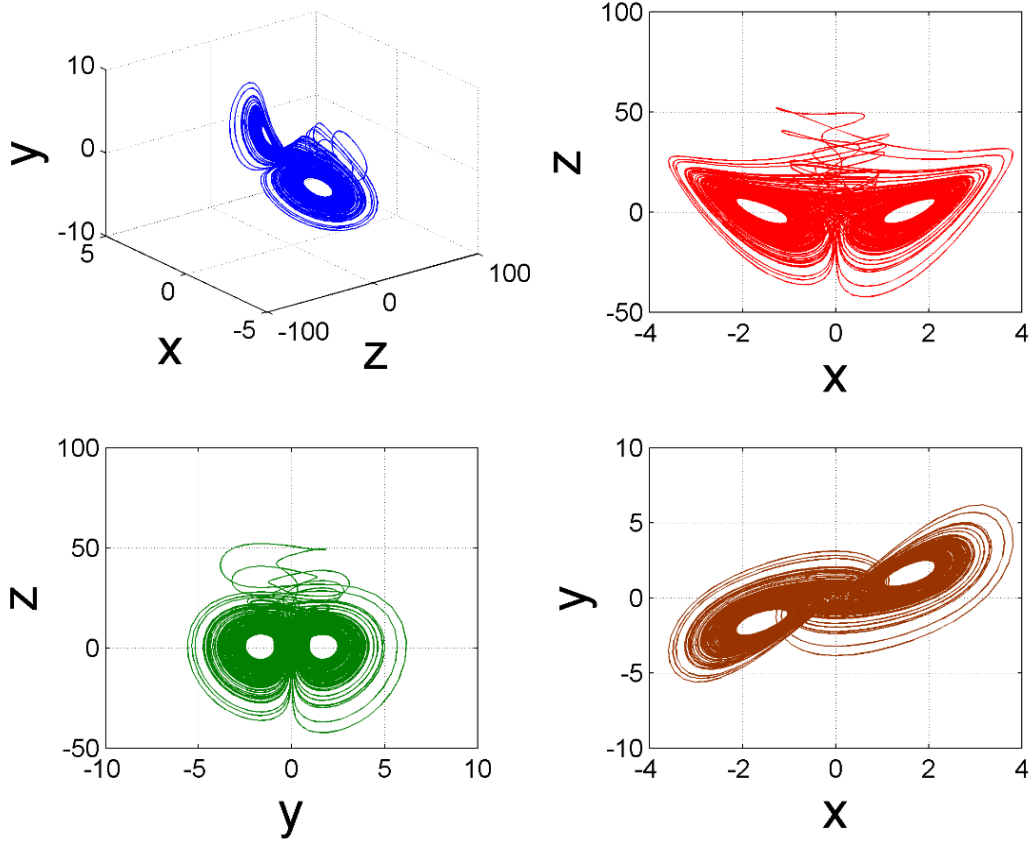

*Figure 47: Phase space dynamics of Lorenz-YZ13*

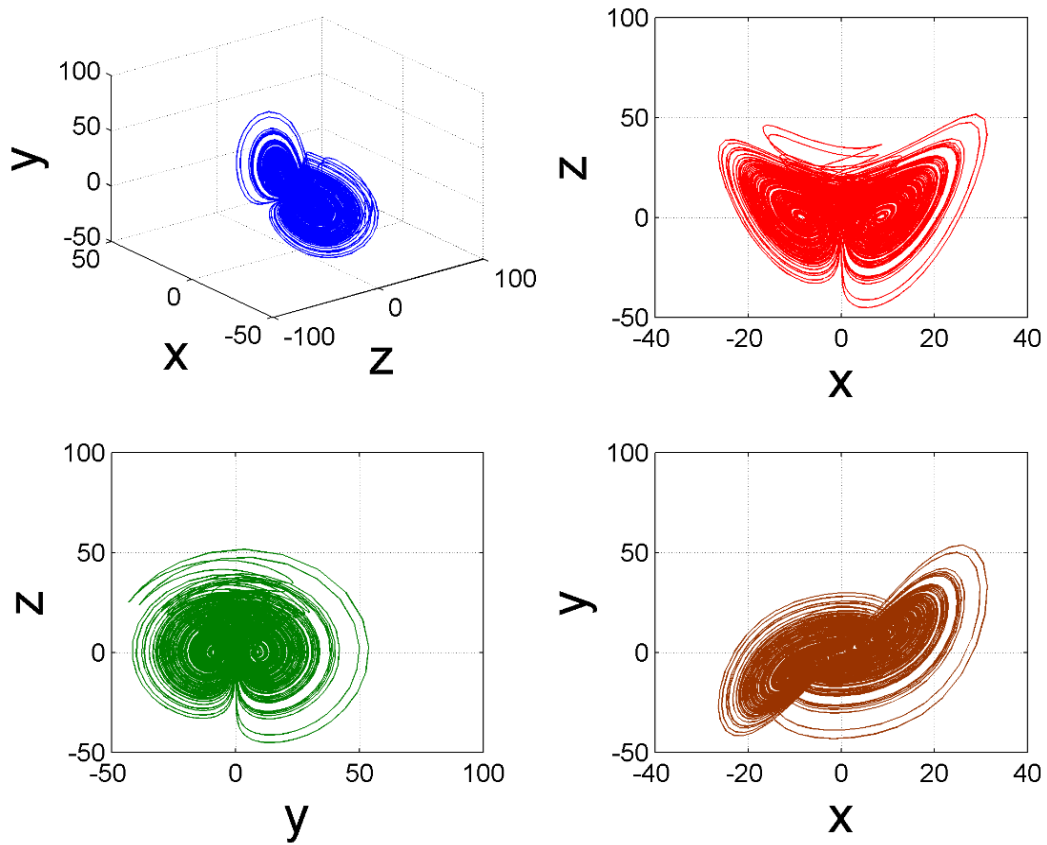

*Figure 48: Phase space dynamics of Lorenz-YZ14*

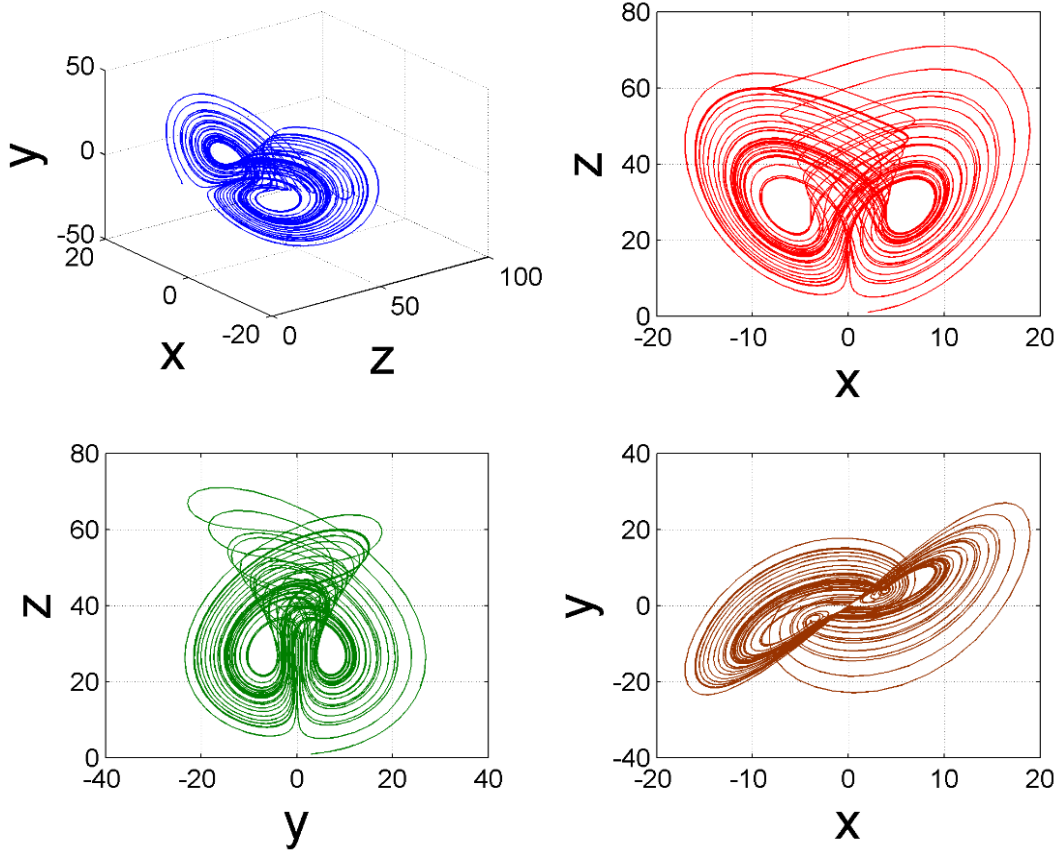

*Figure 49: Phase space dynamics of Lorenz-YZ15*

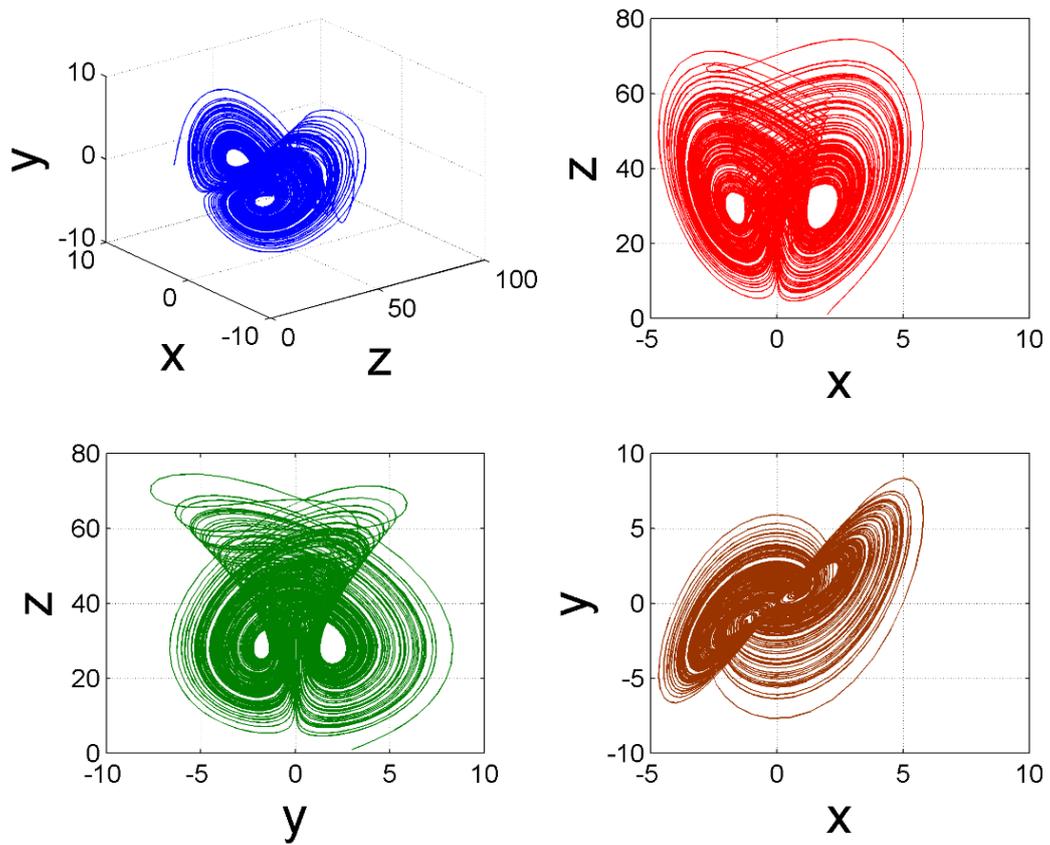

*Figure 50: Phase space dynamics of Lorenz-YZ16*

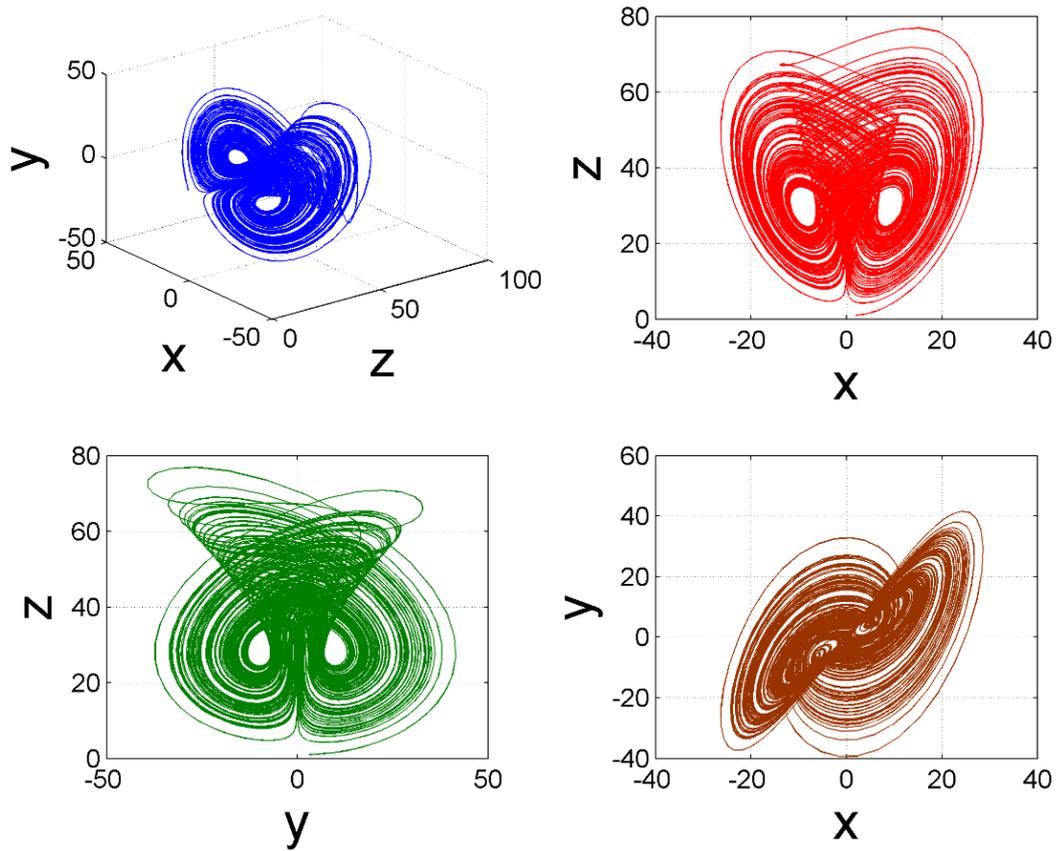

*Figure 51: Phase space dynamics of Lorenz-YZ17*

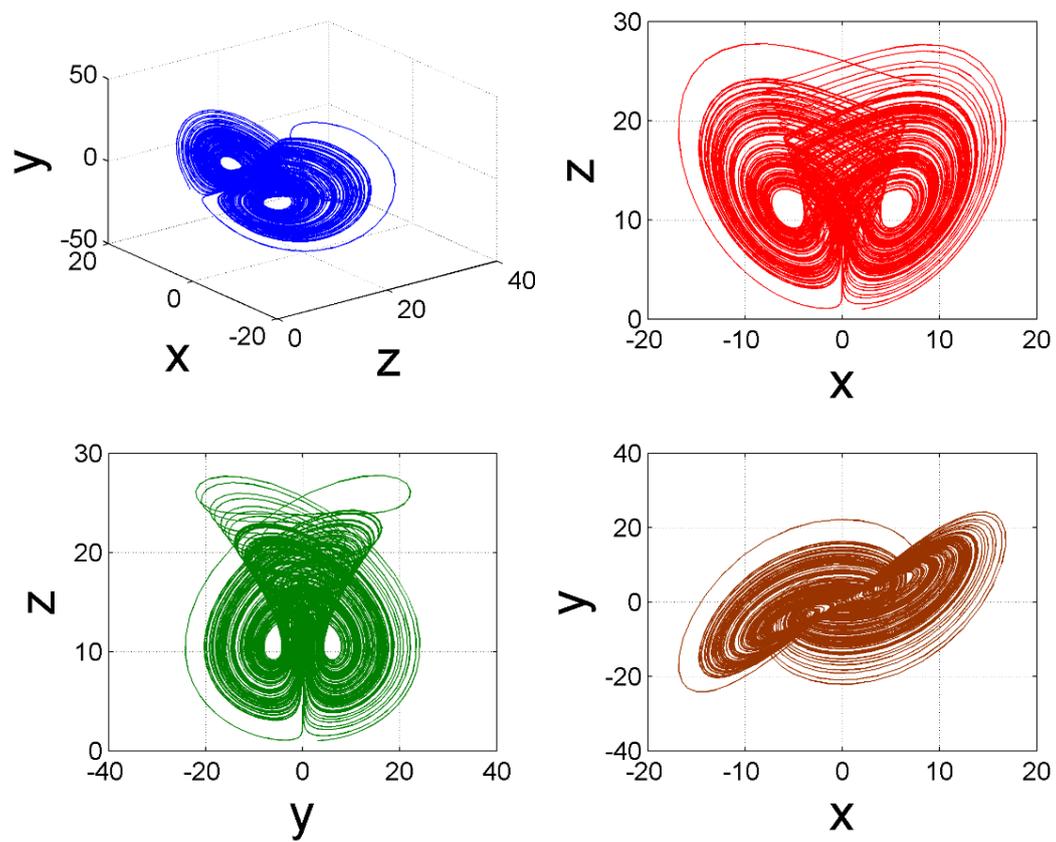

*Figure 52: Phase space dynamics of Lorenz-YZ18*

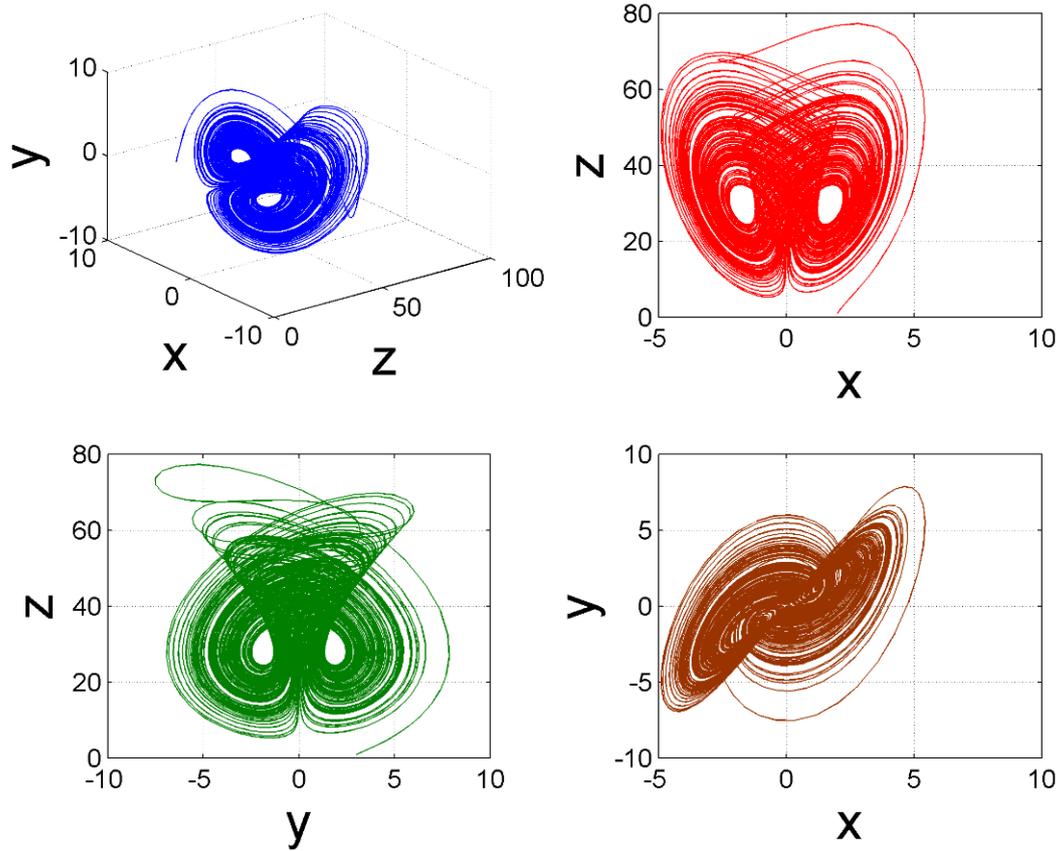

*Figure 53: Phase space dynamics of Lorenz-YZ19*

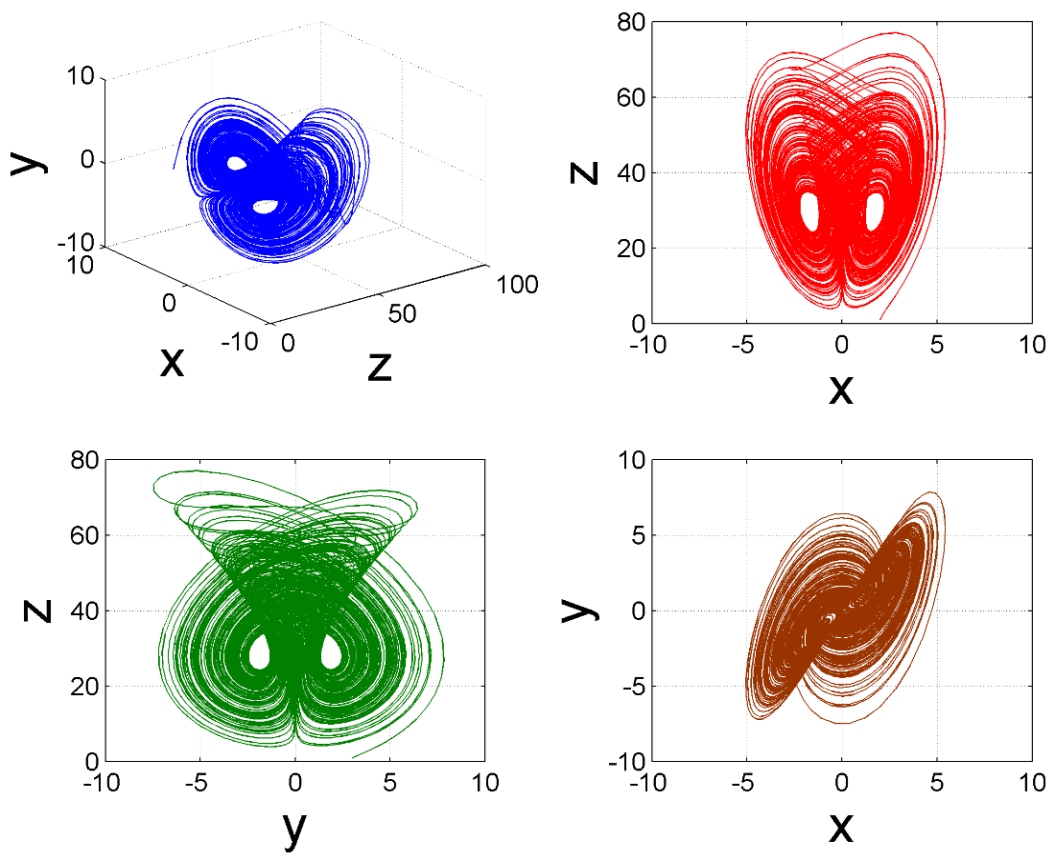

*Figure 54: Phase space dynamics of Lorenz-YZ20*

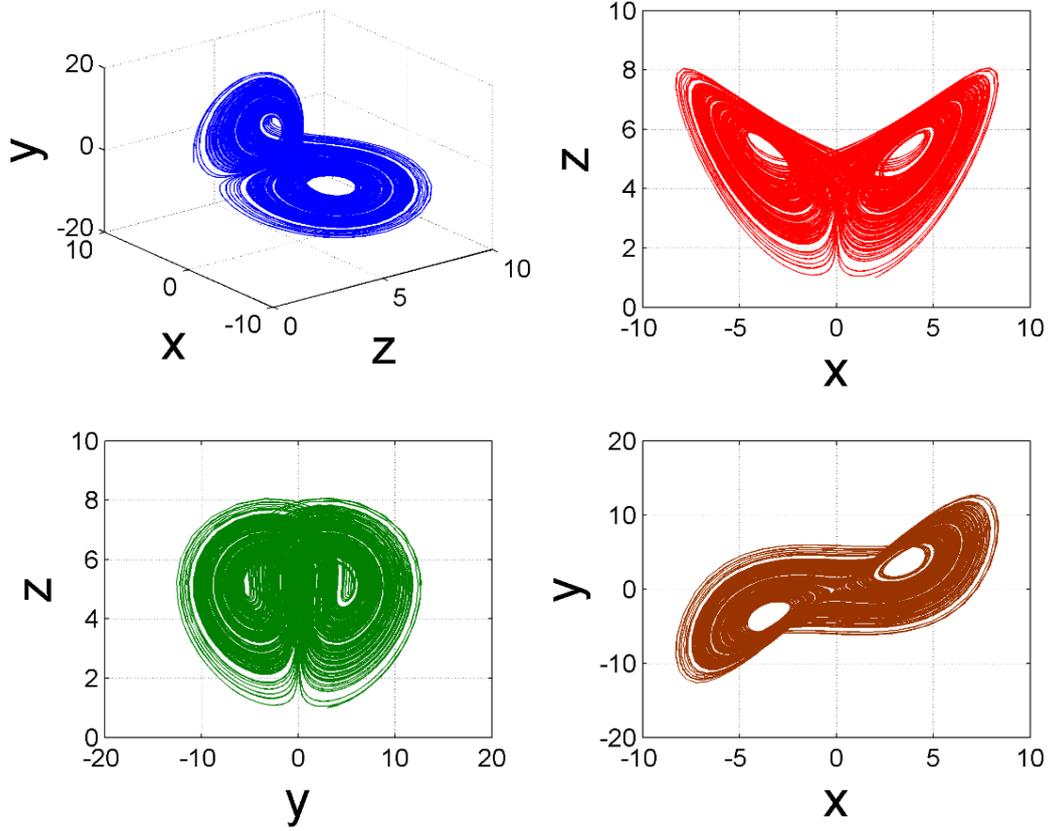

*Figure 55: Phase space dynamics of Lorenz-YZ21*

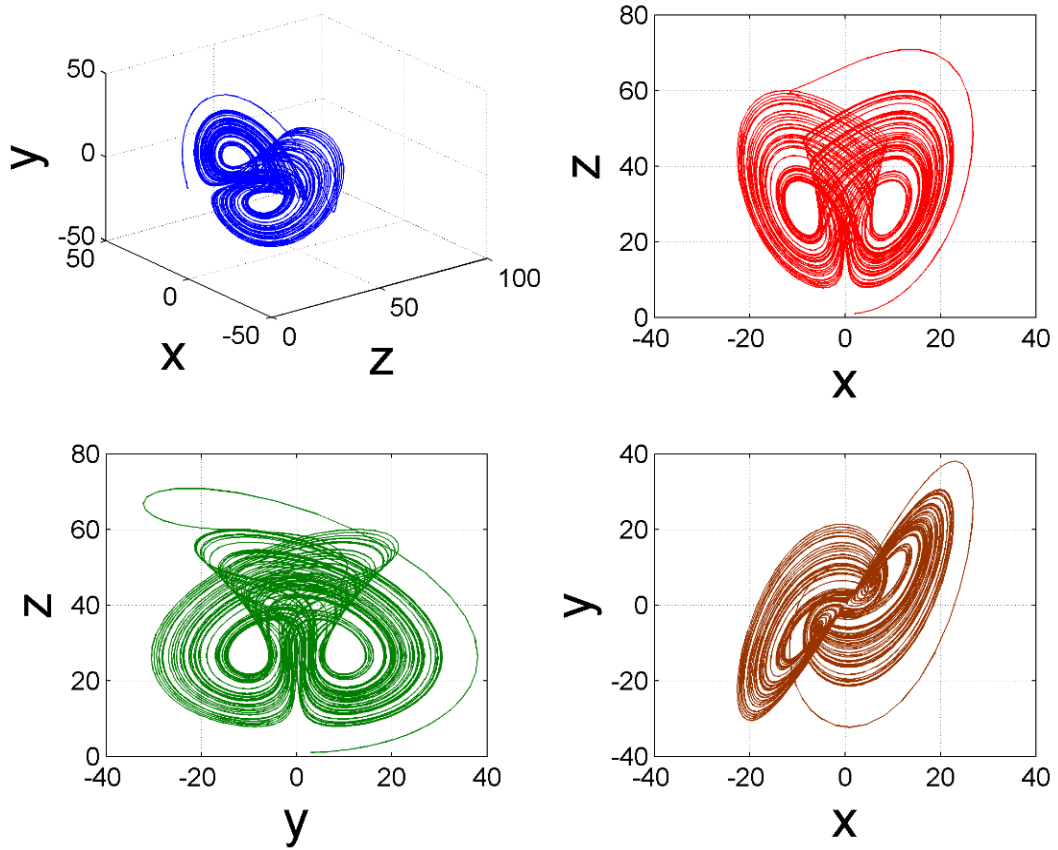

*Figure 56: Phase space dynamics of Lorenz-YZ22*

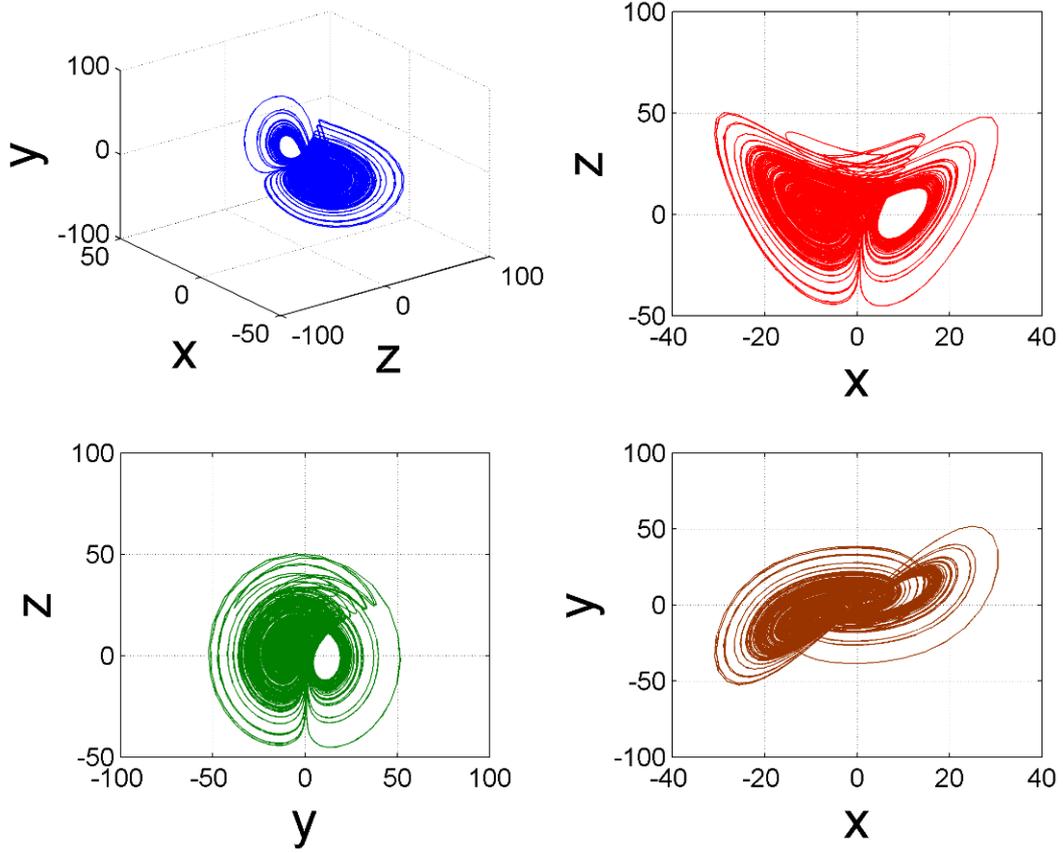

*Figure 57: Phase space dynamics of Lorenz-YZ24*

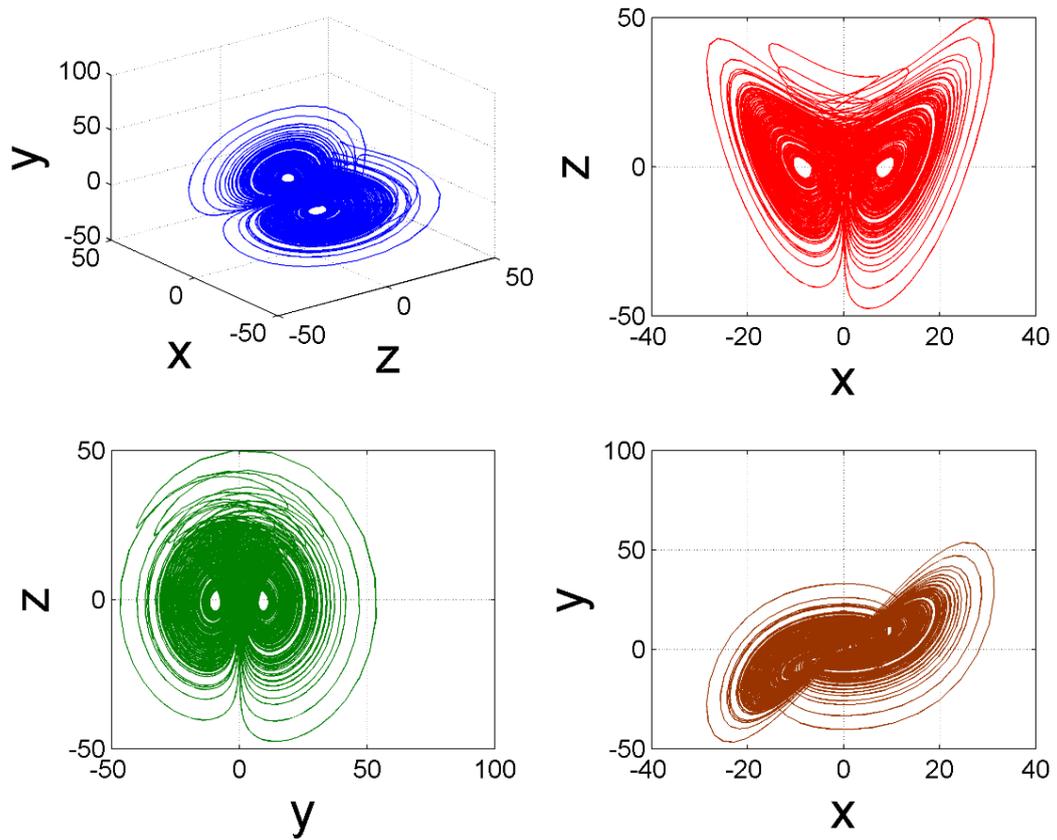

*Figure 58: Phase space dynamics of Lorenz-YZ25*

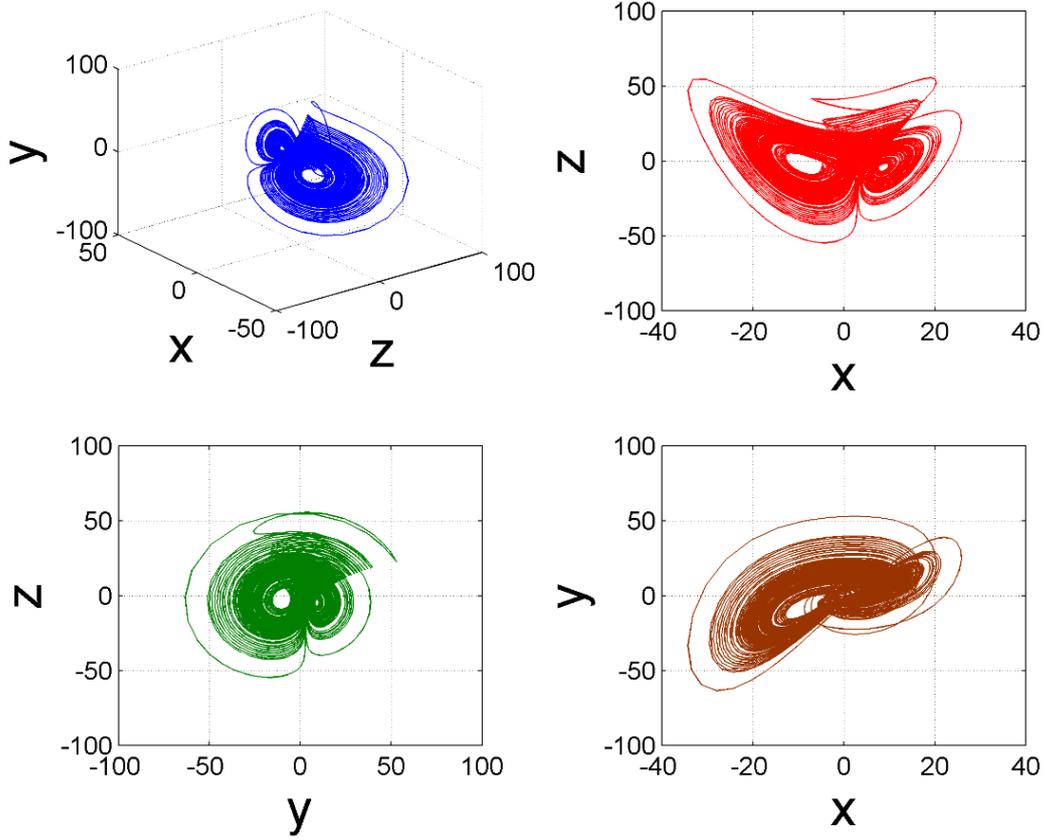

*Figure 59: Phase space dynamics of Lorenz-YZ26*

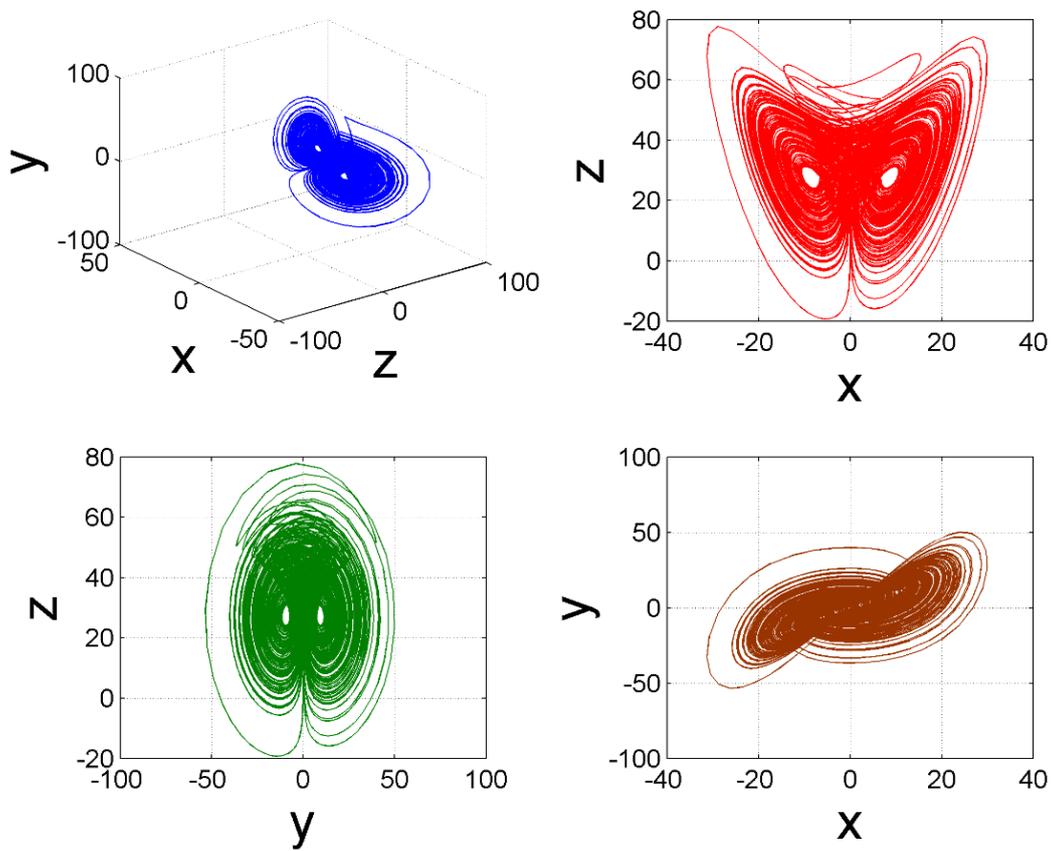

*Figure 60: Phase space dynamics of Lorenz-YZ27*

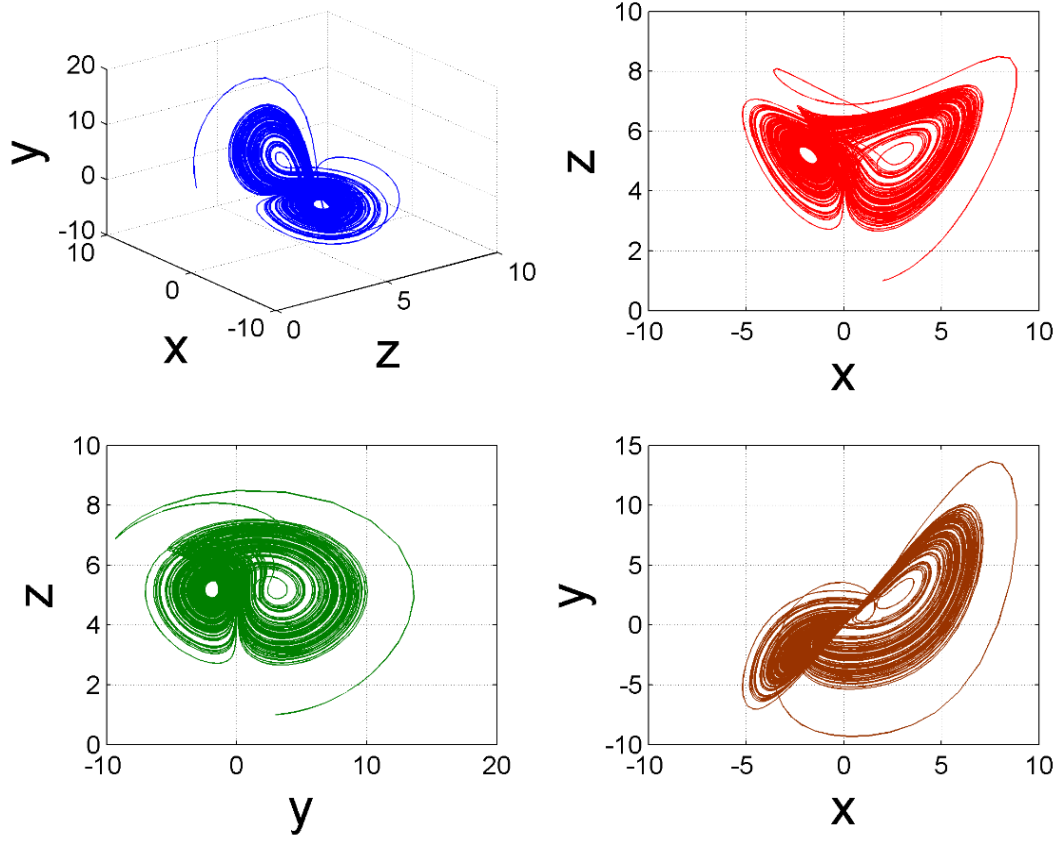

*Figure 61: Phase space dynamics of Lorenz-YZ28*

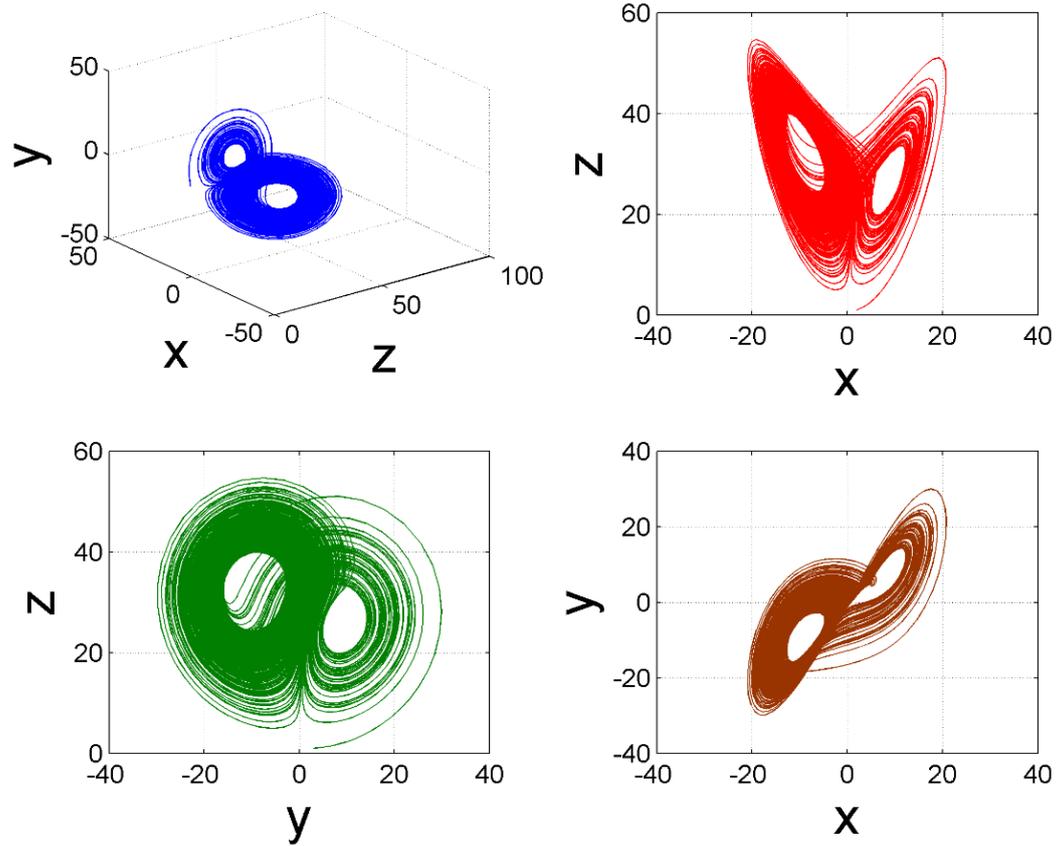

*Figure 62: Phase space dynamics of Lorenz-YZ29*

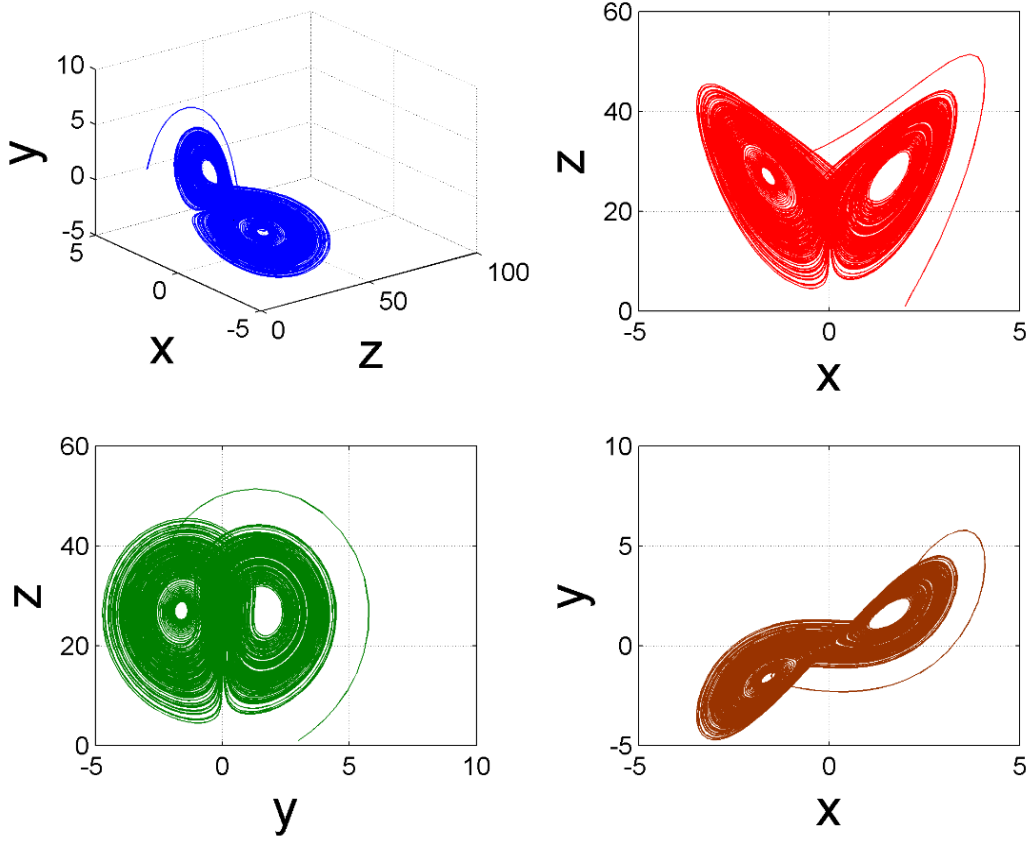

*Figure 63: Phase space dynamics of Lorenz-YZ30*

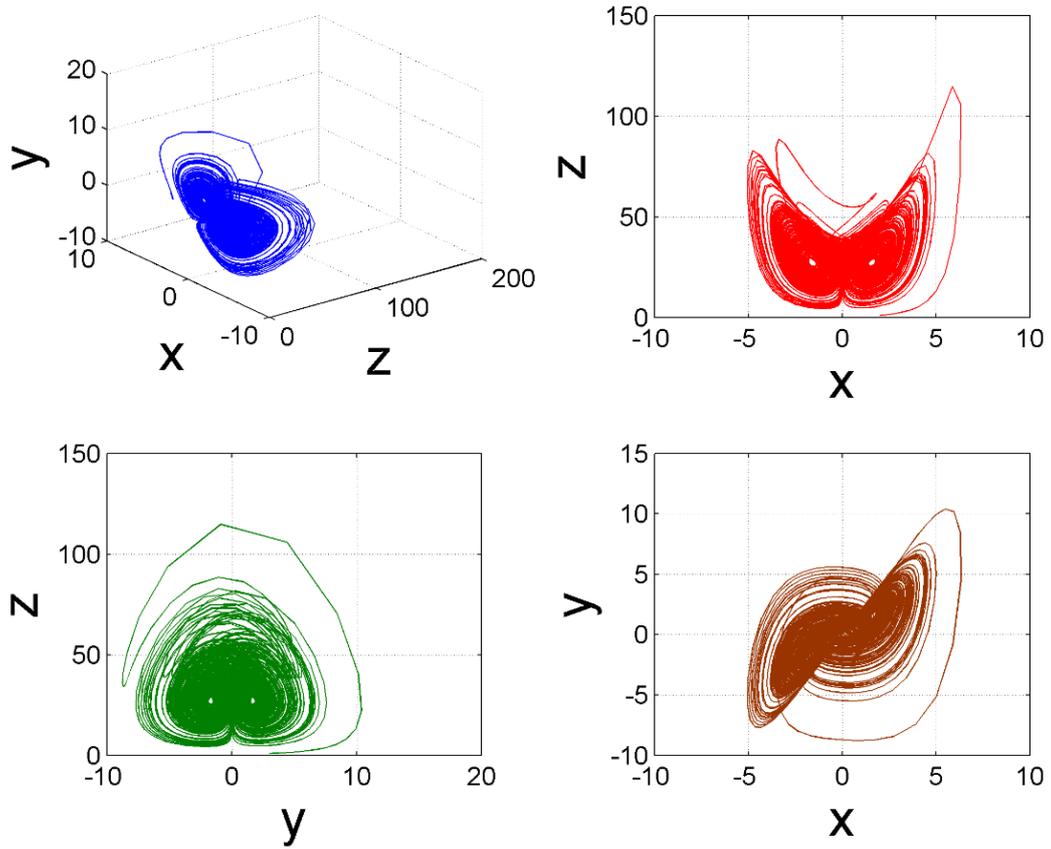

*Figure 64: Phase space dynamics of Lorenz-YZ31*

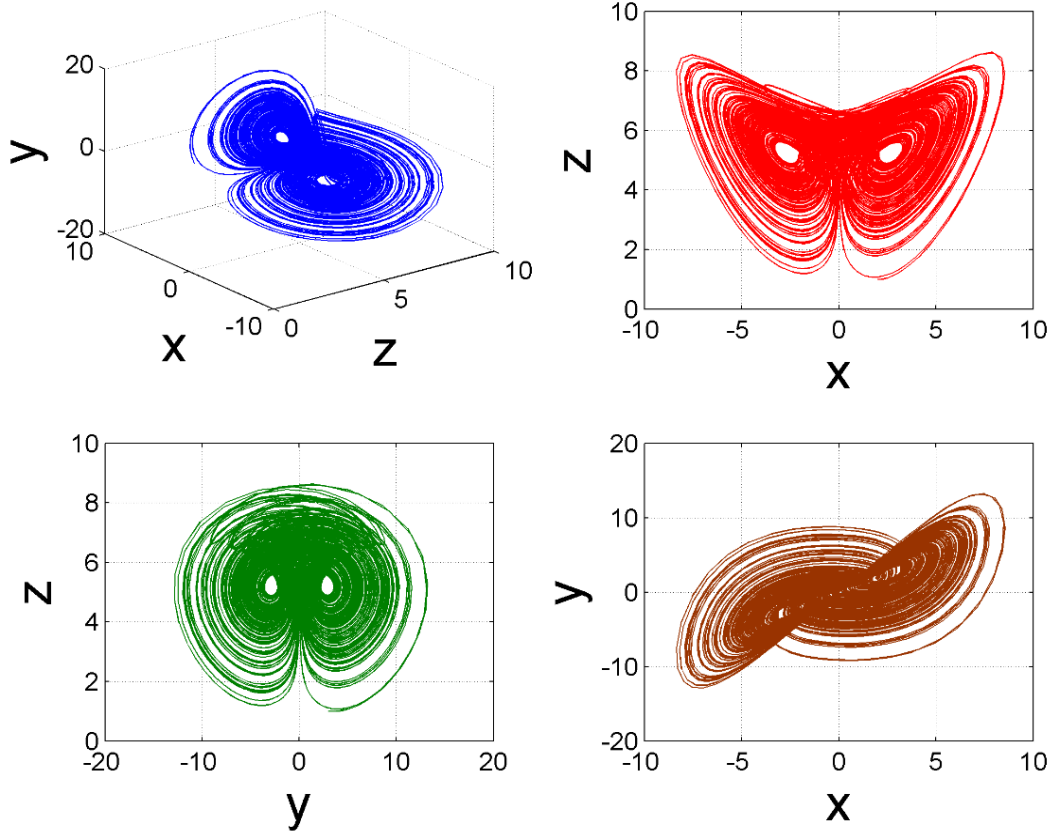

*Figure 65: Phase space dynamics of Lorenz-YZ33*

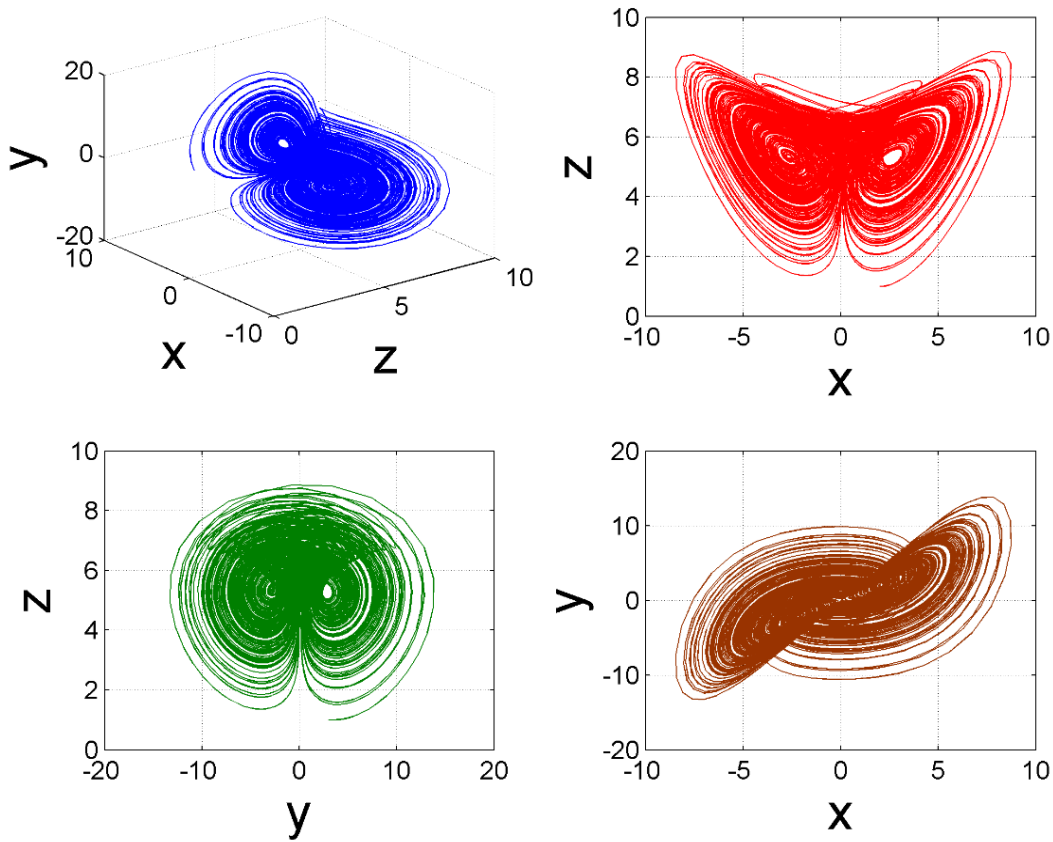

*Figure 66: Phase space dynamics of Lorenz-YZ34*

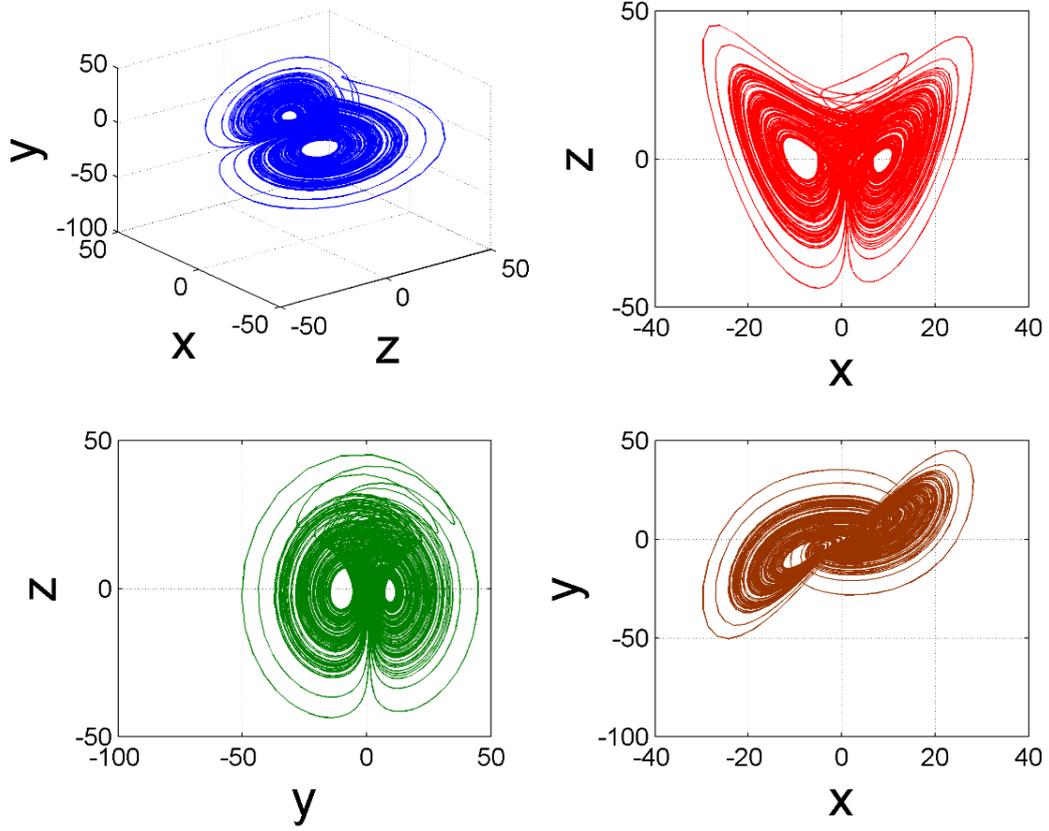

*Figure 67: Phase space dynamics of Lorenz-YZ35*

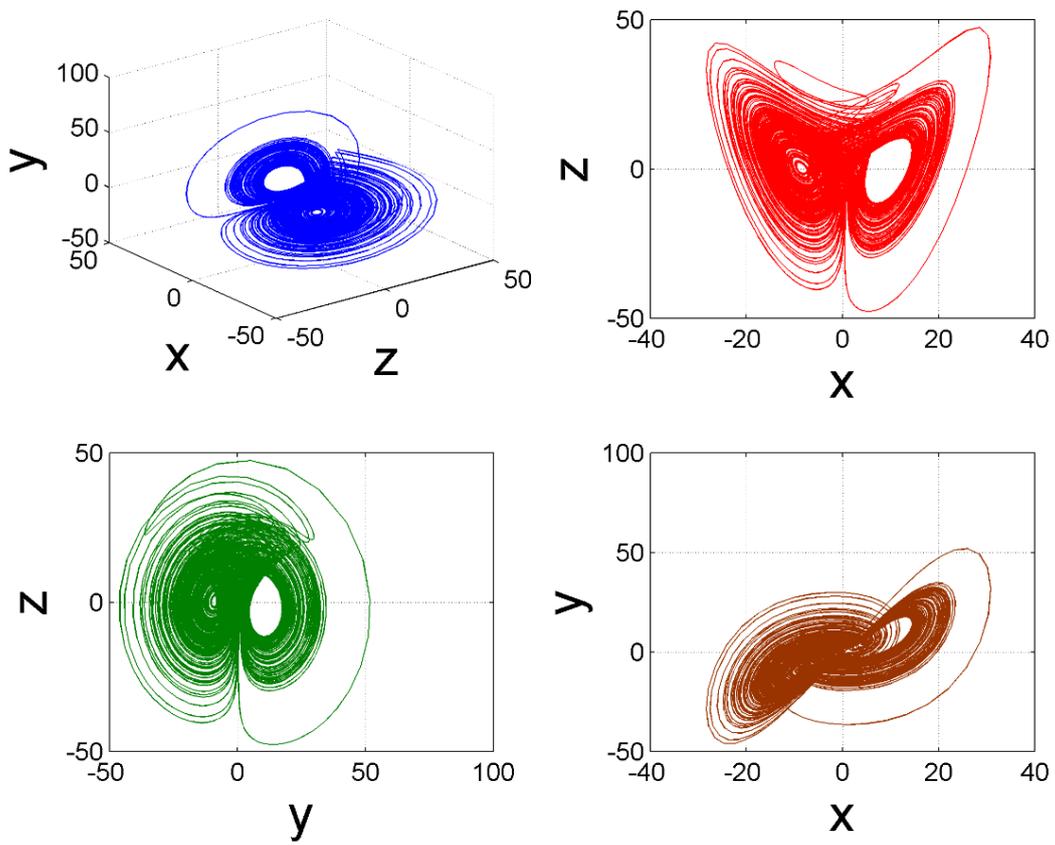

*Figure 68: Phase space dynamics of Lorenz-YZ36*

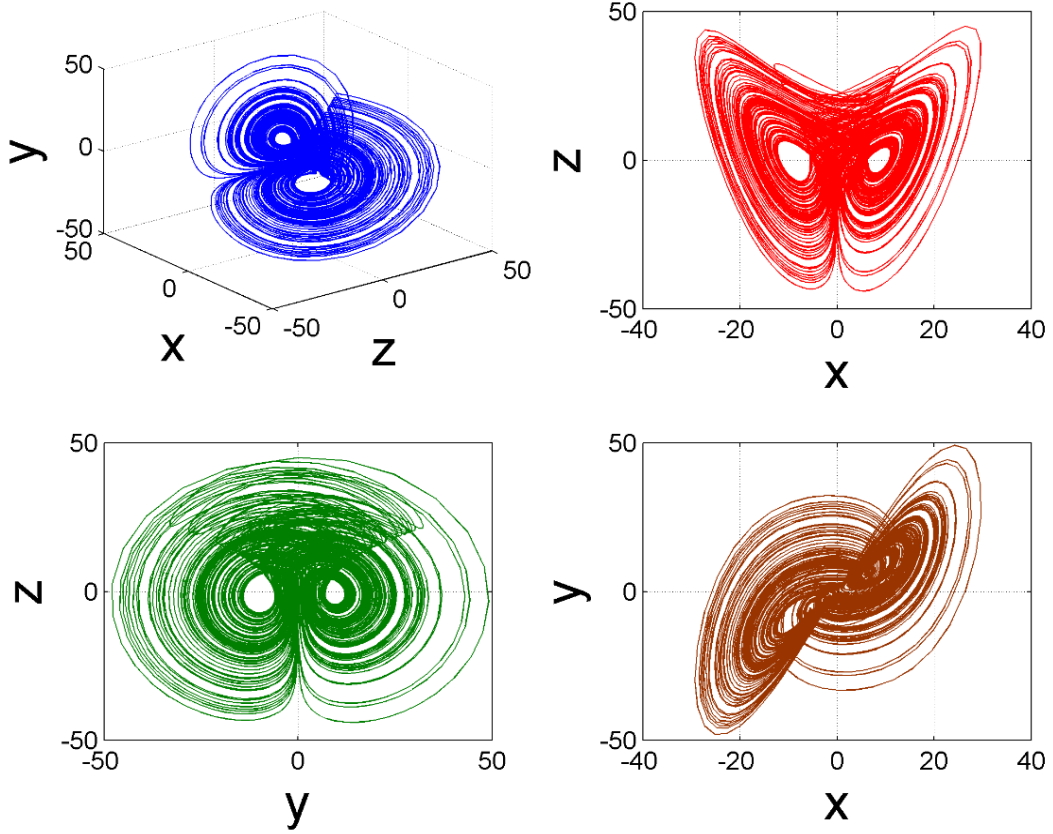

*Figure 69: Phase space dynamics of Lorenz-YZ37*

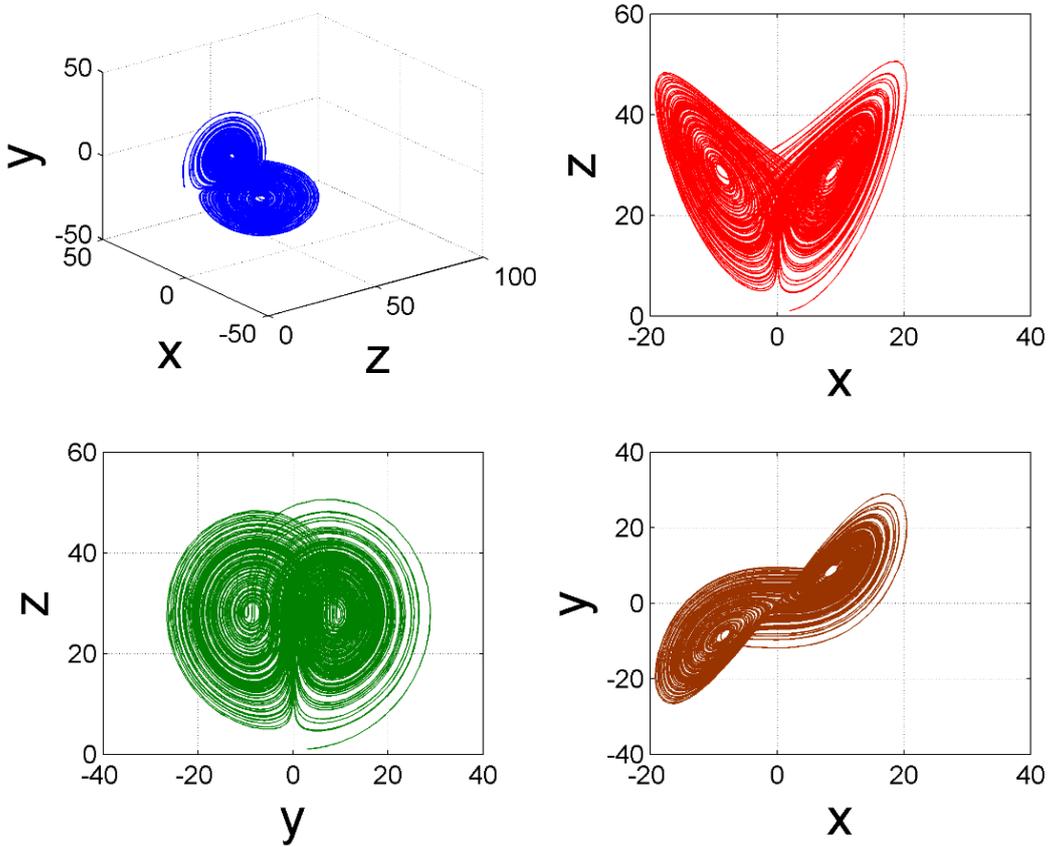

*Figure 70: Phase space dynamics of Lorenz-YZ38*

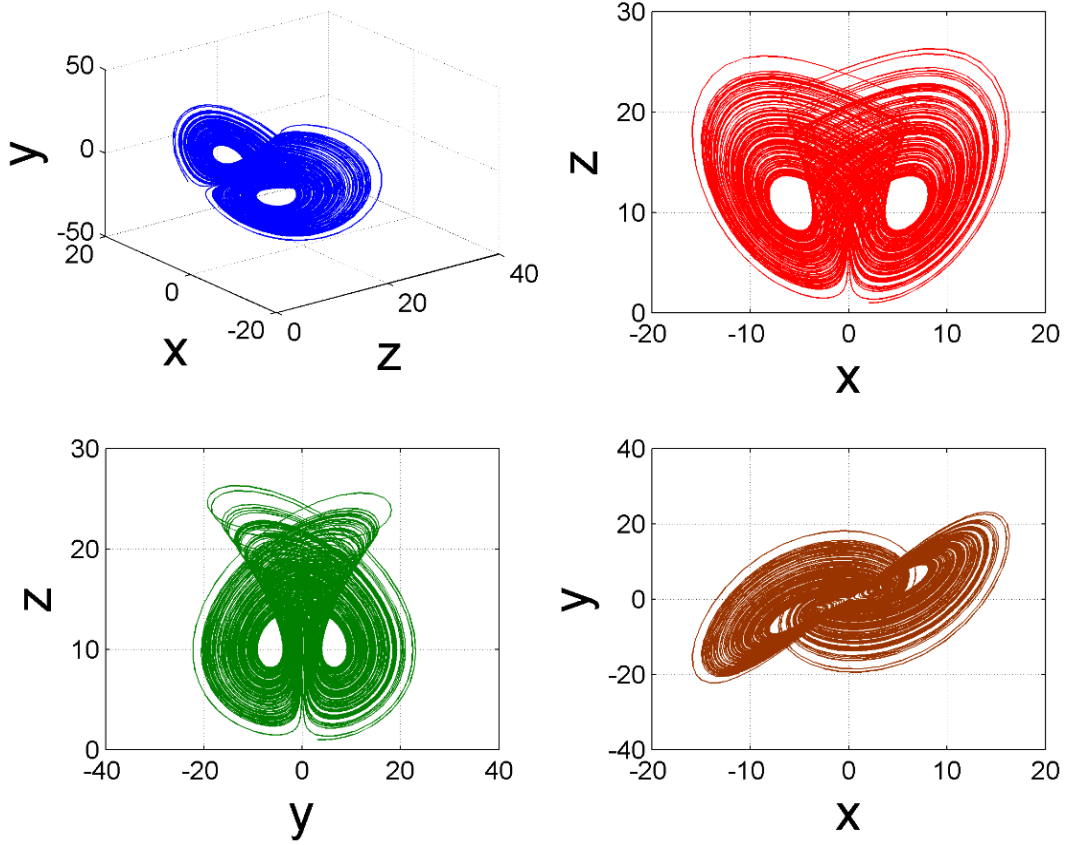

*Figure 71: Phase space dynamics of Lorenz-YZ39*

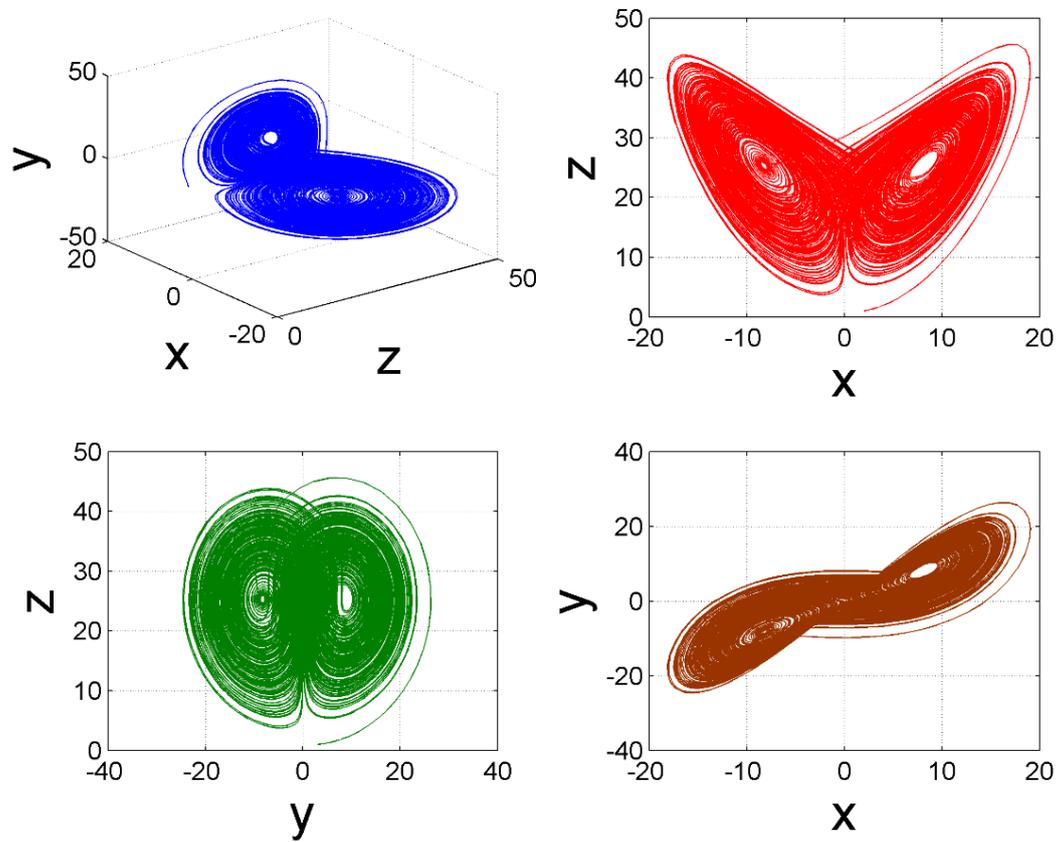

*Figure 72: Phase space dynamics of Lorenz-YZ40*

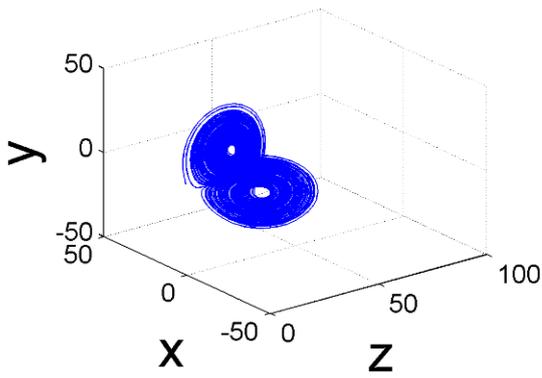
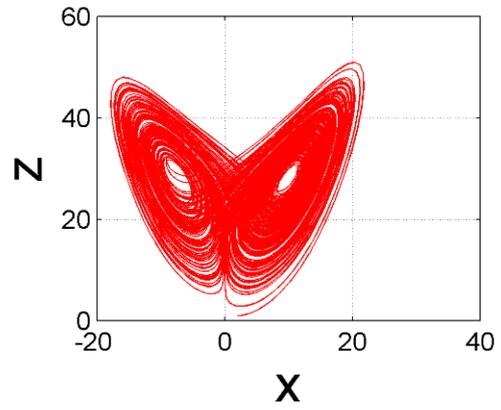
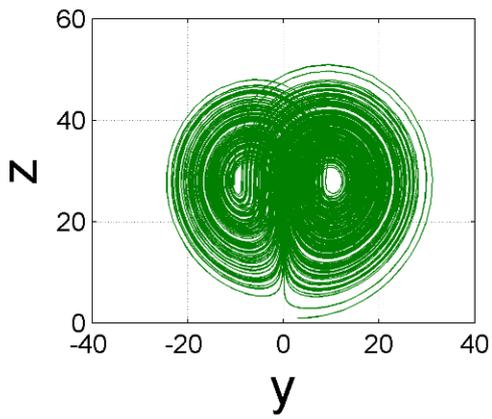
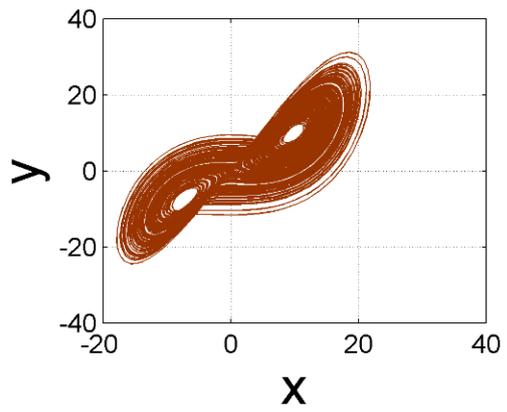

*Figure 73: Phase space dynamics of Lorenz-YZ41*

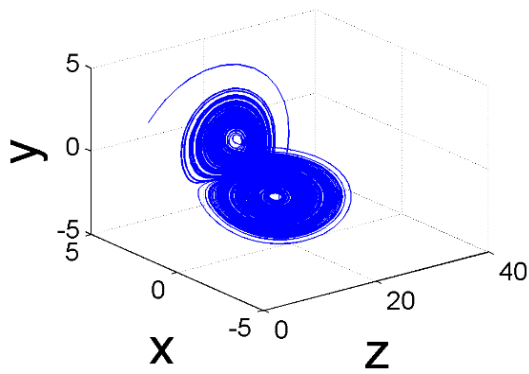
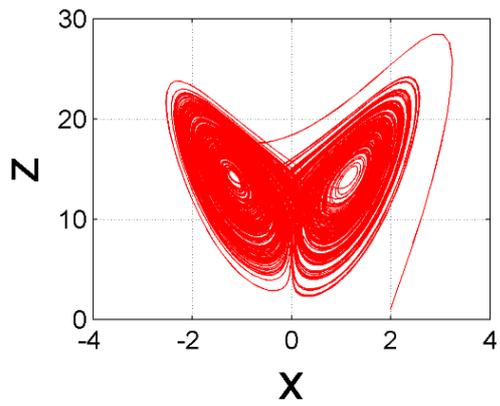
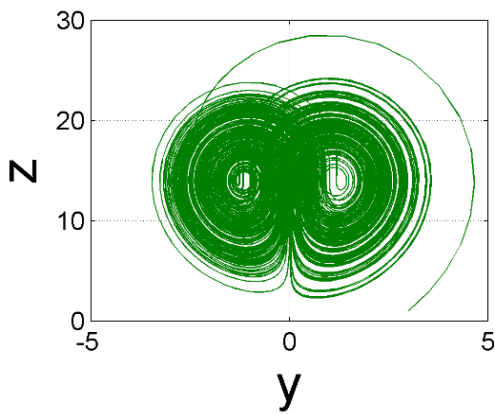
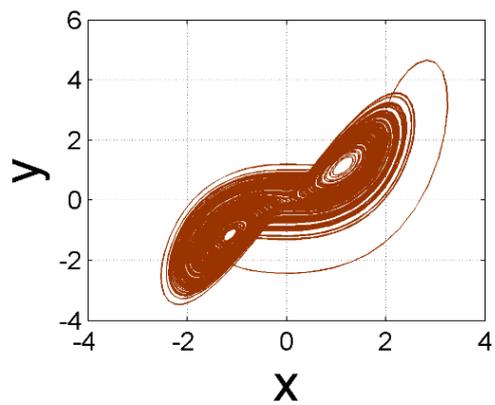

*Figure 74: Phase space dynamics of Lorenz-YZ42*

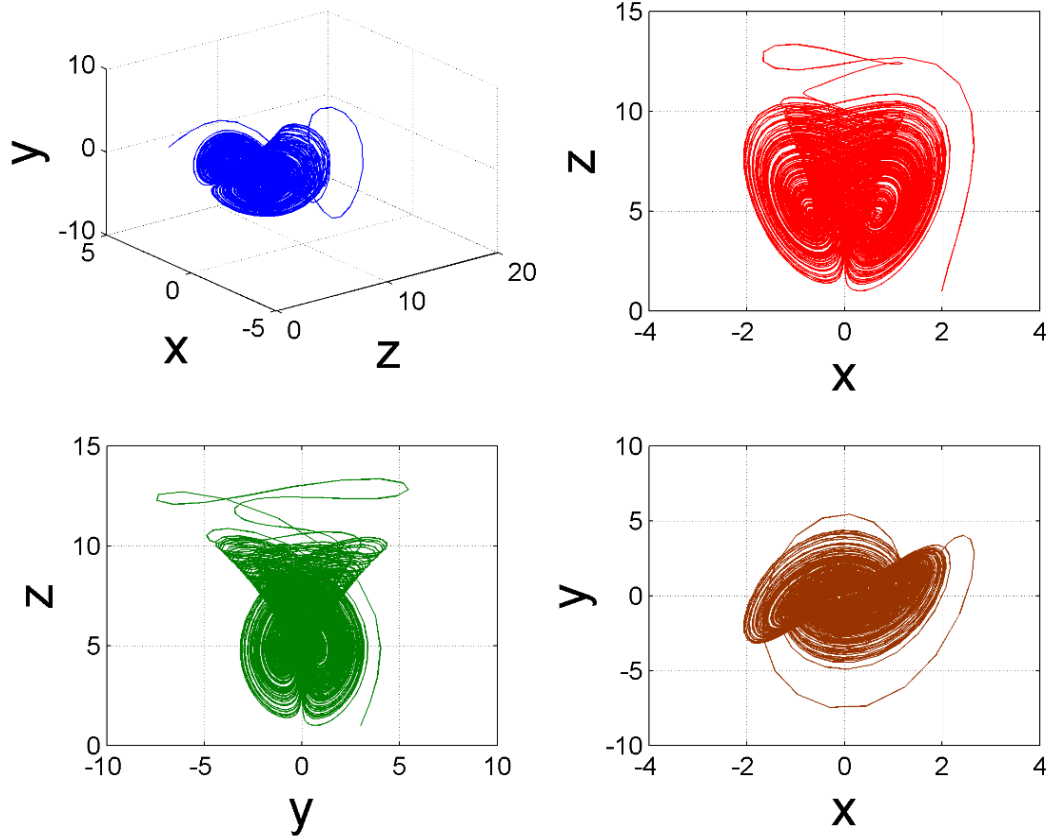

*Figure 75: Phase space dynamics of Lorenz-YZ43*

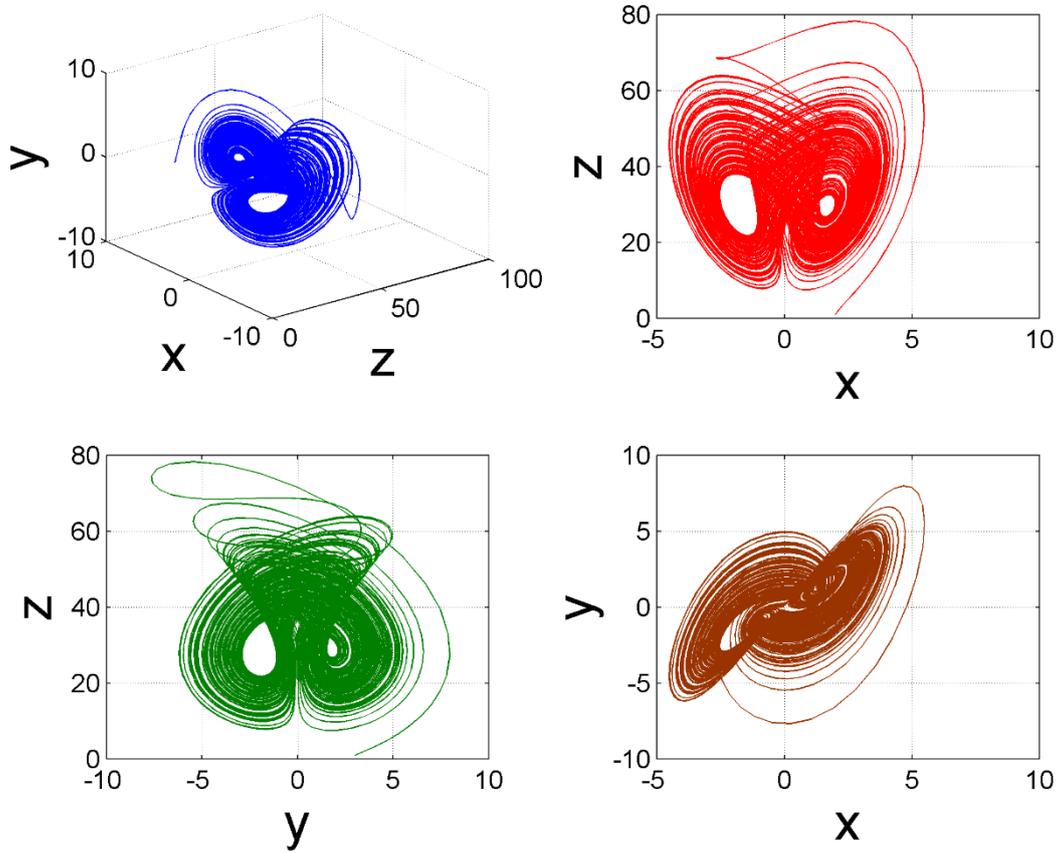

*Figure 76: Phase space dynamics of Lorenz-YZ44*

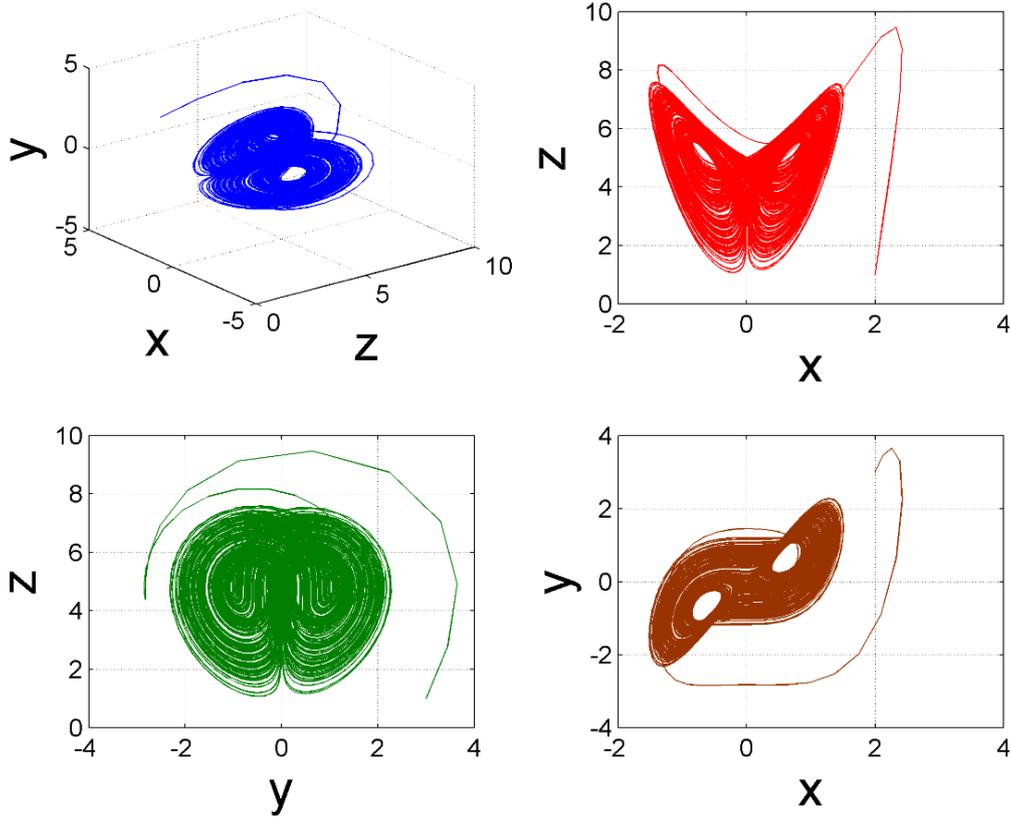

*Figure 77: Phase space dynamics of Lorenz-YZ45*

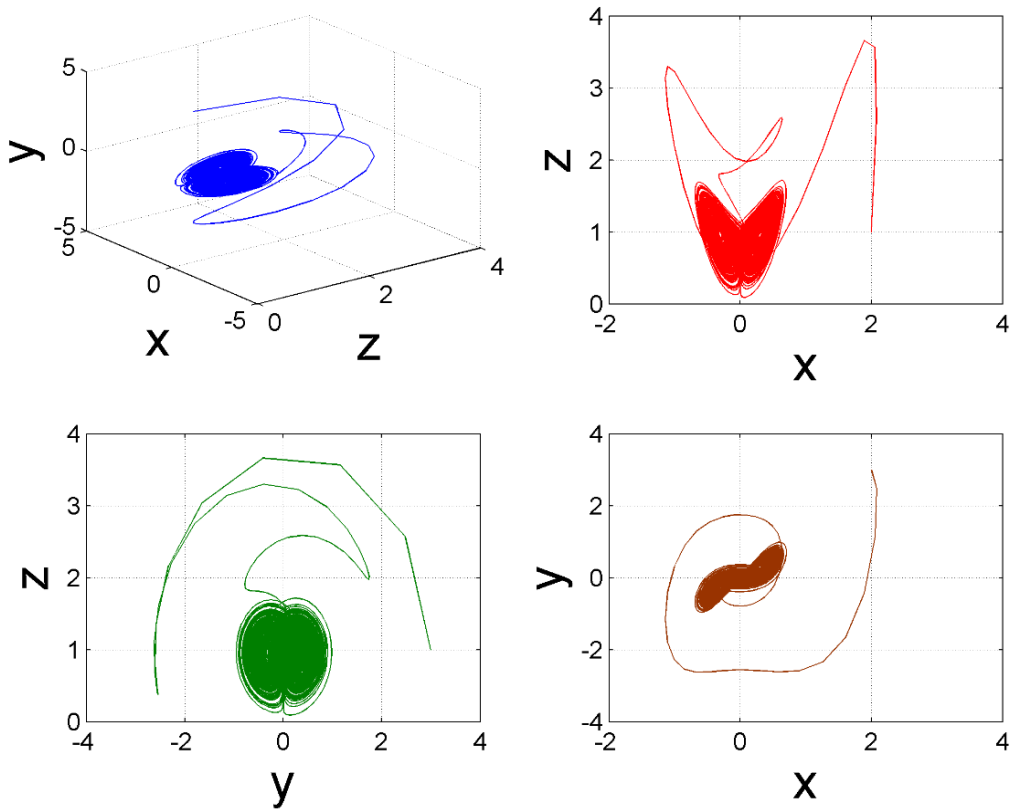

*Figure 78: Phase space dynamics of Lorenz-YZ46*

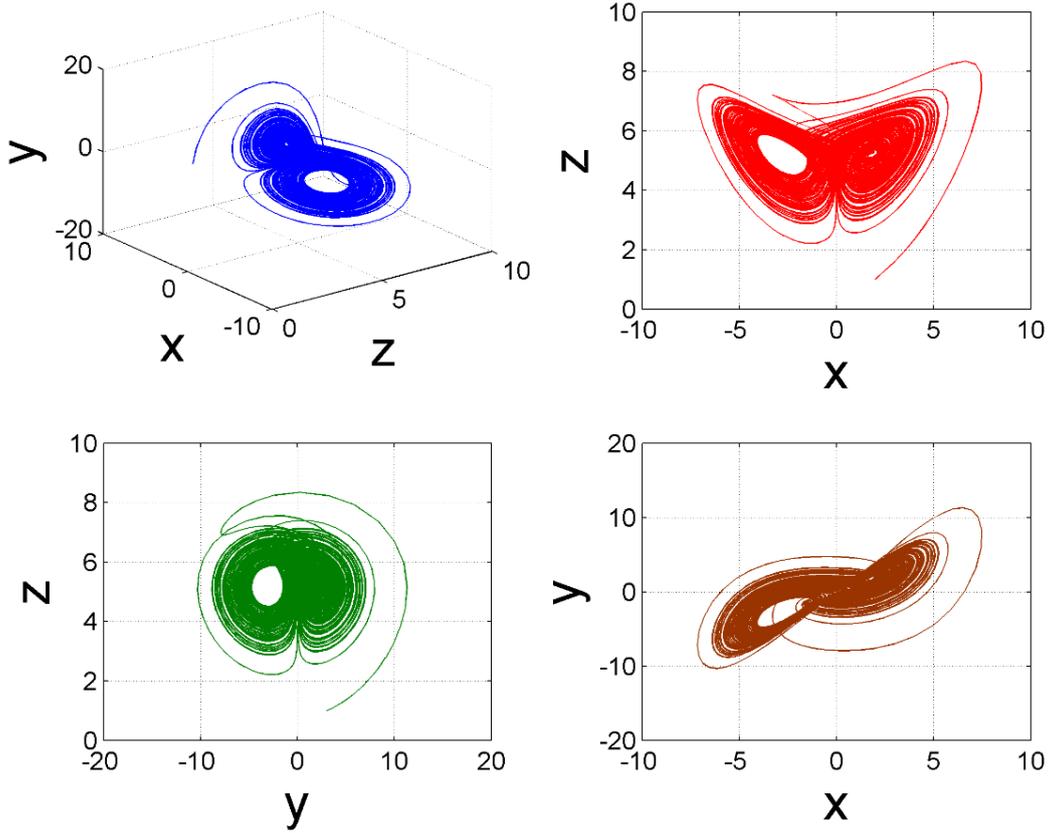

*Figure 79: Phase space dynamics of Lorenz-YZ47*

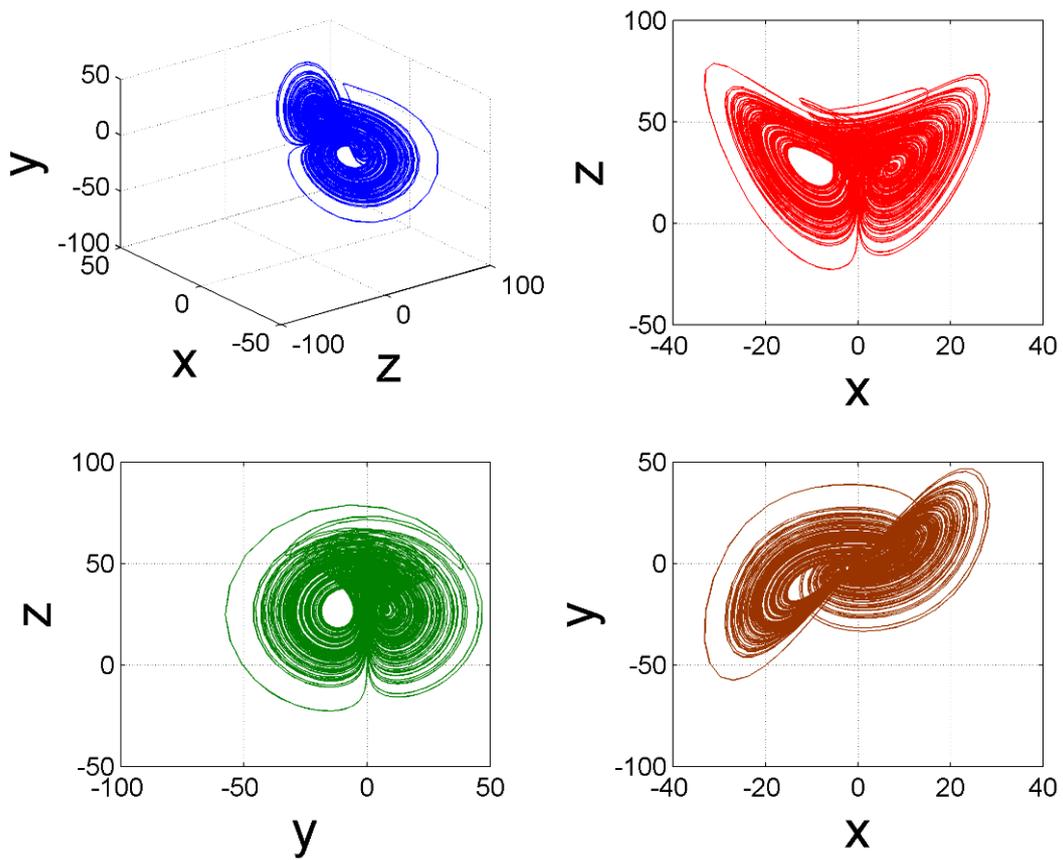

*Figure 80: Phase space dynamics of Lorenz-YZ48*

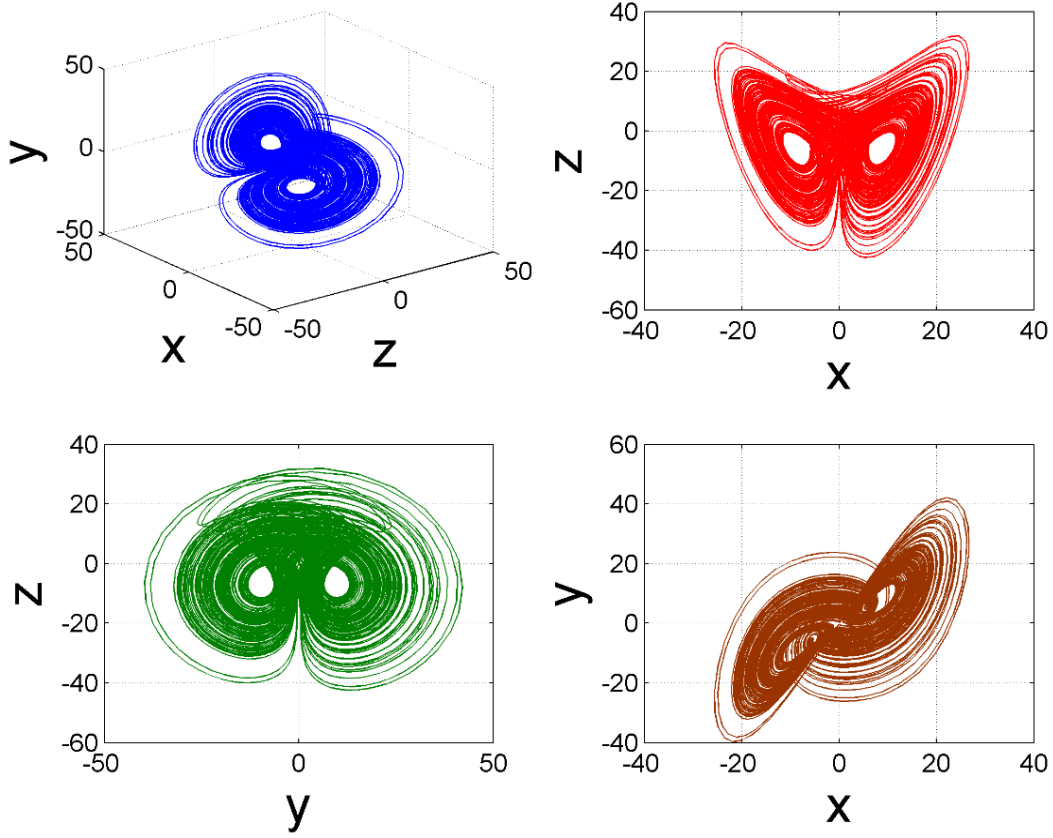

*Figure 81: Phase space dynamics of Lorenz-YZ49*

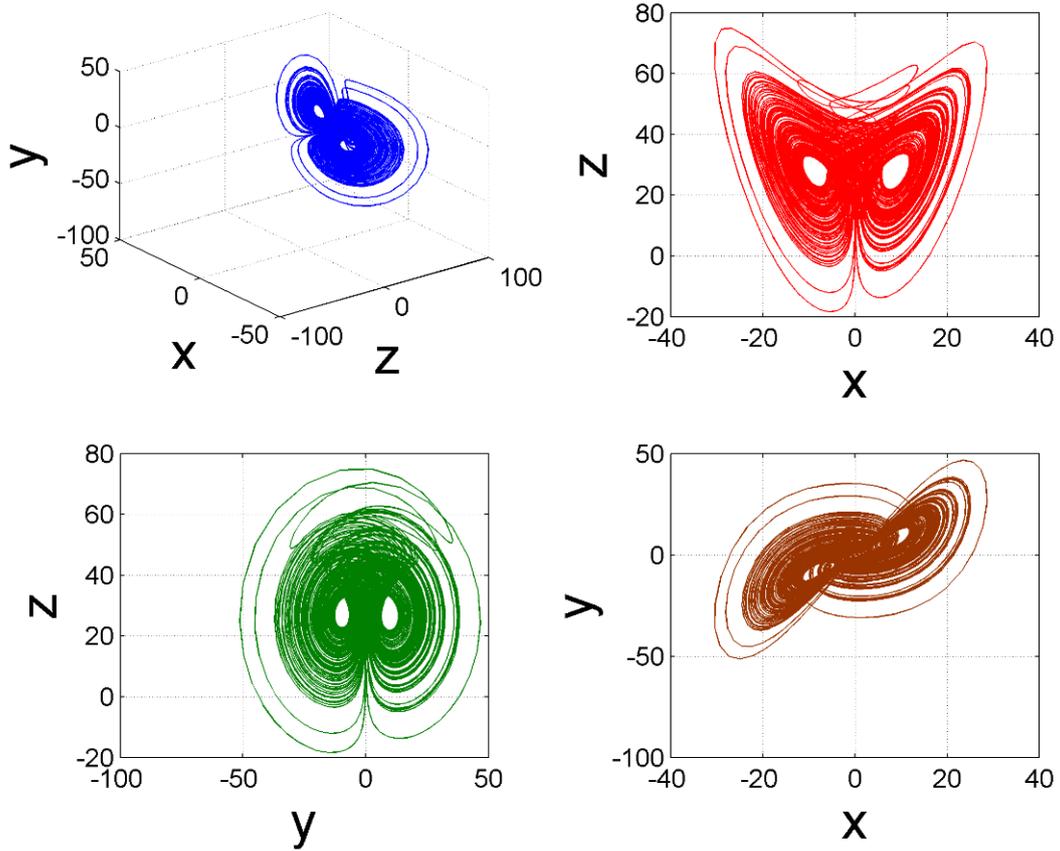

*Figure 82: Phase space dynamics of Lorenz-YZ50*

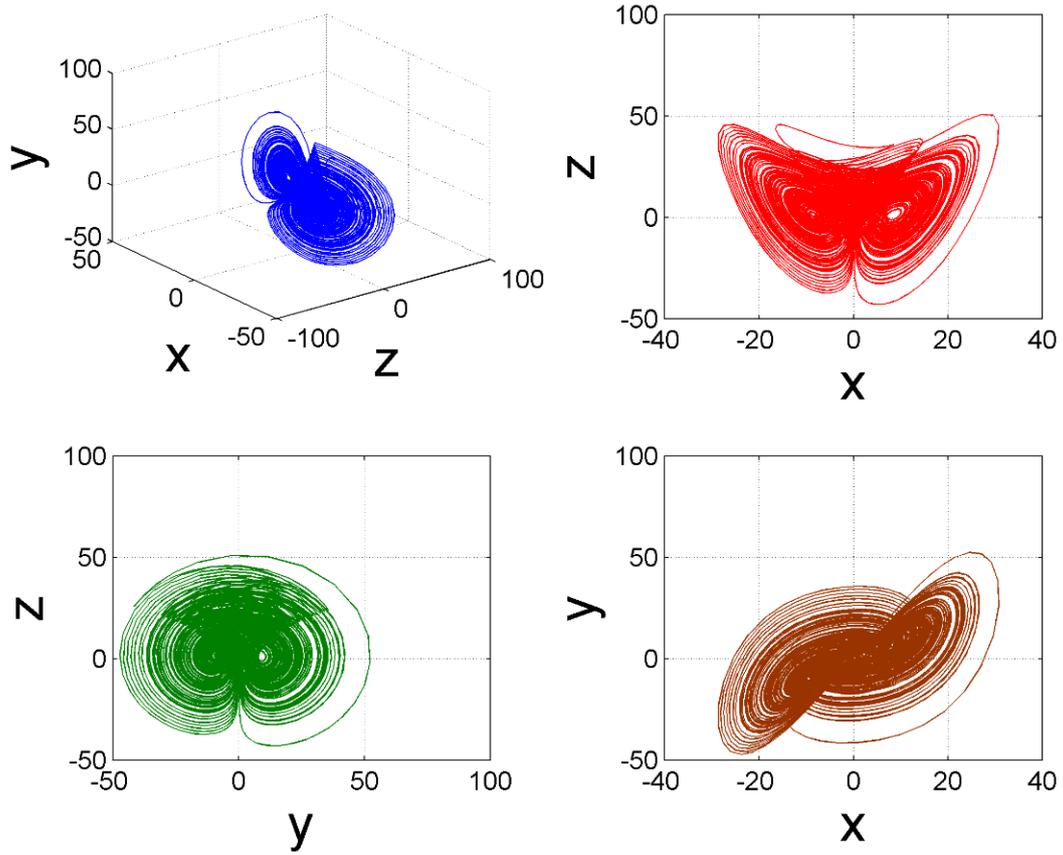

*Figure 83: Phase space dynamics of Lorenz-YZ51*

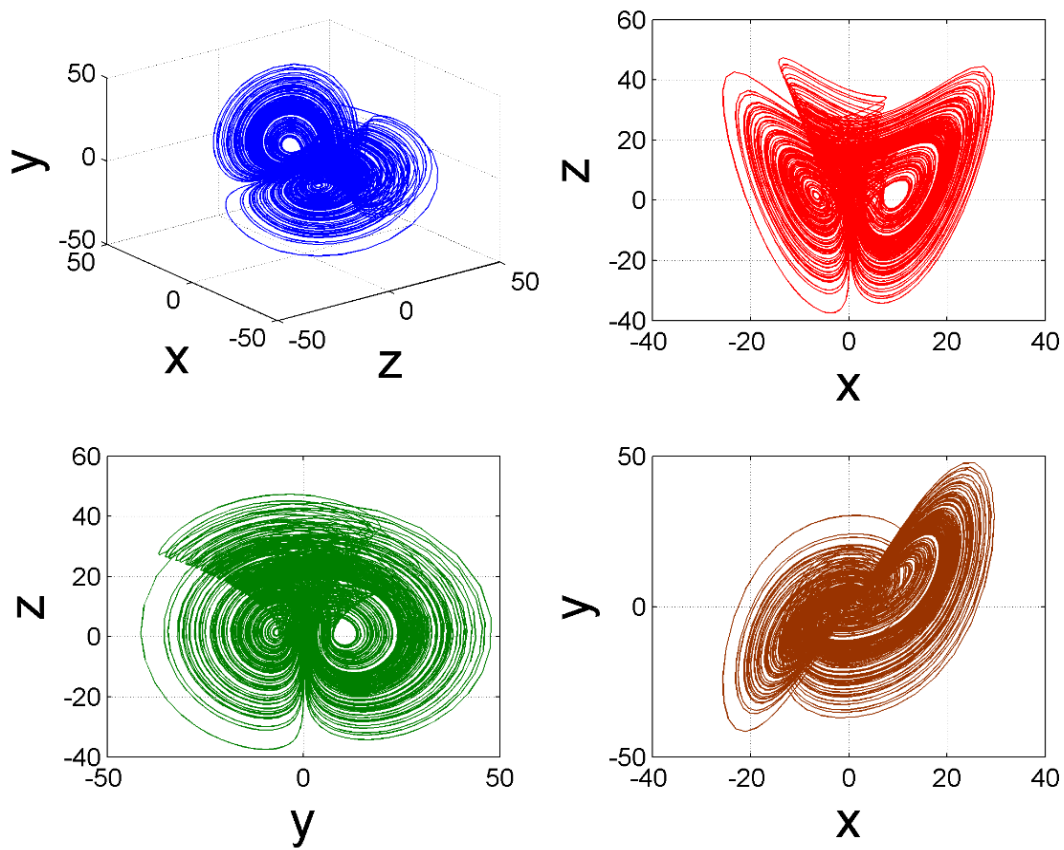

*Figure 84: Phase space dynamics of Lorenz-YZ52*

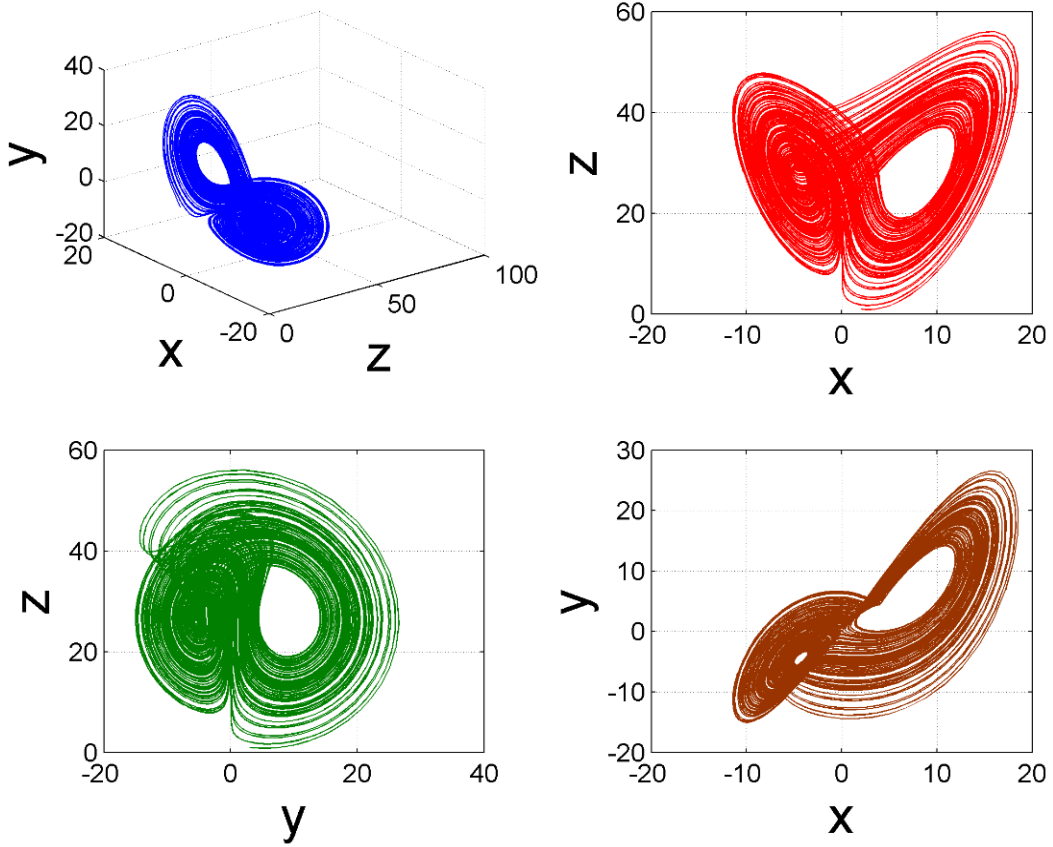

*Figure 85: Phase space dynamics of Lorenz-YZ53*

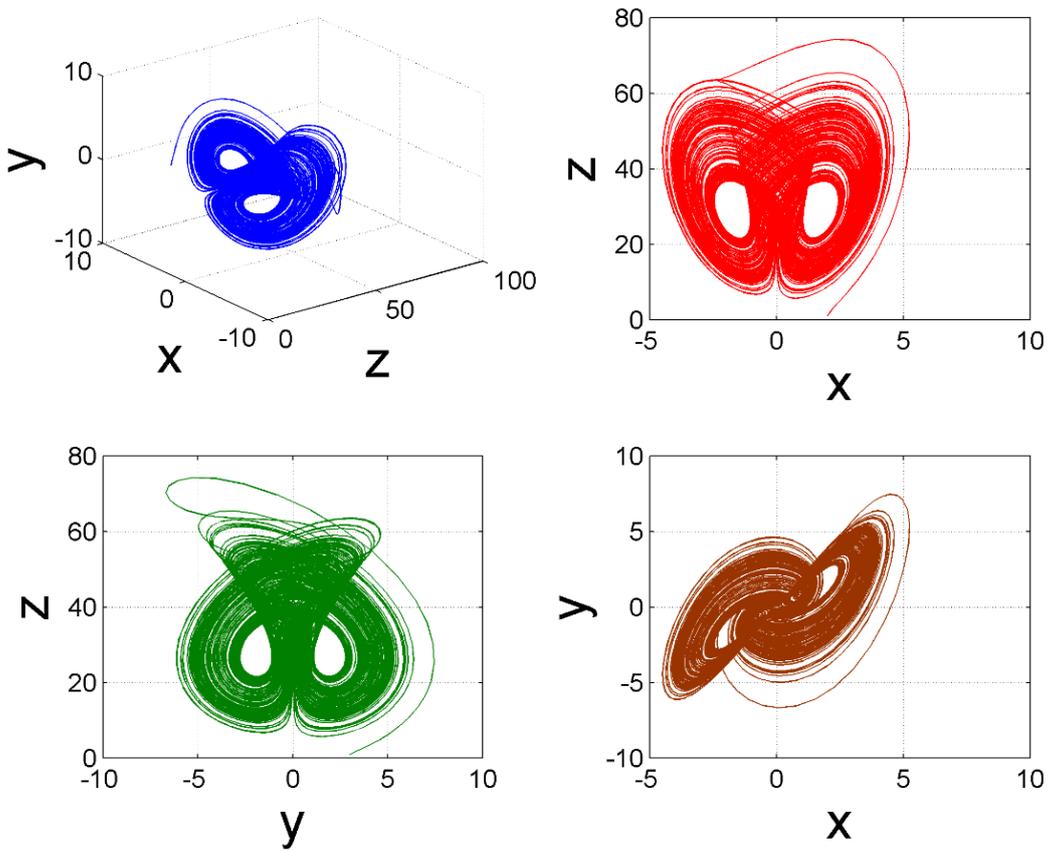

*Figure 86: Phase space dynamics of Lorenz-YZ54*

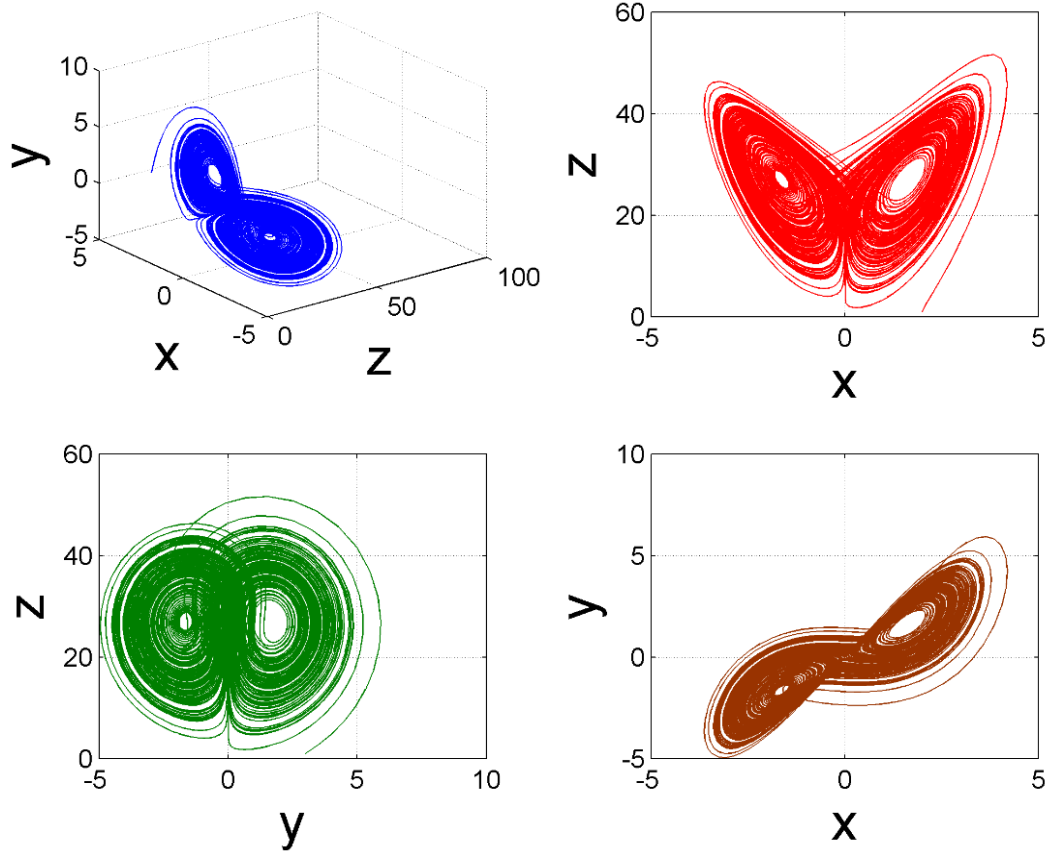

*Figure 87: Phase space dynamics of Lorenz-YZ55*

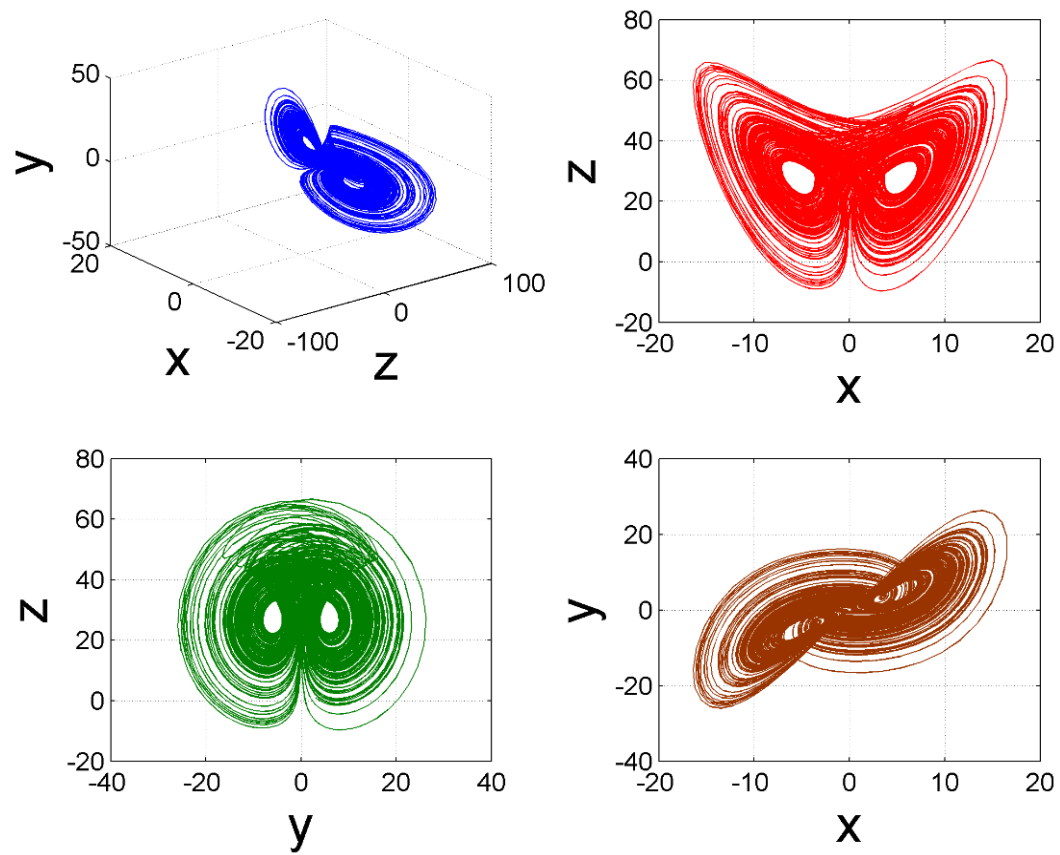

*Figure 88: Phase space dynamics of Lorenz-YZ56*

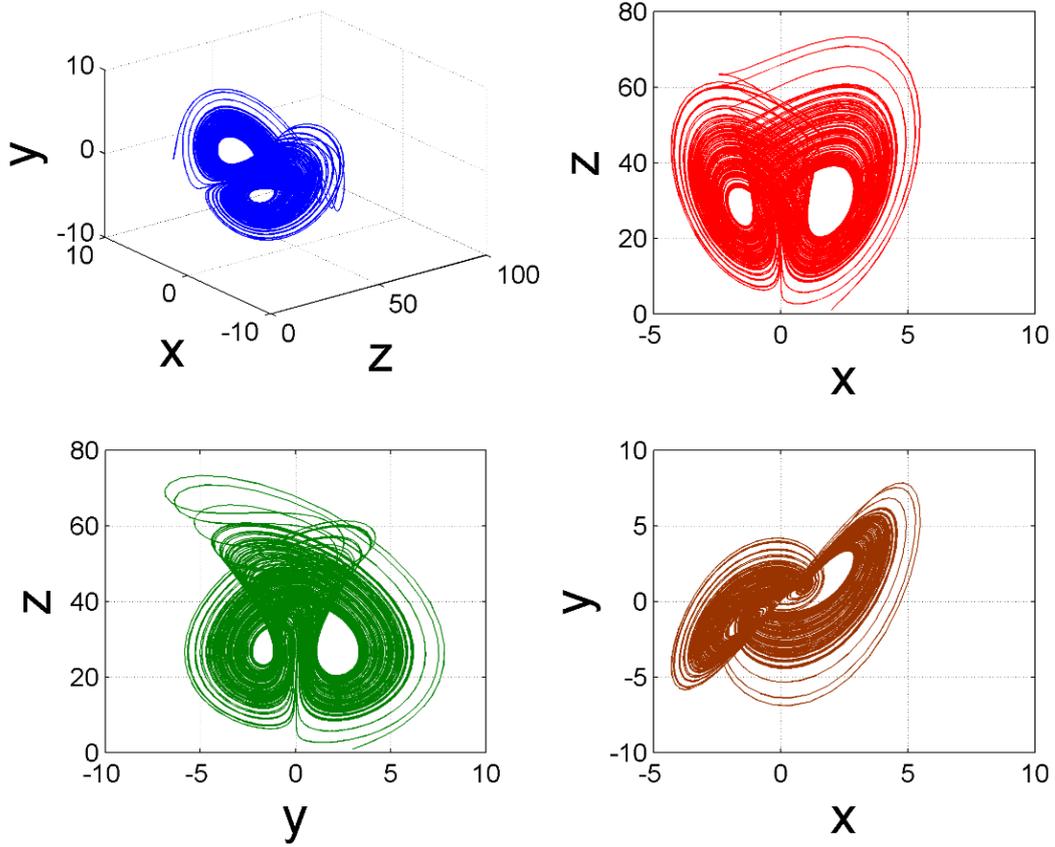

*Figure 89: Phase space dynamics of Lorenz-YZ57*

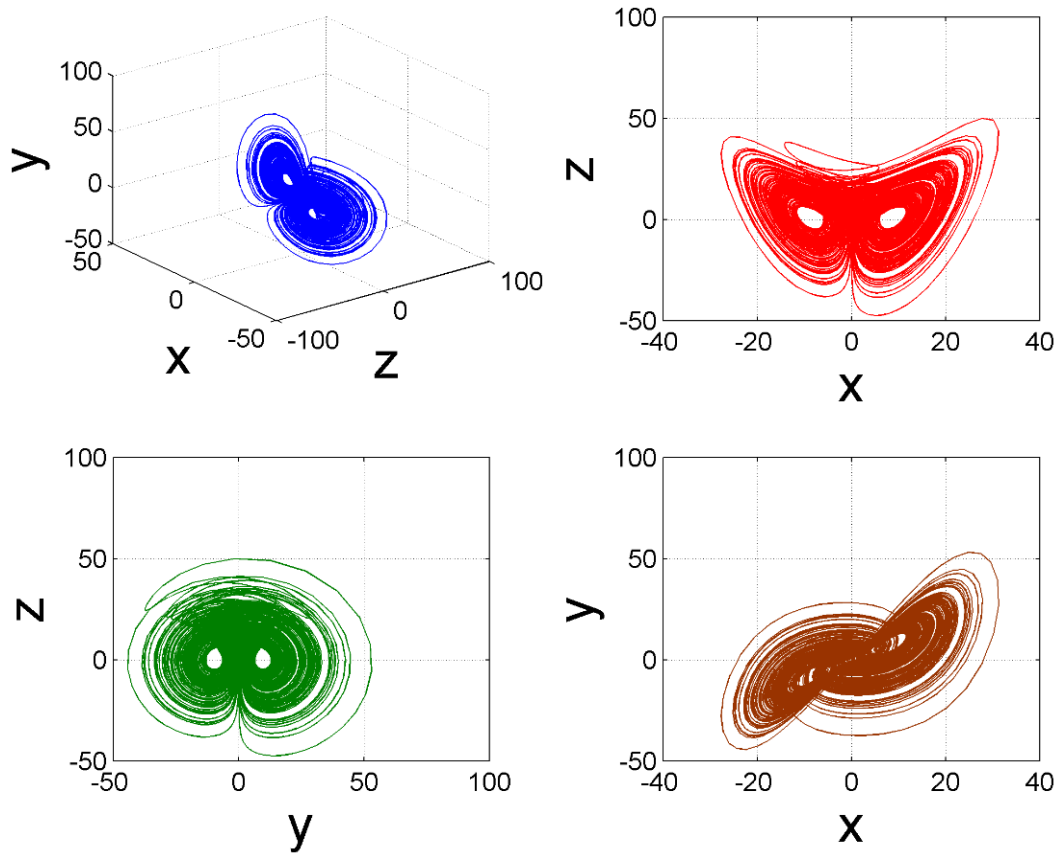

*Figure 90: Phase space dynamics of Lorenz-YZ58*

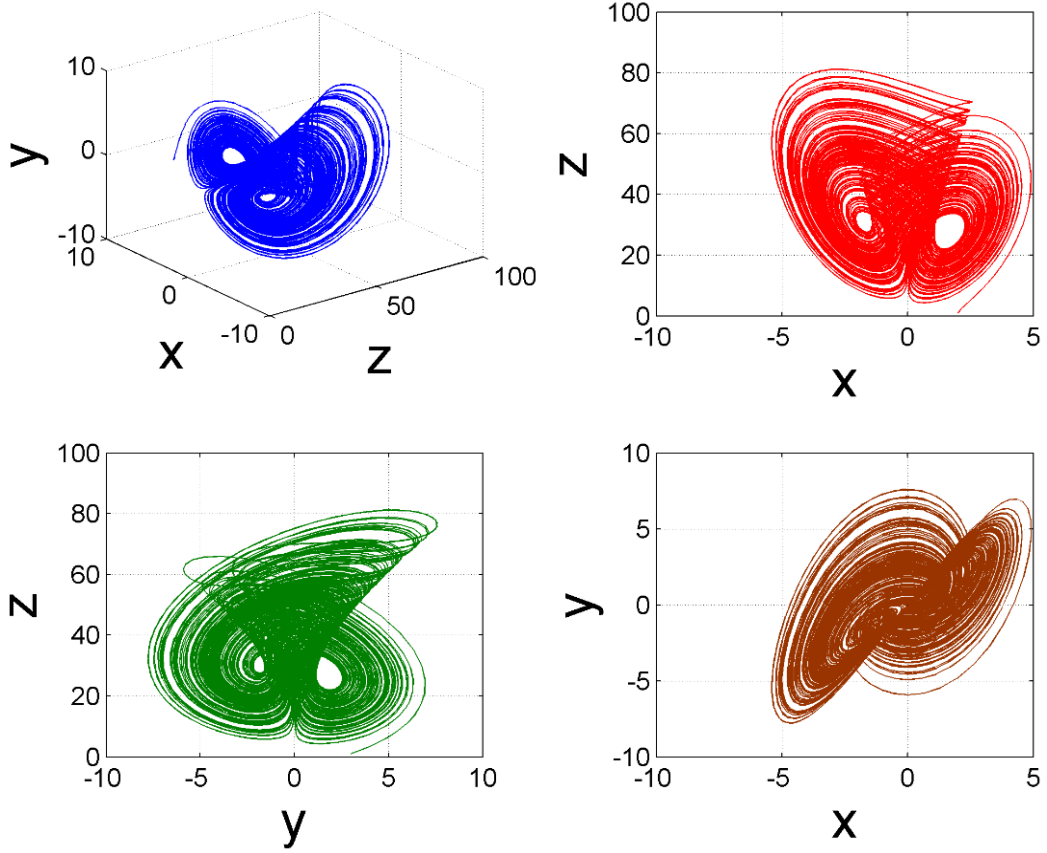

*Figure 91: Phase space dynamics of Lorenz-YZ59*

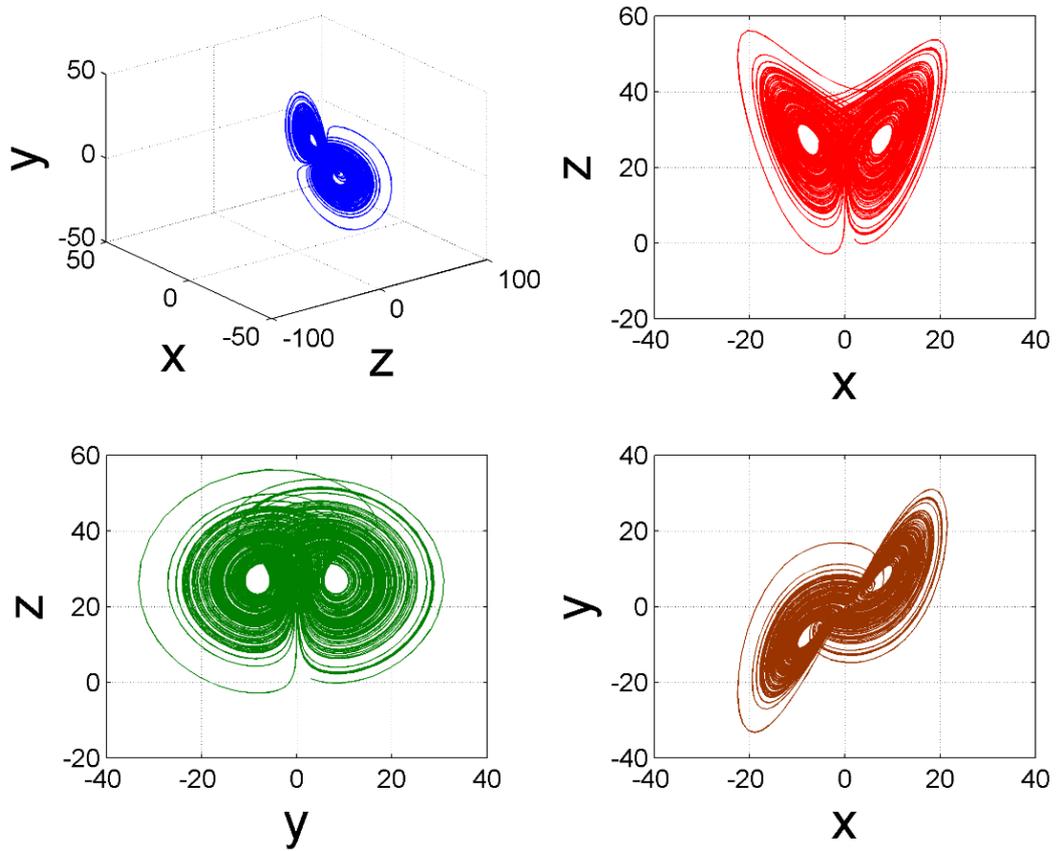

*Figure 92: Phase space dynamics of Lorenz-YZ60*

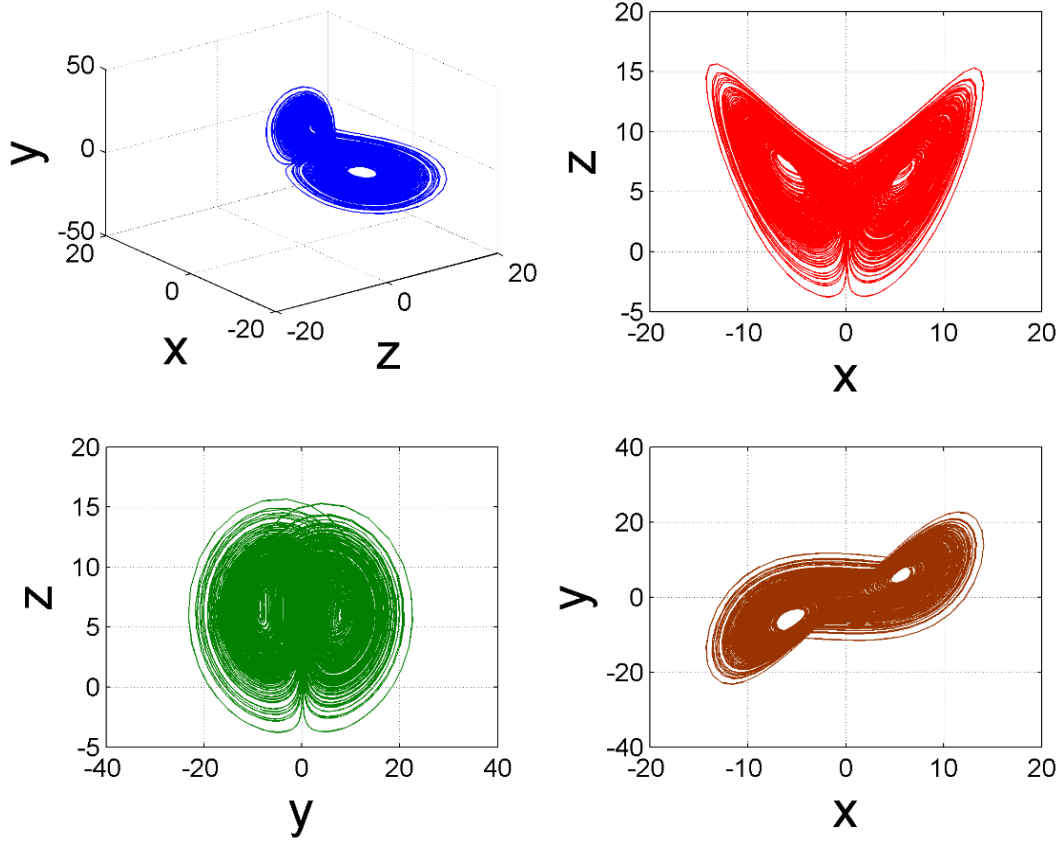

*Figure 93: Phase space dynamics of Lorenz-YZ61*

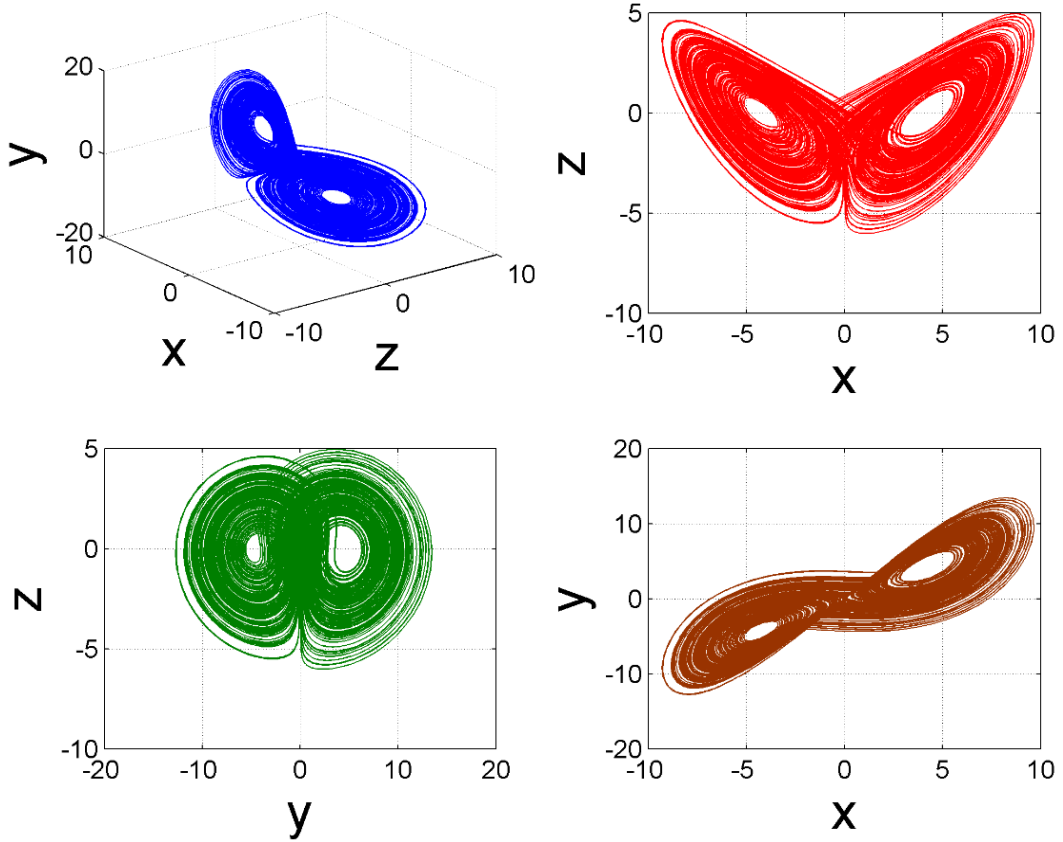

*Figure 94: Phase space dynamics of Lorenz-YZ62*

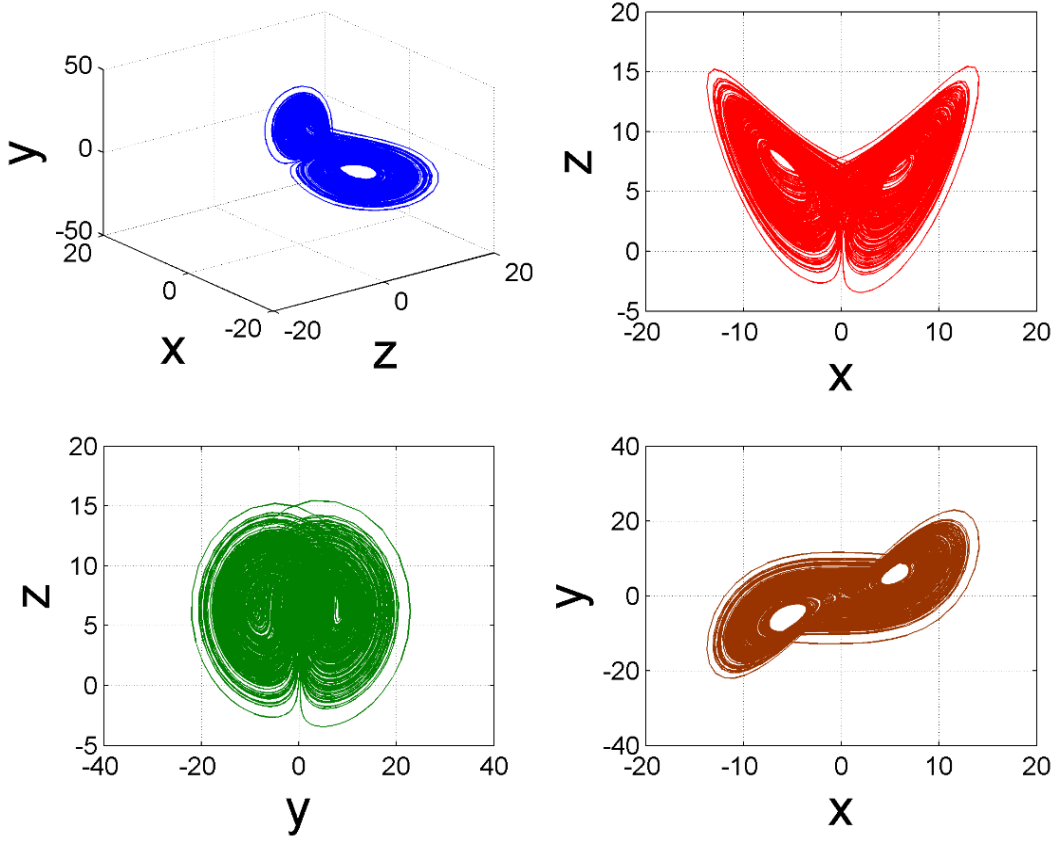

*Figure 95: Phase space dynamics of Lorenz-YZ63*

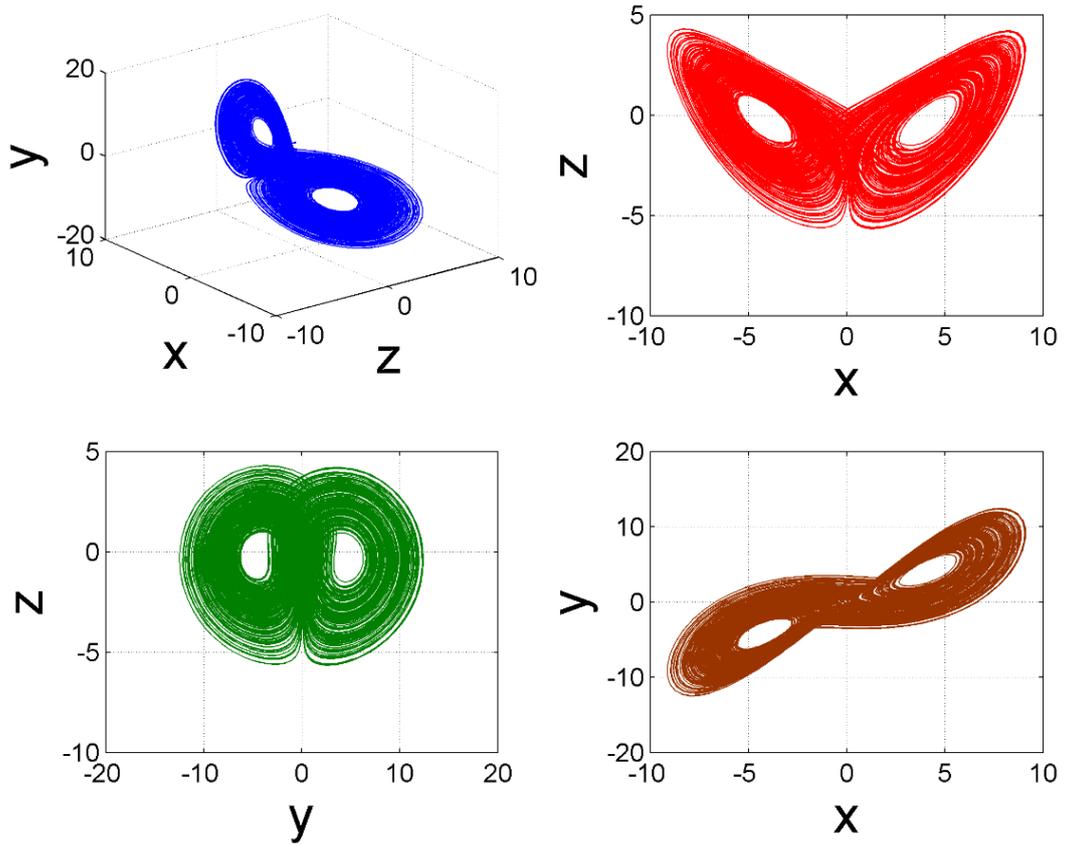

*Figure 96: Phase space dynamics of Lorenz-YZ64*

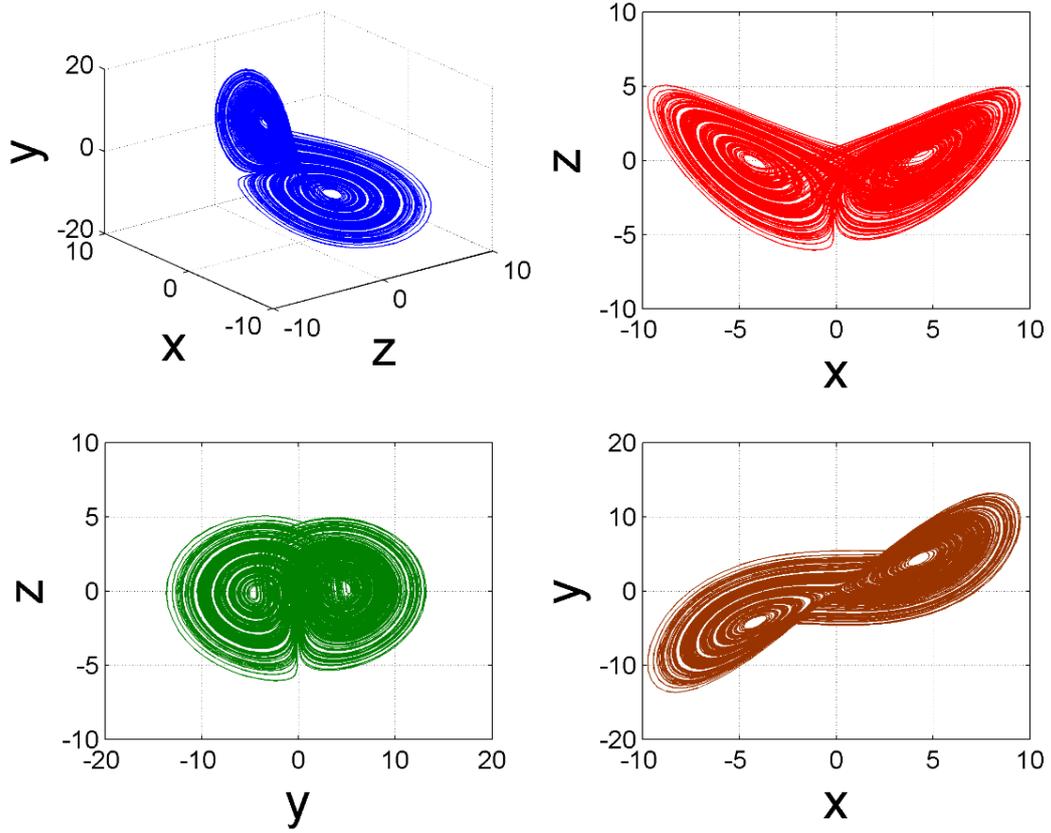

*Figure 97: Phase space dynamics of Lorenz-YZ65*

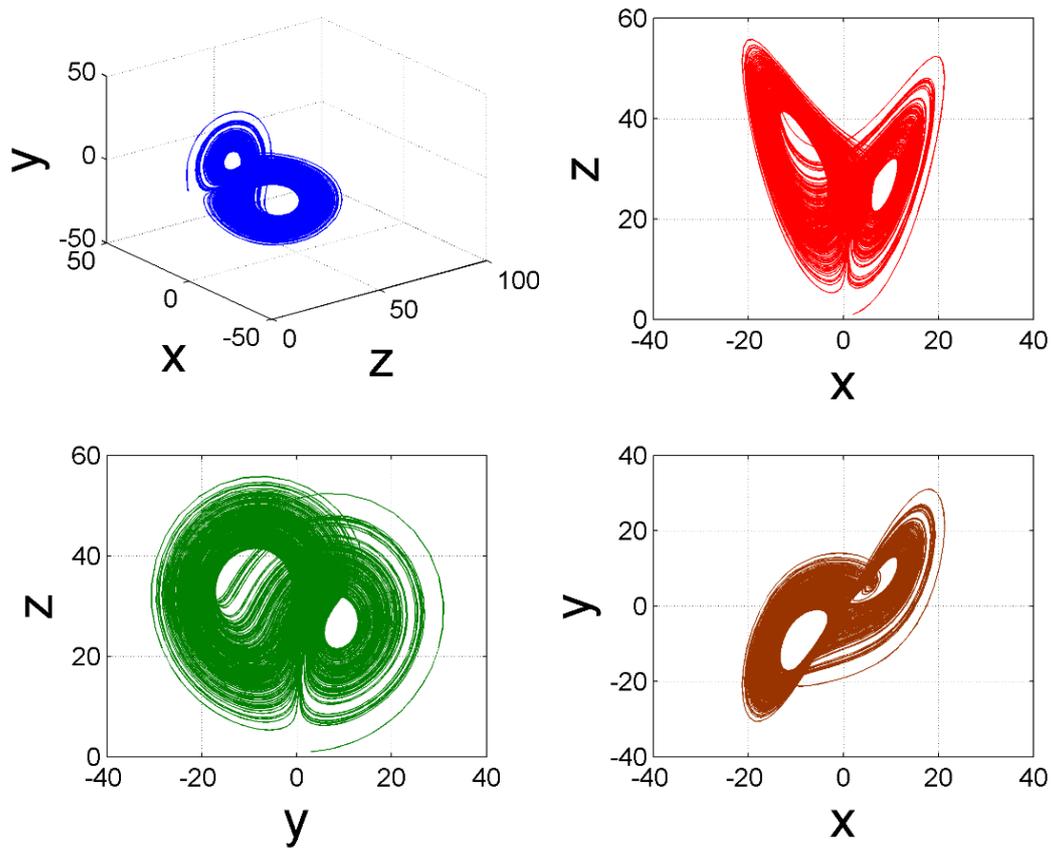

*Figure 98: Phase space dynamics of Lorenz-YZ66*

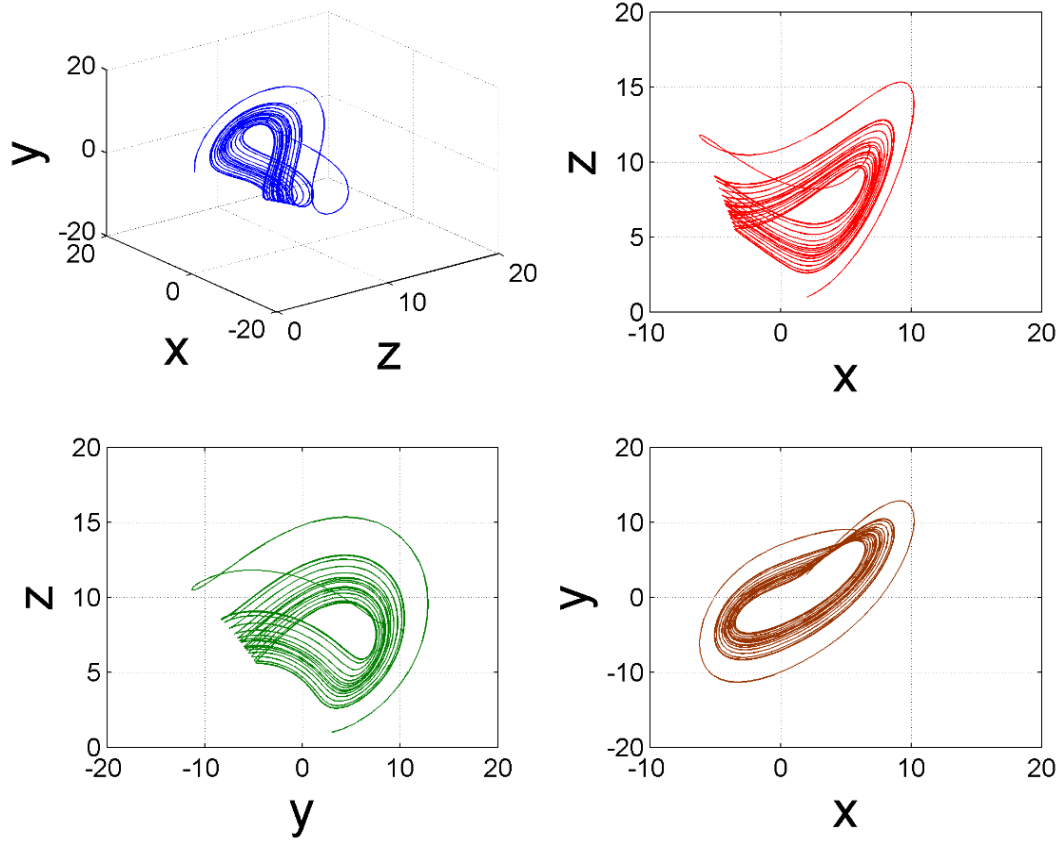

*Figure 99: Phase space dynamics of Lorenz-YZ67*

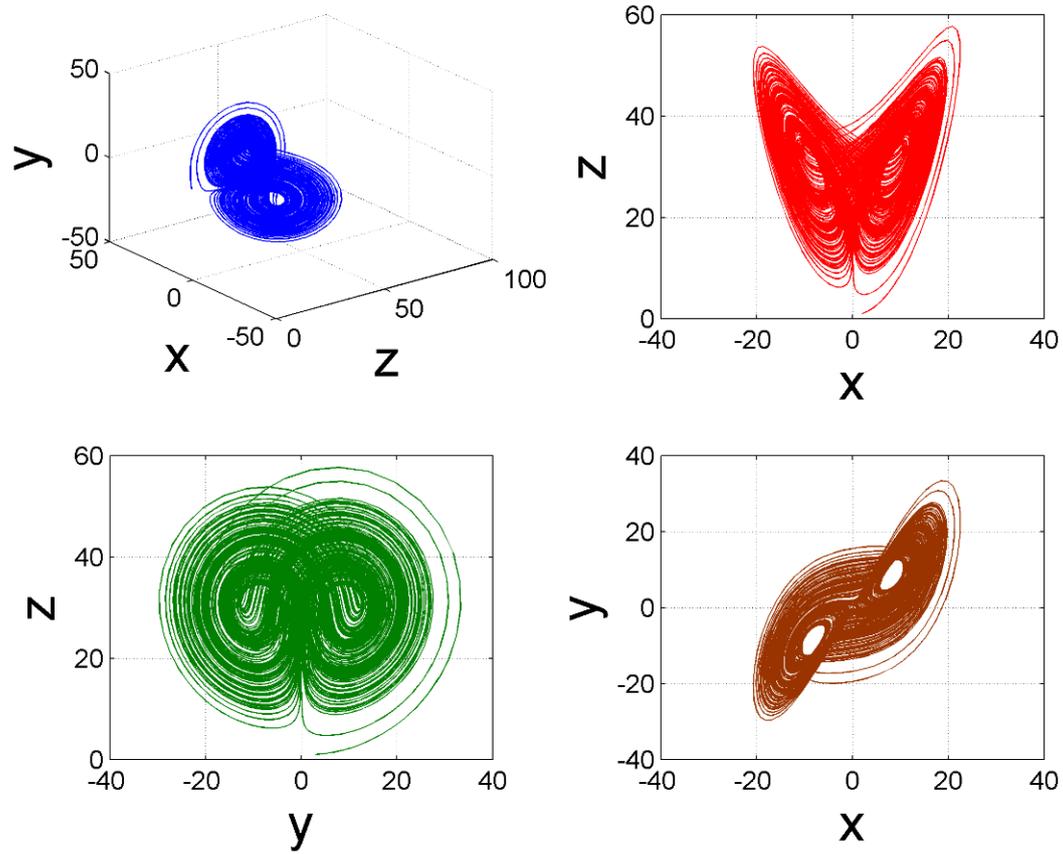

*Figure 100: Phase space dynamics of Lorenz-YZ68*

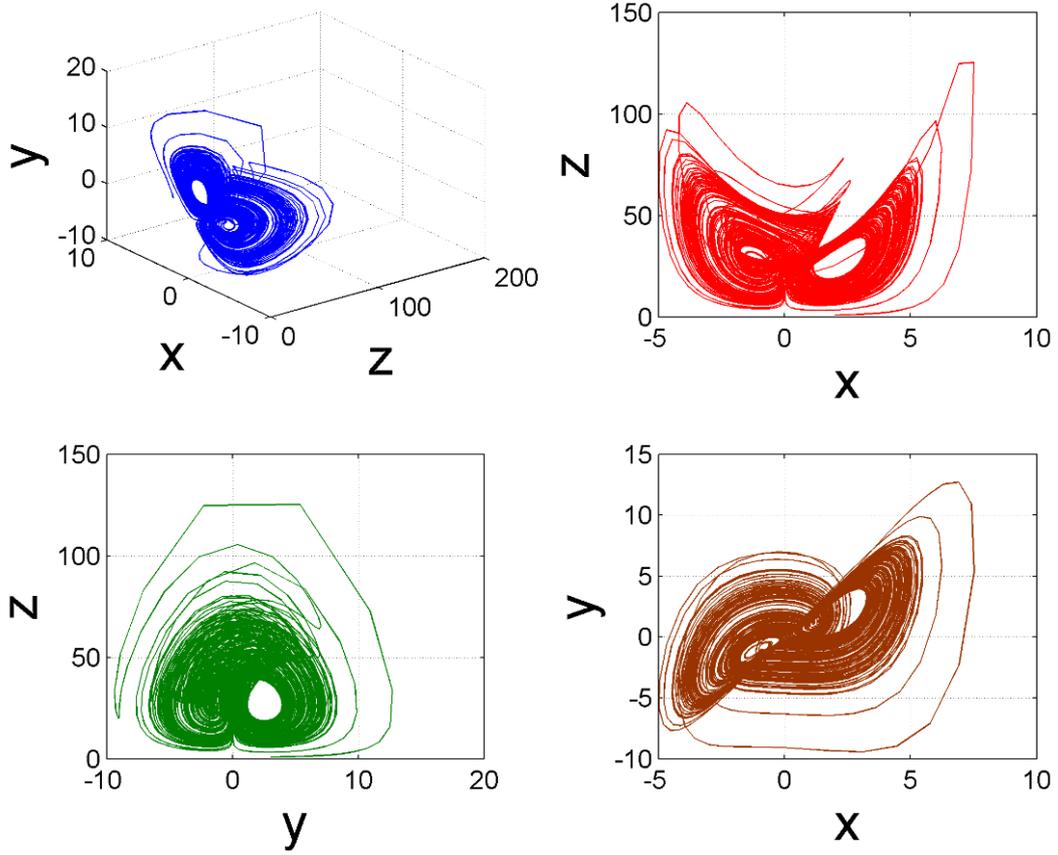

*Figure 101: Phase space dynamics of Lorenz-YZ69*

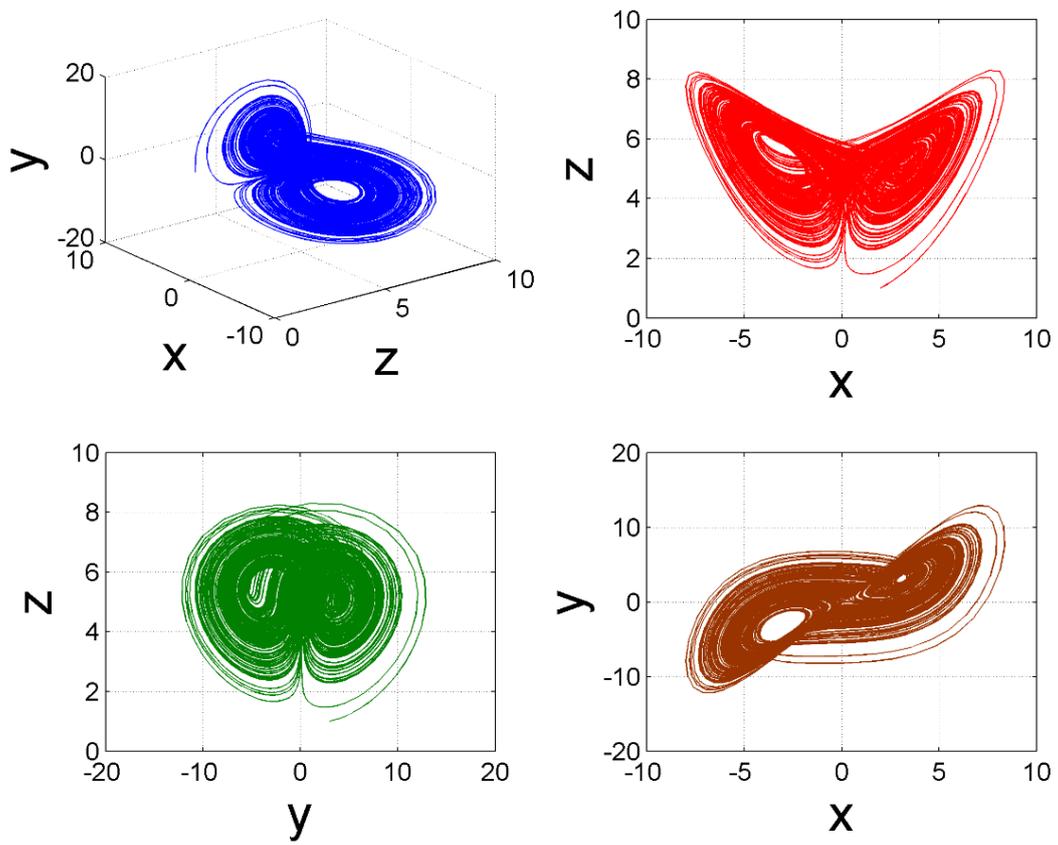

*Figure 102: Phase space dynamics of Lorenz-YZ70*

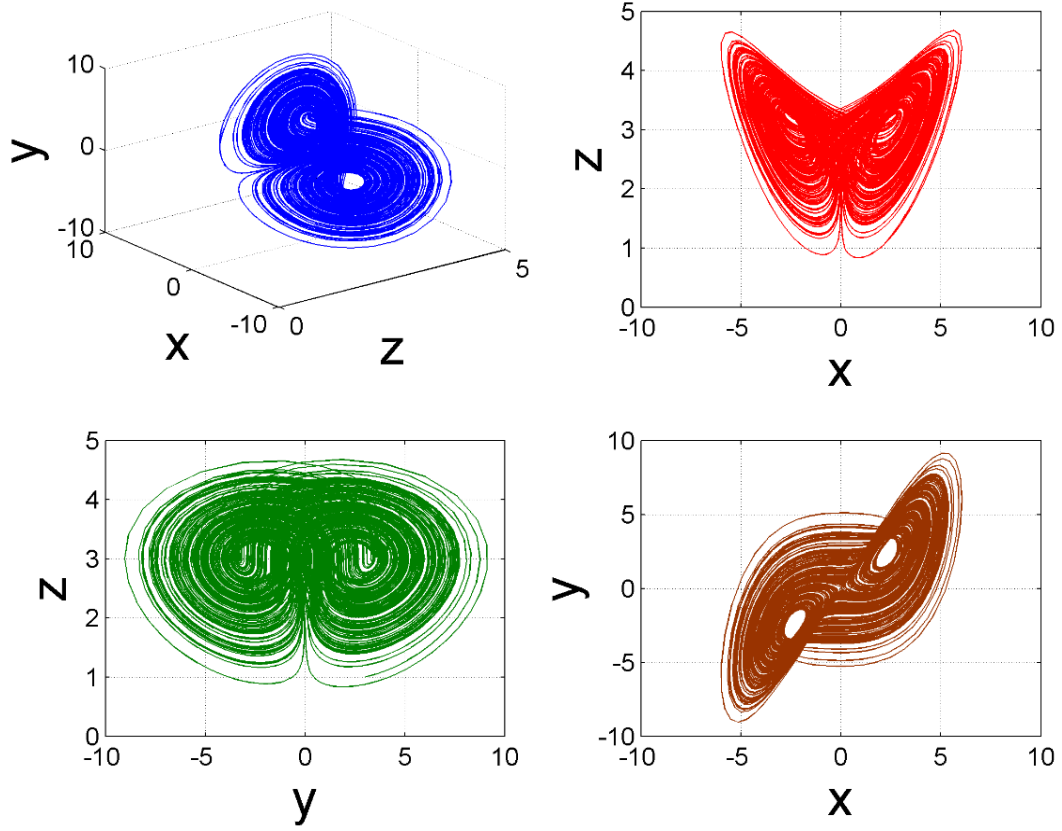

*Figure 103: Phase space dynamics of Lorenz-YZ71*

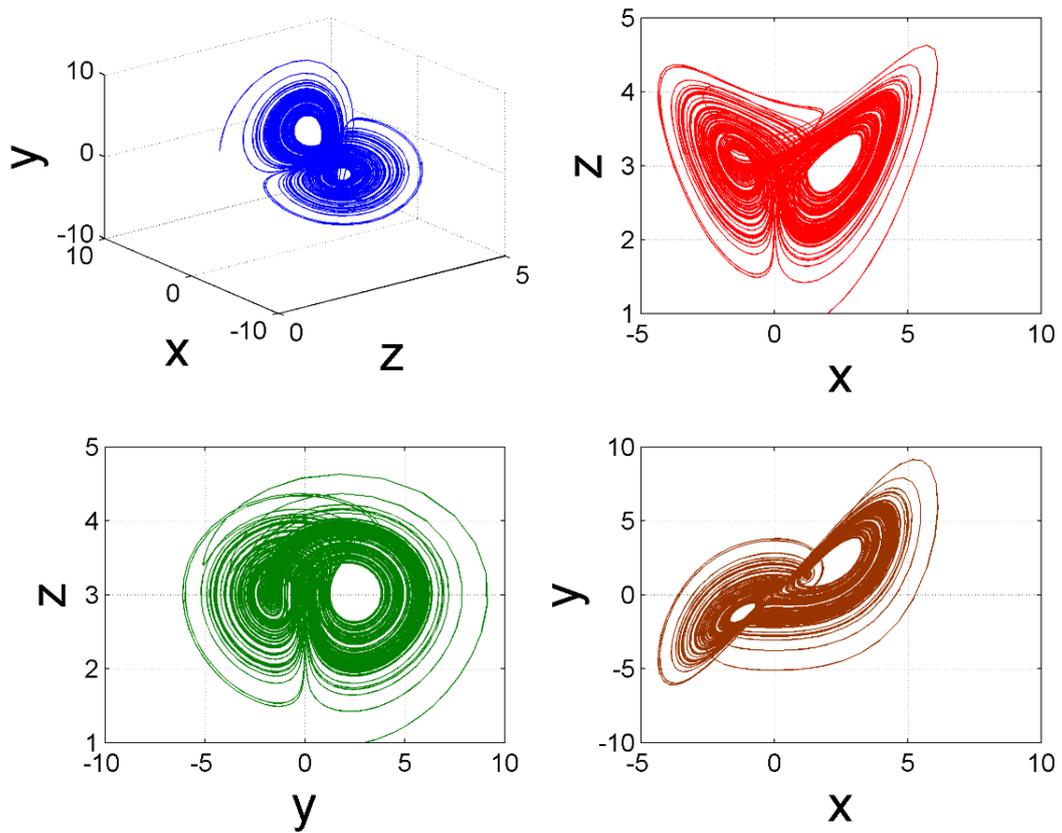

*Figure 104: Phase space dynamics of Lorenz-YZ72*

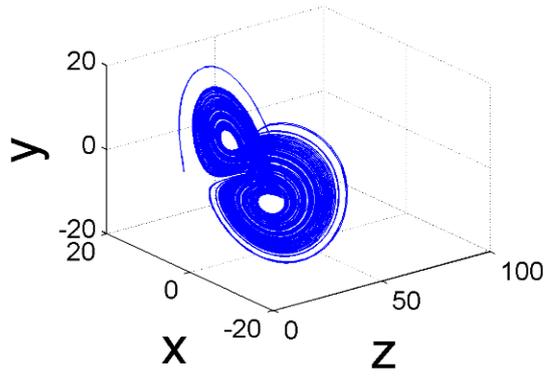
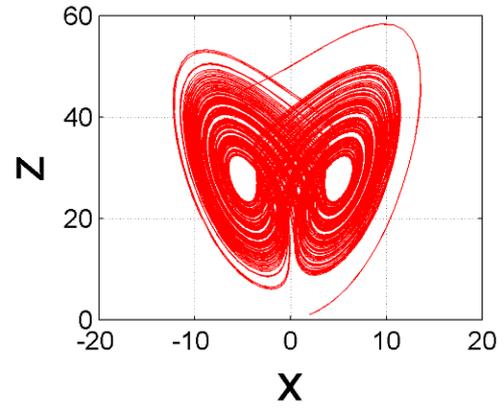
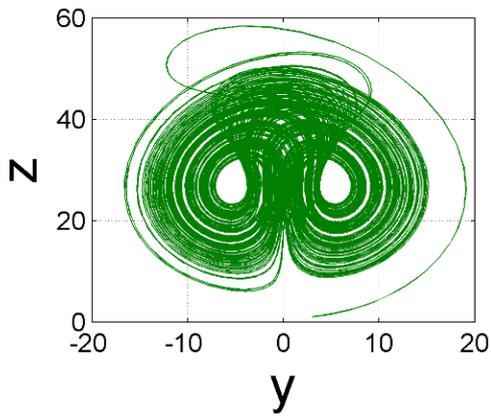
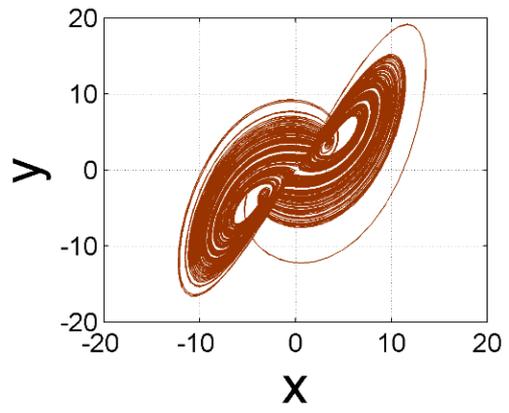

*Figure 105: Phase space dynamics of Lorenz-YZ73*

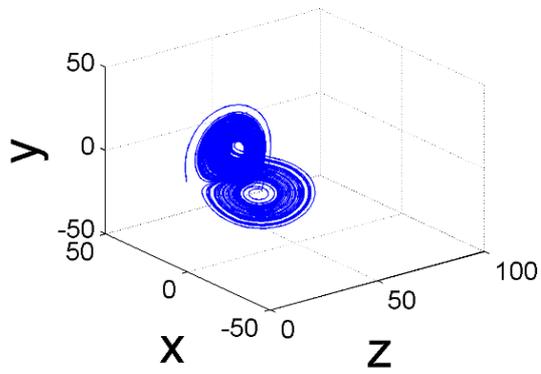
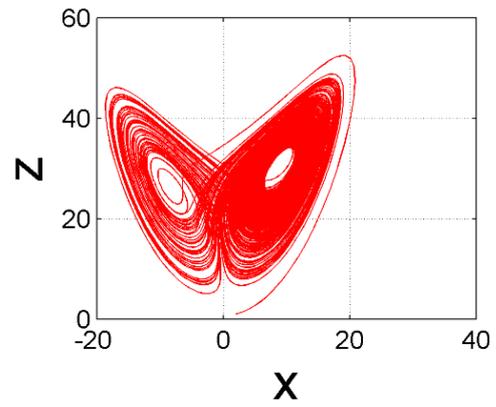
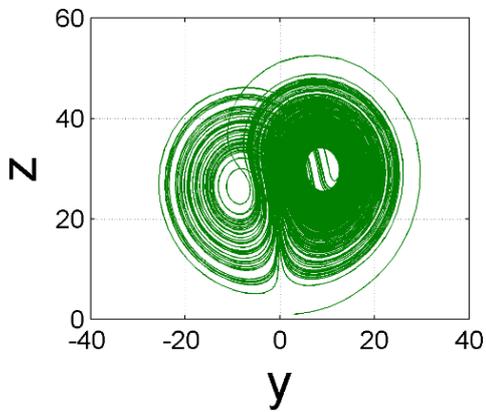
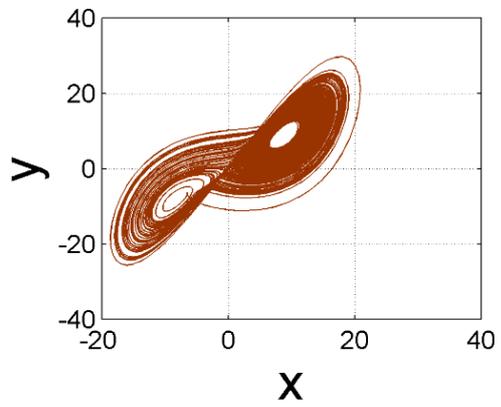

*Figure 106: Phase space dynamics of Lorenz-YZ74*

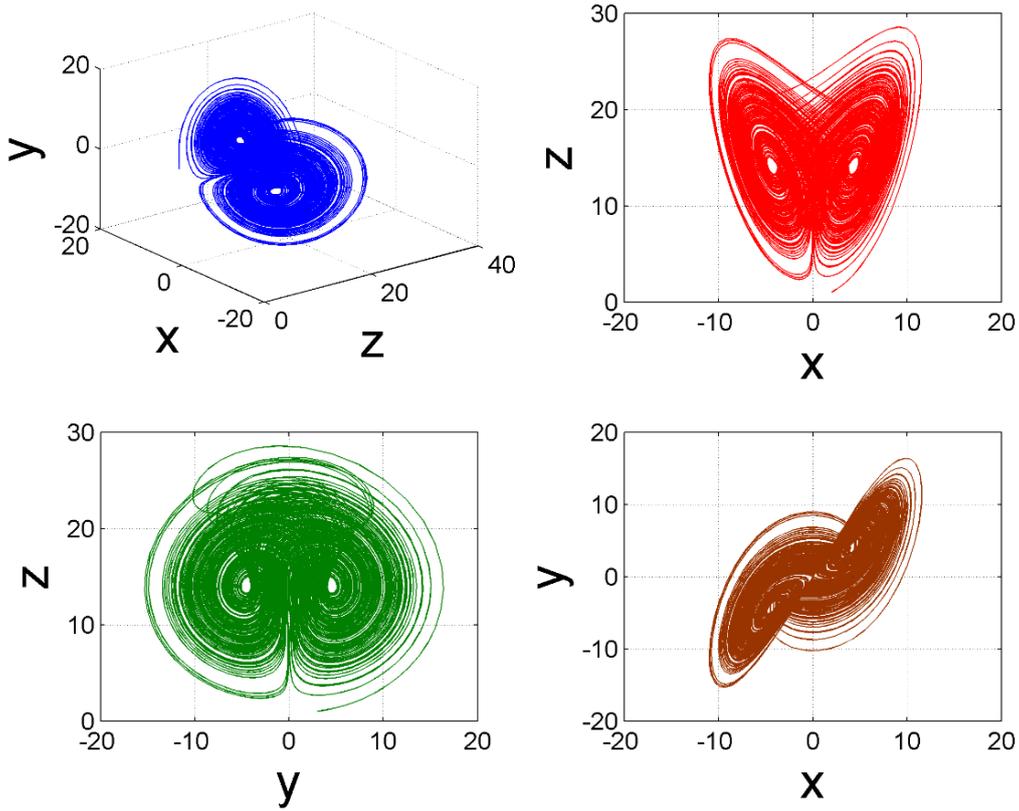

*Figure 107: Phase space dynamics of Lorenz-YZ75*

### 3. Phase portraits of generalized Lorenz-XZ family of attractors

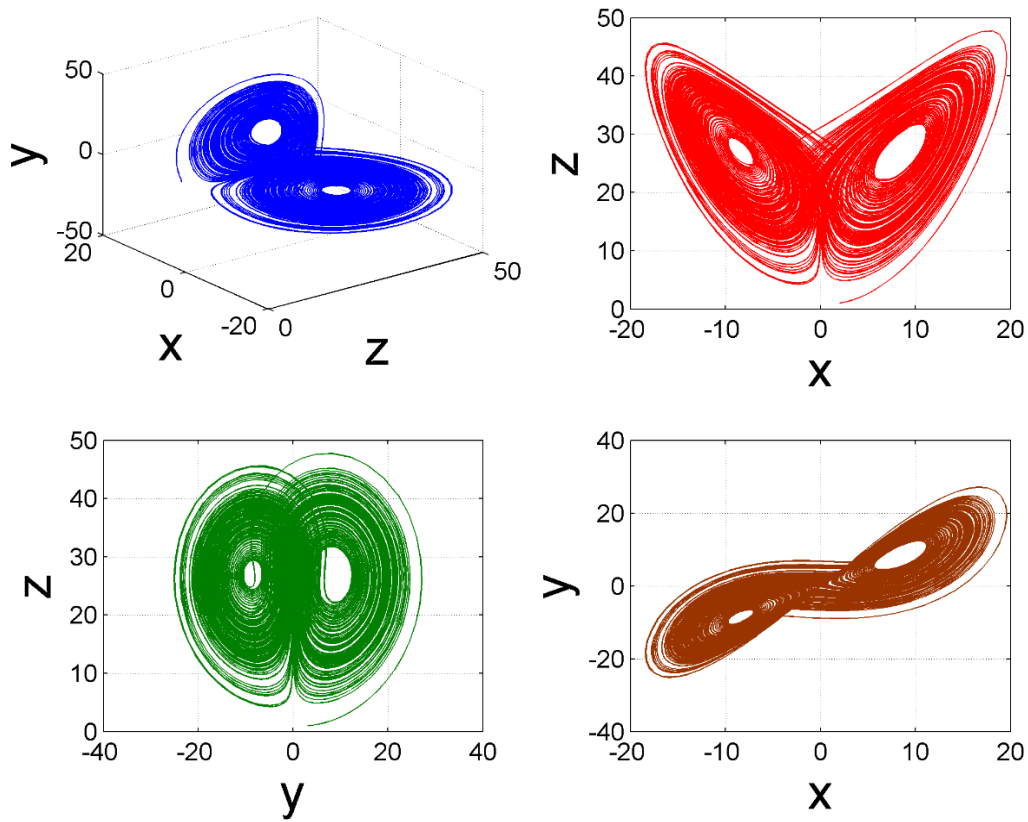

*Figure 108: Phase space dynamics of Lorenz-XZ1*

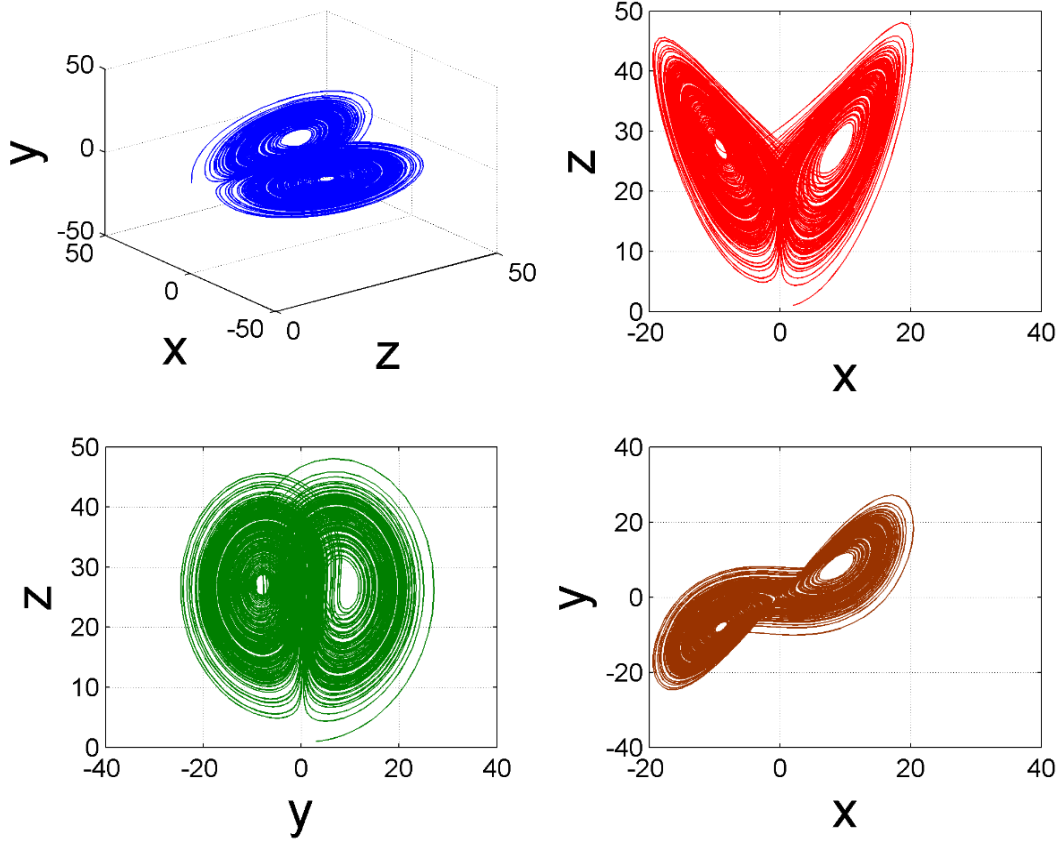

*Figure 109: Phase space dynamics of Lorenz-XZ2*

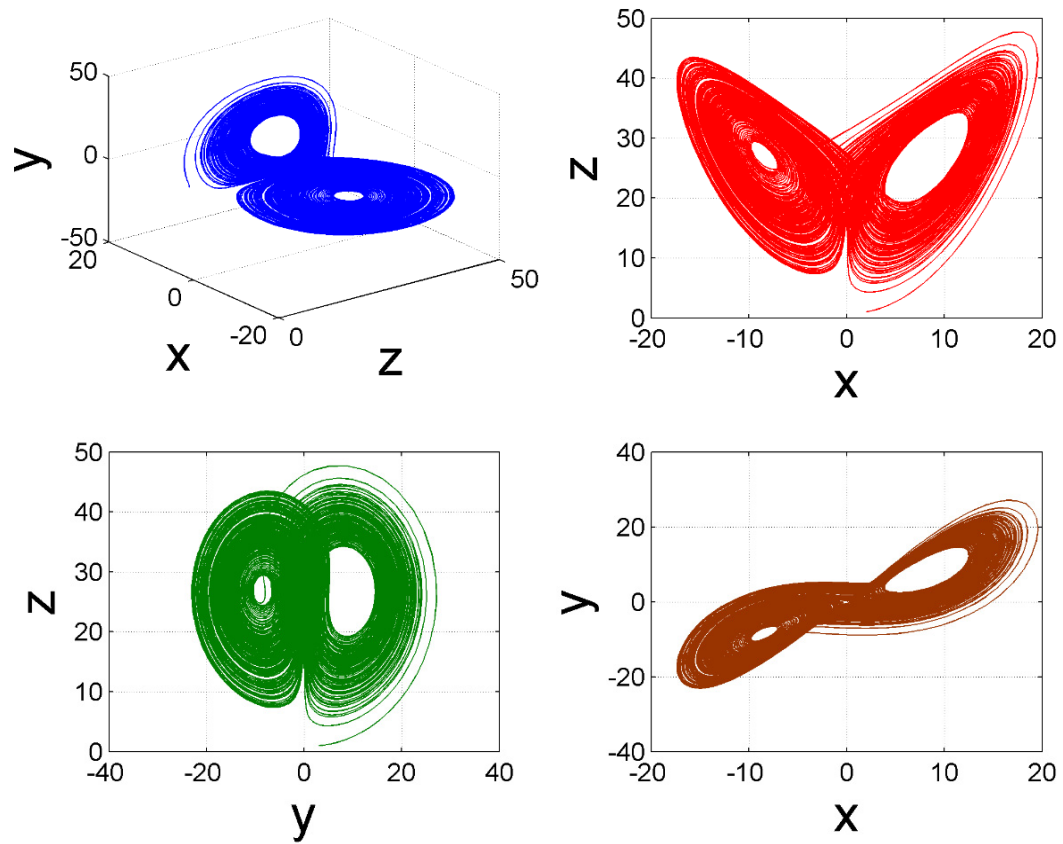

*Figure 110: Phase space dynamics of Lorenz-XZ3*

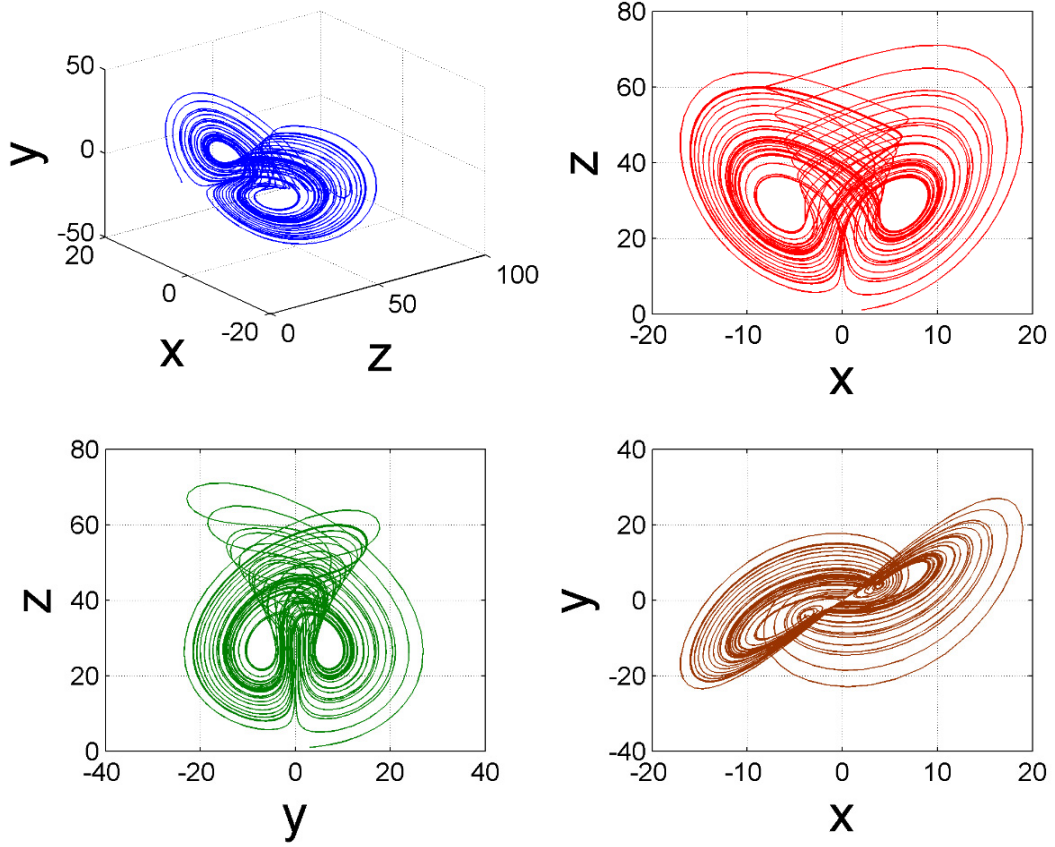

*Figure 111: Phase space dynamics of Lorenz-XZ4*

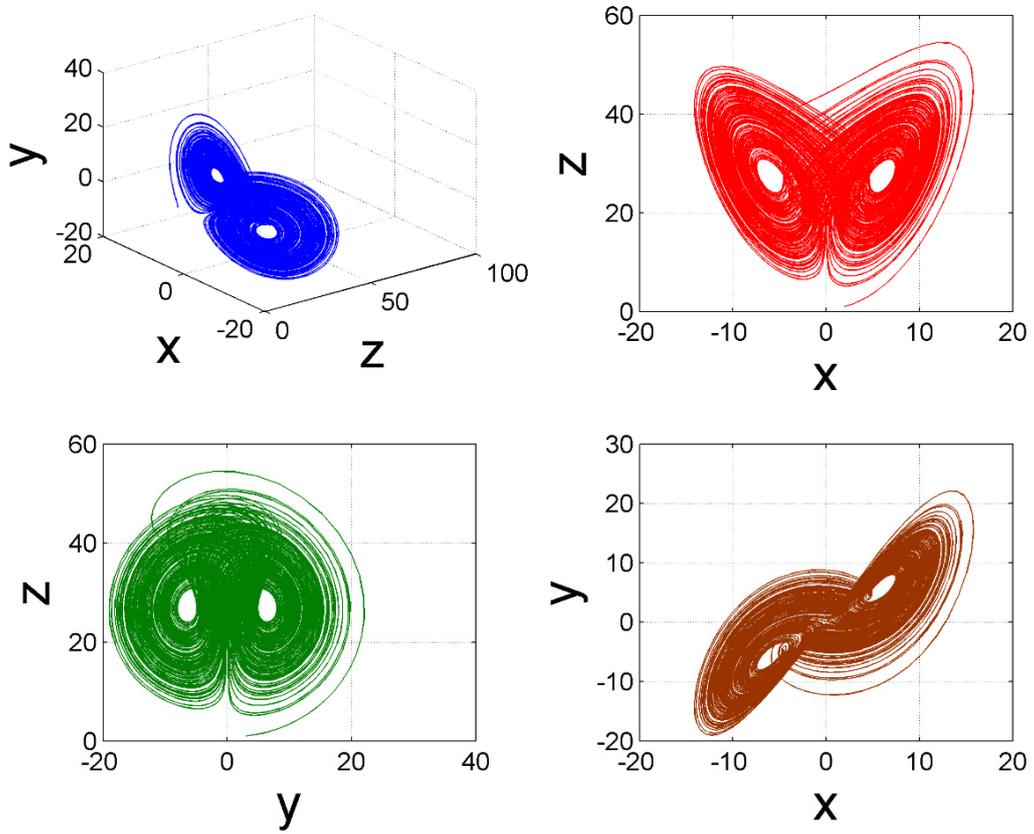

*Figure 112: Phase space dynamics of Lorenz-XZ5*

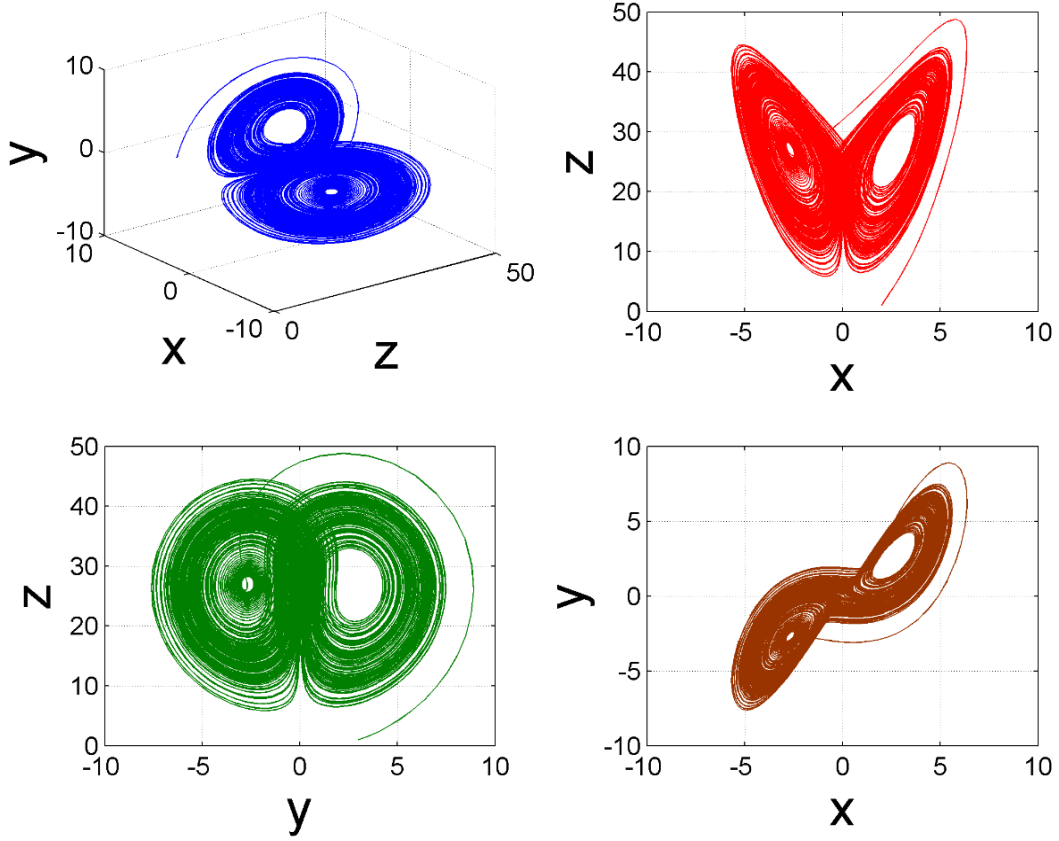

*Figure 113: Phase space dynamics of Lorenz-XZ6*

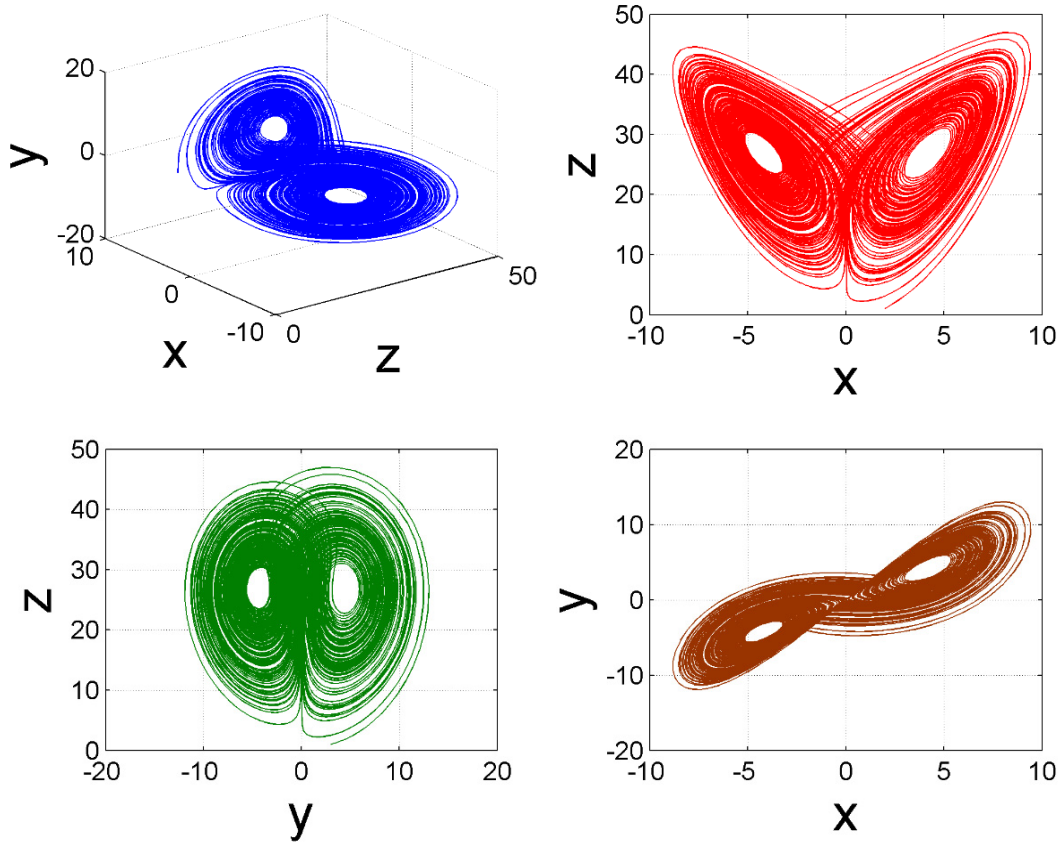

*Figure 114: Phase space dynamics of Lorenz-XZ7*

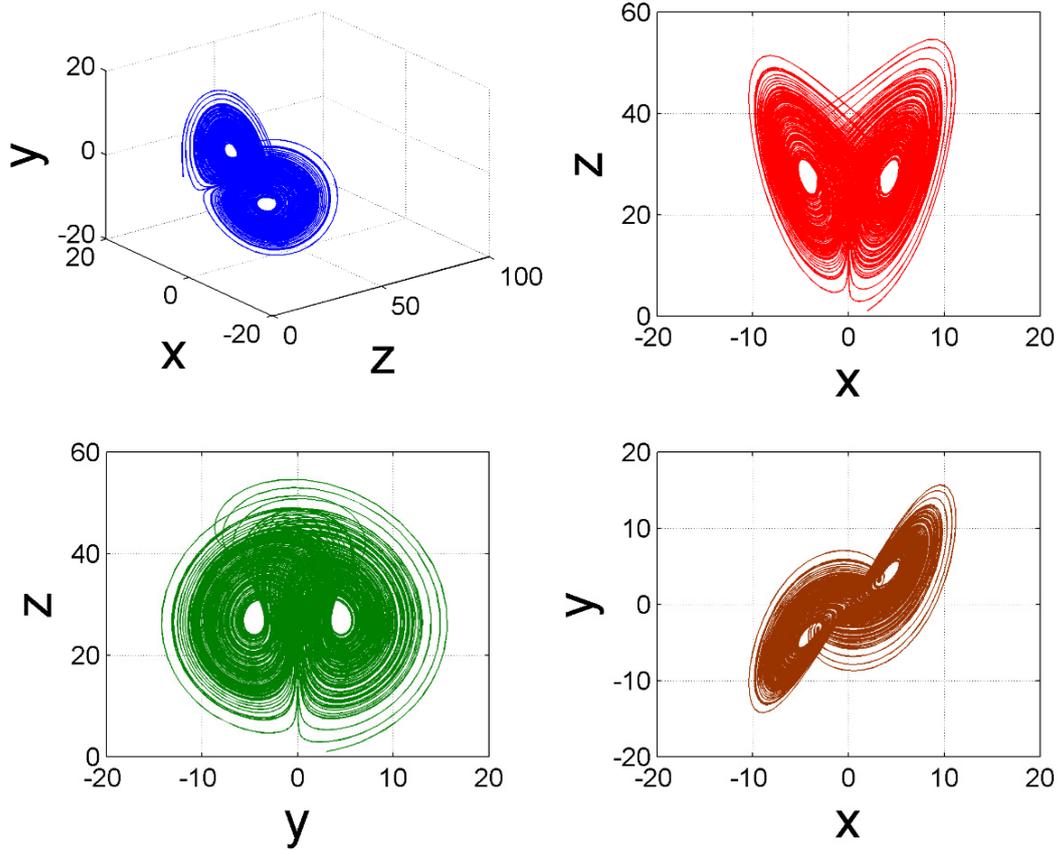

*Figure 115: Phase space dynamics of Lorenz-XZ8*

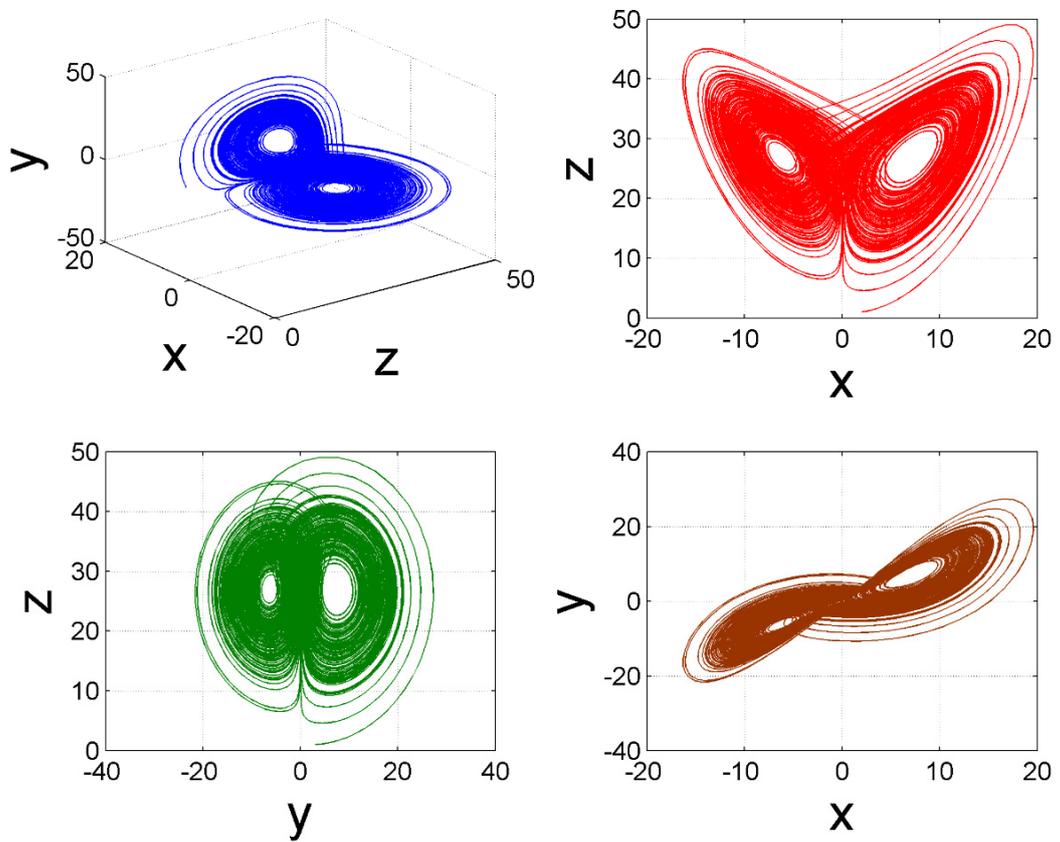

*Figure 116: Phase space dynamics of Lorenz-XZ9*

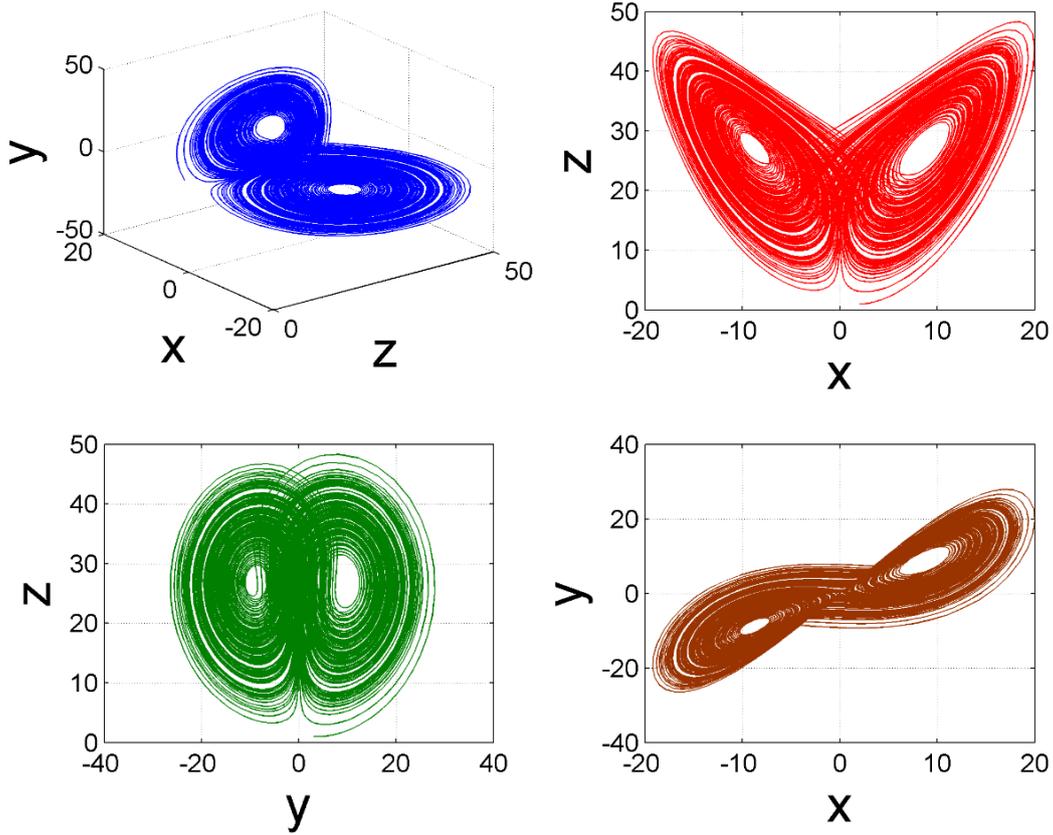

*Figure 117: Phase space dynamics of Lorenz-XZ10*

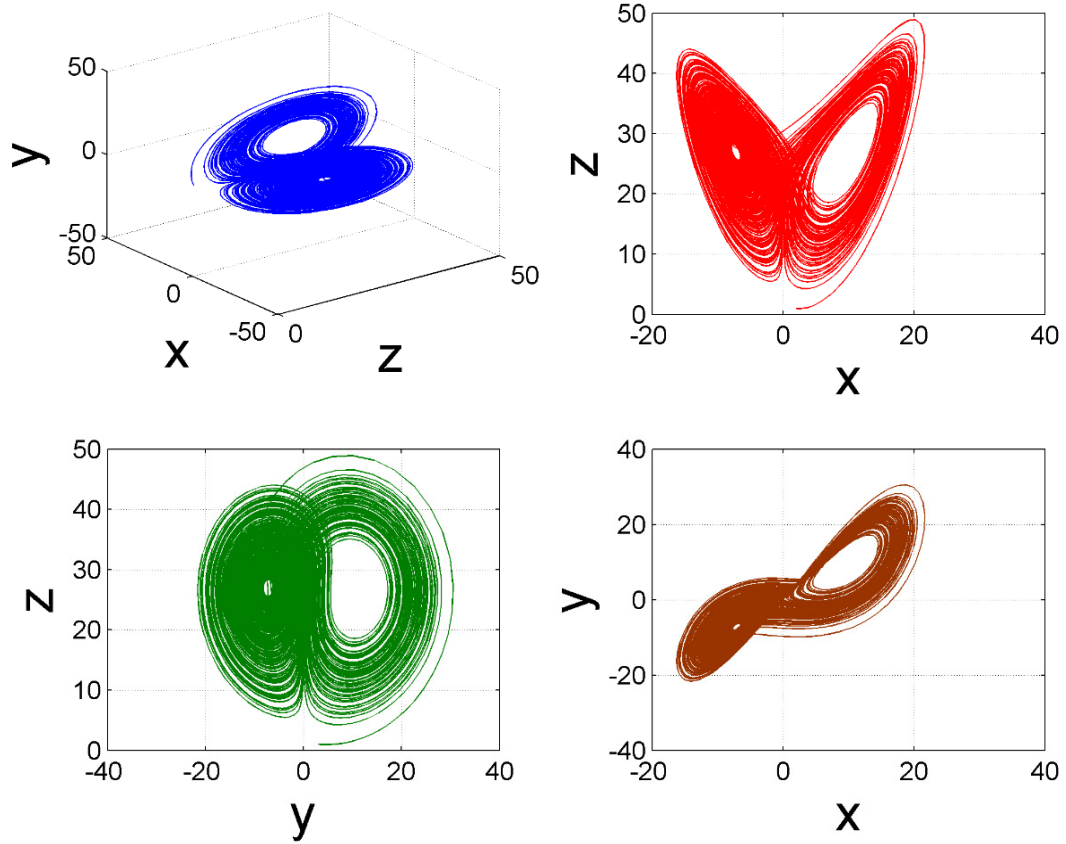

*Figure 118: Phase space dynamics of Lorenz-XZ12*

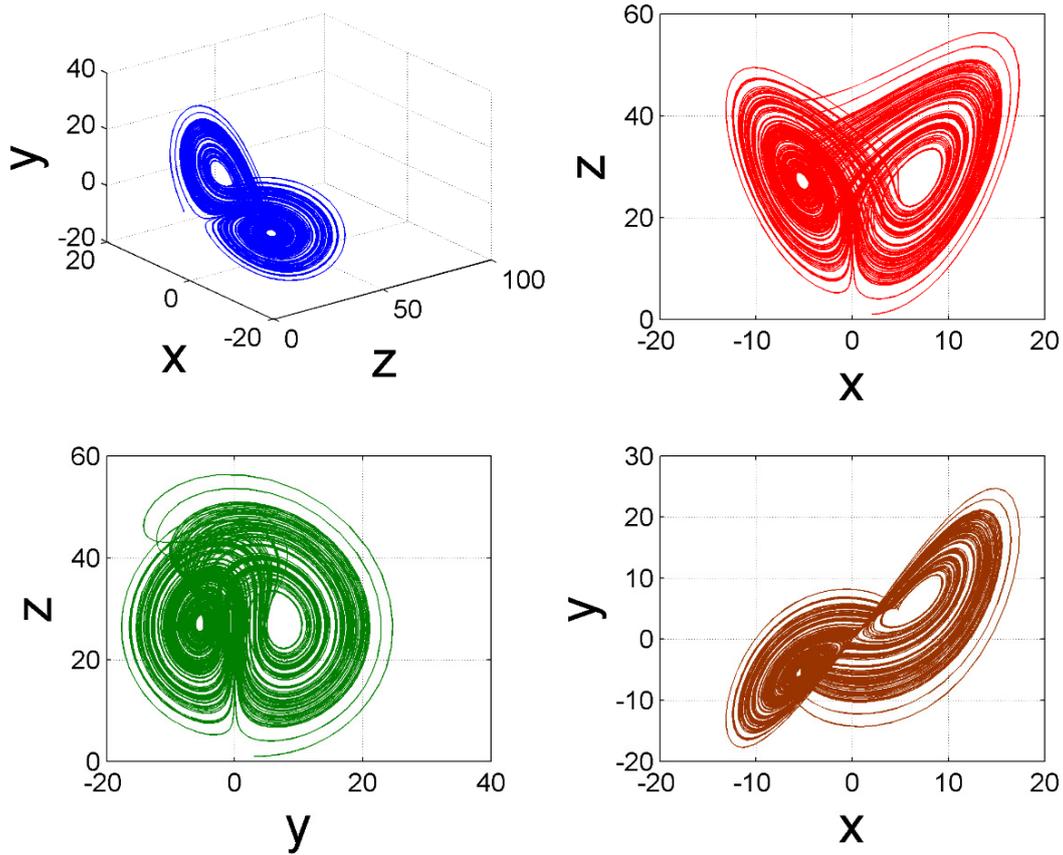

*Figure 119: Phase space dynamics of Lorenz-XZ13*

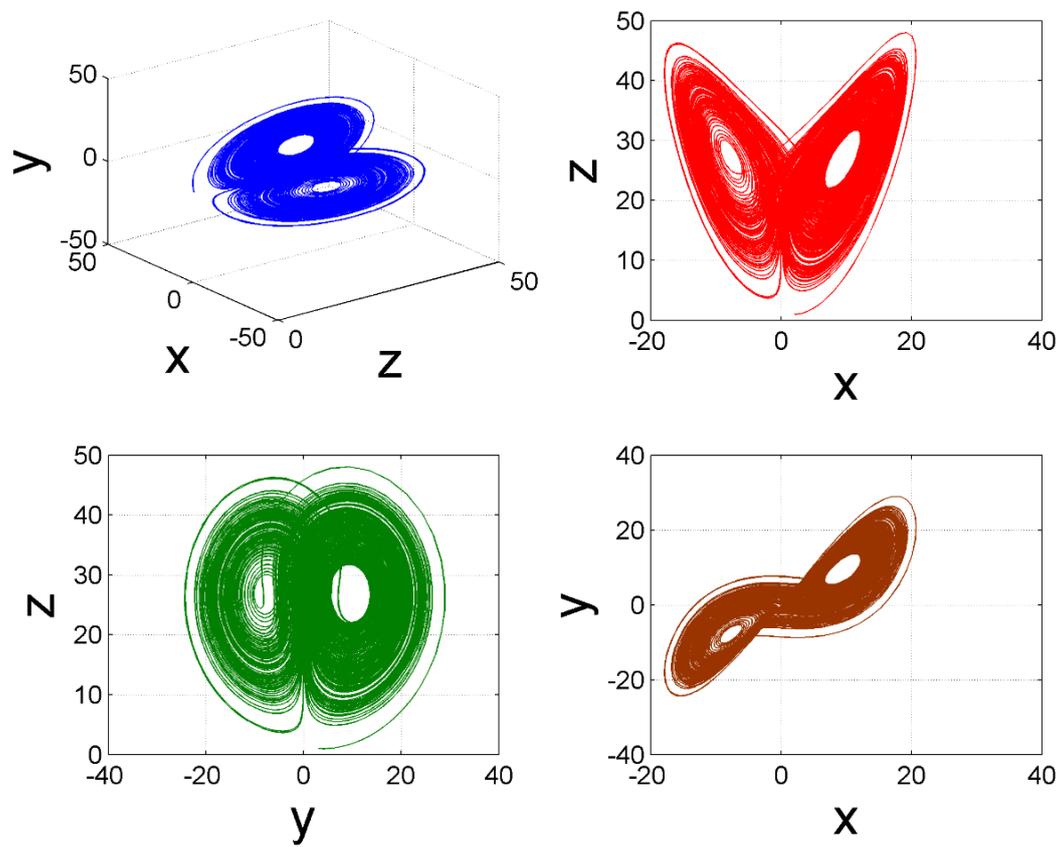

*Figure 120: Phase space dynamics of Lorenz-XZ14*

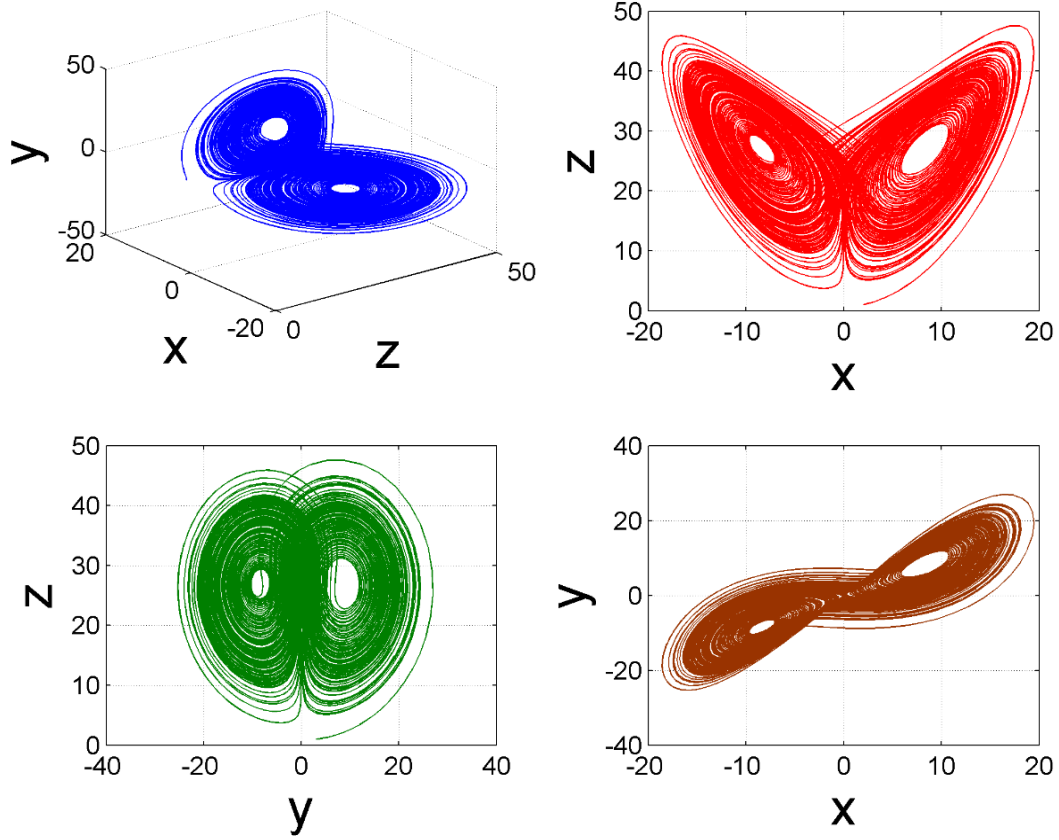

*Figure 121: Phase space dynamics of Lorenz-XZ15*

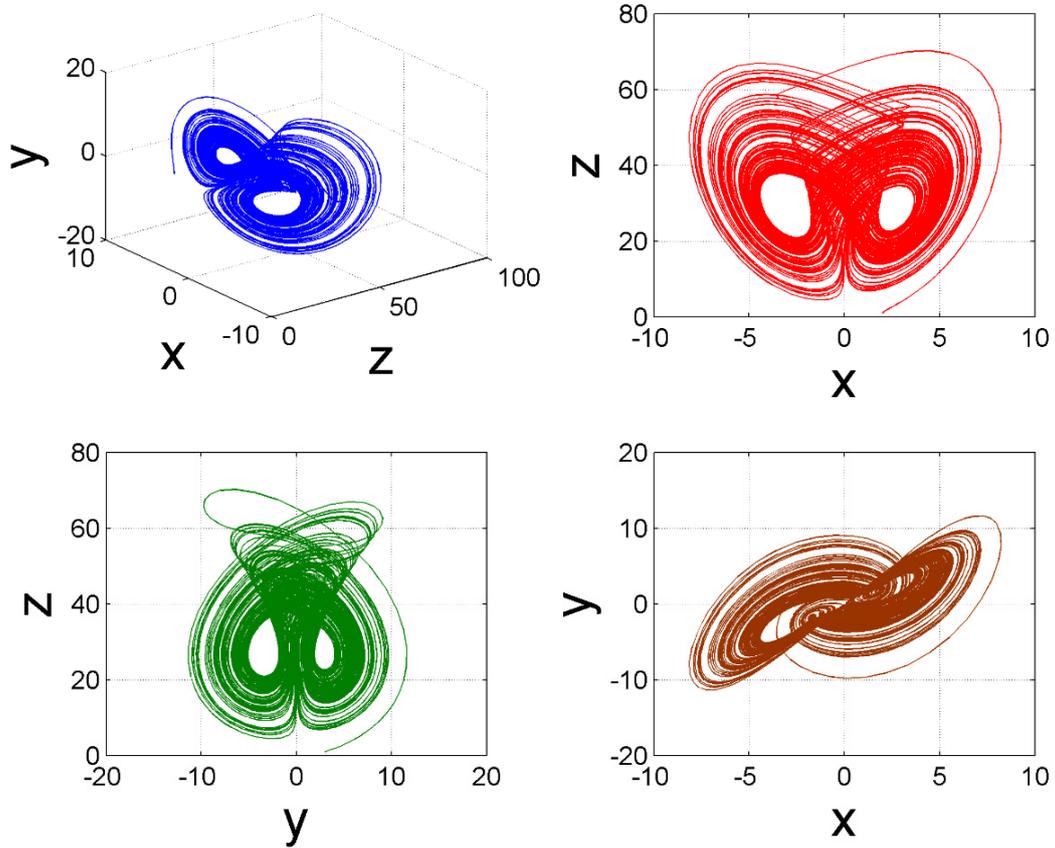

*Figure 122: Phase space dynamics of Lorenz-XZ17*

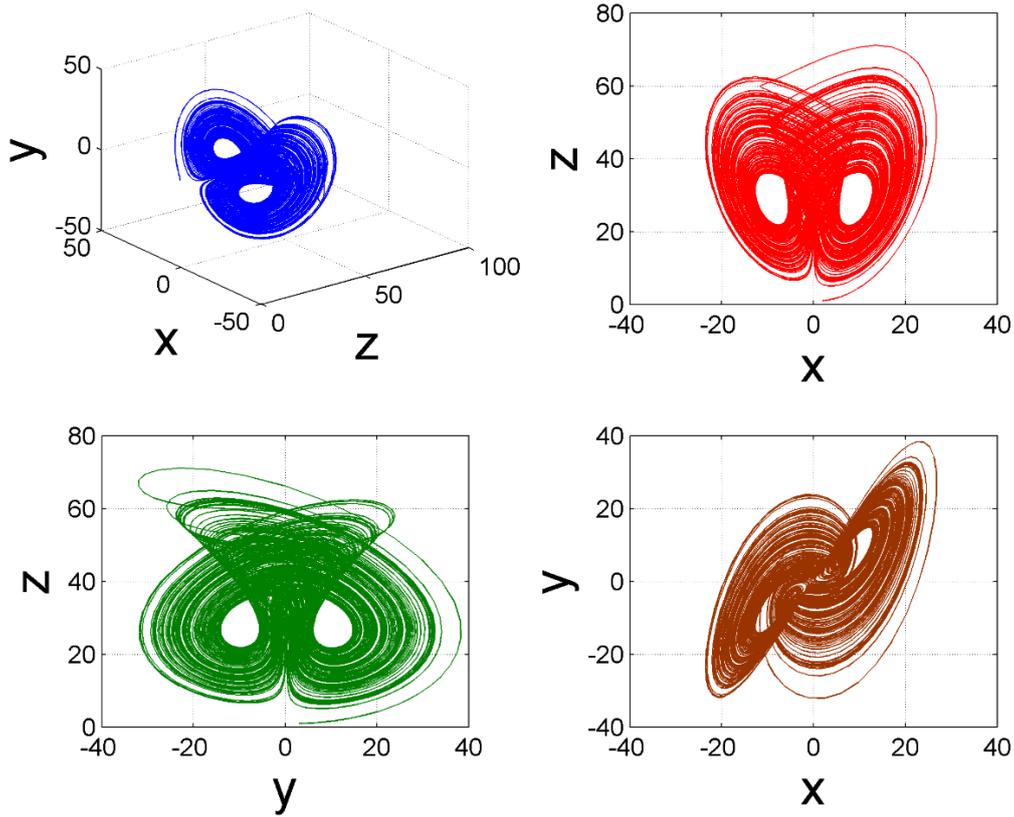

*Figure 123: Phase space dynamics of Lorenz-XZ18*

### 4. Phase portraits of generalized Lorenz-XYZ family of attractors

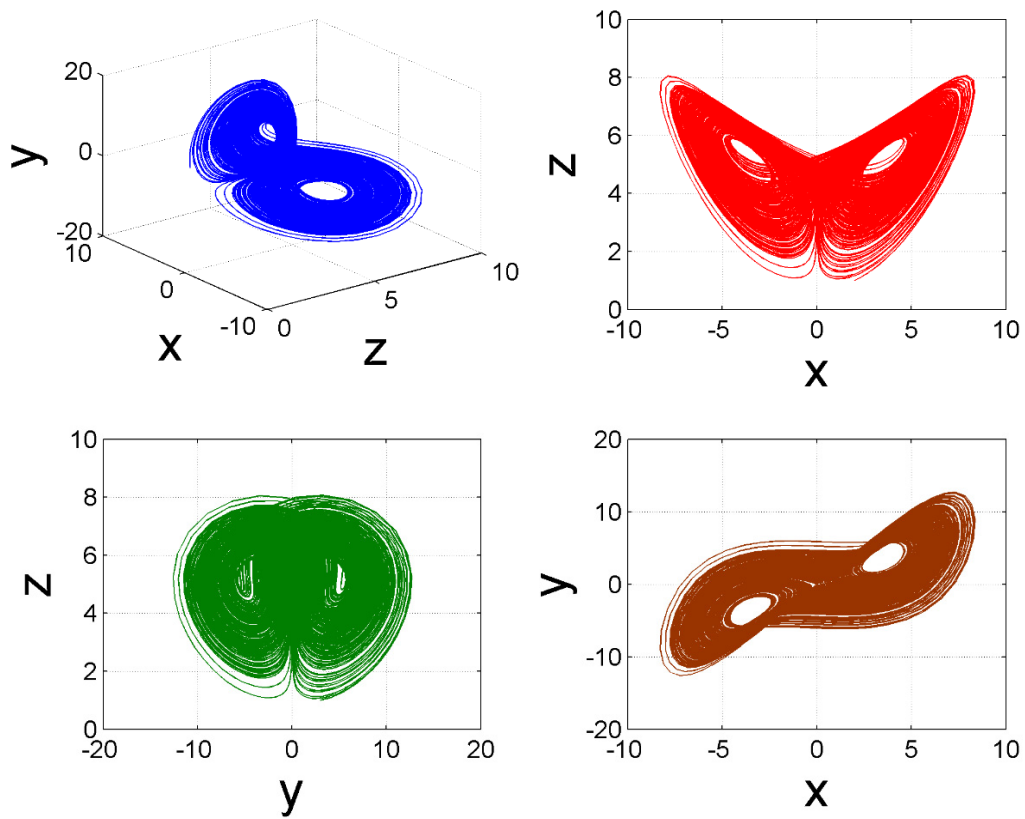

*Figure 124: Phase space dynamics of Lorenz-XYZ1*

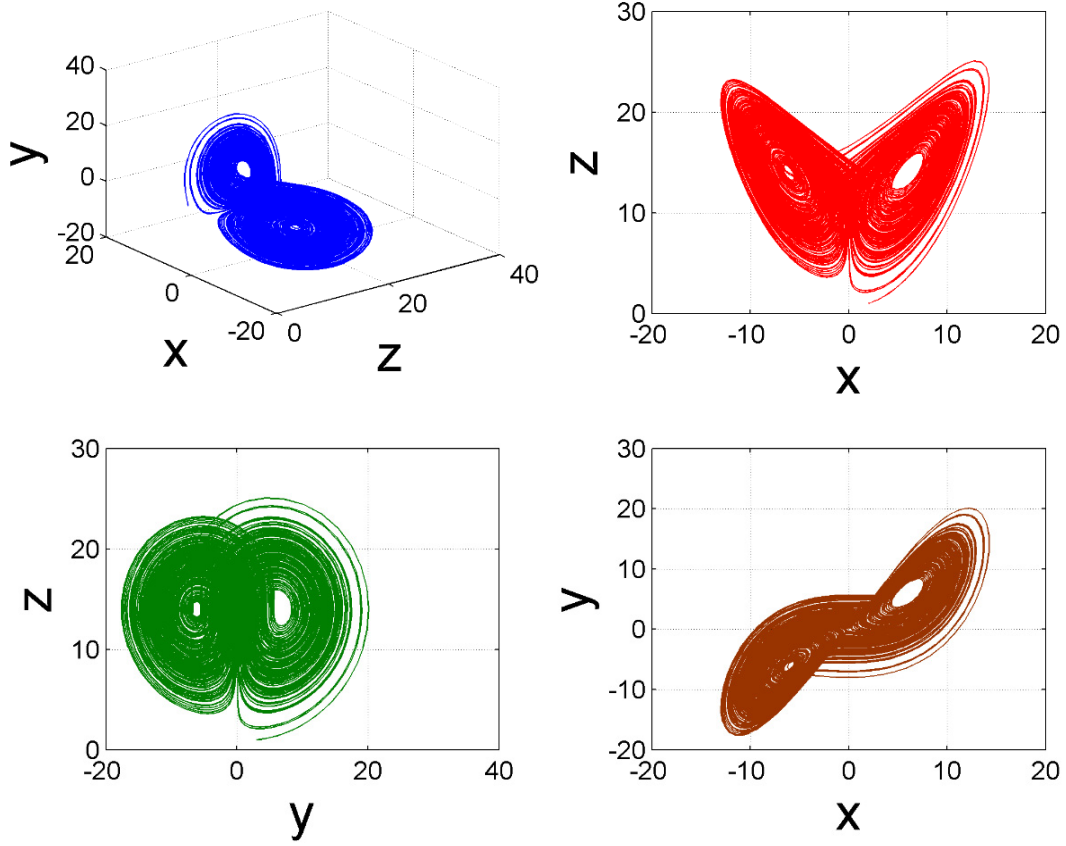

*Figure 125: Phase space dynamics of Lorenz-XYZ2*

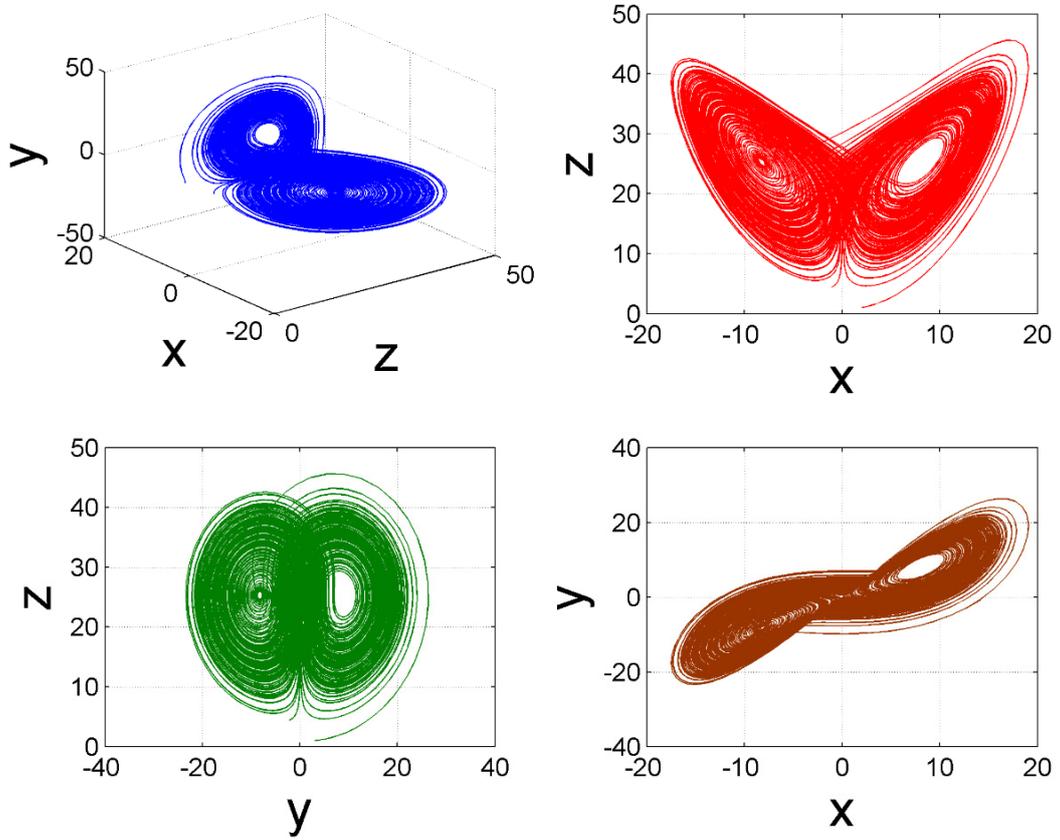

*Figure 126: Phase space dynamics of Lorenz-XYZ3*

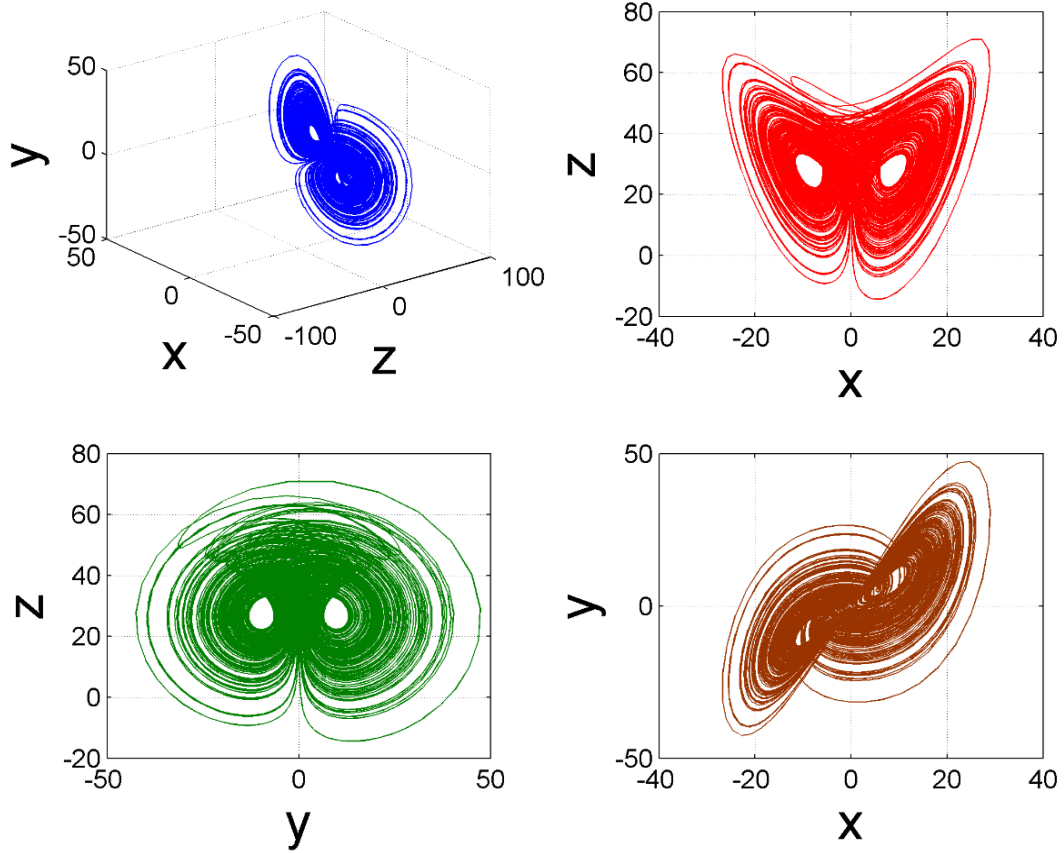

*Figure 127: Phase space dynamics of Lorenz-XYZ6*

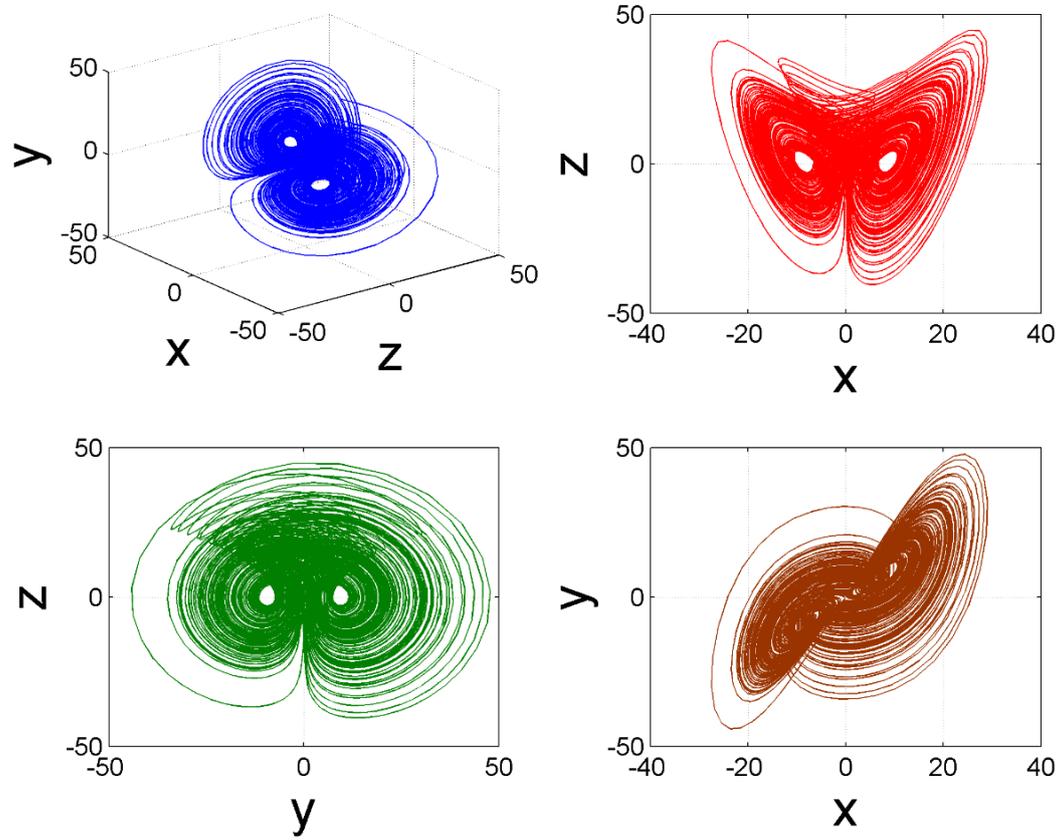

*Figure 128: Phase space dynamics of Lorenz-XYZ7*

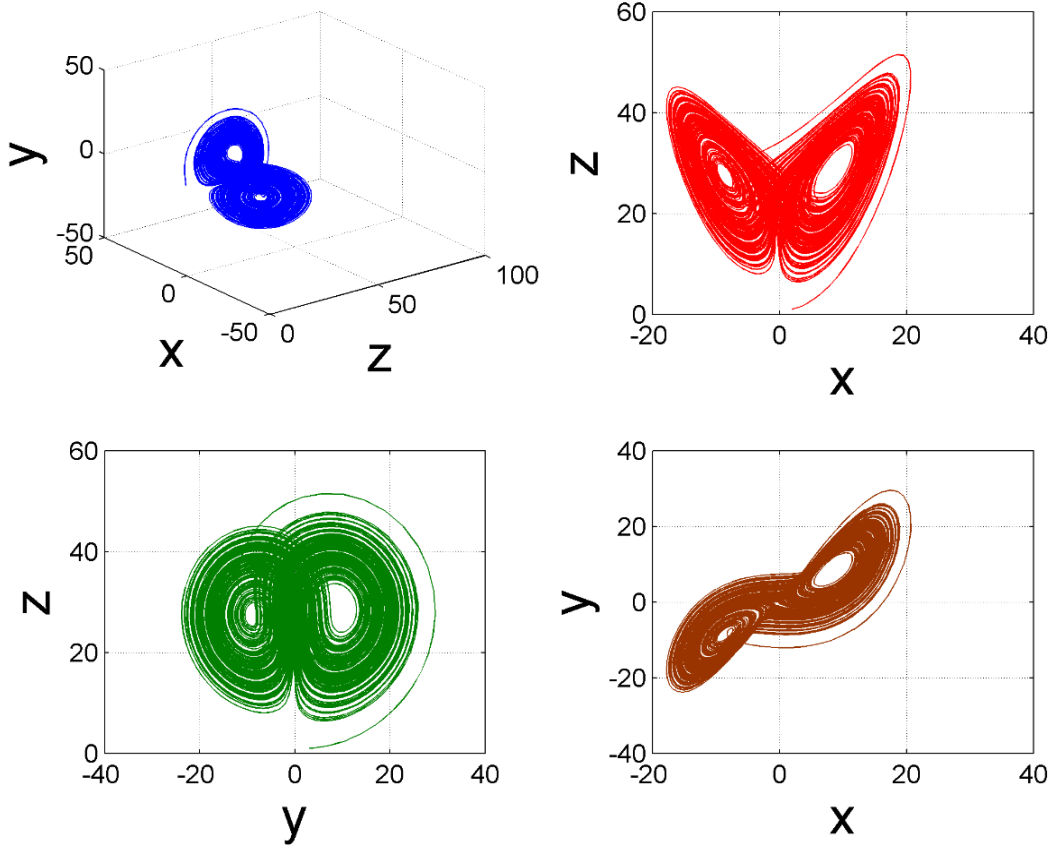

*Figure 129: Phase space dynamics of Lorenz-XYZ8*

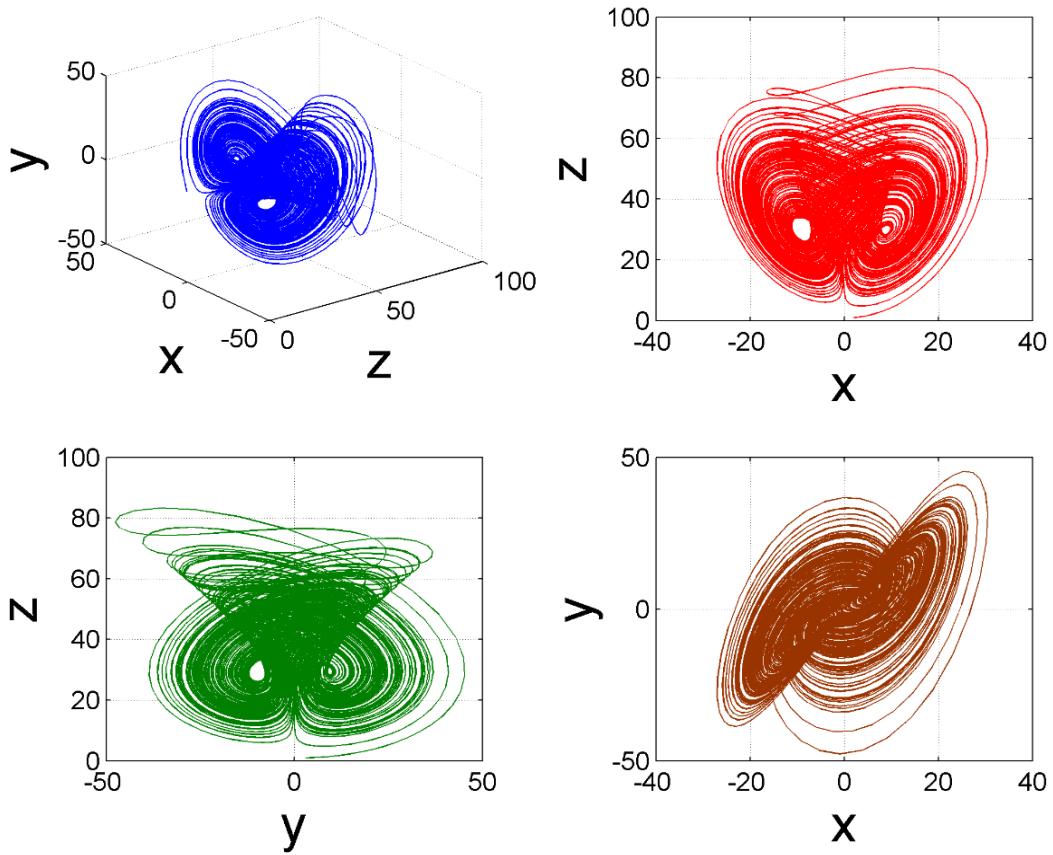

*Figure 130: Phase space dynamics of Lorenz-XYZ9*

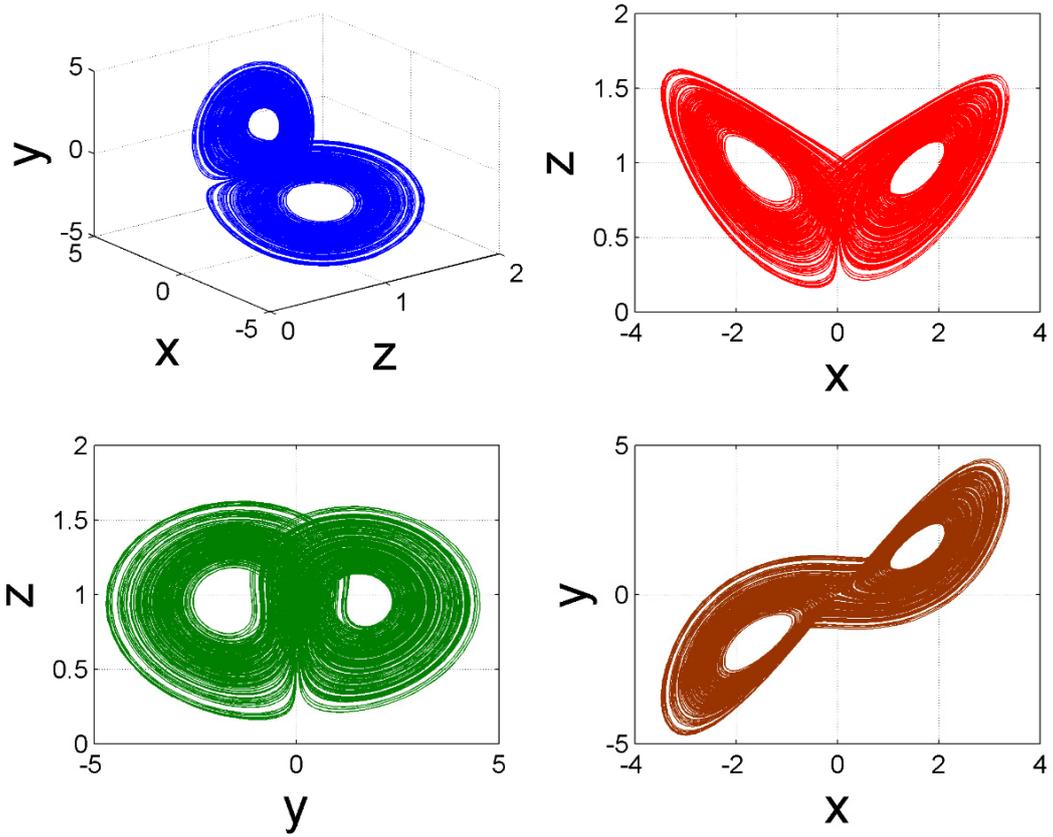

*Figure 131: Phase space dynamics of Lorenz-XYZ10*

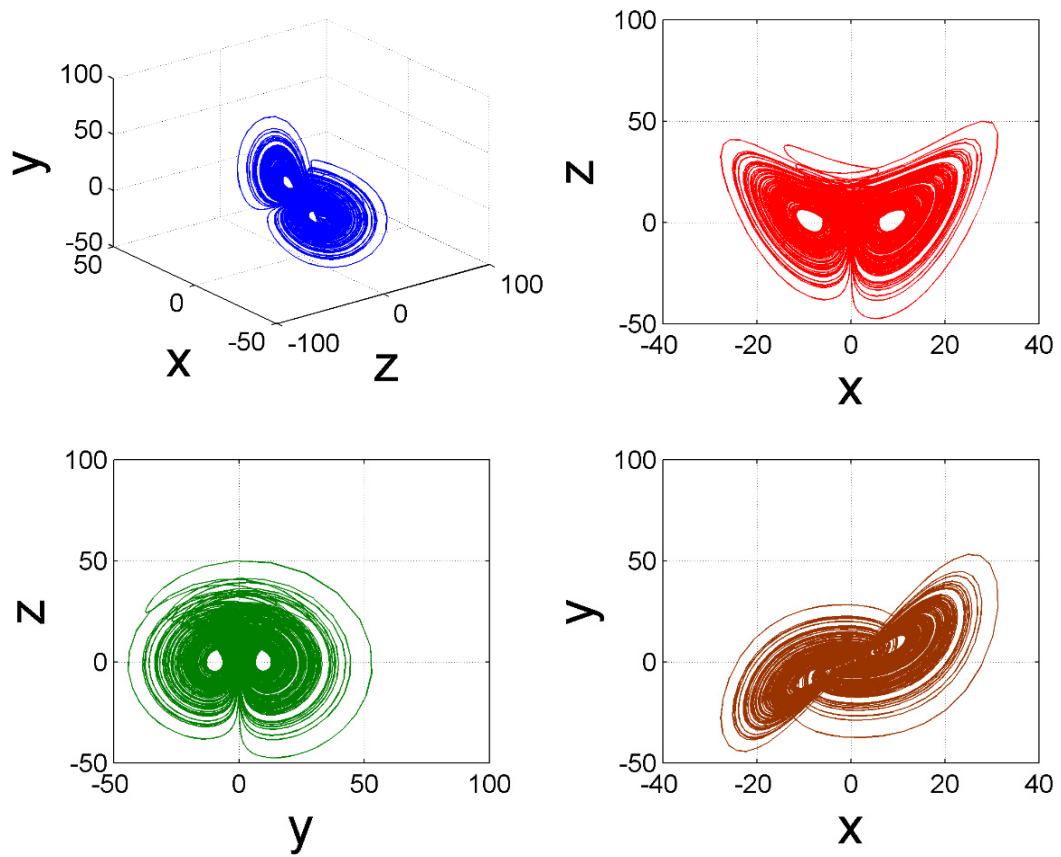

*Figure 132: Phase space dynamics of Lorenz-XYZ11*

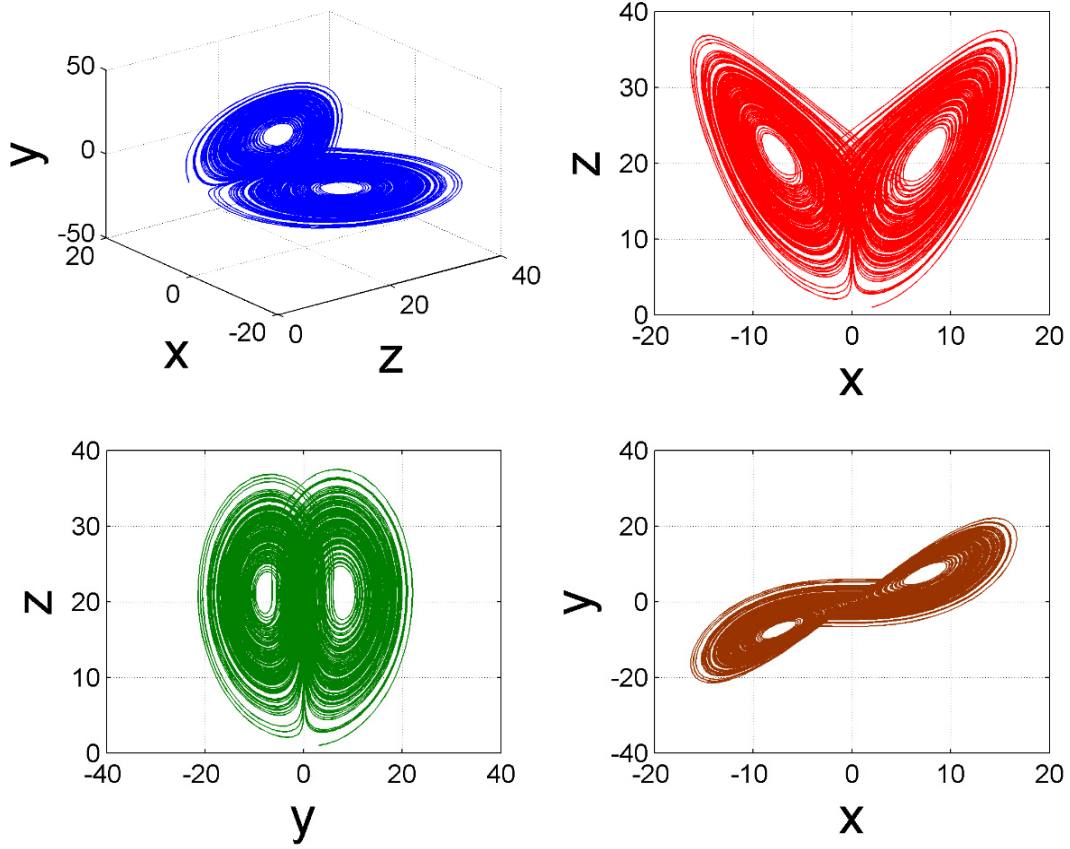

*Figure 133: Phase space dynamics of Lorenz-XYZ12*

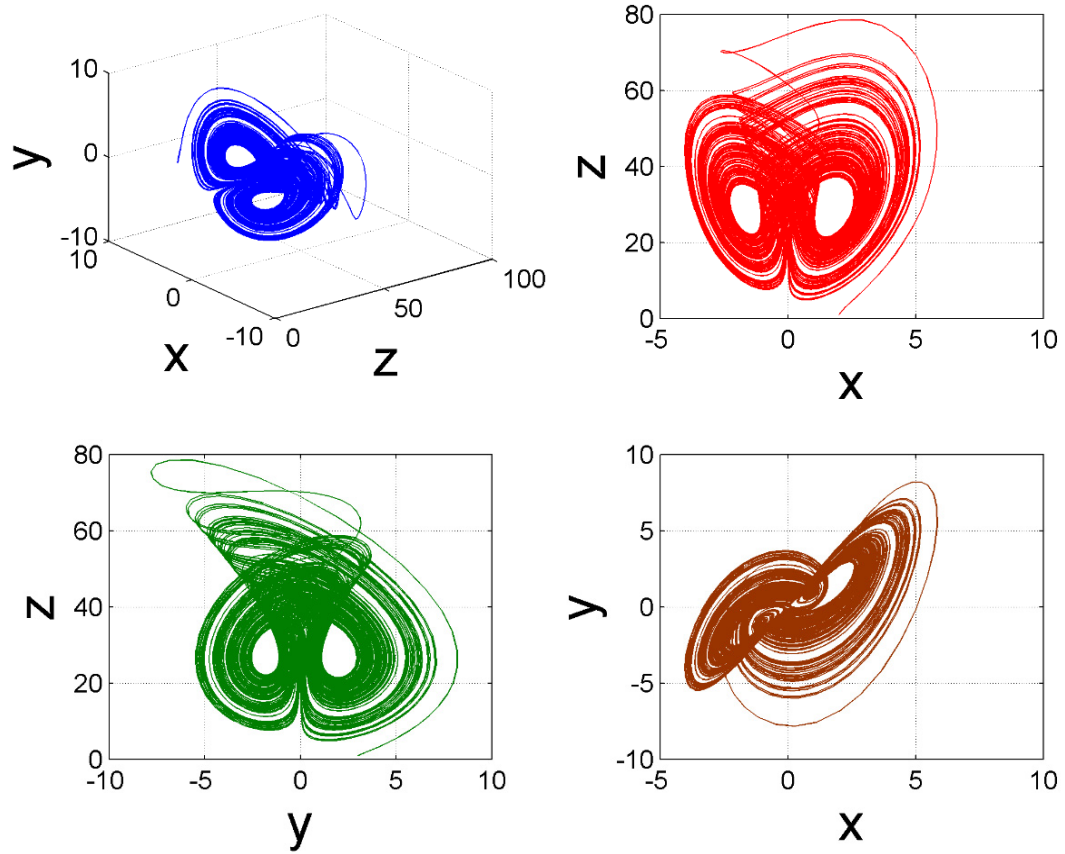

*Figure 134: Phase space dynamics of Lorenz-XYZ13*

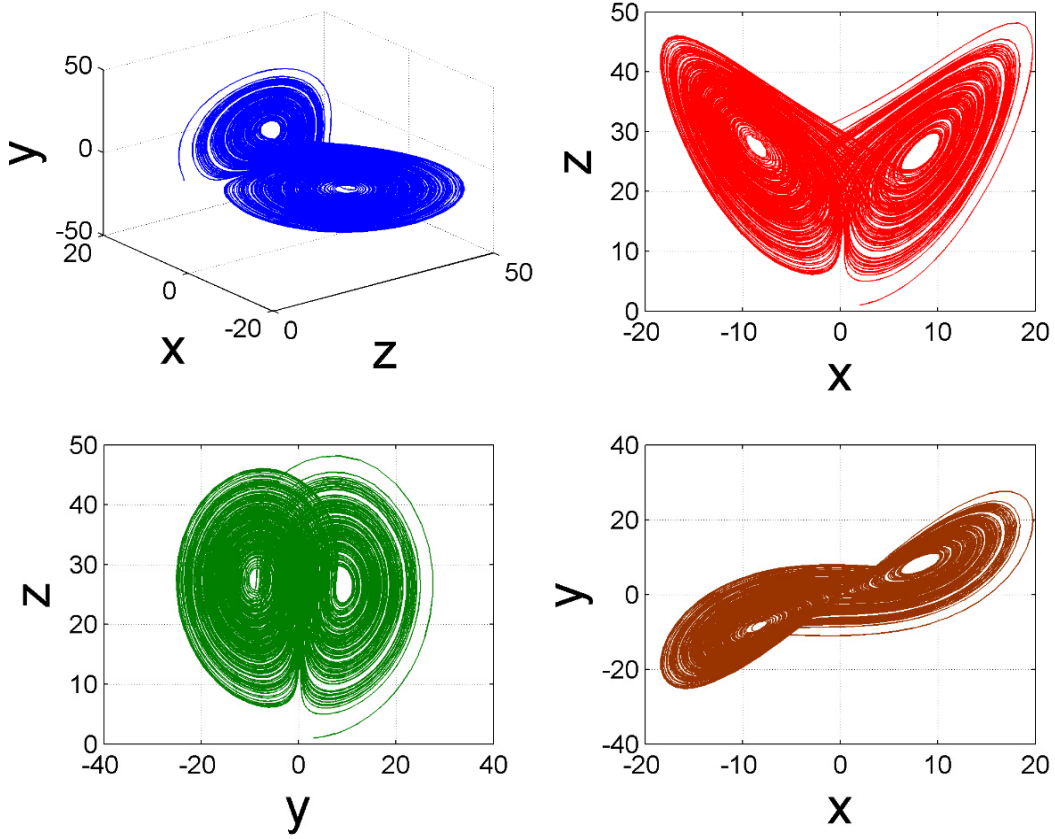

*Figure 135: Phase space dynamics of Lorenz-XYZ14*

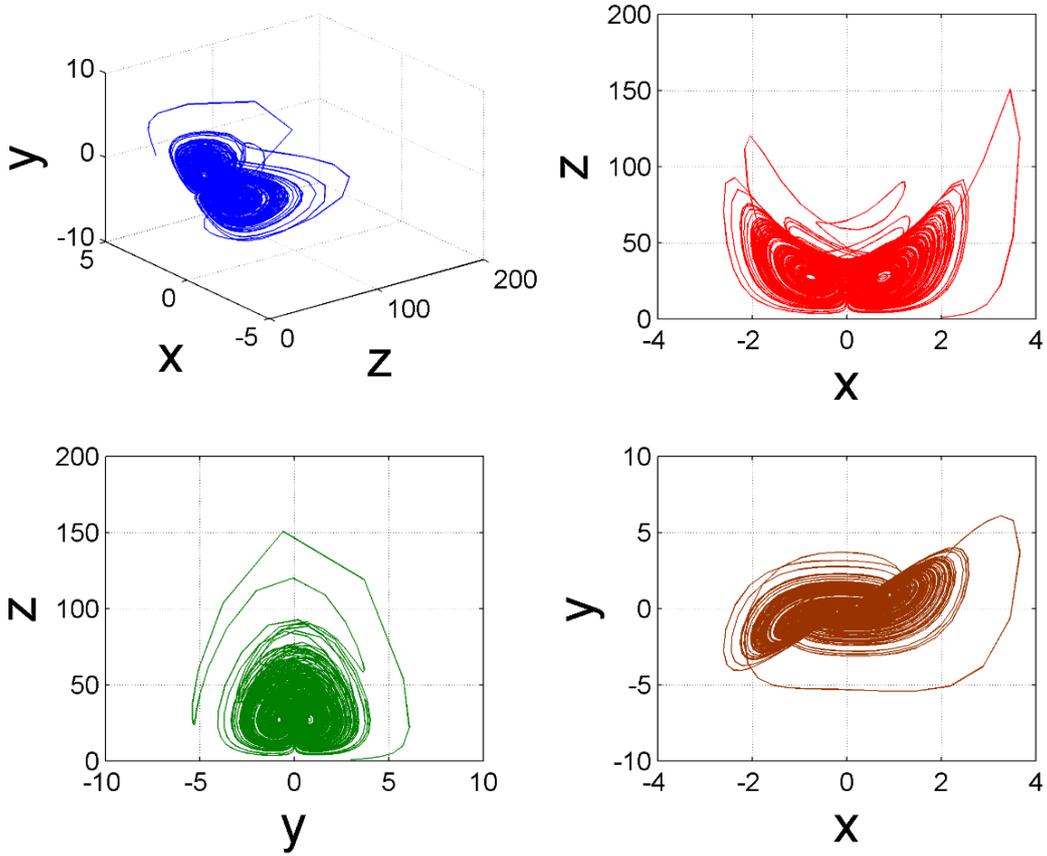

*Figure 136: Phase space dynamics of Lorenz-XYZ15*

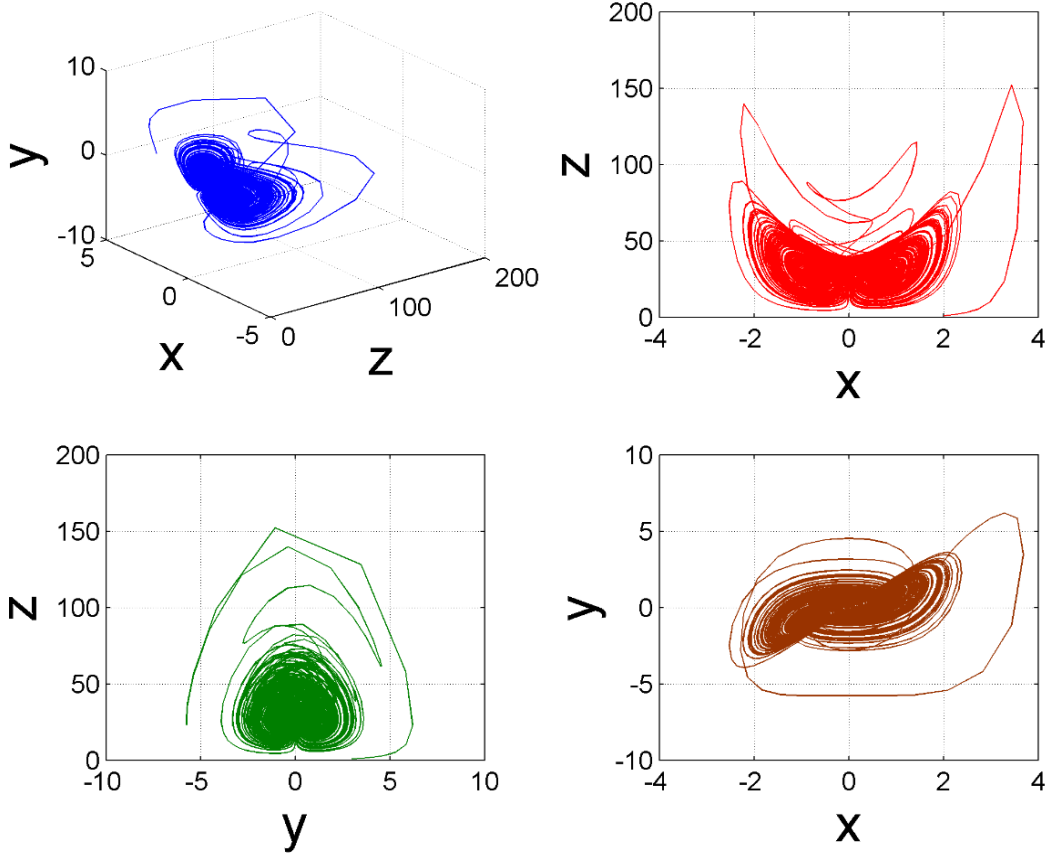

*Figure 137: Phase space dynamics of Lorenz-XYZ16*

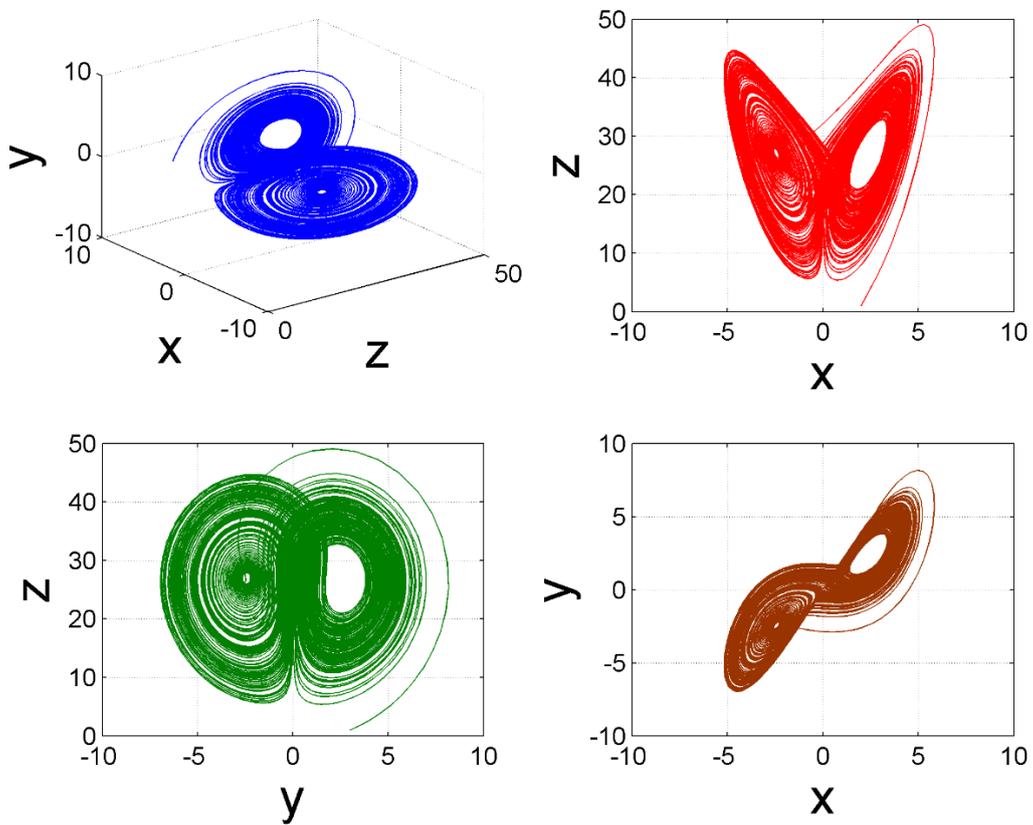

*Figure 138: Phase space dynamics of Lorenz-XYZ17*

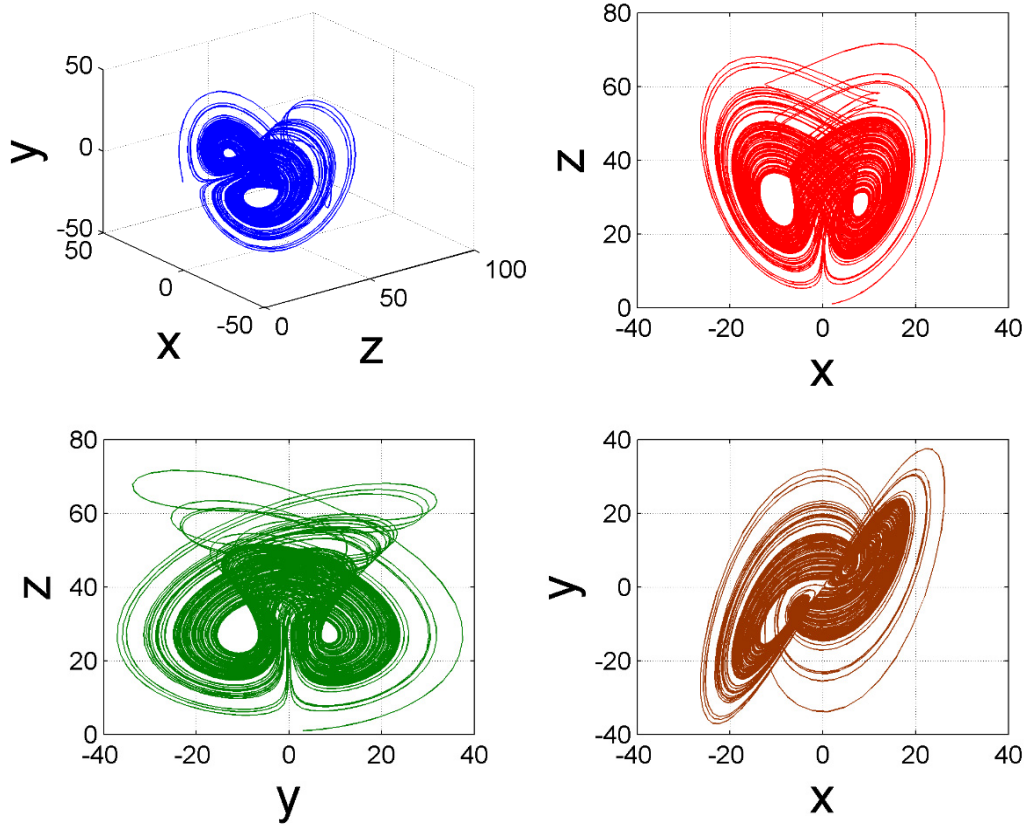

*Figure 139: Phase space dynamics of Lorenz-XYZ18*

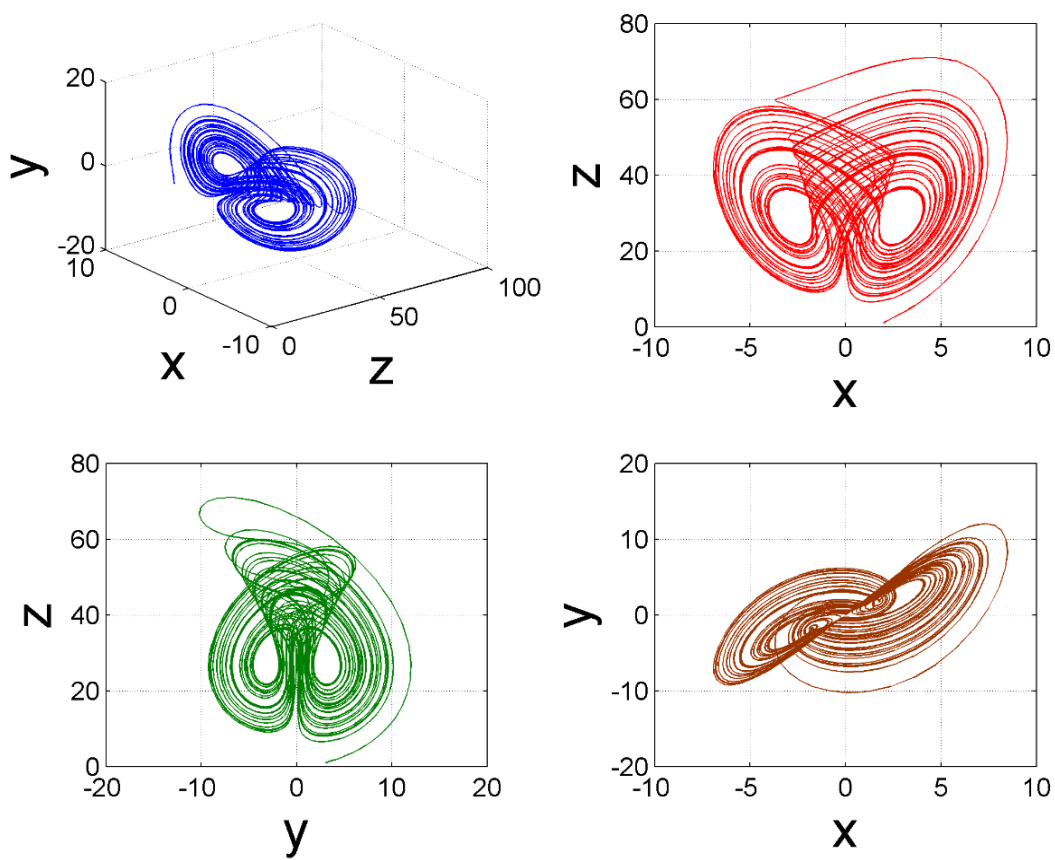

*Figure 140: Phase space dynamics of Lorenz-XYZ19*

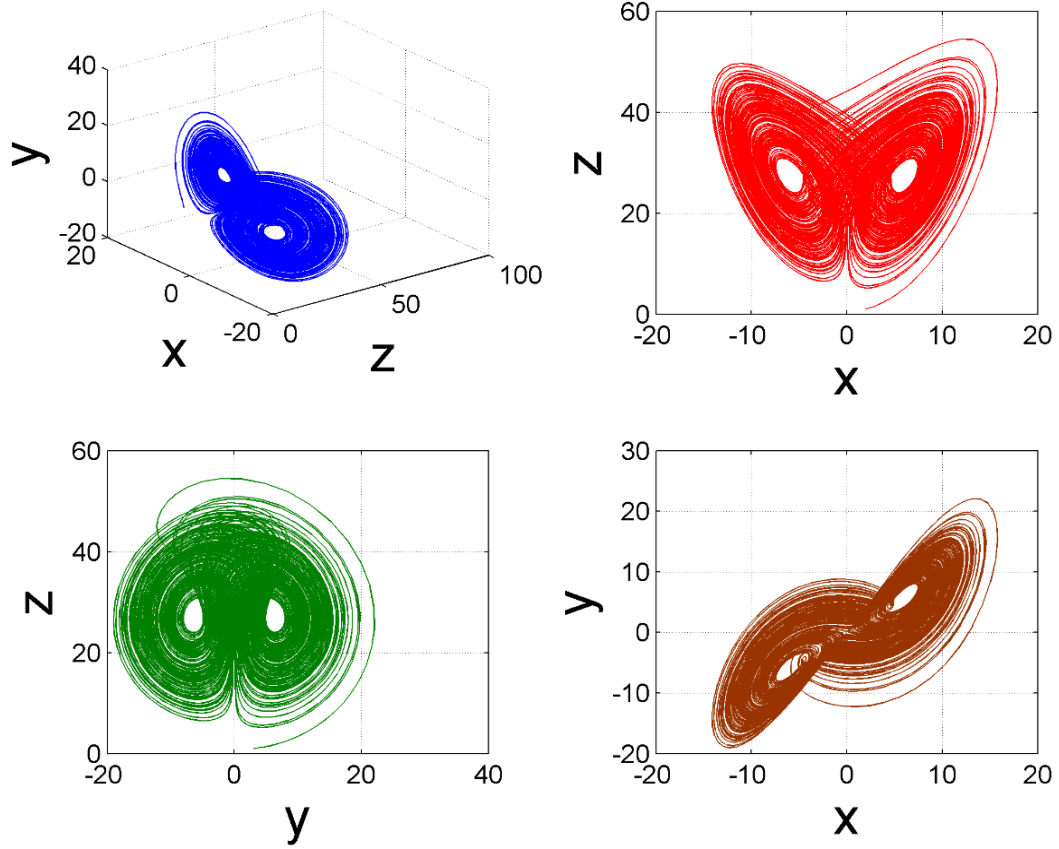

*Figure 141: Phase space dynamics of Lorenz-XYZ20*

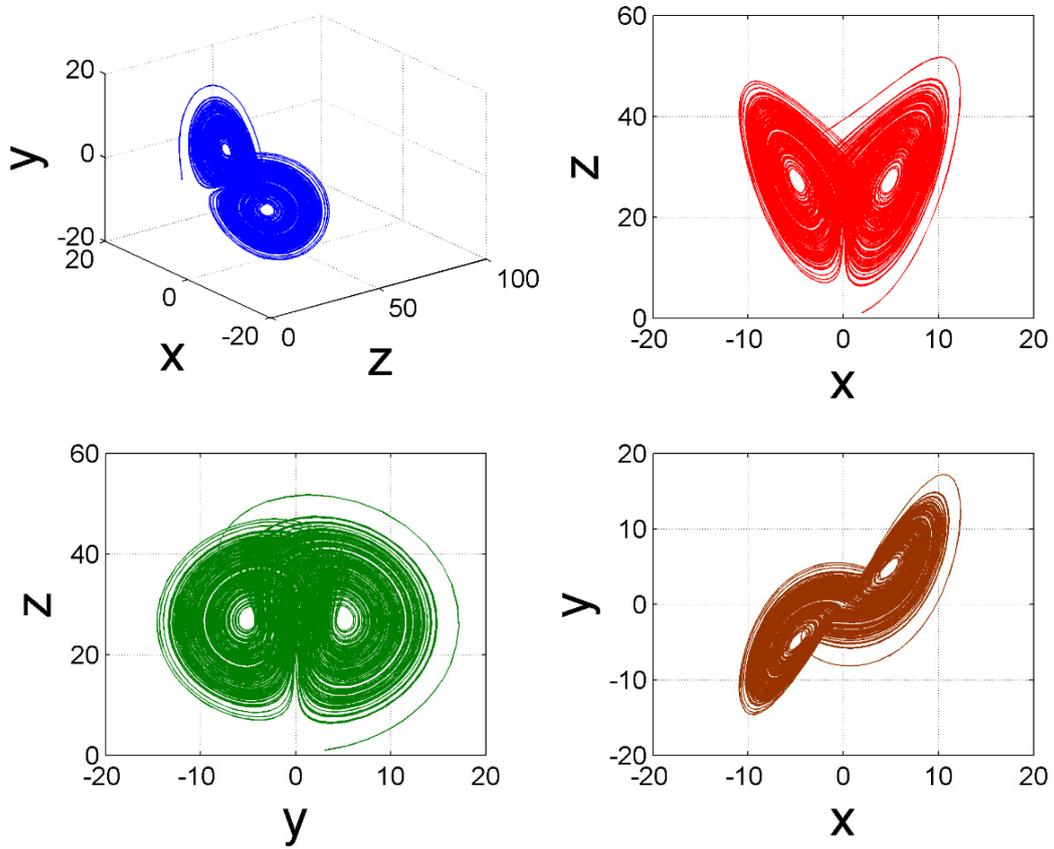

*Figure 142: Phase space dynamics of Lorenz-XYZ21*

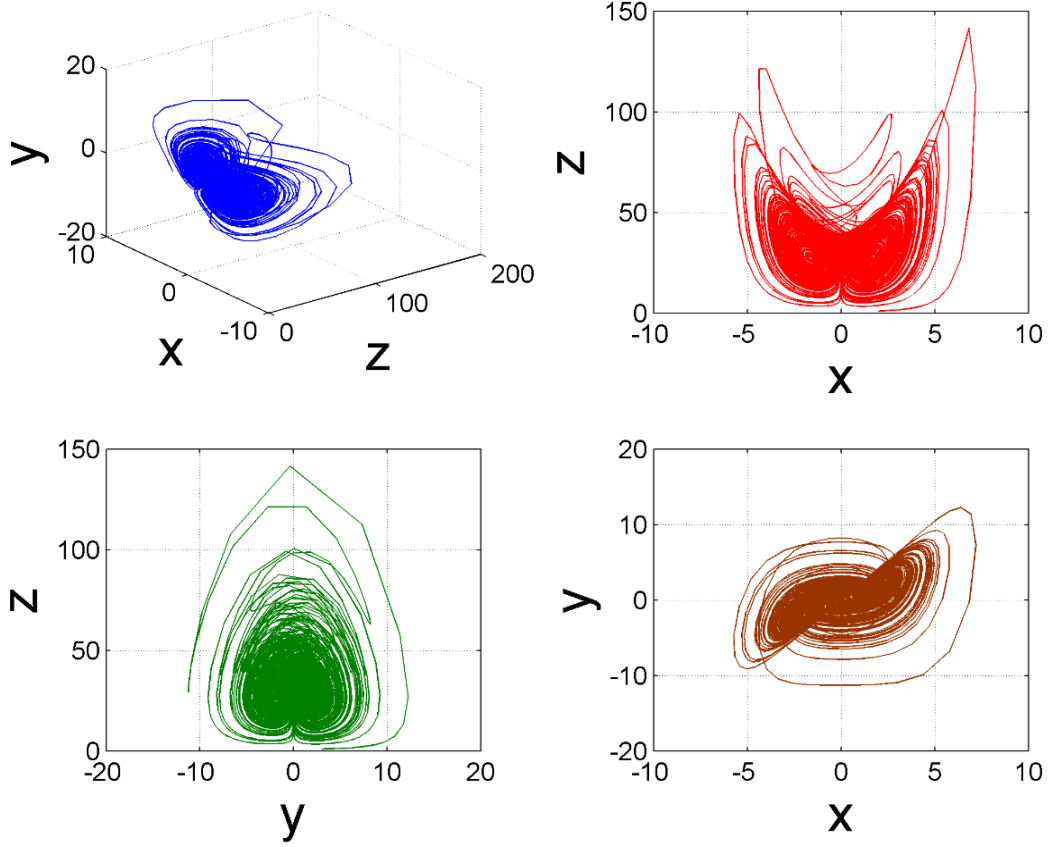

*Figure 143: Phase space dynamics of Lorenz-XYZ22*

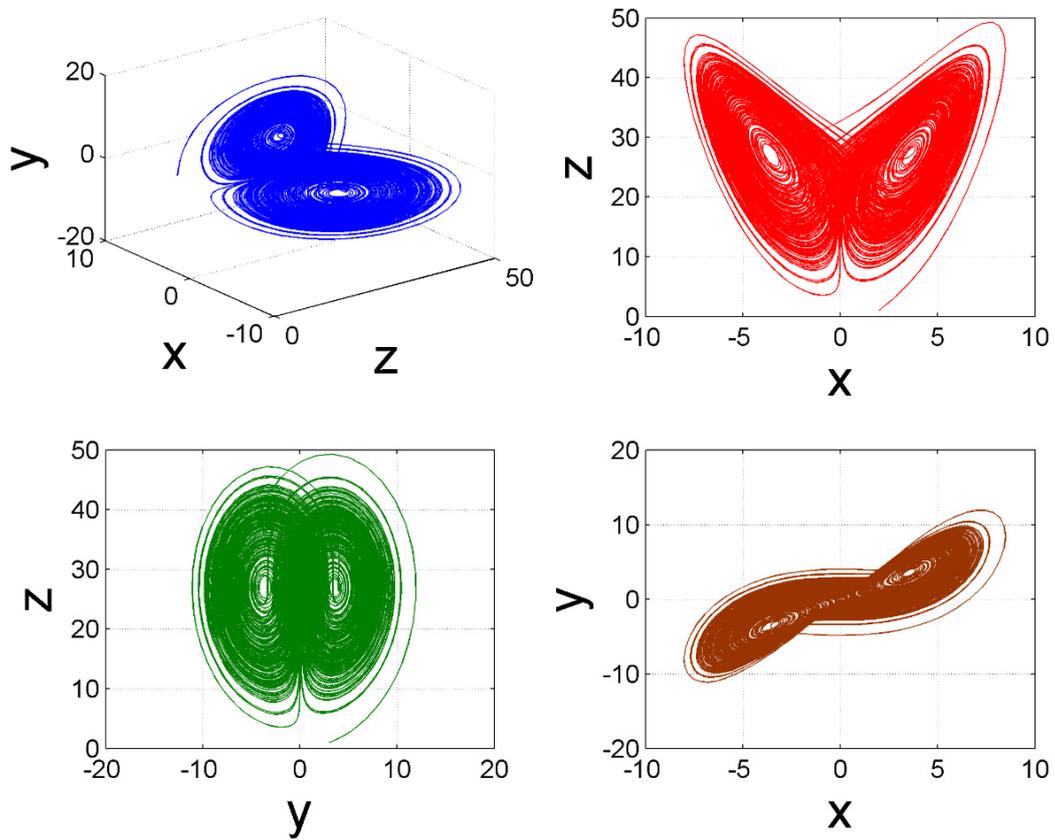

*Figure 144: Phase space dynamics of Lorenz-XYZ23*

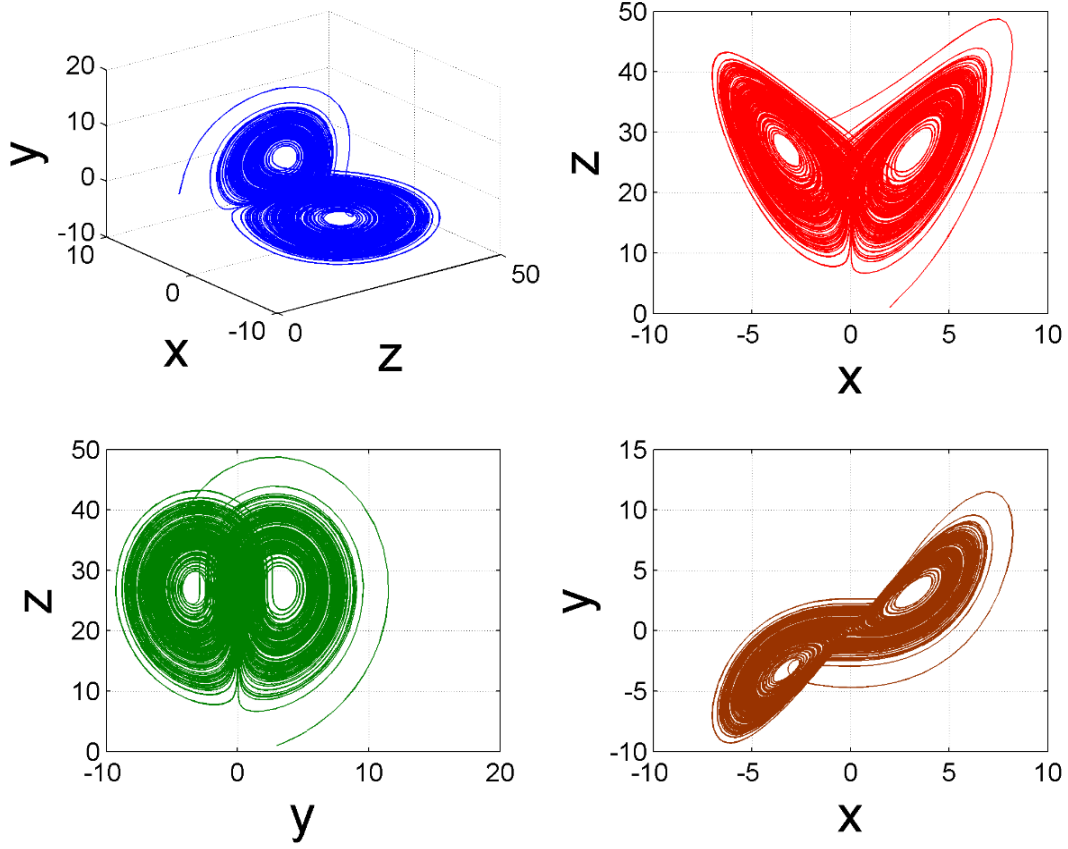

*Figure 145: Phase space dynamics of Lorenz-XYZ24*

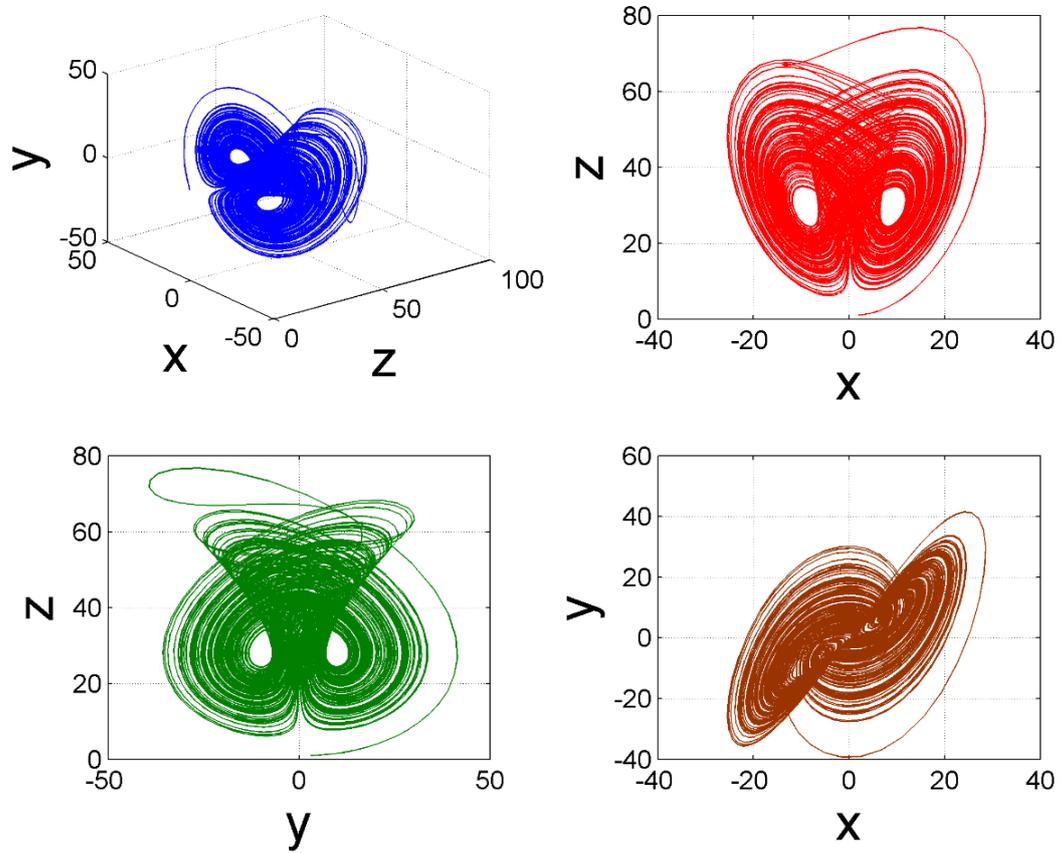

*Figure 146: Phase space dynamics of Lorenz-XYZ25*

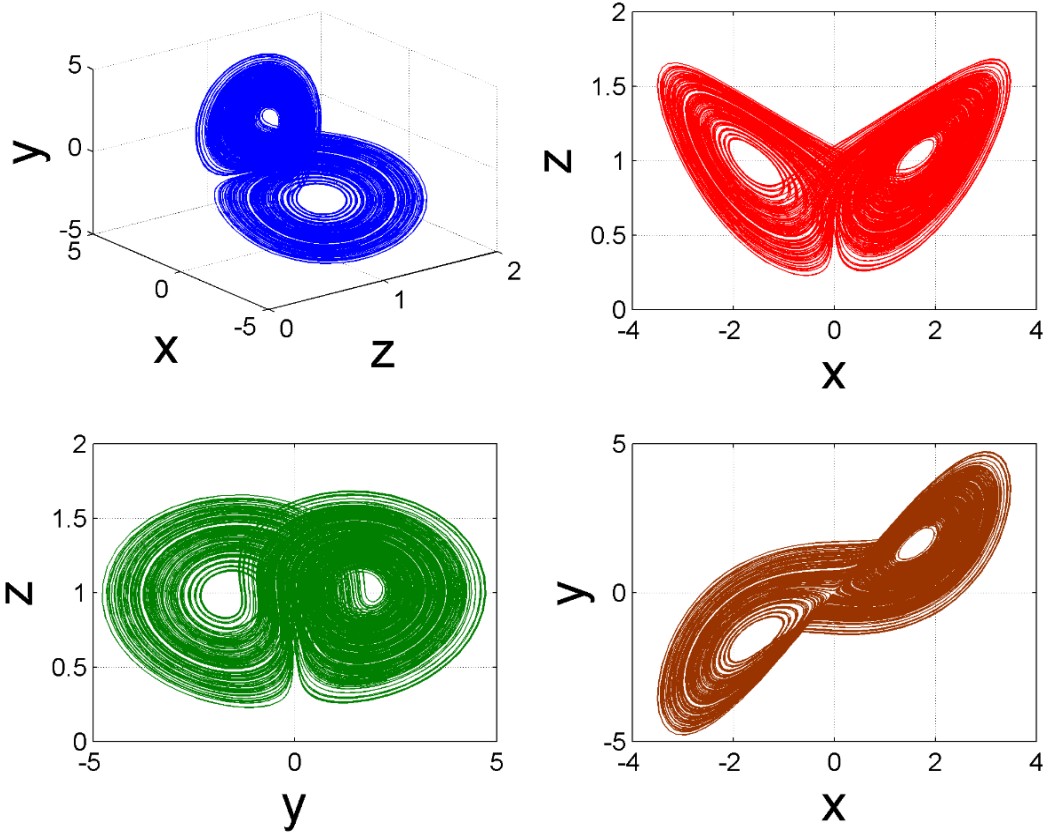

*Figure 147: Phase space dynamics of Lorenz-XYZ26*

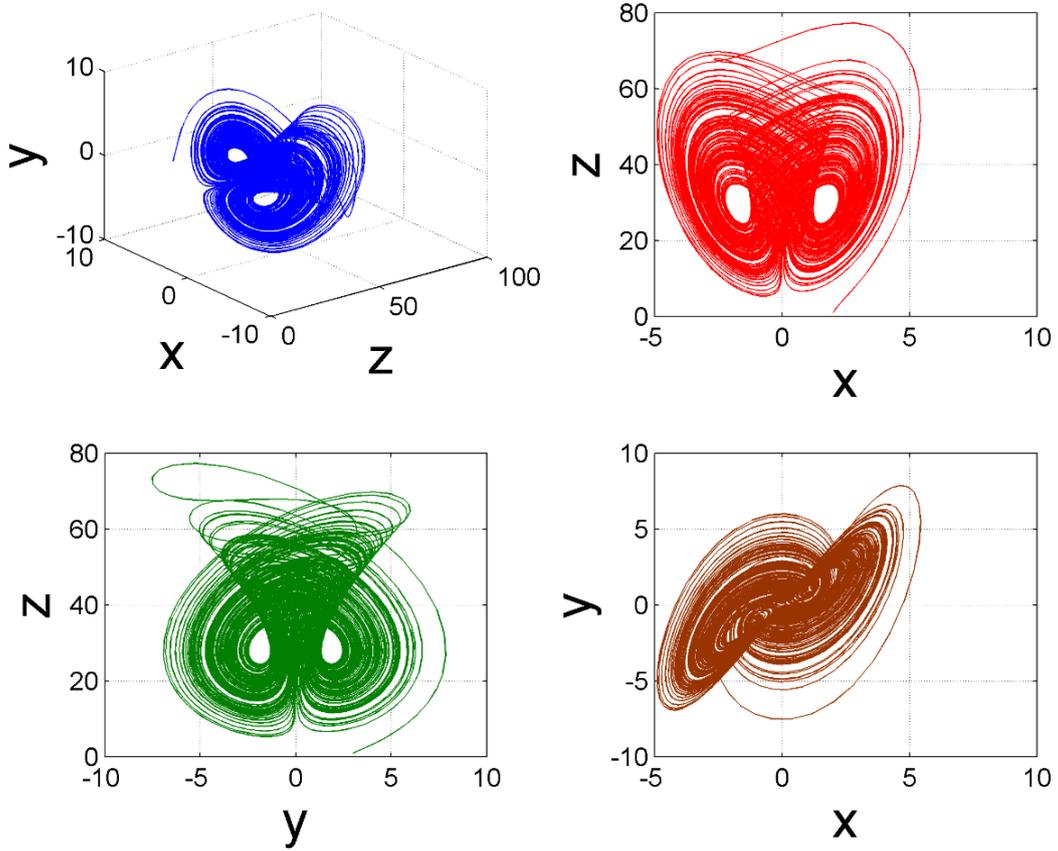

*Figure 148: Phase space dynamics of Lorenz-XYZ27*

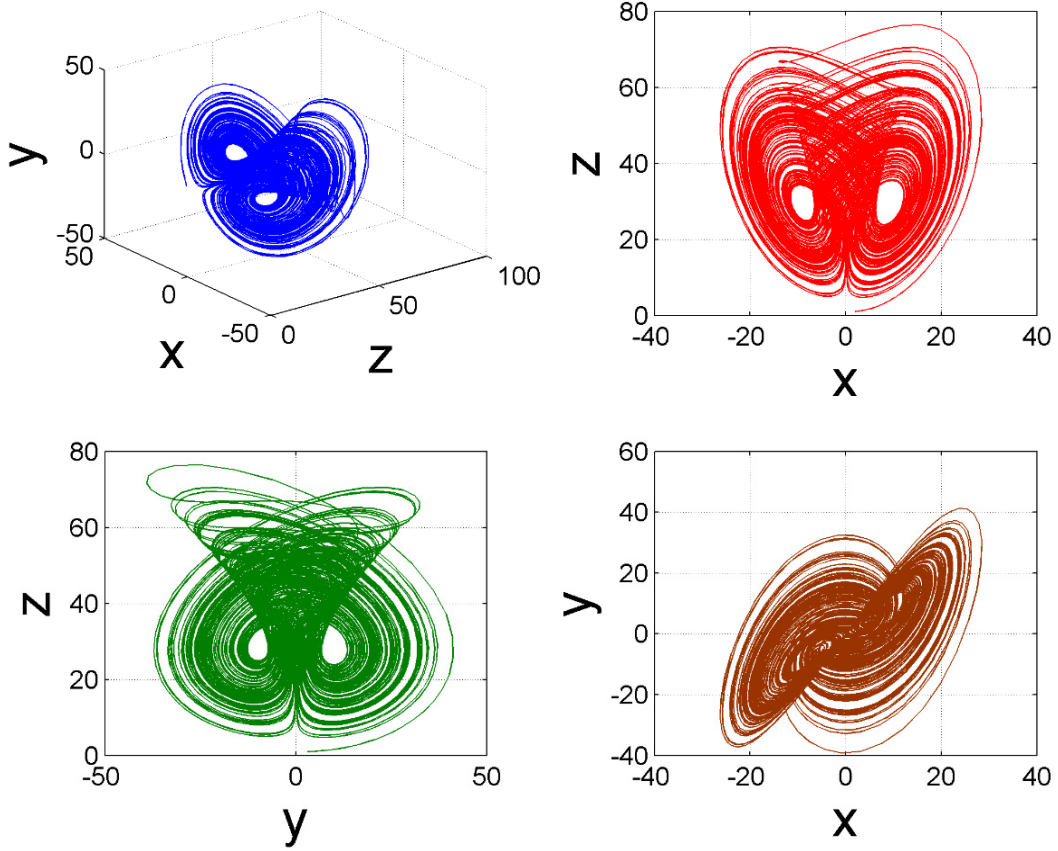

*Figure 149: Phase space dynamics of Lorenz-XYZ28*

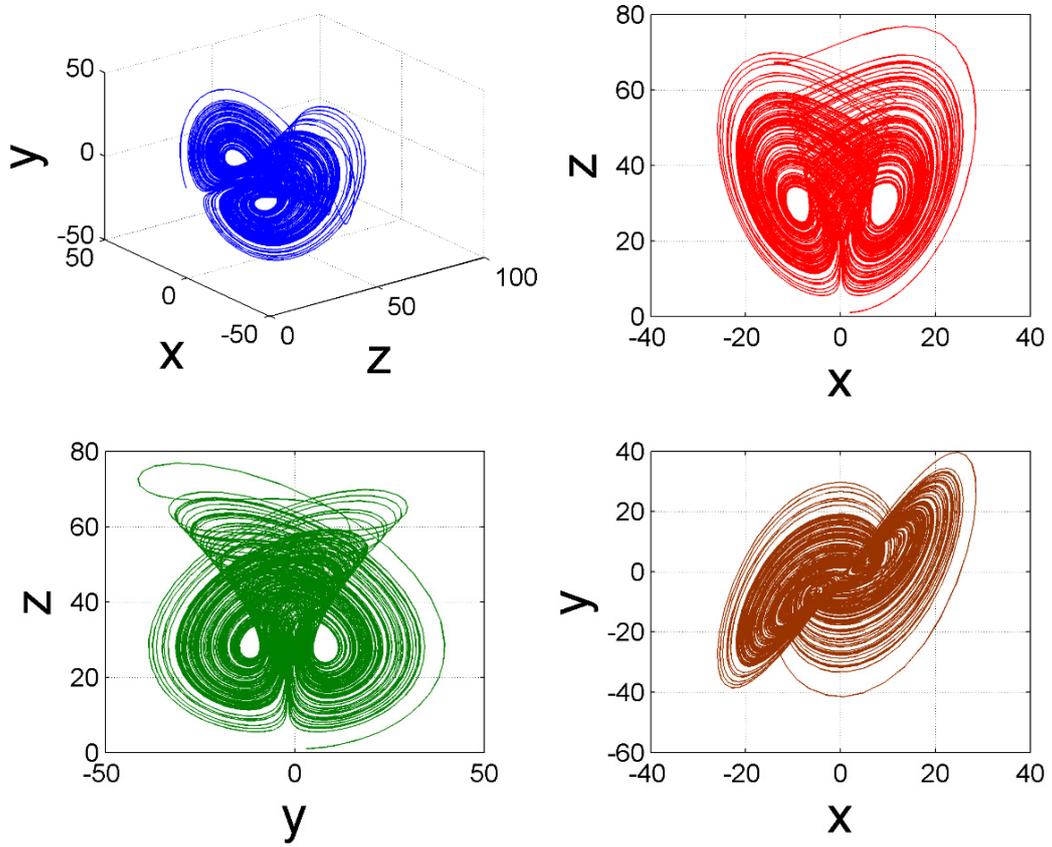

*Figure 150: Phase space dynamics of Lorenz-XYZ29*

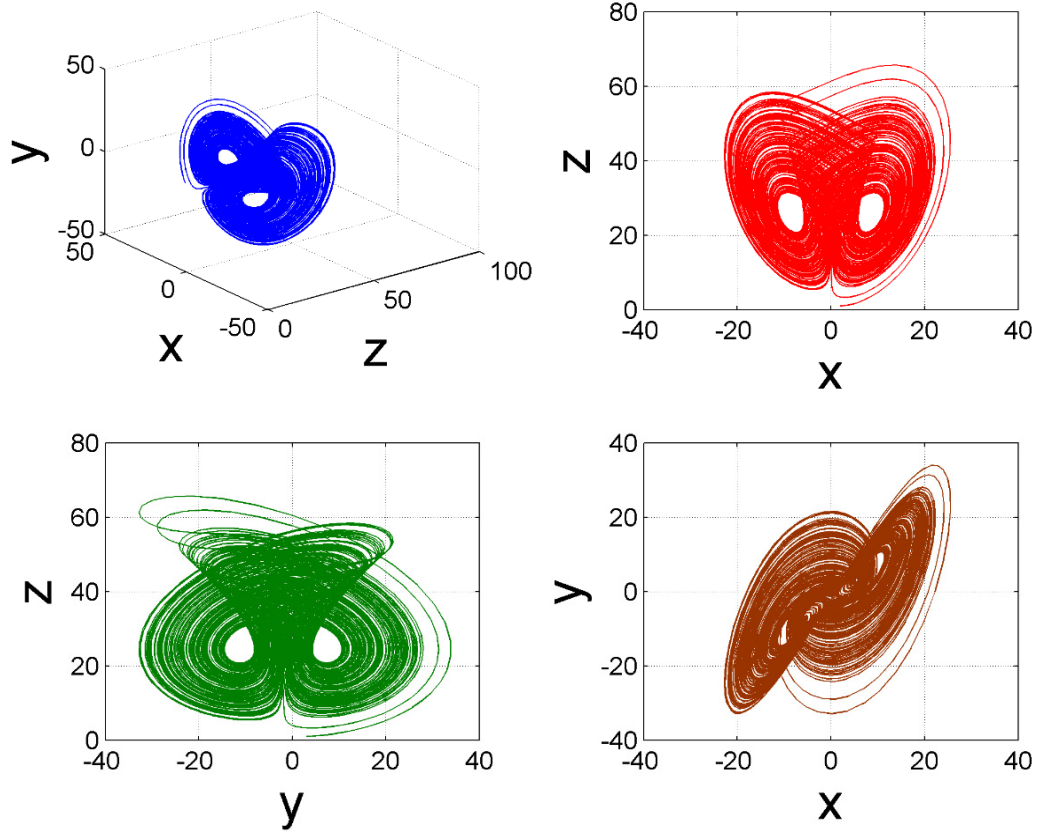

*Figure 151: Phase space dynamics of Lorenz-XYZ30*

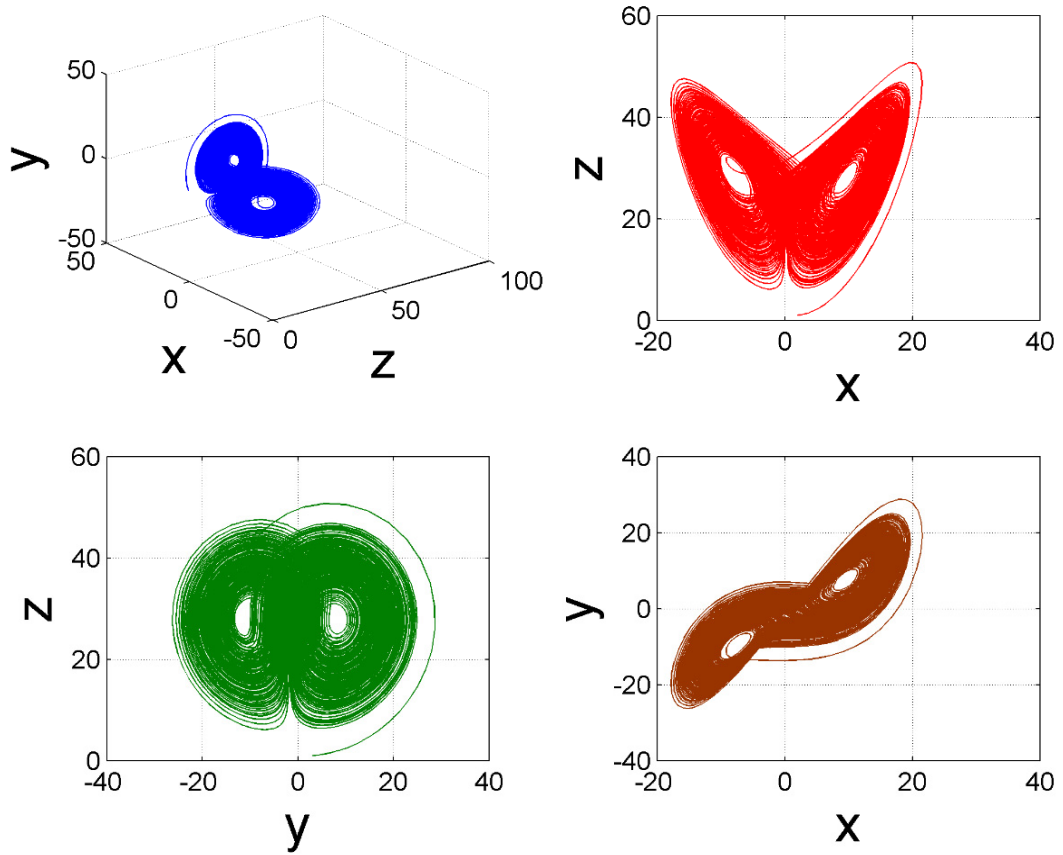

*Figure 152: Phase space dynamics of Lorenz-XYZ31*

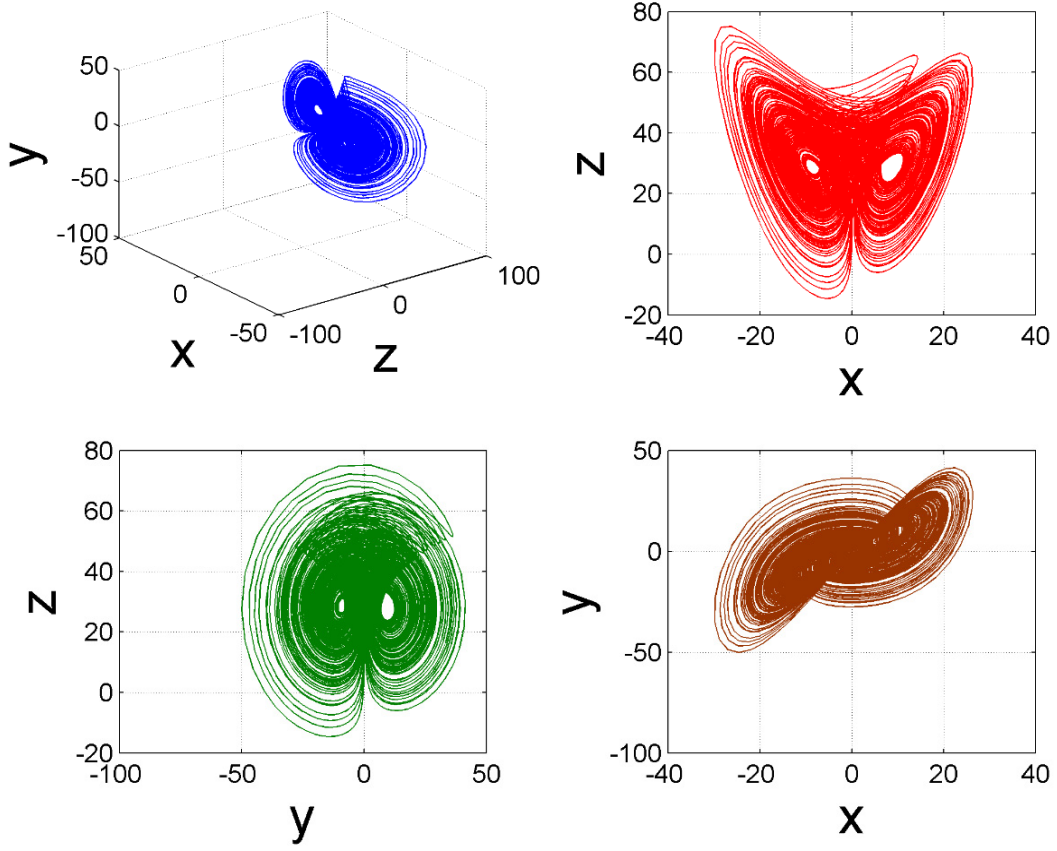

*Figure 153: Phase space dynamics of Lorenz-XYZ32*

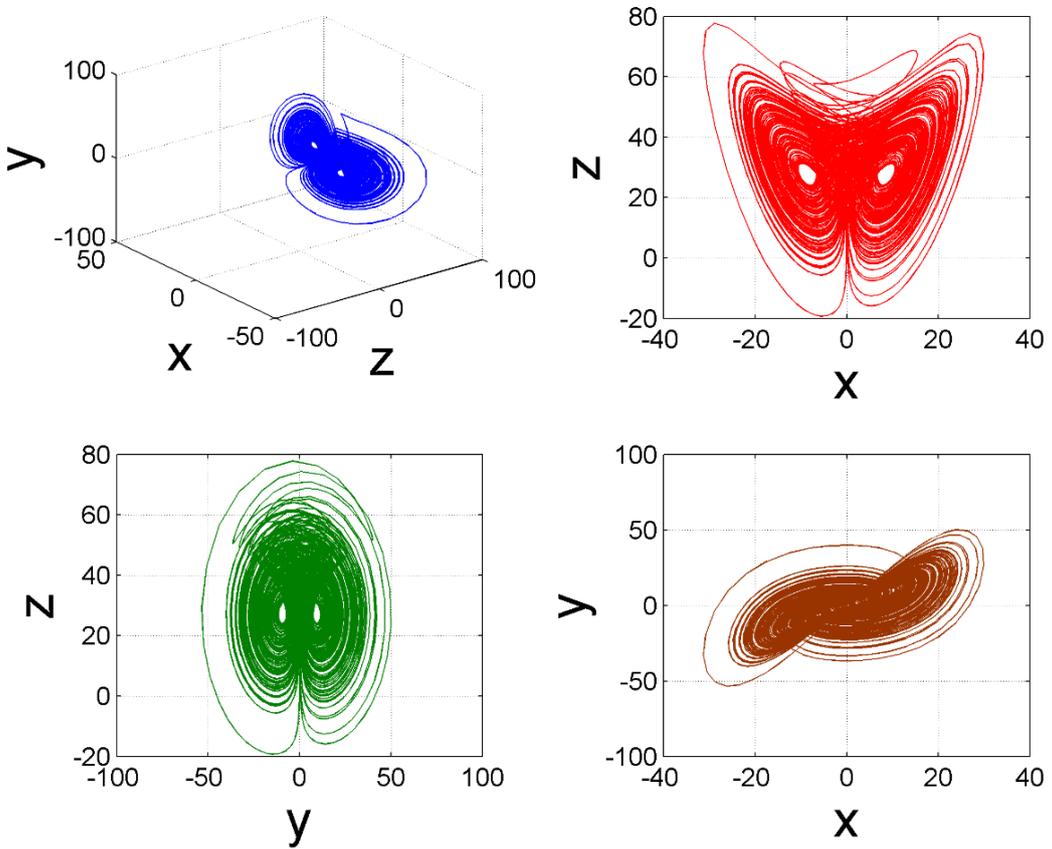

*Figure 154: Phase space dynamics of Lorenz-XYZ33*

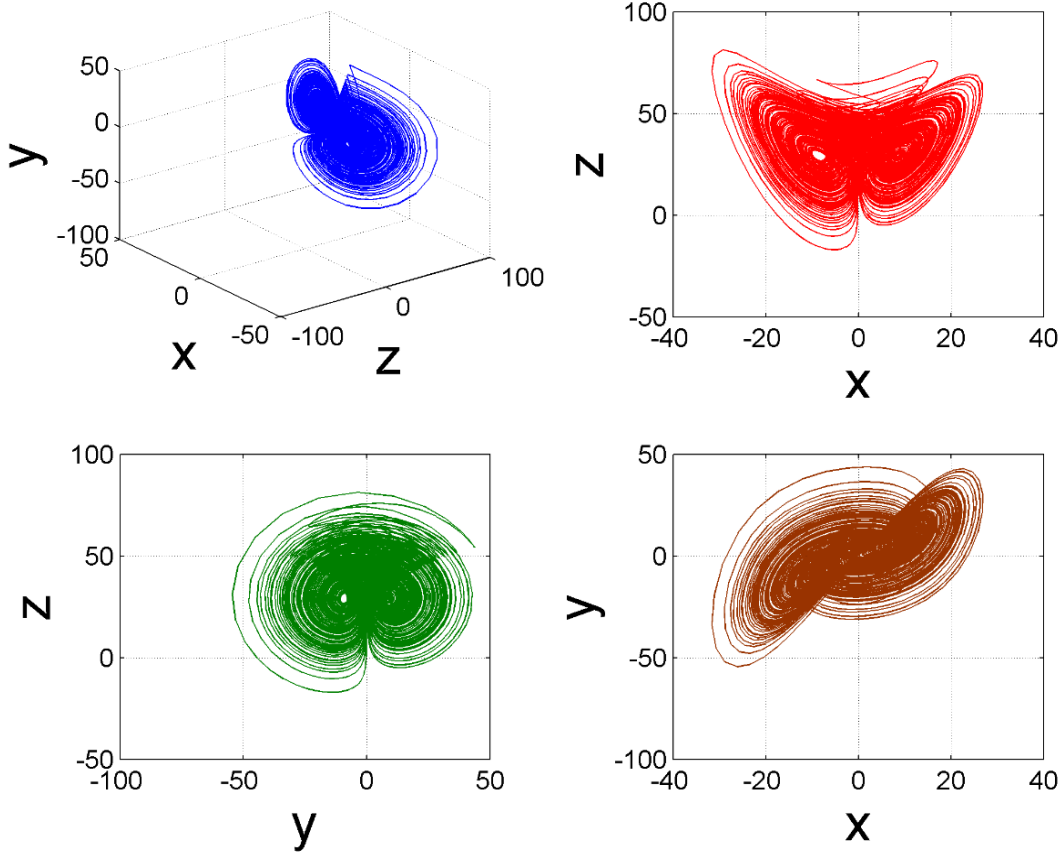

*Figure 155: Phase space dynamics of Lorenz-XYZ34*

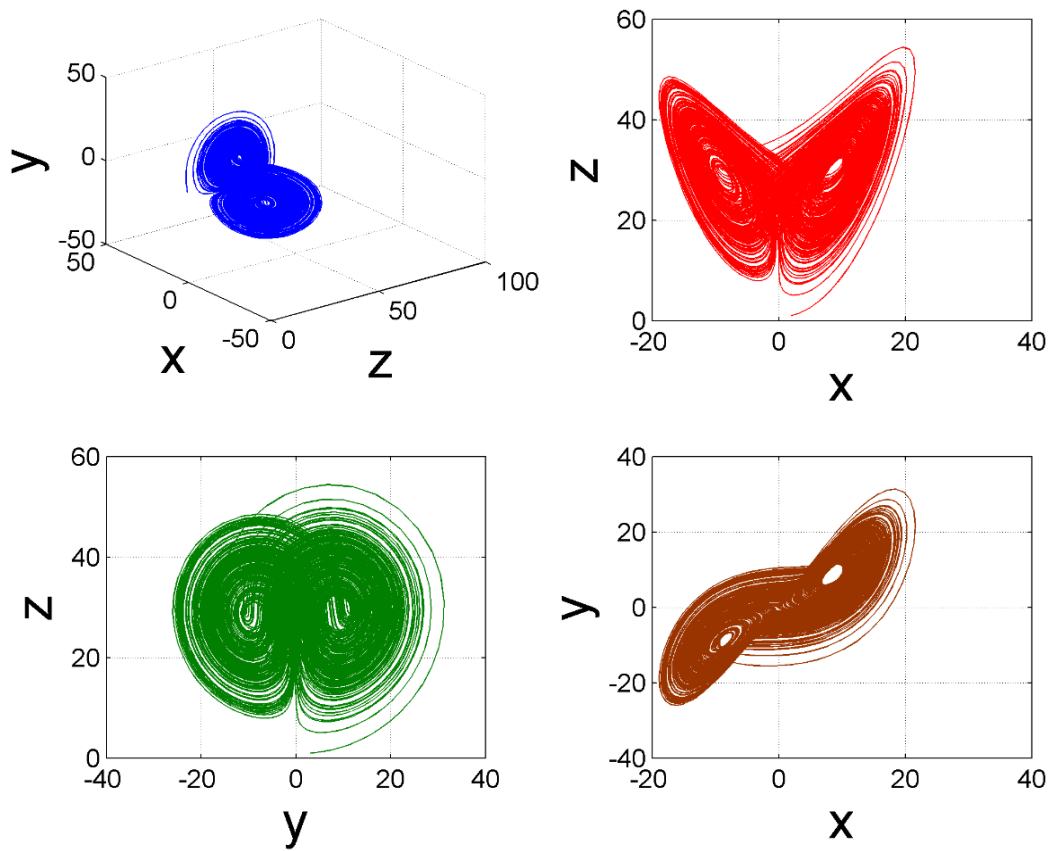

*Figure 156: Phase space dynamics of Lorenz-XYZ35*

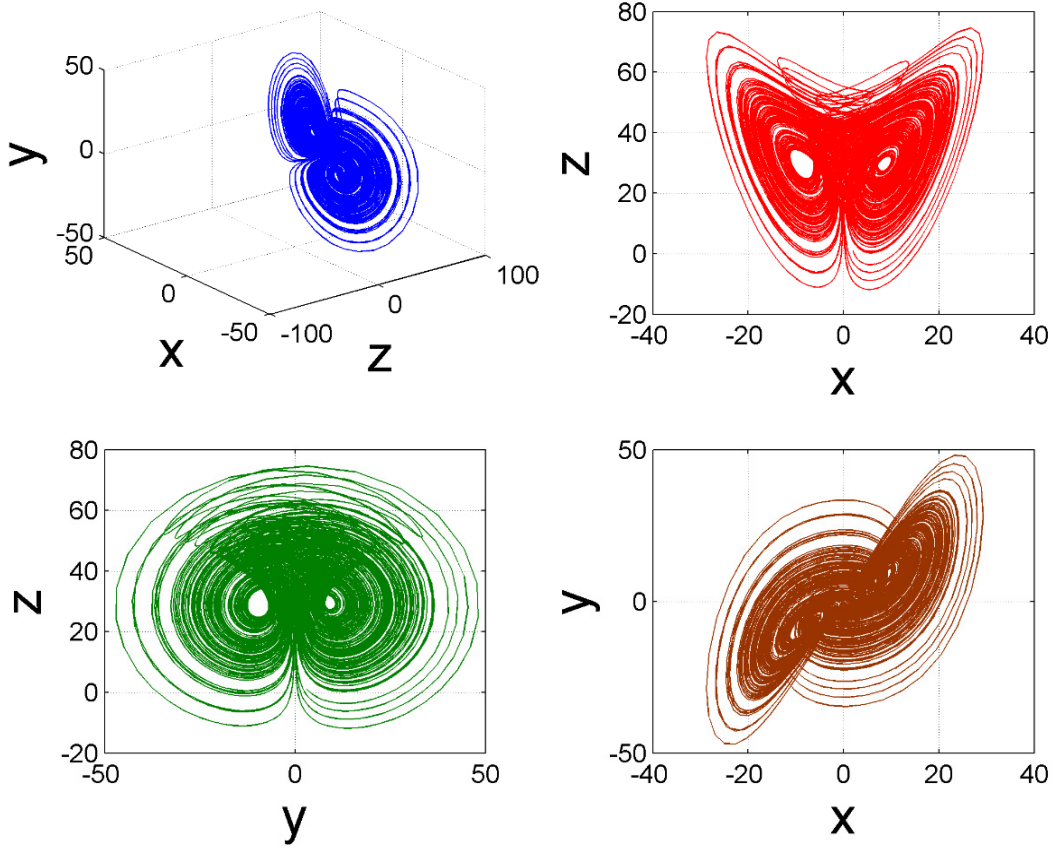

*Figure 157: Phase space dynamics of Lorenz-XYZ36*

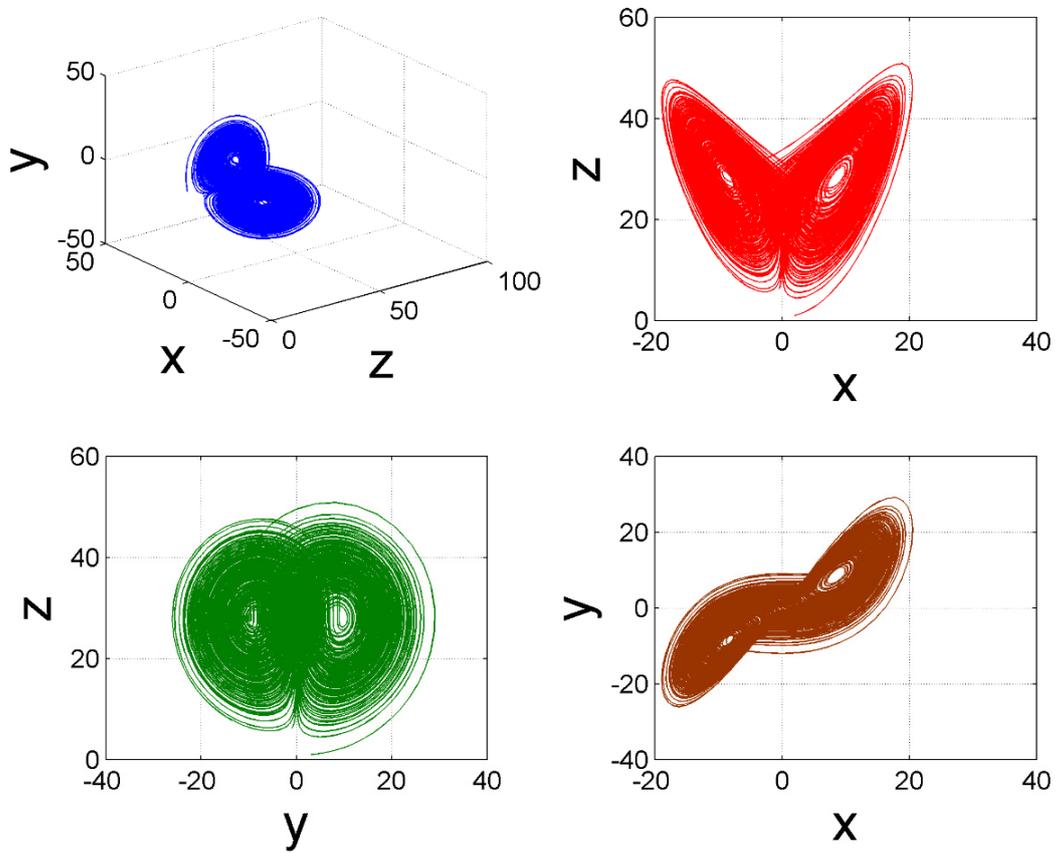

*Figure 158: Phase space dynamics of Lorenz-XYZ37*